\documentclass[a4paper,prd,preprint,amsmath,amssymb,
               superscriptaddress,nofootinbib,floatfix,showpacs]{revtex4}
\usepackage{graphicx}

\begin{document}

\title{%
Patterns of flavor signals in supersymmetric models
}
\date{February 13, 2008}

\author{Toru Goto}
\email{tgoto@post.kek.jp}
\affiliation{%
Theory Group, KEK, Tsukuba, Ibaraki 305-0801, Japan
}
\thanks{Present address.}
\affiliation{%
YITP, Kyoto University, Kyoto 606-8502, Japan
}
\author{Yasuhiro Okada}
\email{yasuhiro.okada@kek.jp}
\affiliation{%
Theory Group, KEK, Tsukuba, Ibaraki 305-0801, Japan
}
\affiliation{%
Department of Particle and Nuclear Physics, The Graduate University for
Advanced Studies (Sokendai), Tsukuba, Ibaraki 305-0801, Japan
}
\author{Tetsuo Shindou}
\email{tetsuo.shindou@desy.de}
\affiliation{%
DESY Theory Group, Notkestrasse 85 D22607 Hamburg, Germany
}
\thanks{Present address.}
\affiliation{%
SISSA/ISAS, via Beirut 2-4 I-34014 Trieste, Italy
}
\author{Minoru Tanaka}
\email{tanaka@phys.sci.osaka-u.ac.jp}
\affiliation{%
Department of Physics, Graduate School of Science, Osaka University,
Toyonaka, Osaka 560-0043, Japan
}

\begin{abstract}
Quark and lepton flavor signals are studied in four
supersymmetric models, namely the minimal supergravity model, 
the minimal supersymmetric standard model with right-handed neutrinos, 
SU(5) supersymmetric grand unified theory with right-handed neutrinos
and the minimal supersymmetric standard model with U(2) flavor symmetry.
We calculate $b\to s(d)$ transition observables in $B_d$ and $B_s$
decays, taking the constraint from the $B_s-\bar{B_s}$ mixing recently
observed at Tevatron into account.
We also calculate lepton flavor violating processes $\mu\to e\gamma$,
$\tau\to \mu\gamma$ and $\tau\to e\gamma$ for the models with
right-handed neutrinos.
We investigate possibilities to distinguish the flavor structure of
the supersymmetry breaking sector with use of patterns of various flavor
signals which are expected to be measured in
experiments such as MEG, LHCb and a future Super $B$ Factory.
\end{abstract}

\pacs{
12.60.Jv, 
14.40.Nd, 
12.15.Hh, 
11.30.Er  
}
\keywords{}

\preprint{KEK-TH-1198}
\preprint{DESY 07-201}
\preprint{OU-HET-590-2007}
\preprint{arXiv:0711.2935~[hep-ph]}

\maketitle

\section{Introduction}

The problem of flavors is one of the interesting aspects of particle
physics.
Results obtained at $B$ factory experiments so far indicate that the
Cabibbo-Kobayashi-Maskawa (CKM) mixing \cite{ref:CKM} is the main
mechanism of flavor mixing phenomena in the quark sector, although there
still remains some room for new physics beyond the standard model (SM).
On the other hand, in the lepton sector, neutrino experiments unveil
large flavor mixings quite different from the quark sector
\cite{ref:sol_atm_nu,ref:K2K,ref:KamLAND,ref:reactor_neutrino}.
These mixings in the lepton sector are certainly beyond the
SM, suggesting a new mechanism of flavor mixing.
It is clear that flavor physics is a clue to new physics.

In the coming years, we expect new experimental results from the energy
frontier, that is, CERN Large Hadron Collider (LHC) \cite{ref:LHC}.
LHC experiments will provide us with invaluable information on new
physics.
Among several candidates of new physics, supersymmetry (SUSY) is the
most attractive and widely discussed \cite{ref:SUSY}.
It is possible that some of the superparticles are discovered in
the early stage of LHC experiments.

One of the key questions in realistic SUSY models is to identify
the mechanism of SUSY breaking.
The SUSY breaking mechanism can be explored by determining the SUSY mass
spectrum in LHC experiments at the energy frontier.
On the other hand, the whole flavor structure of the SUSY breaking cannot
be determined by the energy frontier experiment alone.
There is no a priori reason to exclude flavor changing soft SUSY
breaking terms in the squark and the slepton sectors, and some of them are
already strongly restricted by the existing low energy experimental
data \cite{ref:SUSY_flavor_problem,Ellis:1982tk}.
It means that we can extract important aspects of the SUSY breaking
mechanism from flavor physics.

Two new flavor experiments are under construction, and several others
are proposed.
The MEG experiment \cite{ref:MEG}, which intends to search for the
lepton flavor violating (LFV) process $\mu\to e\gamma$ at a branching
ratio down to $10^{-13}$, will start data taking soon.
The LHCb experiment \cite{ref:LHCb,ref:Nakada:SUSY2010s} 
is another dedicated flavor
experiment under construction and will be ready by the LHC startup in
2008.
It is designed to observe several rare decays and CP violations in $B$
and $B_s$ decays.
There are plans of future Super $B$ Factories under discussion
\cite{ref:SuperKEKB,ref:SuperBFrascati,Hewett:2004tv}.
In addition to measure several $B$ decay observables with higher
statistics, it is expected to search for LFV processes in tau decays at
a branching ratio of $10^{-9}$.
These new flavor experiments themselves and their interplay with LHC
experiments at the energy frontier will augment our knowledge on flavors
and eventually new physics.

Several strategies are possible in order to study the implication of the
past and present experimental data on SUSY models and predictions of
flavor signals in future experiments.
One of them is a model-independent method based on the 
mass insertion \cite{ref:massinsertion,%
ref:massinsertion-flavour,ref:massinsertion-bsbs}.
In this approach, a general set of off-diagonal matrix elements (mass
insertions) of the squarks and the sleptons is assumed, and one (or two)
of the elements is (are) activated in order to obtain a bound from a
specific experiment.
Repeating this procedure for every relevant experiment, a list of bounds
for the possible mass insertions is obtained.
This list is used to evaluate flavor signals in future experiments.
As an opposite way, a model specific analysis is 
possible
\cite{Bouquet:1984pp,Hall:1985dx,Barbieri:1994pv,Bertolini:1991if,%
Baek:2000sj,Moroi:2000mr,ref:flavoursignalSUSYGUT,Calibbi:2006nq}.
In this approach, one specifies a SUSY model with a well-defined SUSY
breaking sector and analyzes one (or more) selected flavor signal(s).
In this way one can make definite predictions on observable quantities
in flavor changing processes provided that the relevant model parameters
are given.

In our previous works \cite{Goto:2002xt,Goto:2003iu}, 
we adopted a different
approach.
We selected three well-motivated SUSY models: the minimal supergravity
(mSUGRA), the SU(5) SUSY grand unified theory (GUT) with right-handed
neutrinos, and the minimal supersymmetric standard model (MSSM) with
U(2) flavor symmetry.
Each of these models has a distinct flavor structure in its SUSY
breaking sector at the electroweak scale.
Then, we investigated various flavor signals in these models in a
unified fashion.
This approach allows us to evaluate flavor signals definitely and to
discuss the possibility to distinguish several different flavor
structures in the SUSY breaking sector in future flavor experiments.
The quark flavor signals which we studied are the CP violation parameter
$\varepsilon_K$ in the $K^0-\bar K^0$ mixing, the $B_d-\bar B_d$ and the
$B_s-\bar B_s$ mass splittings
 ($\Delta m_{B_d}$ and $\Delta m_{B_s}$
respectively), CP asymmetries in $B\to J/\psi K_S$ and related modes,
the direct and the mixing-induced CP asymmetries in $b\to s\gamma$, and
the CP asymmetry in $B\to\phi K_S$.
An LFV process $\mu\to e\gamma$ was studied in addition.
Comparing predictions of the models with each other, we showed that the
study of quark flavor signals at low energies could discriminate several
SUSY models that have different flavor structures in their SUSY breaking
sectors.

In the present work, we extend our previous works.
New features and improvements of the present work are the followings.
\begin{itemize}
\item[(i)]
In addition to the three models, we consider the MSSM with right-handed
neutrinos and the seesaw mechanism without GUT.
\item[(ii)]
Three cases of the low energy neutrino mass spectrum and
three types of Ans\"atze for the neutrino Yukawa coupling matrix are
studied.
\item[(iii)]
New and up-to-date experimental data are incorporated.
In particular $\Delta m_{B_s}$ measured by the CDF and D\O\ experiments
at Fermilab Tevatron affects predictions of several $B$ decay modes
\cite{ref:massinsertion-bsbs,ref:BsBs}.
\item[(iv)] 
LFV tau decays and their implications are examined.
\item[(v)] 
As computational improvements, two-loop renormalization group equations
for the MSSM (with right-handed neutrinos) parameter running and
one-loop threshold corrections at the electroweak scale are implemented.
\end{itemize}
With these new features and improvements, we pursue the possibility to
distinguish the flavor structure of the SUSY breaking sector by low
energy flavor experiments and to understand the SUSY breaking mechanism
consequently.

A brief summary of our analysis is as follows.
We expect significant flavor signals in the lepton sector
for the models with right-handed neutrinos if the neutrino Yukawa
coupling is $O(1)$.
In the MSSM with right-handed neutrinos, depending on the texture of the
neutrino Yukawa coupling matrix, some of the LFV processes, 
$\mu\rightarrow e\gamma$, $\tau\rightarrow\mu\gamma$ and 
$\tau\rightarrow e\gamma$ could be discovered in near future.
In the SU(5) SUSY GUT with right-handed neutrinos, 
in addition to the above texture dependent signals, 
$\mu\rightarrow e\gamma$ can be close to the present experimental
bound due to GUT interactions. As for the quark flavor signals,
CP violating asymmetries in $b\rightarrow s$ and $b\rightarrow d$
transitions can be significant in the SU(5) SUSY GUT with right-handed
neutrinos and in the U(2) model. Enhanced modes vary according to
the texture of the neutrino Yukawa coupling matrix in 
the SU(5) SUSY GUT with right-handed neutrinos.
Our analysis indicates that clarifying a pattern of the quark and lepton
flavor signals is an important step to determine the correct SUSY model.

This paper is organized as follows. In
Sec.~\ref{sec:models-and-parameters}, the models are presented and the
relevant SUSY parameters are introduced.
Our numerical analysis with the experimental inputs and the outline of
computational procedure are shown in Sec.~\ref{sec:numerical-analysis}.
Conclusions are given in Sec.~\ref{sec:conclusions}.

\section{Models and SUSY parameters}
\label{sec:models-and-parameters}

\subsection{Models}

In this section, we give a brief description of the models considered in
this paper.
They are well-motivated examples of SUSY models, and are chosen as
representatives that have distinct flavor signals.
Every model is reduced to the MSSM at low energy scale, which is an 
$\text{SU(3)}_C\times \text{SU(2)}_L\times \text{U(1)}_Y$ 
supersymmetric gauge theory with the SUSY being softly broken.
The MSSM matter contents are the following chiral superfields:
\begin{align}
& Q_i(3,2,\frac{1}{6})\;,\quad
  \bar{U}_i(\bar{3},1,-\frac{2}{3})\;,\quad
  \bar{D}_i(\bar{3},1,\frac{1}{3})\;,
\nonumber\\
& L_i(1,2,-\frac{1}{2})\;,\quad
  \bar{E}_i(1,1,1)\;,\quad (i=1,2,3)\nonumber\\
& H_1(1,2,-\frac{1}{2})\;,\quad H_2(1,2,\frac{1}{2})\;,
\label{eq:matterMSSM}
\end{align}
where the gauge quantum numbers are shown in parentheses.
The MSSM superpotential can be written as
\begin{align}
\mathcal{W}_{\text{MSSM}}=y_D^{ij}\bar{D}_iQ_jH_1+y_U^{ij}\bar{U}_iQ_jH_2
+y_E^{ij}\bar{E}_iL_jH_1+\mu H_1H_2\;,
\label{eq:superpotentialMSSM}
\end{align}
with an assumption of $R$-parity conservation and renormalizability.
The SUSY breaking effect is described by the following soft SUSY
breaking terms in the Lagrangian.
\begin{align}
-\mathcal{L}_{\text{soft}}^{\text{MSSM}}=&
(m_Q^2)^{ij}\tilde{q}^{\dagger}_i\tilde{q}_j
+(m_U^2)^{ij}\tilde{u}^{\dagger}_i\tilde{u}_j
+(m_D^2)^{ij}\tilde{d}^{\dagger}_i\tilde{d}_j
+(m_L^2)^{ij}\tilde{l}^{\dagger}_i\tilde{l}_j
+(m_E^2)^{ij}\tilde{e}^{\dagger}_i\tilde{e}_j\nonumber\\
&+m_{H_1}^2h_1^{\dagger}h_1
+m_{H_2}^2h_2^{\dagger}h_2
-(B\mu h_1h_2+\text{H.c.})\nonumber\\
&+\left(
A_U^{ij}\tilde{u}_i^{\dagger}\tilde{q}_j h_2
+A_D^{ij}\tilde{d}_i^{\dagger}\tilde{q}_j h_1
+A_E^{ij}\tilde{e}_i^{\dagger}\tilde{l}_j h_1
+\text{H.c.}\right)\nonumber\\
&
+\frac{M_3}{2}\bar{\tilde{g}}\tilde{g}
+\frac{M_2}{2}\bar{\tilde{W}}\tilde{W}
+\frac{M_1}{2}\bar{\tilde{B}}\tilde{B}\;,
\label{eq:softtermMSSM}
\end{align}
where $\tilde{q}_i$, $\tilde{u}_i^{\dagger}$, $\tilde{d}_i^{\dagger}$,
$\tilde{l}_i$, $\tilde{e}_i^{\dagger}$, $h_1$, and $h_2$ are the
corresponding scalar components of the chiral superfields given in
Eq.~(\ref{eq:matterMSSM}), and $\tilde{g}$, $\tilde{W}$, and $\tilde{B}$
are $\text{SU(3)}_C$, $\text{SU(2)}_L$ and $\text{U(1)}_Y$ gauge
fermions, respectively.

\subsubsection{The minimal supergravity model}
\label{sec:mSUGRA}

The mSUGRA consists of the MSSM sector and a hidden sector where 
the SUSY is assumed to be spontaneously broken.
Only a gravitational interaction interconnects these two sectors.
This gravitational interaction mediates the SUSY breaking effect from
the hidden sector to the observable MSSM sector, and the soft breaking
terms  in Eq.~(\ref{eq:softtermMSSM}) are induced in the following
manner:
\begin{align}
  & (m_Q^2)_{ij} = (m_U^2)_{ij} = (m_D^2)_{ij} = (m_L^2)_{ij} =
    (m_E^2)_{ij} = m_0^2\delta_{ij} \;,
\nonumber\\
  & m_{H_1}^2 = m_{H_2}^2=m_0^2 \;,
\nonumber\\
  & A_U^{ij} = m_0 A_0 y_U^{ij} \;,\quad
    A_D^{ij} = m_0 A_0 y_D^{ij} \;, \quad
    A_E^{ij} = m_0 A_0 y_E^{ij} \;,
\label{eq:mSUGRAboundaryconditions}
\\
  & M_1 = M_2 = M_3 = m_{1/2} \;,
\nonumber
\end{align}
where we assume the GUT relation among the gaugino masses.
The above relations are applied at the energy scale where the soft
breaking terms are induced by the gravitational interaction.
We identify this scale with the GUT scale 
($\mu_G\simeq 2\times 10^{16}$GeV) for simplicity.
Thus the soft breaking terms are specified at $\mu_G$ by the universal
scalar mass, $m_0$, the universal gaugino mass, $m_{1/2}$, and the
universal trilinear coupling, $A_0$.
The soft breaking terms at the electroweak scale are determined by
solving renormalization group equations.

In this model, the only source of flavor mixings is the CKM matrix.
New flavor mixings in the squark sector at the electroweak scale
come from the CKM matrix through radiative corrections
\cite{Bouquet:1984pp,Hall:1985dx}.
In addition to the CP phase in the CKM matrix, there can be two
physically independent CP phases.
We take the complex phase of the $\mu$ term ($\phi_{\mu}\equiv\arg\mu$)
and the phase of $A_0$ ($\phi_A\equiv\arg A_0$) as the new CP phases
while we take the gaugino mass $m_{1/2}$ as real and positive by
convention.
These CP phases contribute to the neutron and electron electric dipole
moments (EDMs) \cite{Ellis:1982tk,ref:EDM-1,ref:EDM-2,Chang:1998uc,%
Falk:1999tm} and experimental constraints on these phases are very
severe.

We assume that the generation mechanism of neutrino masses in this model
does not affect the flavor mixing in the SUSY sector.
For example, in the mSUGRA model with right-handed neutrinos, which is
described bellow, the effect of the neutrino mass on the flavor mixing
in the SUSY sector is negligible in a small right-handed neutrino mass
limit.

\subsubsection{The MSSM with right-handed neutrinos}
\label{sec:MSSMRN}

Recent developments of neutrino experiments have established the
existence of small finite masses of neutrinos.
A simple extension of the mSUGRA model for giving small finite 
masses of neutrinos is introducing gauge singlet right-handed Majorana
neutrino superfields, $\bar{N}_i$ ($i=1,2,3$).
This is known as the type I seesaw mechanism \cite{ref:seesaw}.
The superpotential can be written as
\begin{align}
  \mathcal{W}_{\text{MSSM}\nu_R} = \mathcal{W}_{\text{MSSM}}
  + (y_N)^{ij} \bar{N}_i L_j H_2
  + \frac{1}{2}(M_N)^{ij}\bar{N}_i\bar{N}_j\;,
\label{eq:superpotentialMSSMnuR}
\end{align}
which leads to the following higher dimensional term
\begin{align}
  \Delta \mathcal{W}_{\nu} =
  -\frac{1}{2}K_{\nu}^{ij}(L_iH_2)(L_jH_2) \;, \quad
  K_{\nu}=(y_N^T)^{ik}(M_N^{-1})^{kl}(y_N)^{lj} \;,
\label{eq:spMSSMnu}
\end{align}
after heavy fields $\bar{N}_{1,2,3}$ are integrated out at the energy scale 
below the Majorana mass scale ($\equiv \mu_R$).
This higher dimensional term yields the neutrino mass matrix by the 
electroweak symmetry breaking as
\begin{align}
(m_{\nu})^{ij}=(K_{\nu})^{ij}\langle h_2\rangle^2\;.
\end{align}
Taking the basis in which the charged lepton mass matrix is diagonal,
one can obtain the observable neutrino mass eigenvalues and
the Pontecorvo-Maki-Nakagawa-Sakata (PMNS) mixing matrix \cite{ref:PMNS}
as
\begin{align}
  (m_{\nu})^{ij} =
  (V_{\text{PMNS}}^*)^{ik}m_{\nu_k}(V_{\text{PMNS}}^{\dagger})^{kj}\;.
\end{align}
From the neutrino oscillation experiments, it is known that there is a
hierarchy among the two squared mass differences as
$|m_{\nu_3}^2-m_{\nu_2}^2|\gg |m_{\nu_2}^2-m_{\nu_1}^2|$.
We define $\nu_1$ and $\nu_2$ so that $m_{\nu_2} > m_{\nu_1}$.
Therefore there are two possibilities for the neutrino mass hierarchy
when the mass of the lightest neutrino is much smaller than the mass
splittings.
\begin{itemize}
\item Normal hierarchy: $m_{\nu_3}\gg m_{\nu_2}>m_{\nu_1}$;
\item Inverted hierarchy: $m_{\nu_2}>m_{\nu_1}\gg m_{\nu_3}$.
\end{itemize}
When the overall mass scale is much larger than the mass splittings,
all the three neutrinos are nearly degenerate in mass.
In the present analysis we take
\begin{itemize}
\item Degenerate: $m_{\nu_3}\gtrsim m_{\nu_2}\gtrsim m_{\nu_1}$.
\end{itemize}
For numerical calculations, we consider three sets of low energy
neutrino parameters corresponding to the above three cases.

As for the soft breaking terms, scalar mass terms, $A$ terms and $B$ terms 
of sneutrinos, $\tilde{\nu}^{\dagger}_i$, are added as
\begin{align}
  -\mathcal{L}_{\text{soft}}^{\text{MSSM}\nu_R} =
  -\mathcal{L}_{\text{soft}}^{\text{MSSM}}
  +(m_N^2)^{ij}\tilde{\nu}_i^{\dagger}\tilde{\nu}_j
  +
  \left(
    A_N^{ij} \tilde{\nu}_i \tilde{l}_j h_2
    + (\tilde{m}_N^2)^{ij} \tilde{\nu}_i^{\dagger} \tilde{\nu}_j^{\dagger}
    + \text{H.c.}
  \right)\;.
\end{align}
We assume that the soft breaking terms are generated in a universal
fashion at $\mu_G$, {\it i.e.}
\begin{align}
  (m_N^2)^{ij} = m_0^2 \delta^{ij} \;,\quad
  A_N^{ij} = m_0 A_0 y_N^{ij} \;.
\label{eq:MSSMnuboundaryconditions}
\end{align}
We neglect the $\tilde{m}_N^2$ terms in the present work.
These terms can significantly affect the EDMs,
while contributions to the lepton flavor violation 
processes are sub-dominant \cite{ref:BnuR}.

The new flavor mixing in the scalar lepton sector comes from the 
neutrino mixing through the renormalization group running between 
$\mu_G$ and $\mu_R$.
In the leading logarithmic approximation, they are given as
\begin{subequations}
\label{eq:yNeffectlog}
\begin{align}
(m_L^2)^{ij}\simeq & -\frac{1}{8\pi^2}m_0^2(3+|A_0|^2)
(y_N^{\dagger}y_N)^{ij}\ln \frac{\mu_G}{\mu_R}\;,\\
(m_E^2)^{ij}\simeq & 0\;,\\
(A_E)^{ij}\simeq &
-\frac{3}{8\pi^2} m_0 A_0 y_e^i (y_N^{\dagger}y_N)^{ij}
\ln \frac{\mu_G}{\mu_R}\;,
\end{align}
\end{subequations}
for $i\neq j$.
We numerically solve full RGEs in the actual analysis given in
Sec.~\ref{sec:numerical-analysis}.
Consequences of these mixings on lepton flavor
violating processes have been investigated from various aspects.
Lepton flavor violating processes such as $\mu\to e\gamma$ are 
sensitive to the off-diagonal elements of $y_N^{\dagger}y_N$
\cite{ref:meg}.

As we have discussed in the end of the last subsection, this model
reduces to the mSUGRA model in the limit of $y_N\to 0$.
For instance, the effect of (\ref{eq:yNeffectlog}) is negligible if
$\mu_R \ll 10^{12} \mathrm{GeV}$.

We consider three typical structures of the neutrino Yukawa couplings.
\begin{itemize}
\item degenerate $\nu_R$ case
\begin{equation}
y_N=\frac{\sqrt{\hat{M}_N}}{\langle h_2\rangle}
\begin{pmatrix}
\sqrt{m_{\nu_1}}&0&0\\
0&\sqrt{m_{\nu_2}}&0\\
0&0&\sqrt{m_{\nu_3}}
\end{pmatrix}
V_{\text{PMNS}}^{\dagger}\;.
\label{eq:yNdegenerate}
\end{equation}
This is a case that all the masses of
the right-handed neutrinos 
are the same and there are no CP phases in the heavy neutrino sector.
In Eq.~(\ref{eq:yNdegenerate}), $\hat{M}_N$ denotes the eigenvalue of the
right-handed neutrino mass matrix, {\it i.e.}
$(M_N)^{ij}=\hat{M}_N\delta^{ij}$.
In this simplest case, the mixing in $y_N$ should be identified with the
PMNS mixing because there is no flavor structure in $M_N$.
The large mixing in the PMNS matrix leads to large off-diagonal
elements of $y_N^{\dagger}y_N$, which enhance the $\mu\to e\gamma$
branching ratio.
As we will see later, the SUSY breaking parameter space is
strongly constrained by the present experimental limit in the normal
hierarchy case.
\item non-degenerate $\nu_R$ (I)
\begin{align}
y_N=
\begin{pmatrix}
y_{11} & 0 & 0\\
0 & y_{22} & y_{23}\\
0 & y_{32} & y_{33}
\end{pmatrix}\;.
\label{eq:Y-non-degenerate-I}
\end{align}
In this case, the PMNS mixing arises from the above $y_N$ and 
a non-degenerate mass matrix of right-handed neutrinos, $M_N$, as is
described in Sec.~\ref{sec:inputparameters}. Because $y_N^\dagger y_N$
has the same texture as $y_N$ in Eq.~(\ref{eq:Y-non-degenerate-I}),
$\mu\to e\gamma$ is suppressed enough to satisfy the present
experimental bound. As for other LFV processes, $\tau\to e\gamma$ is
also suppressed, while $\tau\to \mu\gamma$ is not.
The specific structure in Eq.~(\ref{eq:Y-non-degenerate-I}) could be an
implication of electron-number conservation which works above the
right-handed neutrino mass scale, $\mu_R$, and is broken by the
right-handed neutrino mass matrix, $M_N$.
\item non-degenerate $\nu_R$ (II)
\begin{align}
y_N=
\begin{pmatrix}
y_{11} & 0 & y_{13}\\
0 & y_{22} & 0\\
y_{31} & 0 & y_{33}
\end{pmatrix}\;.
\label{eq:Y-non-degenerate-II}
\end{align}
This case is similar to the non-degenerate $\nu_R$ (I) case, except that
the first and the second generations are interchanged in $y_N$. 
Accordingly, $\mu\to e\gamma$ and $\tau\to \mu\gamma$
are suppressed, while we expect a larger branching ratio of 
$\tau\to e\gamma$.
\end{itemize}

\subsubsection{The SU(5) SUSY GUT with right-handed neutrinos}
\label{sec:SU5RN}

The idea of supersymmetric grand unification is supported by the precise
determination of three gauge coupling constants at LEP and other
experiments in the last decade.
In view of this, we consider SU(5) SUSY GUT with right-handed neutrinos
as an extension of the MSSM with right-handed neutrinos.
Here we follow the analysis of
Ref.~\cite{Baek:2000sj,Goto:2002xt} and we give a brief description of
the model.

This model is defined by the following superpotential:
\begin{eqnarray}
  \mathcal{W}_{\text{SU(5)}\nu_R} &=&
  \frac{1}{8}\epsilon_{abcde}(\lambda_U)^{ij}(T_i)^{ab}(T_j)^{cd}H^e
  +(\lambda_D)^{ij}(\bar{F}_i)_a(T_j)^{ab}\bar{H}_b
\nonumber\\&&
 +(\lambda_N)^{ij}\bar{N}_i(\bar{F}_j)_a H^a
 +\frac{1}{2}(M_N)^{ij}\bar{N}_i\bar{N}_j
 + \mathcal{W}_H
 + \Delta\mathcal{W}_{\text{SU(5)}\nu_R}
 \;,
\label{eq:SU5RN}
\end{eqnarray}
where $i$ and $j$ are generation indexes, while $a,\,b,\,c,\,d$ and
$e$ are SU(5) indices.
$\epsilon_{abcde}$ denotes the totally antisymmetric tensor of
the SU(5). $T_i$, $\bar{F}_i$, and $\bar{N}_i$ are
${\bf 10}$, ${\bf \bar{5}}$, and ${\bf 1}$ representations of
the SU(5) gauge group, respectively.
$Q_i$, $\bar{U}_i$, and $\bar{E}_i$ are embedded in 
$T_i$, $\bar{F}_i$ consists of $\bar{D}_i$ and $L_i$, and
$\bar{N}_i$ is identified with the right-handed heavy Majorana neutrinos.
$H$ ($\bar{H}$) denotes a Higgs superfield in ${\bf 5}$
(${\bf \bar{5}}$) representation and includes
$H_C({\bf 3,1},-\frac{1}{3})$ and $H_2$
($\bar{H}_C({\bf \bar{3},1},\frac{1}{3})$ and $H_1$).
$(\lambda_U)^{ij}$, $(\lambda_D)^{ij}$, and $(\lambda_N)^{ij}$
are the Yukawa coupling matrices, and $(M_N)^{ij}$ is
the Majorana mass matrix.
The superpotential for Higgs superfields, $\mathcal{W}_H$, contains terms
with $H$, $\bar{H}$, and ${\Sigma^a}_b$ which is a {\bf 24} representation 
of the SU(5) gauge group.
It is assumed that a vacuum expectation value (VEV) of ${\Sigma^a}_b$, 
$\langle \Sigma^a_{~b} \rangle=
\text{diag}(\frac{1}{3},\frac{1}{3},\frac{1}{3},
-\frac{1}{2},-\frac{1}{2})v_G$ breaks the SU(5) symmetry to 
$\text{SU(3)}_{\text{C}}\times
\text{SU(2)}_{\text{L}}\times\text{U(1)}_{\text{Y}}$
at $\mu_G$.
$\Delta\mathcal{W}_{\text{SU(5)}\nu_R}$ is a dimension five operator,
which is introduced in order to reproduce the realistic mass relations
between the down-type quarks and charged leptons, as explained in
Ref.~\cite{Goto:2002xt}.

The supermultiplets with the masses of order of the GUT scale 
such as $H_C$ and $\bar{H}_C$ are integrated out at $\mu_G$
and the effective theory below $\mu_G$ is the MSSM with the
right-handed neutrinos described by the superpotential 
$\mathcal{W}_{\text{MSSM}\nu_R}$ in Eq.~(\ref{eq:spMSSMnu}).
The Yukawa coupling matrices in Eq.~(\ref{eq:spMSSMnu}) are related 
to those in Eq.~(\ref{eq:SU5RN}) as $(y_U)^{ij}=(\lambda_U)^{ij}$ and
$(y_{N})^{ij}=(\lambda_N)^{ij}$.
$(\lambda_D)^{ij}$ is determined from $(y_D)^{ij}$ and $(y_E)^{ij}$, taking
$O(\mu_G/\mu_P)$ corrections from
$\Delta\mathcal{W}_{\text{SU(5)}\nu_R}$ into account \cite{Goto:2002xt}.

There are additional degrees of freedom in the matching relations
between $y_{U,D,E,N}$ and  $\lambda_{U,D,N}$, which cannot be determined
from the quark and lepton masses and the CKM and PMNS matrices at low
energy \cite{Ellis:1979hy,Moroi:2000mr,Baek:2000sj}.
In the present analysis, we introduce only two relative phase parameters
for simplicity, which corresponds to $\widehat{\Theta}_L$ in
Ref.~\cite{Moroi:2000mr}.

The SU(5) invariant and renormalizable soft breaking terms are written as
\begin{align}
  -\mathcal{L}^{\text{SU(5)}}_{\text{soft}} &=
  \frac{1}{2} (m^2_T)^{ij} (\tilde{T}_i^{*})_{ab} (\tilde{T}_j)^{ab}
  + (m^2_{\bar{F}})^{ij}
    (\tilde{\bar{F}}_i^{*})^a (\tilde{\bar{F}}_j)_a
  + (m^2_{\bar{N}})^{ij}
    \tilde{\bar{N}}_i^{*}\tilde{\bar{N}}_j
\nonumber\\&
  +(m^2_{H}) H^*_{~a} H^a
  +(m^2_{\bar{H}}) \bar{H}^{*a} \bar{H}_a
  +\left(\frac{3}{2}B_H\lambda_Hv_G\bar{H}_aH^a+\text{H.c.}\right)
  +(\text{terms with ${\Sigma^a}_b$})
\nonumber\\&
  +\left\{
      \frac{1}{8}\epsilon_{abcde}(\tilde{\lambda}_U)^{ij}
      (\tilde{T}_i)^{ab}(\tilde{T}_j)^{cd} H^e
    + (\tilde{\lambda}_D)^{ij}(\tilde{\bar{F}}_i)_a
      (\tilde{T}_j)^{ab} \bar{H}_b
    \right.
\nonumber\\&\phantom{+}
    \left.
    + (\tilde{\lambda}_N)^{ij}
      \tilde{\bar{N}}_i(\tilde{\bar{F}}_j)_a H^a
    + \frac{1}{2}(\tilde{M}_N)^{ij}
      \tilde{\bar{N}}_i\tilde{\bar{N}}_j
      + \text{H.c.} \right\}
\nonumber\\&
    +\frac{1}{2}M_5\bar{\tilde{G}}_5\tilde{G}_5,
\label{eq:SUSY-Breaking-SU5RN}
\end{align}
where $\tilde{T}_i$, ${\tilde{\bar{F}}}_i$, and
${\tilde{\bar{N}}}_i$ are the scalar components of $T_i$,
$\bar{F}_i$, and $\bar{N}_i$, respectively;
$H$ and $\bar{H}$ stand for the corresponding scalar components
of the superfields denoted by the same symbols; and
$\tilde{G}_5$ represents the SU(5) gaugino.
We assume that the soft breaking terms are generated in a universal
fashion at the Planck scale, $\mu_P$, {\it i.e.}
\begin{align}
  (m^2_T)^{ij}= &
  (m^2_{\bar{F}})^{ij} =
  (m^2_{\bar{N}})^{ij} =
  m_0^2\delta^{ij}\;,\nonumber\\
  (\widetilde{\lambda})^{ij} = & m_0 A_0 (\lambda)^{ij}\;,\quad
  ( \lambda = \lambda_U, \lambda_D, \lambda_N )\;, \nonumber\\
  M_5 = & m_{1/2}\;.
  \label{eq:bcSU5GUT}
\end{align}
We solve the RG equations of the SU(5) SUSY GUT from $\mu_P$
to $\mu_G$ with Eq.~(\ref{eq:bcSU5GUT}) as boundary conditions
at $\mu_P$, then those of MSSM with right-handed neutrinos
between the $\mu_G$ and $\mu_R$.
Finally, the squark and slepton mass matrices are obtained by the RG
equations of the MSSM below $\mu_R$.

Unlike the previous two models, a large flavor mixing in the 
neutrino sector can affect the right-handed down type squark sector
because the lepton doublets and the down quarks are embedded in the
same representation of SU(5).
For a similar reason, the CKM mixing in the quark sector contributes to
the mixing in the right-handed charged slepton sector
\cite{Hall:1985dx,Barbieri:1994pv}.
For instance, the correction to $m_E^2$ is given in the leading
logarithmic approximation as
\begin{equation}
  (m_E^2)^{ij}
  \simeq -\frac{3}{8\pi^2}m_0^2(3+|A_0|^2)
  (\lambda_U^\dagger \lambda_U)^{ij} \ln \frac{\mu_P}{\mu_G}\;.
\end{equation}
Quark flavor signals in models with a grand unification have been studied 
in literature
\cite{Moroi:2000mr,Baek:2000sj,ref:flavoursignalSUSYGUT}.
It is shown in these papers that large contributions to
$\epsilon_K$ and the $\mu\to e\gamma$ decay can arise from the new 
source of flavor mixing in the neutrino sector.

We study the same patterns of neutrino Yukawa couplings as those in MSSM
with right-handed neutrinos, {\it i.e.} degenerate $\nu_R$,
non-degenerate $\nu_R$ (I) and non-degenerate $\nu_R$ (II) cases.

\subsubsection{A model with U(2) flavor symmetry}
\label{sec:U(2)}

There is a class of models which are intended to solve the flavor
problem of the MSSM by introducing appropriate symmetry structure.
U(2) flavor symmetry \cite{ref:U2-1,ref:U2-2} is a typical example of such
models. 
We consider the model given in Ref.~\cite{ref:U2-2}.
In this model, the quark and lepton supermultiplets in the first and
the second generations transform as doublets under the U(2) flavor
symmetry, and the third generation and the Higgs supermultiplets are
singlets under the U(2).
In addition to the ordinary matter fields, we introduce the following
superfields: a doublet $\Phi^i(-1)$, a symmetric tensor $S^{ij}(-2)$,
and an antisymmetric tensor $A^{ij}(-2)$, where $i$ and $j$ run over the
first two generations and the numbers in the parentheses represent the 
charge of the U(1) subgroup.

The U(2) invariant superpotential relevant to the quark
Yukawa couplings is given as follows:
\begin{align}
\mathcal{W}_{\text{U(2)}} =&
  Y_U
  \left(
    \bar{U}_3 Q_3 H_2
    +\frac{b_U}{M_F}\Phi^i\bar{U}_i Q_3 H_2
    +\frac{c_U}{M_F}\bar{U}_3\Phi^i Q_i H_2
  \right.
\nonumber
\\&
  \left.
    +\frac{d_U}{M_F}S^{ij}\bar{U}_i Q_j H_2
    +\frac{a_U}{M_F}A^{ij}\bar{U}_i Q_j H_2\right)
\nonumber
\\& +
  Y_D
  \left(
    \bar{D}_3 Q_3 H_1
    +\frac{b_D}{M_F}\Phi^i\bar{D}_i Q_3 H_1
    +\frac{c_D}{M_F}\bar{D}_3\Phi^i Q_i H_1
  \right.
\nonumber\\&
  \left.
    +\frac{d_D}{M_F}S^{ij}\bar{D}_i Q_j H_1
    +\frac{a_D}{M_F}A^{ij}\bar{D}_i Q_j H_1
  \right),
\label{U2superpotential}
\end{align}
where $M_F$ is the scale of the flavor symmetry, and
$Y_Q$, $a_Q$, $b_Q$, $c_Q$, and $d_Q$ ($Q=U,D$) are dimensionless
coupling constants. Dimension five and higher dimensional operators 
are neglected in the superpotential in 
Eq.~(\ref{U2superpotential}).
Absolute values of the above dimensionless coupling constants
except for $Y_D$ are supposed to be of $O(1)$.

The breaking pattern of the U(2) symmetry is assumed to be
\begin{align}
\text{U(2)}\to \text{U(1)}\to \text{no symmetry}\;,
\label{eq:U2breaking}
\end{align}
in order to reproduce the preferable quark Yukawa coupling matrices 
which can explain the mass eigenvalues and the mixing of quarks.
The first breaking is induced by VEV's of $\Phi^i$ and $S^{ij}$,
and the second one by a VEV of $A^{ij}$.
These VEV's are given as
\begin{equation}
\frac{\langle \Phi^i \rangle}{M_F}=\delta^{i2}\,\epsilon,\;
\frac{\langle S^{ij} \rangle}{M_F}=\delta^{i2}\,\delta^{j2}\epsilon,\;
\frac{\langle A^{ij} \rangle}{M_F}=\epsilon^{ij}\,\epsilon',
\end{equation}
where $\epsilon$ and $\epsilon'$ are taken to be real without loss of
generality. 
Because $\epsilon$ and $\epsilon'$ are order parameters of the U(2) 
and U(1) symmetry breaking respectively, they satisfy
$\epsilon'\ll \epsilon$.
Note that $\langle S^{ij}\rangle$ is chosen
to leave the U(1) unbroken.
With the breaking pattern given in Eq.~(\ref{eq:U2breaking}), 
we obtain the quark Yukawa coupling matrix $y_Q$
as
\begin{align}
y_Q^{ij}=Y_Q
\begin{pmatrix}
0&a_Q\epsilon'&0\\
-a_Q\epsilon'&d_Q\epsilon&b_Q\epsilon\\
0&c_Q\epsilon&1
\end{pmatrix}\;,
\quad Q=U,D\;.
\label{eq_u2_yukawa}
\end{align}

The U(2) symmetry controls not only the superpotential but also 
the soft breaking terms. After the U(2) broken with the pattern 
in Eq.~(\ref{eq:U2breaking}), 
the squark mass matrices $m_X^2$ can be obtained as
\begin{align}
m_X^2=(m_0^{X})^2
\begin{pmatrix}
1&0&0\\
0&1+r_{22}^{X}\epsilon^2&r_{23}^X\epsilon\\
0&r_{23}^{X*}\epsilon&r_{33}^X
\end{pmatrix}\;,
X=Q,U,D\;,
\label{eq_u2_susy_para}
\end{align}
where $r^X_{ij}$ are dimensionless parameters of $\mathcal{O}(1)$.
As for the squark $A$ terms, they have the same structure as the quark
Yukawa coupling matrices:
\begin{align}
  A_Q^{ij}=A_Q^0Y_Q
  \begin{pmatrix}
    0 & \tilde{a}_{Q} \epsilon' & 0\\
-\tilde{a}_{Q}\epsilon'&\tilde{d}_Q\epsilon&
\tilde{b}_Q\epsilon\\
0&\tilde{c}_Q\epsilon&1
\end{pmatrix}\;,\quad
Q=U,D\;.
\label{eq:U2-Aterm}
\end{align}
In general, though being of $\mathcal{O}(1)$,
$\tilde{a}_Q$, $\tilde{b}_Q$, $\tilde{c}_Q$, and
$\tilde{d}_Q$ take different values from the
corresponding parameters in Eq.~(\ref{eq_u2_yukawa}), and we expect no
exact universality of the $A$ terms in this model.

With the help of the U(2) symmetry, the masses of the first and second
generation squarks are naturally degenerate.
On the other hand, the mass of the third generation squarks may be
separated from the others.
There exist flavor mixings of $\mathcal{O}(\epsilon)$ between the second
and the third generations of squarks.
These are new sources of flavor mixing besides the CKM matrix.

There are several efforts to explain the observed neutrino masses and
mixings in SUSY models with the U(2) flavor symmetry (or its discrete
relatives) \cite{ref:U2-neutrino}.
Unlike the quark sector, application of the U(2) symmetry to the lepton
sector is not straightforward because of the large mixings of the
neutrinos.
Therefore we focus on the quark sector in the following analysis, taking
the same boundary conditions as Eq.~(\ref{eq:mSUGRAboundaryconditions})
for the slepton sector.

\subsection{Treatments of radiative breaking of the electroweak symmetry}

\begin{figure}
\includegraphics{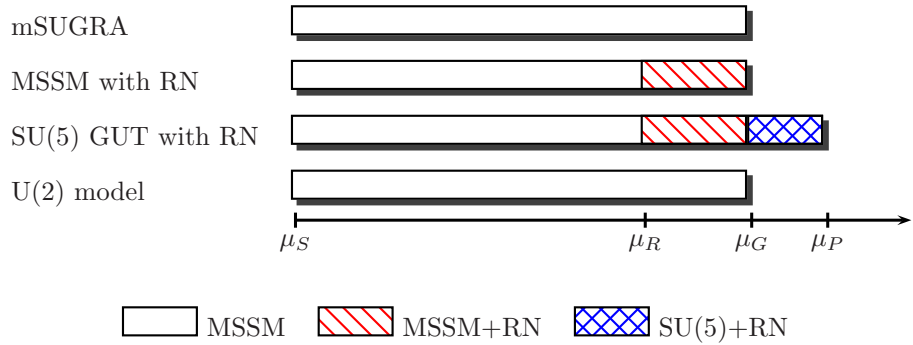}
\caption{%
The cut-off scales and the models. RN stands for right-handed neutrinos.
}
\label{fig:cutoff-and-models}
\end{figure}

In the models we consider, SUSY parameters such as $m_0$, $m_{1/2}$,
$A_0$ {\it etc.} are given at the high energy cut-off scale.
In order to analyze flavor signals, we need to connect the parameters at
the cut-off scale and those at the electroweak scale with help of the
renormalization group equations.
In the present work, we adopt the following procedure to determine the
parameters at the electroweak scale.
\begin{enumerate}
\item
The masses of quarks and leptons and the mixings (the CKM and PMNS matrices)
are given as inputs at the electroweak scale, $\mu_W = M_Z$.
These masses are running masses in the standard model.
The Yukawa couplings and the coupling matrix of the dimension five
operator $K_\nu$ in Eq.~(\ref{eq:spMSSMnu})
are determined by these masses and another input parameter
$\tan\beta\equiv \langle h_2\rangle /\langle h_1\rangle$.
\item\label{item:runup}
Two-loop RGEs for the Yukawa couplings and $K_{\nu}$, as well as
the gauge coupling constants, are solved up to a high energy cut-off
scale with the boundary conditions given at $\mu_W$.
The cut-off scale is taken as the GUT scale, $\mu_G$, for the mSUGRA,
the MSSM with right-handed neutrinos, and the U(2) model and the Planck
scale, $\mu_P$, for the SU(5) SUSY GUT with right-handed neutrinos.
By this procedure, we calculate the parameters in the superpotential at
the cut-off scale.
A schematic picture of the cut-off scale involved in these models 
is displayed in Fig.~\ref{fig:cutoff-and-models}.
Here, the $\overline{\text{DR}}'$ scheme \cite{ref:DRbarRGE}  is adopted
as a renormalization scheme.

In the MSSM with right-handed neutrinos and SU(5) SUSY GUT with 
right-handed neutrinos, we decompose $K_{\nu}$ to $y_N$ and $M_N$ at 
the $\mu_R$ scale so that they satisfy the seesaw relation,
Eq.~(\ref{eq:spMSSMnu}).
In the SU(5) SUSY GUT with right-handed neutrinos the parameters in 
$\mathcal{W}_{\text{SU(5)}\nu_R}$ are matched with the MSSM with
right-handed neutrinos at the GUT scale, $\mu_G$.
\item
The boundary conditions for the soft SUSY breaking parameters are set at
the cut-off scale as Eq.~(\ref{eq:mSUGRAboundaryconditions}) for the
mSUGRA, Eq.~(\ref{eq:mSUGRAboundaryconditions}) and
(\ref{eq:MSSMnuboundaryconditions}) for the MSSM with right-handed
neutrinos, Eq.~(\ref{eq:bcSU5GUT}) for the SU(5) SUSY GUT with
right-handed neutrinos, and Eq.~(\ref{eq_u2_susy_para}) for the U(2)
model.
We take the same boundary conditions for the $A$ parameters in the
U(2) model as the mSUGRA case for simplicity.
\item
With help of two-loop RGEs, we evaluate the soft breaking terms at a
typical SUSY breaking scale, $\mu_S = 1$TeV, and calculate the SUSY
masses and mixings at the leading order which are considered as
$\overline{\text{DR}}'$ masses.
For the masses of the Higgs bosons, the one loop corrections are
included.

Then we set the value of $\mu$ and $B$ so that the tadpole diagrams of
the Higgs bosons up to one loop level vanish.

Then running the $\mu$ to the electroweak scale $M_Z$, we obtain the
$\mu$ at $M_Z$. 
\item\label{item:mssmTH}
The SUSY threshold corrections to the gauge couplings and the masses of
quarks and leptons are evaluated in order to  determine
$\overline{\text{DR}}'$ gauge couplings, $\overline{\text{DR}}'$ Higgs
vev, and $\overline{\text{DR}}'$ masses of the matter fermions in the
MSSM which lead to the $\overline{\text{DR}}'$ Yukawa couplings,
according to Ref.~\cite{Pierce:1996zz}.
\item 
We iterate from \ref{item:runup} to \ref{item:mssmTH} in the above list
until the numerical behavior converges.
\item 
The physical mass spectrum of SUSY particles is calculated at the $M_Z$
scale up to one loop level \cite{Pierce:1996zz}.
The flavor observables are also calculated with the parameters
determined at the $M_Z$ scale.
\end{enumerate}

In comparison with the previous work, two-loop RGEs for the 
running of SUSY parameters are used and the one loop SUSY threshold corrections 
at the electroweak scale are included in the calculation of this work.

\section{Numerical analysis}
\label{sec:numerical-analysis}

\subsection{Flavor observables}

The flavor observables considered in the following are the 
$K^0-\bar{K}^0$, $B_d-\bar{B}_d$ and $B_s-\bar{B}_s$ mixings,
both the direct and mixing induced CP asymmetries of $b\to s\gamma$ and
$b\to d\gamma$, and the time dependent CP asymmetry of $B\to \phi K_S$.
The branching ratios of the lepton flavor violating decay processes 
$\mu\to e\gamma$, $\tau\to \mu\gamma$ and $\tau\to e\gamma$ are also
evaluated in the MSSM with right-handed neutrinos and SU(5) SUSY GUT
with right-handed neutrinos.
As mentioned in Sec.~\ref{sec:U(2)}, we do not consider the flavor
signals in the lepton sector for the U(2) model.
Here we show the calculation methods of the flavor observables briefly.
Detail on the calculation methods of the flavor observables is available
in Ref.~\cite{Baek:2000sj,Goto:2002xt,Goto:2003iu}.

\subsubsection{$K-\bar{K}$ and $B-\bar{B}$ mixings}

The $K^0-\bar{K}^0$, $B_d-\bar{B}_d$ and $B_s-\bar{B}_s$ mixings 
are described by the effective Lagrangian of the following form:
\begin{align}
\mathcal{L}_{\Delta F=2}=&
C_{LL}(\bar{q}_L^{\alpha} \gamma^\mu Q_{L\alpha})
(\bar{q}_L^{\beta } \gamma_\mu Q_{L\beta })
+ C_{RR}(\bar{q}_R^{\alpha} \gamma^\mu Q_{R\alpha})
(\bar{q}_R^{\beta } \gamma_\mu Q_{R\beta })
\nonumber\\&
+ C_{LR}^{(1)}
(\bar{q}_R^{\alpha} Q_{L\alpha})
(\bar{q}_L^{\beta } Q_{R\beta })
+ C_{LR}^{(2)}
(\bar{q}_R^{\alpha} Q_{L\beta })
(\bar{q}_L^{\beta } Q_{R\alpha})
\nonumber\\&
+ \tilde{C}_{LL}^{(1)}
(\bar{q}_R^{\alpha} Q_{L\alpha})
(\bar{q}_R^{\beta } Q_{L\beta })
+ \tilde{C}_{LL}^{(2)}
(\bar{q}_R^{\alpha} Q_{L\beta })
(\bar{q}_R^{\beta } Q_{L\alpha})
\nonumber\\&
+ \tilde{C}_{RR}^{(1)}
(\bar{q}_L^{\alpha} Q_{R\alpha})
(\bar{q}_L^{\beta } Q_{R\beta })
+ \tilde{C}_{RR}^{(2)}
(\bar{q}_L^{\alpha} Q_{R\beta })
(\bar{q}_L^{\beta } Q_{R\alpha}),
\label{eq:lagDB2}
\end{align}
where $(q,\,Q)=(d,\,b)$, $(s,\,b)$ and $(d,\,s)$ for the $B_d-\bar{B}_d$,
$B_s-\bar{B}_s$ and $K^0-\bar{K}^0$ mixings, respectively, and
the suffices $\alpha$ and $\beta$ denote color indices.
New physics contributions to the Wilson coefficients $C$'s, as well as
the SM ones, are obtained by calculating relevant box diagrams.
Explicit formulae of the coefficients are found in 
{\it e.g.} Ref.~\cite{Baek:2000sj}.
The mixing matrix elements $M_{12}(B_d)$, $M_{12}(B_s)$, and $M_{12}(K)$
are given as
\begin{equation}
  M_{12}(P) = -\frac{1}{2m_P}
  \langle P | \mathcal{L}_{\Delta F=2} | \bar{P} \rangle,
\end{equation}
where $P=B_d,\,B_s,\,K^0$.

In the evaluation of the matrix elements 
$\langle P|\mathcal{L}_{\Delta F=2}|\bar{P}\rangle$,
we parameterize the matrix elements of the operators in
Eq.~(\ref{eq:lagDB2}) as
\begin{subequations}
\begin{align}
  \langle K^0 |
  (\bar{d}_L^{\alpha} \gamma^\mu s_{L\alpha})
  (\bar{d}_L^{\beta } \gamma_\mu s_{L\beta })
  | \bar{K}^0 \rangle
  =&
  \frac{2}{3}m_K^2 f_K^2 B_K,
\\
  \langle K^0 |
  (\bar{d}_R^{\alpha} s_{L\alpha})
  (\bar{d}_L^{\beta } s_{R\beta })
  | \bar{K}^0 \rangle
  =&
  \frac{1}{2} \left( \frac{m_K}{m_s+m_d} \right)^2
  m_K^2 f_K^2 B_K^{LR(1)},
\\
  \langle K^0 |
  (\bar{d}_R^{\alpha} s_{L\beta })
  (\bar{d}_L^{\beta } s_{R\alpha})
  | \bar{K}^0 \rangle
  =&
  \frac{1}{6} \left( \frac{m_K}{m_s+m_d} \right)^2
  m_K^2 f_K^2 B_K^{LR(2)},
\\
  \langle K^0 |
  (\bar{d}_L^{\alpha} s_{R\alpha})
  (\bar{d}_L^{\beta } s_{R\beta })
  | \bar{K}^0 \rangle
  =&
  -\frac{5}{12} \left( \frac{m_K}{m_s+m_d} \right)^2
  m_K^2 f_K^2 \tilde{B}_K^{RR(1)},
\\
  \langle K^0 |
  (\bar{d}_L^{\alpha} s_{R\beta })
  (\bar{d}_L^{\beta } s_{R\alpha})
  | \bar{K}^0 \rangle
  =&
  \frac{1}{12} \left( \frac{m_K}{m_s+m_d} \right)^2
  m_K^2 f_K^2 \tilde{B}_K^{RR(2)},
\end{align}
\label{eq kk matrix element}
\end{subequations}
where $B_K$, $B_K^{LR(1,2)}$, and $\tilde{B}_K^{RR(1,2)}$ are bag
parameters.
$B-\bar{B}$ mixing matrix elements are also defined in the same way.
The bag parameters of $B$ and $K$ mesons and the decay constants 
of the $B$ mesons are
evaluated by the lattice QCD method \cite{ref:LatticeQCD}.
We list the numerical values used in our calculation in
Table~\ref{tab:f-and-bag}.
\begin{table}
\begin{center}
\begin{tabular}{ccccccc}
\hline
$P$ & $f_P$(MeV) & $B_P$ & $B_P^{LR(1)}$ & $B_P^{LR(2)}$
  & $\tilde{B}_P^{RR(1)}$ & $\tilde{B}_P^{RR(2)}$
\\
\hline
$K$ & 159.8 & 0.63 & 1.03 & 0.77 & 0.59 & 0.85 \\
$B_d$ & 198 & 0.87 & 1.15 & 1.72 & 0.79 & 0.92 \\
$B_s$ & 239 & 0.87 & 1.16 & 1.75 & 0.80 & 0.94 \\
\hline
\end{tabular}
\end{center}
\caption{%
Decay constants and bag parameters for the $B^0-\bar{B}^0$ and
the $K^0-\bar{K}^0$ mixing matrix elements \cite{ref:LatticeQCD} used in
the numerical calculation.
Here $f_K$ is the experimental value.
}
\label{tab:f-and-bag}
\end{table}

The observables $\varepsilon_K$, $\Delta m_{B_d}$ and $\Delta m_{B_s}$ are
expressed in terms of $M_{12}$ as
\begin{align}
\varepsilon_K =& \frac{e^{i\pi/4}\mathrm{Im} M_{12}(K) }{ \sqrt{2}\Delta m_K },
\\
\Delta m_{B_d}=& 2\left| M_{12}(B_d) \right|,
\\
\Delta m_{B_s}=& 2\left| M_{12}(B_s) \right|.
\end{align}
%

\subsubsection{CP asymmetries in $B$ meson decays}

The time-dependent CP asymmetry in the $B_d$ decays to a CP eigenstate
$f_{\text{CP}}$ is given by
\begin{align}
\frac{\Gamma(\bar{B}_d(t)\to f_{\text{CP}})-\Gamma(B_d(t)\to f_{\text{CP}})}
{\Gamma(\bar{B}_d(t)\to f_{\text{CP}})+\Gamma(B_d(t)\to f_{\text{CP}})}
= & A_{\text{CP}}(B_d\to f_{\text{CP}})\cos\Delta m_{B_d} t
\nonumber\\
 & +S_{\text{CP}}(B_d\to f_{\text{CP}})\sin \Delta m_{B_d} t\;,
\end{align}
where $A_{\text{CP}}$ and $S_{\text{CP}}$ are direct and indirect
(mixing-induced) CP violation parameters, respectively.

For $f_{\text{CP}}= J/\psi K_S$, the $b\to c\bar{c}s$ decay amplitude is
assumed to be dominated by the tree level Standard Model contribution.
Consequently, the direct CP asymmetry $A_{\text{CP}}(B_d\to J/\psi K_S)$
is negligibly small.
The weak phase of the $b\to c\bar{c}s$ decay amplitude comes from a
product of the CKM matrix elements $V_{cb}V_{cs}^*$, which is almost
real by convention.
Therefore we can write
\begin{align}
\frac{\Gamma(\bar{B}_d(t)\to J/\psi K_S) - \Gamma(B_d(t)\to J/\psi K_S)}
     {\Gamma(\bar{B}_d(t)\to J/\psi K_S) + \Gamma(B_d(t)\to J/\psi K_S)}
&=
S_{\text{CP}}(B_d\to J/\psi K_S)\,\sin\Delta m_{B_d} t\;,
\\
S_{\text{CP}}(B_d\to J/\psi K_S) &= \sin\phi_M\;,
\end{align}
with $\phi_M$ being $e^{i\phi_M} = M_{12}(B_d)/| M_{12}(B_d) |$.
In the Standard Model,
$\phi_M=2\phi_1=2\arg(-V_{cb}^*V_{cd}/(V_{tb}^*V_{td}))$.
Experimentally, $\sin\phi_M$ can be determined by combining decay
modes with the $b \to c\overline{c}s$ transition such as
$B_d \to J/\psi K_S$, $B_d \to J/\psi K_L$, and $B_d \to \psi' K_S$.

The time-dependent CP asymmetry in the $B_s$ decay is formulated in the
same way.
$B_s \to J/\psi \phi$ is the $b\to c\bar{c}s$ mode of the $B_s$ decay,
which corresponds to $B_d \to J/\psi K_S$.
The mixing-induced CP violation parameter
$S_{\text{CP}}(B_s \to J/\psi\phi)$ is written as
$S_{\text{CP}}(B_s \to J/\psi\phi)=\sin\phi_{M_s}$ where $\phi_{M_s}$ is
defined as $e^{i\phi_{M_s}} = M_{12}(B_s)/|M_{12}(B_s)|$.
In actual extraction, the angular analysis is needed to separate
CP odd and even contribution
\cite{ref:BtoVVangularanalysis,ref:Nakada:SUSY2010s}.
The Standard Model prediction is given as
$\left.\sin\phi_{M_s}\right|_{\text{SM}} \simeq -0.04$.

We also consider the decay mode $B_d\to \phi K_S$, which is supposed to
be a pure $b\to s\bar{s}s$ process.
The mixing-induced CP asymmetry $S_{\text{CP}}(B_d\to \phi K_S)$, 
is given as
\begin{align}
S_{\text{CP}}(B_d\to \phi K_S)=
\frac{2\mathrm{Im}(e^{-i\phi_M}
\bar{\mathcal{A}}\mathcal{A})}
{|\mathcal{A}|^2+|\bar{\mathcal{A}}|^2}\;,
\end{align}
where $\mathcal{A}$ and $\bar{\mathcal{A}}$ denote
decay amplitudes of $B_d\to \phi K$ and $\bar{B}_d\to \phi \bar{K}$
respectively.
This quantity is expected to coincide with $S_{\text{CP}}(B_d\to J/\psi K_S)$
within the SM. If there is sizable deviation, this will be an evidence of 
new physics beyond the SM in $b\to s$ transition.
The calculation of the decay amplitude involves sizable uncertainty.
Here we use a method based on the naive factorization.
Details of the calculation of $\mathcal{A}$ are given in
Refs.~\cite{Moroi:2000mr,ref:CalcBphiK}.

As for the $b\to q \gamma$ ($q=s,d$) decays, both direct and
mixing-induced CP asymmetries are considered, as well as the branching
ratio $\mathrm{B}(b\to s\gamma)$ which provides a significant constraint
on the parameter space.
Relevant effective Lagrangian is given as
\begin{align}
\mathcal{L} =&
C_{2L}\mathcal{O}_{2L} + C_{2L}'\mathcal{O}_{2L}'
-C_{7L}\mathcal{O}_{7L}
-C_{8L}\mathcal{O}_{8L}
+(L\leftrightarrow R)
+\mathcal{L}_{4q}\;.
\end{align}
The operators $\mathcal{O}$'s are
\begin{subequations}
\begin{align}
\mathcal{O}_{2L}=&
(\bar{q}_{\alpha}\gamma^{\mu}c_{L\alpha})
(\bar{c}_{\beta}\gamma^{\mu}b_{L\beta})
\;,\\
\mathcal{O}_{2L}'=&
(\bar{q}_{\alpha}\gamma^{\mu}u_{L\alpha})
(\bar{u}_{\beta}\gamma^{\mu}b_{L\beta})
-(\bar{q}_{\alpha}\gamma^{\mu}c_{L\alpha})
(\bar{c}_{\beta}\gamma^{\mu}b_{L\beta})
\;,\\
\mathcal{O}_{7L}=&\frac{e}{16\pi^2}m_b\bar{q}
\frac{i}{2}[\gamma^{\mu},\gamma^{\nu}]b_RF_{\mu\nu}
\;,\\
\mathcal{O}_{8L}=&\frac{g_3}{16\pi^2}m_b\bar{q}^{\alpha}
\frac{i}{2}[\gamma^{\mu},\gamma^{\nu}]
T_{\alpha\beta}^{(a)}b_R^{\beta}G_{\mu\nu}^{(a)}
\;,
\end{align}
\end{subequations}
where $q$ is $s$ or $d$ for $b\to s\gamma$ or $b\to d\gamma$ decays,
respectively.
$\mathcal{L}_{4q}$ denotes the terms with four-quark operators induced
by loop effects.
The Wilson coefficients $C_{2L}$ and $C_{2L}'$ are
dominated by the contributions from the tree level $W$ boson exchange.
Therefore, $C_{2L}'=\epsilon_u C_{2L}$ is satisfied, where
$\epsilon_u=-V^*_{uq}V_{ub}/(V^*_{tq}V_{tb})$.
The direct CP asymmetry in the inclusive decays $B\to X_q\gamma$ ($q=s,d$)
is given as \cite{Kagan:1998bh}
\begin{align}
A_{\text{CP}}^{\text{dir}}(B\to X_q\gamma)=&
\frac{\Gamma (\bar{B}\to X_q\gamma)-\Gamma(B\to X_{\bar{q}}\gamma)}
{\Gamma (\bar{B}\to X_q\gamma)+\Gamma(B\to X_{\bar{q}}\gamma)}\nonumber\\
=&
-\frac{\alpha_3}{\pi(|C_{7L}|^2+|C_{7R}|^2)}
\Biggl[
-\mathrm{Im}r_2\mathrm{Im}\left[(1-\epsilon_u)C_{2L}C_{7L}^*\right]
+\frac{80}{81}\pi\mathrm{Im}(\epsilon_uC_{2L}C_{7L}^*)\nonumber\\
&+\frac{8}{9}\pi\mathrm{Im}(C_{8L}C_{7L}^*)
-\mathrm{Im}f_{27}
\mathrm{Im}\left[(1-\epsilon_u)C_{2L}C_{7L}^*\right]\nonumber\\
&+\frac{1}{3}\mathrm{Im}f_{27}\mathrm{Im}
\left[(1-\epsilon_u)C_{2L}C_{8L}^*\right]
+(L\leftrightarrow R)
\Biggr]\;,
\label{eq_dircpbsgamma}
\end{align}
where the functions $r_2$ and $f_{27}$ for $B\to X_s\gamma$
are found in Ref.~\cite{ref:bsgamma-NLO}.
The mixing-induced CP asymmetry is defined for an exclusive
$B_d\to M_q \gamma$ decay.
$M_q$ denotes a hadronic CP eigenstate which includes a strange or down
quark such as $K^*$ (for $q=s$) and $\rho$ ($q=d$).
$S_{\text{CP}}(B_d\to M_q\gamma)$ is given as \cite{Atwood:1997zr}
\begin{align}
S_{\text{CP}}(B_d\to M_q\gamma)=
\frac{2\mathrm{Im}(e^{-i\phi_M}C_{7L}C_{7R})}{|C_{7L}|^2+|C_{7R}|^2}\;.
\end{align}

\subsubsection{Lepton flavor violation}

The effective Lagrangian for the lepton flavor violating
$l_j\to l_i\gamma$ decay is written as
\begin{equation}
\mathcal{L}_{\text{LFV}}=
-\frac{e}{16\pi^2}m_{l_j}
\bar{l}_i\frac{i}{2}\left[\gamma^{\mu},\gamma^{\nu}\right]
\left( A_L^{ij} P_R + A_R^{ij} P_L \right)
l_{j}\,F_{\mu\nu}
\;,\quad(i\neq j)\;,
\end{equation}
where $P_R=(1+\gamma_5)/2$ and $P_L=(1-\gamma_5)/2$.
The decay width is given by 
\begin{align}
\Gamma(l_j\to l_i\gamma)=\frac{\alpha}{64\pi^2}m_{l_j}^5
\left(|A_L^{ij}|^2+|A_R^{ij}|^2\right)\;.
\end{align}

\subsubsection{Electric dipole moments}
\label{sec:EDM}

Electric dipole moment $d_f$ of a fermion $f$ is defined as the
coefficient in the effective Lagrangian
\begin{equation}
  \mathcal{L} = \frac{i}{2} d_f \bar{f}
  \frac{i}{2}\left[\gamma^{\mu},\gamma^{\nu}\right]
  \gamma_5 f\,F_{\mu\nu}\;.
\end{equation}
In addition, chromo-electric dipole moments of quarks and the
three-gluon operator \cite{Weinberg:1989dx} are taken into account for
hadronic EDMs.
Relevant effective Lagrangian is written as
\begin{equation}
  \mathcal{L} = \frac{i}{2} d^C_q \bar{q}
  \frac{i}{2}\left[\gamma^{\mu},\gamma^{\nu}\right]
  \gamma_5 T^{(a)} q \,G_{\mu\nu}^{(a)}
  +
  \frac{d^G}{6} f^{(a)(b)(c)} \epsilon^{\mu\nu\lambda\rho}
  G_{\mu\sigma}^{(a)}G_{\nu}^{(b)\sigma}G_{\lambda\rho}^{(a)}
\;.
\end{equation}
We calculate $d_f$ for quarks and leptons and $d^C_q$ with all the
one-loop SUSY contributions \cite{ref:EDM-1,ref:EDM-2} and two-loop
contributions given in Ref.~\cite{Chang:1998uc}.
$d^G$ is calculated according to Ref.~\cite{ref:EDM-2}.

The neutron and the mercury EDMs, $d(n)$ and $d(\mathrm{Hg})$,
respectively, are written as linear combinations of $d_q$, $d^C_q$ and
$d^G$:
\begin{align}
  d(h) &= \sum_{q=u,d,s} \left[ c_q(h) d_q + c^C_q(h) d^C_q \right] +
  c^G(h) d^G\;,
\quad
  h=n,\; \mathrm{Hg}.
\end{align}
Values of the coefficients used in our calculation are given in
Table~\ref{tab:EDMcoeff}.
\begin{table}
\begin{center}
$
\begin{array}{ccccccccc}
\hline
h & c_u & c_d & c_s & c^C_u & c^C_d & c^C_s & c^G & \text{Ref.}
\\
\hline
n(\mathrm{NDA})
 &\displaystyle -\frac{1}{3}
 &\displaystyle \frac{4}{3}
 & 0
 &\displaystyle -\frac{1}{3}\frac{e}{4\pi}
 &\displaystyle \frac{4}{3}\frac{e}{4\pi}
 & 0
 &\displaystyle -\frac{e}{2\sqrt{2}}f_\pi
 & \mbox{\cite{ref:EDM-2}}
\\
n(\mathrm{ChPT})
 & 0
 & 0
 & 0
 & 1.6e
 & 1.3e
 & 0.26e
 & 0
 & \mbox{\cite{ref:EDM-3}}
\\
\mathrm{Hg}
 & 0
 & 0
 & 0
 &  0.0087e
 & -0.0087e
 &  4.4\times 10^{-5}e
 & 0
 & \mbox{\cite{ref:EDM-3}}
\\
\hline
\end{array}
$
\end{center}
\caption{Hadronic factors used in the calculation of EDMs.}
\label{tab:EDMcoeff}
\end{table}

There are large uncertainties in the estimation of the hadronic EDMs.
Here we use the value of the neutron EDM obtained by the formulae based
on the naive dimensional analysis (NDA) \cite{ref:EDM-2}.
On the other hand, it has been pointed out that an evaluation with use
of the chiral perturbation theory (ChPT) may give a much larger value of
the neutron EDM, due to the chromo-electric dipole moment of the strange
quark \cite{ref:EDM-3}.
We later discuss how the numerical results change if the latter is
applied.

\subsection{Input parameters and experimental constraints}
\label{sec:inputparameters}

As input parameters at the low energy, the mass eigenvalues and the
flavor mixing matrices of the quarks and leptons are used.
We take the top quark mass as $m_t(\text{pole})=170.9$GeV.

The CKM matrix elements $V_{us}$, $V_{cb}$, and $|V_{ub}|$ are 
determined by measurements of the processes which are supposed to be
dominated by the SM tree level contributions.
We adopt $V_{us}=0.224$ and $V_{cb}=0.0416$ in the following
calculations.
As for the $|V_{ub}|$, because the uncertainty is relatively large, we
vary $|V_{ub}|$ within a range $3.0<|V_{ub}|/10^{-3}<4.7$.
The CKM phase is not yet determined by tree level processes free from
new physics contributions.
Therefore we vary the CKM phase
$\phi_3\equiv\arg(-V_{ub}^*V_{ud}/V_{cb}^*V_{cd})$ within
$0<\phi_3< 180^{\circ}$. 

In the models with neutrino masses, we need to specify the
parameters in the neutrino sector in addition to the quark Yukawa
coupling constants.
As explained in Sec.~\ref{sec:MSSMRN}, we consider three cases for the
low energy neutrino mass spectrum and three types for the structure of
the neutrino Yukawa coupling matrix.
Among nine possible combinations, we show the results of the following
five cases:
\begin{itemize}
\item Degenerate $\nu_R$, normal hierarchy (D$\nu_R$-NH);
\item Degenerate $\nu_R$, inverted hierarchy (D$\nu_R$-IH);
\item Degenerate $\nu_R$, degenerate (D$\nu_R$-D);
\item Non-degenerate $\nu_R$ (I), normal hierarchy (ND$\nu_R$(I)-NH);
\item Non-degenerate $\nu_R$ (II), normal hierarchy (ND$\nu_R$(II)-NH).
\end{itemize}
In the non-degenerate $\nu_R$ cases, we have found that the results do
not change much when we take other low energy neutrino mass spectrum,
since the neutrino Yukawa coupling matrix is essentially independent of
the low energy neutrino mass spectra.
As for the mass eigenvalues, we fix
$m_{\nu_2}^2-m_{\nu_1}^2=8.0\times 10^{-5}\text{eV}^2$ in all cases.
The values of $m_{\nu_3}^2-m_{\nu_2}^2$ and the lightest neutrino mass
are shown in Table~\ref{tab:numass}.
\begin{table}
\begin{tabular}{lcc}
\hline
& $m_{\nu_3}^2-m_{\nu_2}^2 (\text{eV}^2)$
& lightest $\nu$ \\
\hline  
Normal hierarchy
& $\phantom{-}2.5\times 10^{-3}$
& $m_{\nu_1}=0.003\text{eV}$
\\
Inverted hierarchy
& $-2.5\times 10^{-3}$
& $m_{\nu_3}=0.003\text{eV}$
\\
Degenerate
& $\phantom{-}2.5\times 10^{-3}$
& $m_{\nu_1}=0.1\text{eV}$
\\
\hline  
\end{tabular}
\caption{Input parameters for the low energy neutrino masses.
$1-2$ splittiong is fixed as
$m_{\nu_2}^2-m_{\nu_1}^2=8.0\times 10^{-5}\text{eV}^2$ in all cases.}
\label{tab:numass}
\end{table}
We take the PMNS mixing matrix as
\begin{align}
V_{\text{PMNS}}=
\begin{pmatrix}
 c_{\odot}c_{13} & s_{\odot}c_{13} & s_{13} \\
-s_{\odot}c_{\text{atm}}-c_{\odot}s_{\text{atm}}s_{13} &
 c_{\odot}c_{\text{atm}}-s_{\odot}s_{\text{atm}}s_{13} &
 s_{\text{atm}}c_{13} \\
 s_{\odot}s_{\text{atm}}-c_{\odot}c_{\text{atm}}s_{13} &
-c_{\odot}s_{\text{atm}}-s_{\odot}c_{\text{atm}}s_{13} &
 c_{\text{atm}}c_{13}
\end{pmatrix}\;,
\label{eq:VPMNS}
\end{align}
($c_i=\cos\theta_i,s_i=\sin\theta_i$) with
$\sin^22\theta_{\text{atm}}=1$, $\tan^2\theta_{\odot}=0.4$, and
$\sin^22\theta_{13}=0$.
These mixing angles are consistent with the observed solar and
atmospheric neutrino oscillations \cite{ref:sol_atm_nu}, the K2K
experiment \cite{ref:K2K}, and the KamLAND experiment
\cite{ref:KamLAND}.
Only the upper bound of $\sin^22\theta_{13}$ is obtained by reactor
experiments \cite{ref:reactor_neutrino}, and we take the above value as an
illustration.
We ignore the Dirac and Majorana CP phases in the neutrino sector for
simplicity, though they can affect the analysis of the lepton flavor
violations \cite{Kanemura:2005cq}.

The neutrino Yukawa coupling and the right-handed neutrino mass
matrices have eighteen independent parameters in general.
Nine of these parameters are determined by the low energy neutrino
parameters, namely three masses $m_{\nu_{1,2,3}}$ and $V_{\text{PMNS}}$
(three mixing angles and three phases).
There remain nine free parameters to specify the neutrino Yukawa
coupling and right-handed neutrino mass matrices.
In the degenerate $\nu_R$ case, these parameters are fixed by the assumption
$(M_N)^{ij} = \hat{M}_N\delta^{ij}$, so that $y_N$ is determined as
Eq.~(\ref{eq:yNdegenerate}).
In the non-degenerate $\nu_R$ cases, we take $y_N$ as inputs for the extra nine
parameters.
We generally parametrize the $y_N$ as
\begin{align}
  y_N =& \hat{y}_N V_L \;,
\\
  \hat{y}_N =& \text{diag}(\hat{y}_1,\;\hat{y}_2,\;\hat{y}_3) \;,
\\
  V_L =&
\begin{pmatrix}
 \bar{c}_{12} \bar{c}_{13} &
 \bar{s}_{12} \bar{c}_{13} &
 \bar{s}_{13} e^{-i\bar{\delta}_{13}} \\
 - \bar{s}_{12} \bar{c}_{23}
 - \bar{c}_{12} \bar{s}_{23} \bar{s}_{13} e^{i\bar{\delta}_{13}} &
   \bar{c}_{12} \bar{c}_{23}
 - \bar{s}_{12} \bar{s}_{23} \bar{s}_{13} e^{i\bar{\delta}_{13}} &
   \bar{s}_{23} \bar{c}_{13} \\
   \bar{s}_{12} \bar{s}_{23}
 - \bar{c}_{12} \bar{c}_{23} \bar{s}_{13} e^{i\bar{\delta}_{13}} &
 - \bar{c}_{12} \bar{s}_{23}
 - \bar{s}_{12} \bar{c}_{23} \bar{s}_{13} e^{i\bar{\delta}_{13}} &
   \bar{c}_{23} \bar{c}_{13}
\end{pmatrix}
\nonumber\\ &\times
\text{diag}(e^{i\bar{\psi}_{13}},\; e^{i\bar{\psi}_{23}},\; 1)
e^{-i(\bar{\psi}_{13}+\bar{\psi}_{23})/3}
\;,
\end{align}
where $\bar{s}_{ij} = \sin\bar{\theta}_{ij}$ and
$\bar{c}_{ij} = \cos\bar{\theta}_{ij}$.
The right-handed neutrino mass matrix is written as
\begin{align}
  M_N =& \hat{y}_N V_L K_\nu^{-1} V_L^T \hat{y}_N \;.
\end{align}
After calculating $M_N$, we rescale $\hat{y}_N$ (and $M_N$) such that
$M_N$ satisfies $\det M_N=\mu_R^3$.
Therefore, the nine input parameters are $\hat{y}_1/\hat{y}_3$,
$\hat{y}_2/\hat{y}_3$, $\bar{s}_{12}$, $\bar{s}_{23}$,
$\bar{s}_{13}$, $\bar{\delta}_{13}$, $\bar{\psi}_{13}$,
$\bar{\psi}_{23}$ and $\mu_R$.
We take the input parameters as shown in Table~\ref{tab:inputs-nd},
which provide us appropriate $y_N$ of the structure
(\ref{eq:Y-non-degenerate-I}) and (\ref{eq:Y-non-degenerate-II}).
With use of these input parameters, as well as the low energy neutrino
mass spectrum of the normal hierarchy, we obtain the
eigenvalues of $y_N$ and $M_N$ at $\mu_R$ scale for
$\mu_R=4\times10^{14}$GeV and $\tan\beta=30$ as
$\hat{y}_N=\{0.213,\;0.406,\;0.647\}$ and
$\hat{M}_N=\{2.58,\;9.95,\;2.49\}\times10^{14}$GeV
in Case (I)
and
$\hat{y}_N=\{0.250,\;0.469,\;0.476\}$ and
$\hat{M}_N=\{1.05,\;10.2,\;5.93\}\times10^{14}$GeV
in Case (II).
\begin{table}
\centering
\begin{tabular}{ccccccccc}
\hline
Case & $\hat{y}_1/\hat{y}_3$ & $\hat{y}_2/\hat{y}_3$
 & $\bar{s}_{12}$ & $\bar{s}_{23}$ & $\bar{s}_{13}$
 & $\bar{\delta}_{13}$ & $\bar{\psi}_{13}$ & $\bar{\psi}_{23}$
\\
\hline
(I) & 0.329 & 0.628 & 0 & -0.666 & 0 & 0 & 0 & 0
\\
(II) & 0.534 & 1.014 & 0 & 0 & 0.435 & 0 & 0 & 0
\\
\hline
\end{tabular}
\caption{%
Input parameters for the neutrino Yukawa coupling matrix
in the non-degenerate $\nu_R$ cases.
}
\label{tab:inputs-nd}
\end{table}

As for the SUSY parameters, we take the convention that the unified
gaugino mass $m_{1/2}$ is real.
As already described in Sec.~\ref{sec:mSUGRA},
it is known that $\phi_{\mu}$ is strongly constrained by the upper
bounds of EDMs, while the corresponding constraint on $\phi_A$ is not
so tight \cite{ref:EDM-1,ref:EDM-2}.
Thus we fix $\phi_{\mu}=0^{\circ}$ ($\mu>0$) at the electroweak scale.
We scan the SUSY breaking parameters within the ranges
$0\leq m_0\leq 4\text{TeV}$, $0<m_{1/2}\leq 1.5\text{TeV}$
($0<M_{5}(\mu_G)\leq 1.5\text{TeV}$ for SU(5) SUSY GUT with right-handed
neutrinos), $|A_0|\leq 4$
and $-180^{\circ}<\phi_A\leq 180^{\circ}$.

In the SU(5) SUSY GUT with right-handed neutrinos, we also vary the two
phase parameters, which are introduced at the GUT scale matching as
mentioned in Sec.~\ref{sec:SU5RN}, within the whole range
$\{-180^{\circ},\;180^{\circ}\}$.

In the U(2) model, the flavor symmetry breaking parameters $\epsilon$
and $\epsilon'$ are fixed to be $\epsilon=0.04$ and $\epsilon'=0.008$,
and the parameters in the quark Yukawa coupling matrices are determined
so that the CKM matrix and the quark masses are reproduced.
There are six independent $O(1)$ parameters in the quark Yukawa coupling
matrices of the form (\ref{eq_u2_yukawa}) for given quark masses and the
CKM matrix.
We scan those free parameters as inputs.
For the squark mass matrices (\ref{eq_u2_susy_para}), we make an assumption
\begin{equation}
m_0^{Q2}= m_0^{U2} = m_0^{D2} = m_0^2\;,
\end{equation}
and scan the range of $m_0$ as $0<m_0<4\text{TeV}$.
Dimensionless parameters in Eq.~(\ref{eq_u2_susy_para}) are varied
within the ranges $0.4 \leq r^X_{22}\,,\; r^X_{33}\,,\; |r^X_{23}| \leq
2.5$ and $-180^{\circ}<\arg r^X_{23}\leq 180^{\circ}$.
We assume that the boundary conditions for the $A$ parameters
are the same as the mSUGRA case for simplicity%
\footnote{We have carried out a preliminary analysis of
the flavor signals for the case with non-universal $A$ terms, where $A^0_Q$,
$\tilde{a}_Q$, $\tilde{b}_Q$, $\tilde{c}_Q$ and $\tilde{d}_Q$ in
Eq.~(\ref{eq:U2-Aterm}) are free $O(1)$ parameters with small number of
samples.  We have found that the EDMs become too large in most of the
parameter sets chosen at random.  The result of the flavor signals does
not change much once the EDM constraints are applied.}.

In order to constrain the parameter space, 
we consider the following experimental results:
\begin{itemize}
\item
Lower limits on the masses of SUSY particles and the Higgs
bosons given by direct searches in collider
experiments \cite{ref:directsearch}.
\item
Branching ratio of the $b\to s\gamma$ decay:
$\text{B}(b\to s\gamma) = (3.55\pm 0.24^{+0.09}_{-0.10}\pm 0.03)\times 10^{-4}$
\cite{ref:bsgamma}.
We take the allowed range for the calculated branching ratio as
$2.85\times 10^{-4}<\text{B}(b\to s\gamma)<4.25\times 10^{-4}$, taking
also account of theoretical uncertainties.
\item
Upper bounds of the branching ratios of the $\mu\to e\gamma$,
$\tau\to\mu\gamma$ and $\tau\to e\gamma$ decays for the MSSM with
right-handed neutrinos and SUSY GUT cases:
$\text{B}(\mu\to e\gamma)<1.2\times 10^{-11}$ \cite{Ahmed:2001eh},
$\text{B}(\tau\to\mu\gamma)<6.8\times 10^{-8}$ \cite{Aubert:2005ye} and
$\text{B}(\tau\to e\gamma)<1.1\times 10^{-7}$. \cite{Aubert:2005wa}
\item
Upper bounds of EDMs of $^{199}\text{Hg}$, the neutron and
the electron:
$|d_{\text{Hg}}|<2.1\times 10^{-28}e\cdot\text{cm}$ \cite{Romalis:2000mg}, 
$|d_n|<2.9\times 10^{-26}e\cdot\text{cm}$ \cite{Baker:2006ts}
and
$|d_e|<1.6\times 10^{-27}e\cdot\text{cm}$ \cite{Regan:2002ta}.
\item
The CP violation parameter $\varepsilon_K$ in the 
$K^0-\bar{K}^0$ mixing $|\varepsilon_K|=(2.232\pm 0.007)\times 10^{-3}$
and the $B_d-\bar{B}_d$ and the $B_s-\bar{B}_s$ 
mixing parameters $\Delta m_{B_d}=0.507\pm 0.005\text{ps}^{-1}$
\cite{Barberio:2007cr}
and $\Delta m_{B_s}=17.77\pm 0.10\pm 0.07\text{ps}^{-1}$
\cite{Abulencia:2006ze}.
Theoretical uncertainties in these quantities are larger than the
experimental ones.
For the $B-\bar{B}$ mixings, $1\sigma$ uncertainties of the decay
constants $f_{B_{d,s}}$ and of the bag parameters $B_{B_d,s}$ are
evaluated as 10 percent and 8 percent, respectively
\cite{ref:LatticeQCD}.
In the present analysis, we calculate $\Delta m_{B_{d,s}}$ with a fixed
set of hadronic parameters as listed in Table~\ref{tab:f-and-bag} and
allow $\pm 40$ percent deviations from the experimental central values.
We expect that these ranges provide typical $2-3\sigma$ allowed
intervals.
In addition, the ratio of the hadronic parameters
$\xi\equiv f_{B_s}\sqrt{B_{B_s}}/(f_{B_d}\sqrt{B_{B_d}})$ is evaluated
with better accuracy.
The uncertainty of $\xi$ is evaluated as $\pm 4$ percent
\cite{ref:LatticeQCD}.
Therefore we also require that the calculated ratio
$\Delta m_{B_s}/\Delta m_{B_d}$, which is proportional to $\xi^2$, be
within $\pm 20$ percent range of the central value.
For $\varepsilon_K$ we assign $\pm 15$ percent uncertainty.
\item
CP asymmetry in the $B_d\to J/\psi K_S$ decay and related modes observed
at the $B$ factory experiments:
$\sin 2\phi_1|_{c\bar{c}s}=0.678 \pm 0.025$ \cite{Barberio:2007cr}.
We take the allowed range for the calculated value as
$0.628< S_{\text{CP}}(B_d\to J/\psi K_S) <0.728$, which is a simple
$2\sigma$ interval, since the theoretical uncertainty of this asymmetry
is expected to be small.
\end{itemize}

\subsection{Numerical results}

\subsubsection{Allowed parameter region from the radiative electroweak
  symmetry breaking condition}

\begin{figure}[htbp]
\centering
\begin{tabular}{ccc}
\includegraphics[scale=.3]{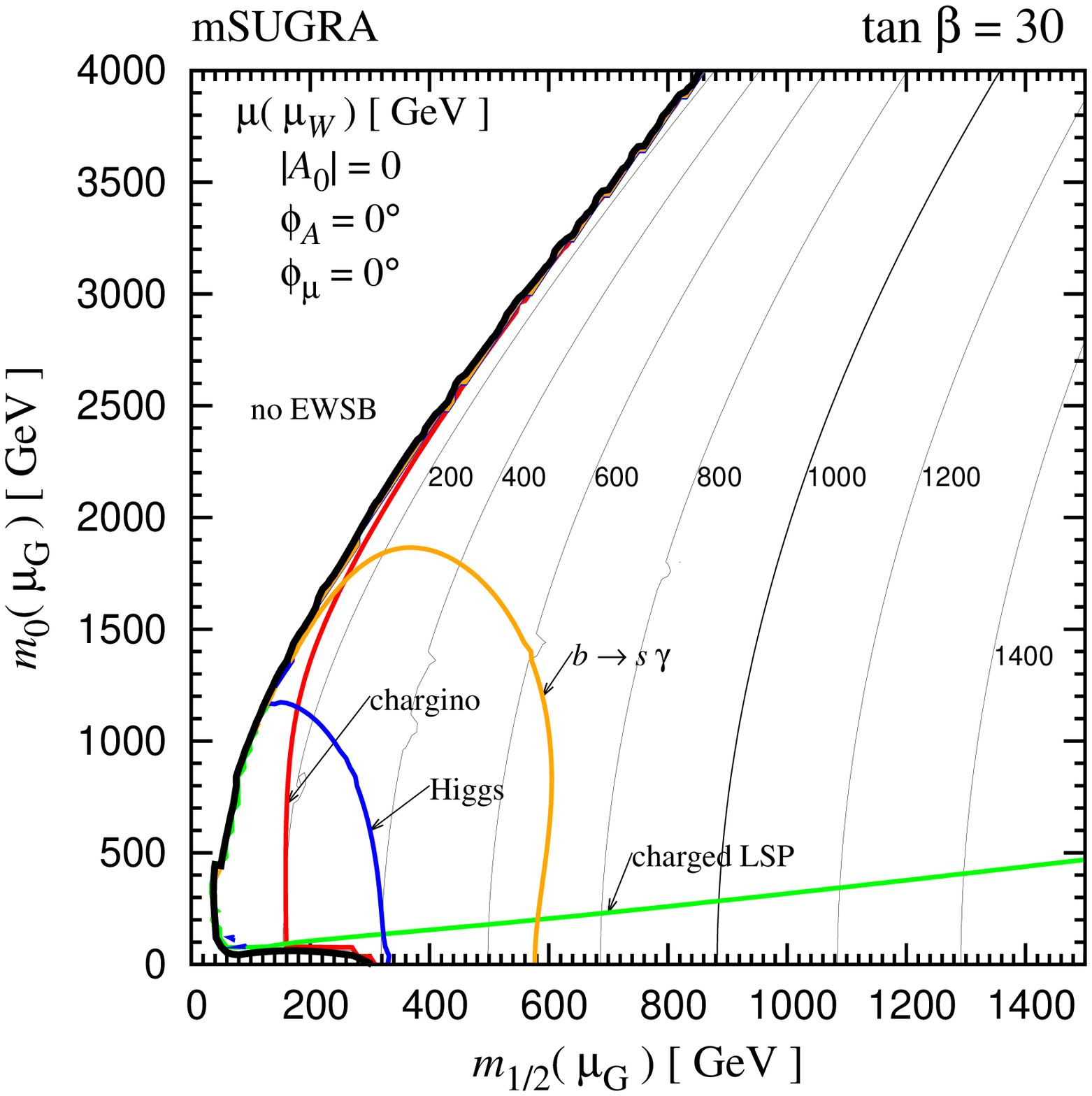} &
\includegraphics[scale=.3]{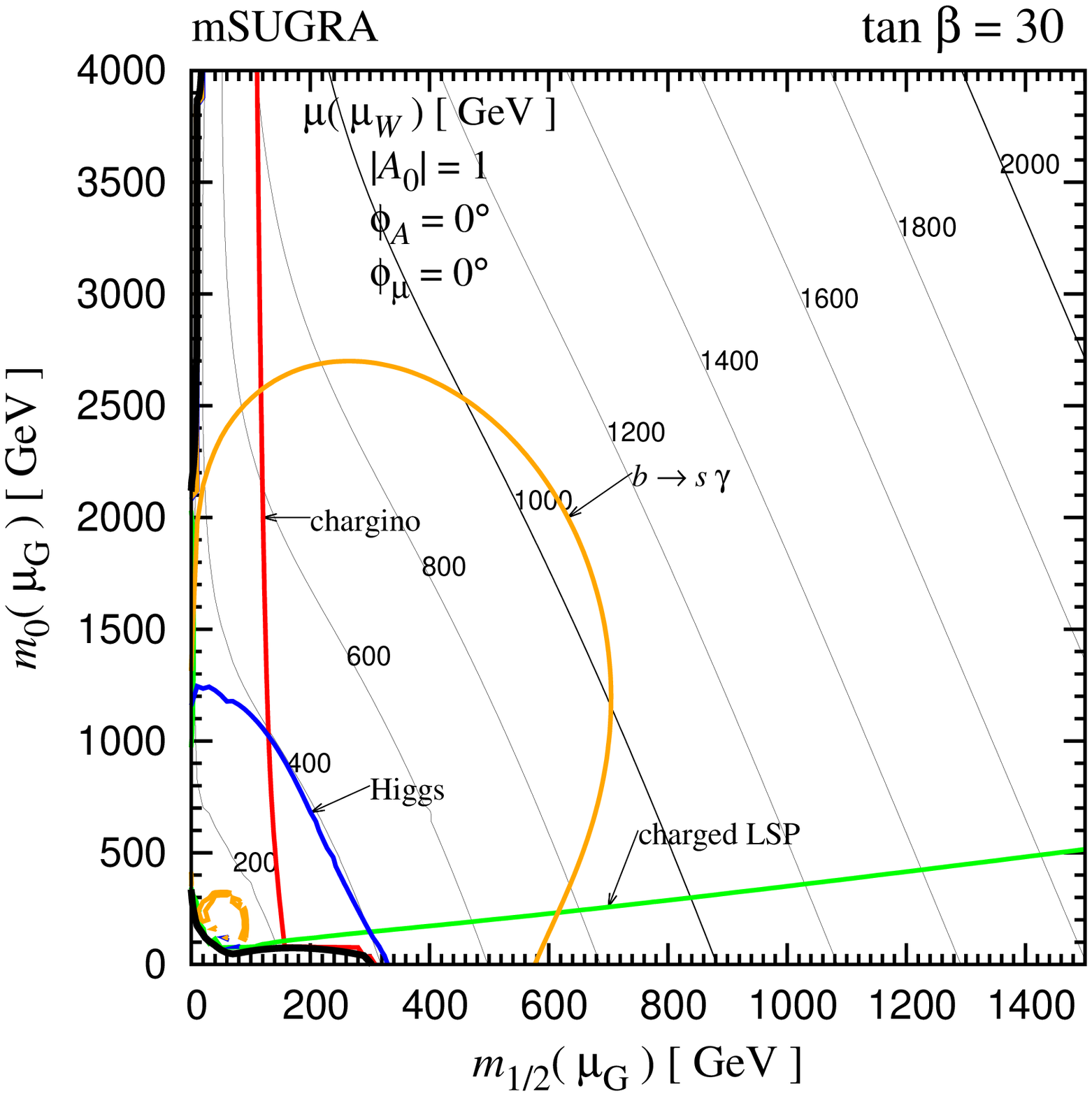} &
\includegraphics[scale=.3]{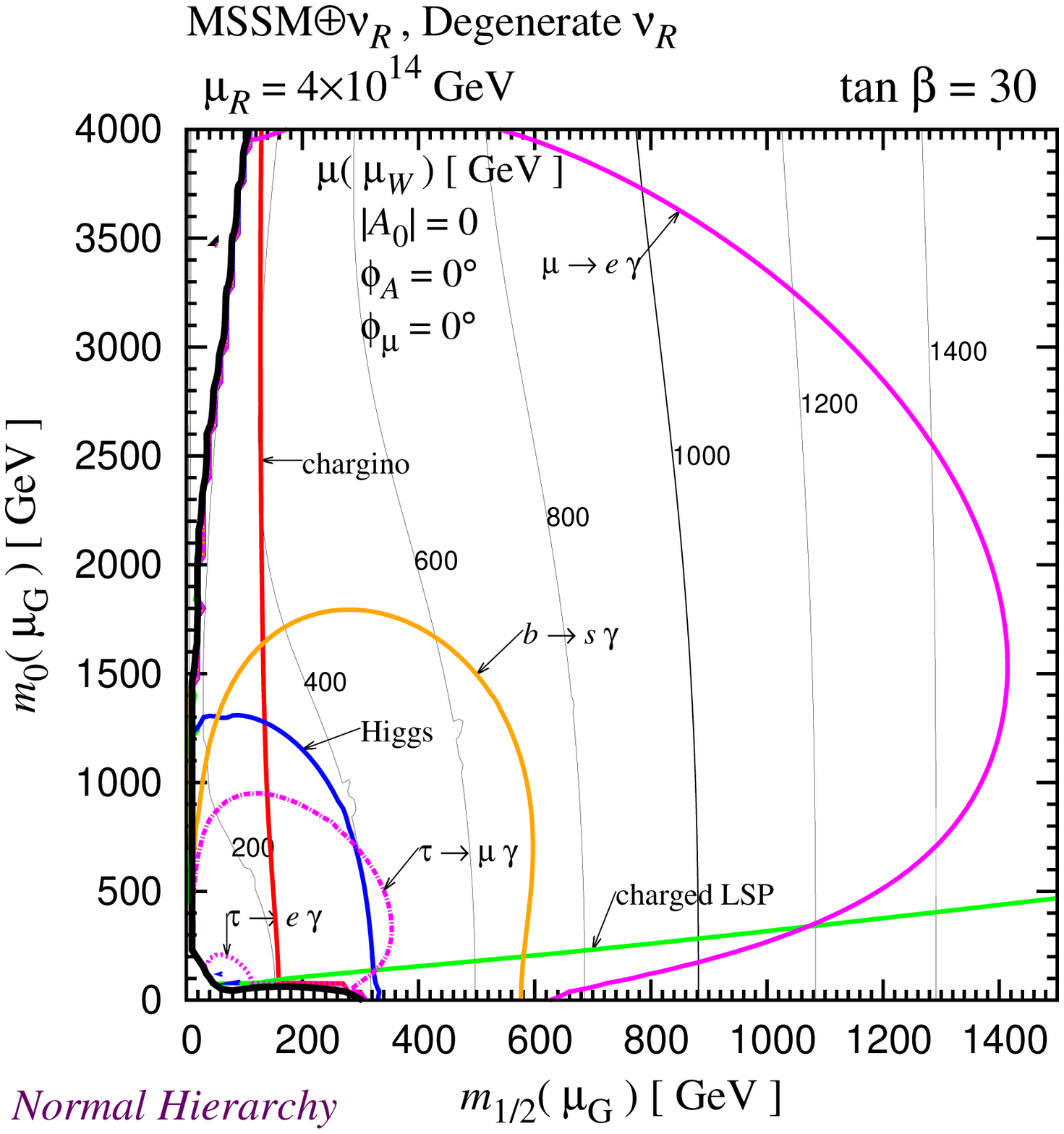} \\
(a) & (b) & (c)\\
\includegraphics[scale=.3]{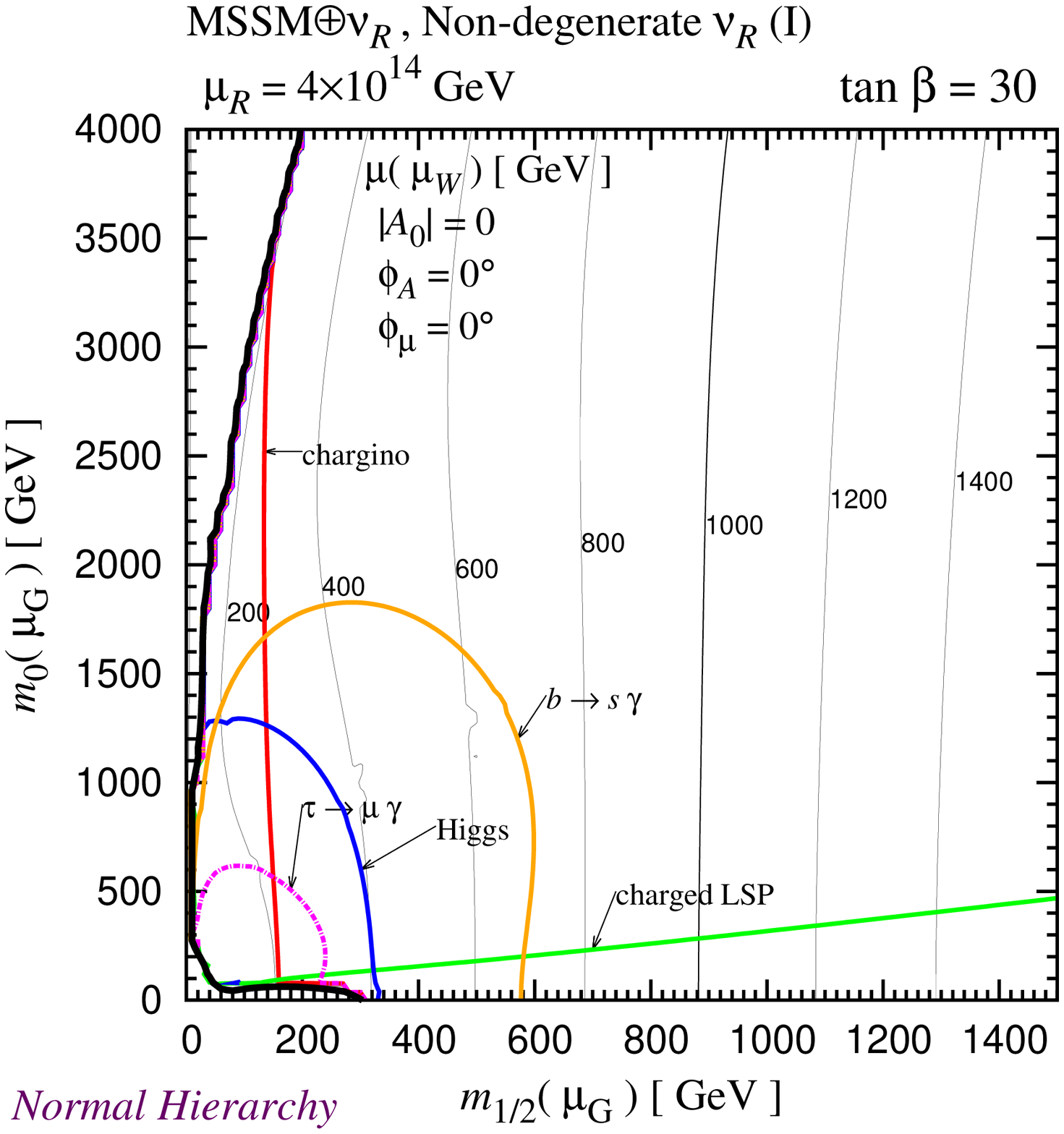} &
\includegraphics[scale=.3]{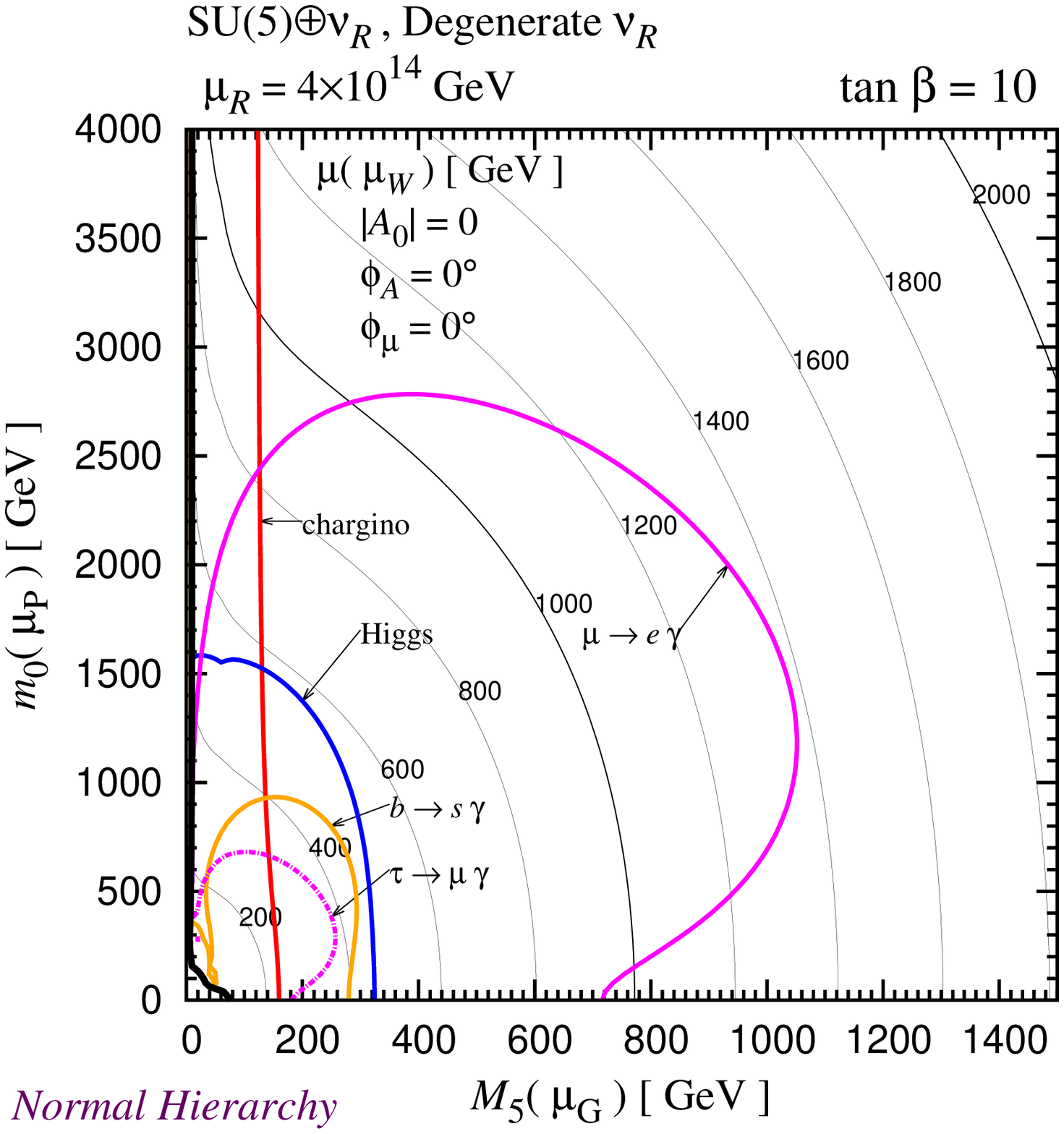} &
\includegraphics[scale=.3]{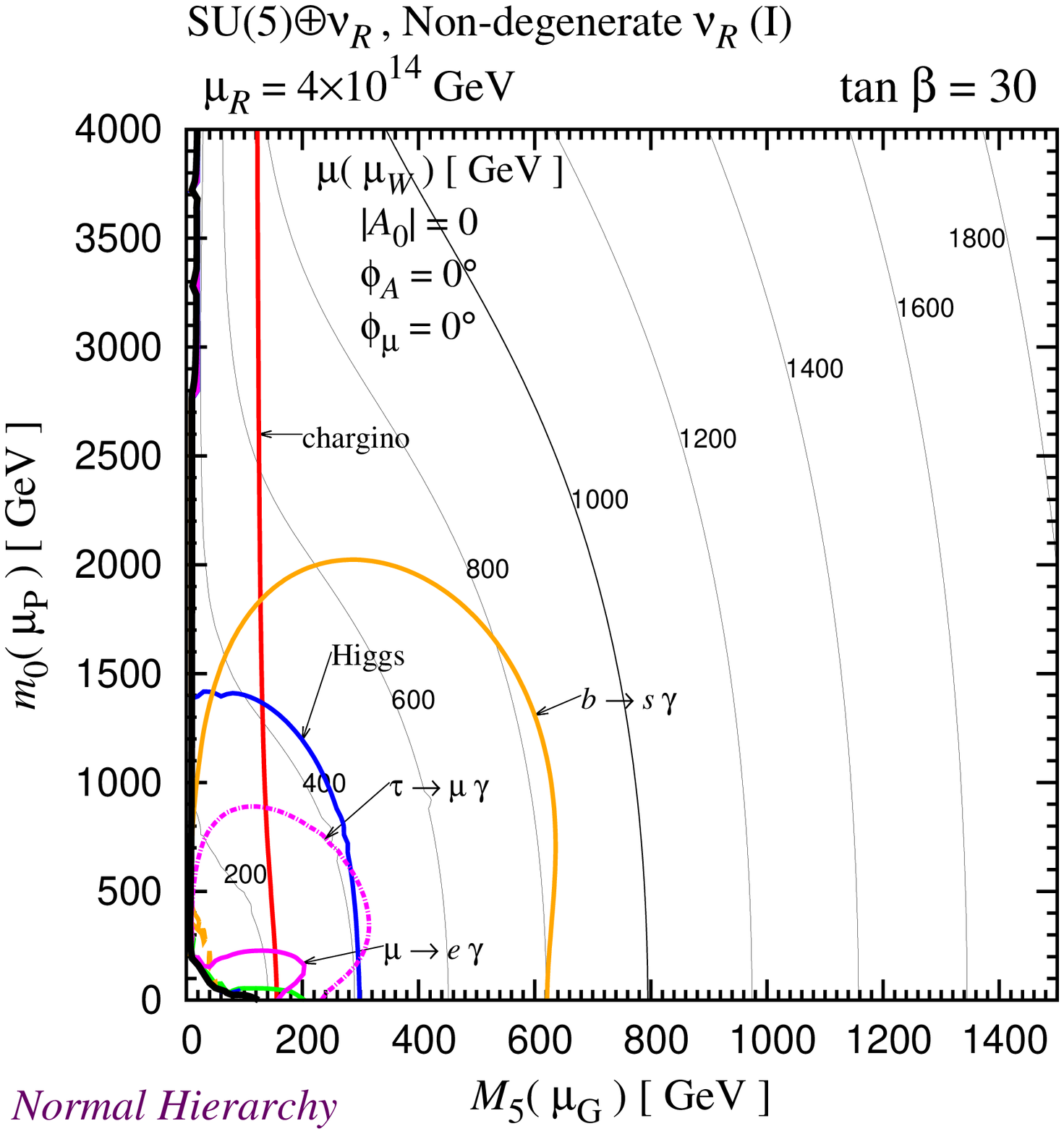} \\
(d) & (e) & (f)
\end{tabular}
\caption{%
(Color online)
Contour plots of the value of $|\mu|$ on $m_0$ and $m_{1/2}$ plane for
fixed $\tan\beta$ and $A_0$ in mSUGRA ((a) and (b)), MSSM with
right-handed neutrinos ((c) and (d)) and SU(5) SUSY GUT with
right-handed neutrinos models ((e) and (f)).
Each thick black line shows the boundary of the region where correct
electroweak symmetry breaking occurs.
In the regions below the lines labeled with ``charged LSP'' (green) in
(a)--(d), the LSP is a charged particle.
Boundaries of excluded regions which come from the chargino mass (red),
the Higgs boson mass (blue) and $\mathrm{B}(b\to s\gamma)$ (orange) are
shown in each plot.
Regions excluded by the lepton flavor violating processes are also shown in
(c)--(f) (magenta).
}
\label{fig:m0-m1/2}
\end{figure}

Before presenting flavor signals, we first discuss the SUSY parameter
space which is allowed by the radiative electroweak symmetry
breaking condition and experimental constraints.

In Fig.~\ref{fig:m0-m1/2}, we show the allowed region in the $m_0$ and
$m_{1/2}$ plane for the mSUGRA, MSSM with right-handed neutrinos, and
SU(5) SUSY GUT with right-handed neutrinos.
Parameters other than $m_0$ and $m_{1/2}$ are fixed as indicated in each
plot.
Contours of $|\mu|$ determined from the electroweak symmetry breaking
condition are also shown.
In mSUGRA, the parameter region is mainly constrained by the lower limit
on the chargino mass, the limit on the lightest Higgs boson mass, the
branching ratio of $b\to s\gamma$ decay, and the requirement that the
lightest supersymmetric particle (LSP) is neutral.
When the neutrino Yukawa couplings are relevant, the lepton flavor
violating decays are enhanced.
As a result, a large portion is excluded due to the experimental upper
limit on the branching ratio of $\mu\to e\gamma$ for MSSM with
right-handed neutrinos and SU(5) SUSY GUT with right-handed neutrinos.
Notice that we take CP violating SUSY phases to be vanishing in these
plots.
A significant potion of the parameter space is excluded due to the
experimental limits on EDMs if we take non-vanishing SUSY CP phases.

In the plot for mSUGRA with $|A_0|=0$ (Fig.~\ref{fig:m0-m1/2}(a)), the
$m_0\gg m_{1/2}$ region is excluded because the electroweak symmetry
breaking cannot be satisfied, namely there is no solution with
$|\mu|^2\geq 0$ for this region.
The allowed region near the boundary corresponds to so-called ``focus
point'' region \cite{Feng:1999zg} where the LSP is the lightest
neutralino with significant higgsino component.
This region is one of the favored regions in the context of the cosmic
dark matter study \cite{ref:DMfocuspoint}.
The pair annihilation of the lightest neutralino into $W$ boson or $Z$
boson pair is enhanced by the gauge interaction of the higgsino
component, so that the relic abundance of the LSP becomes suitable for
the cold dark matter density.
However, in $|A_0|=1$ case (Fig.~\ref{fig:m0-m1/2}(b)), such region
disappears because the $A$-terms affect the running of the Higgs mass
parameter $m_{H_2}^2$ so that a sufficiently large $|\mu|^2$ is
realized.
The ``focus point'' like region disappears also in the cases with
right-handed neutrinos (Fig.~\ref{fig:m0-m1/2}(c)--(f)), since the large
neutrino Yukawa coupling affects the running of $m_{H_2}^2$ in a similar
way.

In the cases of mSUGRA and MSSM with right-handed neutrinos
(Fig.~\ref{fig:m0-m1/2}(a)--(d)), the $m_0\ll m_{1/2}$ region is
excluded because the LSP is the lightest charged slepton.
The allowed region near the boundary provides another dark matter
favored region \cite{Ellis:1998kh}.
The coannihilation effect among the LSP (neutralino) and the next-to-LSP
(slepton), which are nearly degenerate in mass, provides an appropriate
relic abundance of LSP.
On the other hand, in the SU(5) SUSY GUT with right-handed neutrinos
(Fig.~\ref{fig:m0-m1/2}(e), (f)), the running between the Planck and the
GUT scales induces positive contribution to the slepton mass squared,
which makes the charged slepton heavier than the lightest neutralino
even in the $m_0\ll m_{1/2}$ region.
Therefore ``charged LSP'' or ``stau coannihilation'' region disappears
in the SU(5) SUSY GUT with right-handed neutrinos.

The disappearance of the ``focus point'' like region due to the effect
of the neutrino Yukawa couplings and the disappearance of the ``charged
LSP'' region caused by the running between the Planck and the GUT scales
are previously observed in Ref.~\cite{Calibbi:2006nq}, where the SO(10)
SUSY GUT is considered.

\subsubsection{Lepton flavor violating $\mu$ and $\tau$ decays}

\begin{figure}[htbp]
\begin{tabular}{ccc}
\includegraphics[scale=.3]{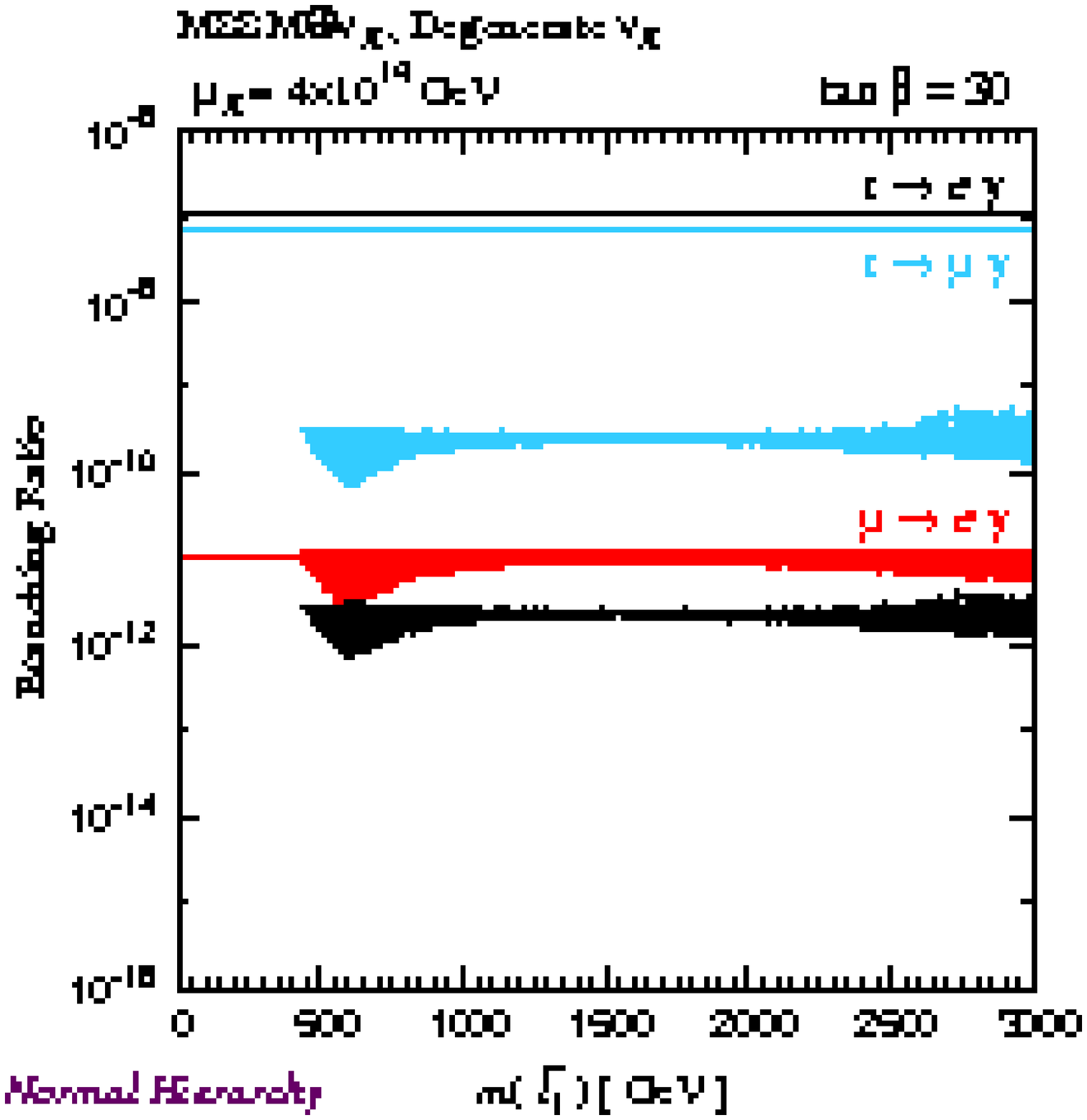} &
\includegraphics[scale=.3]{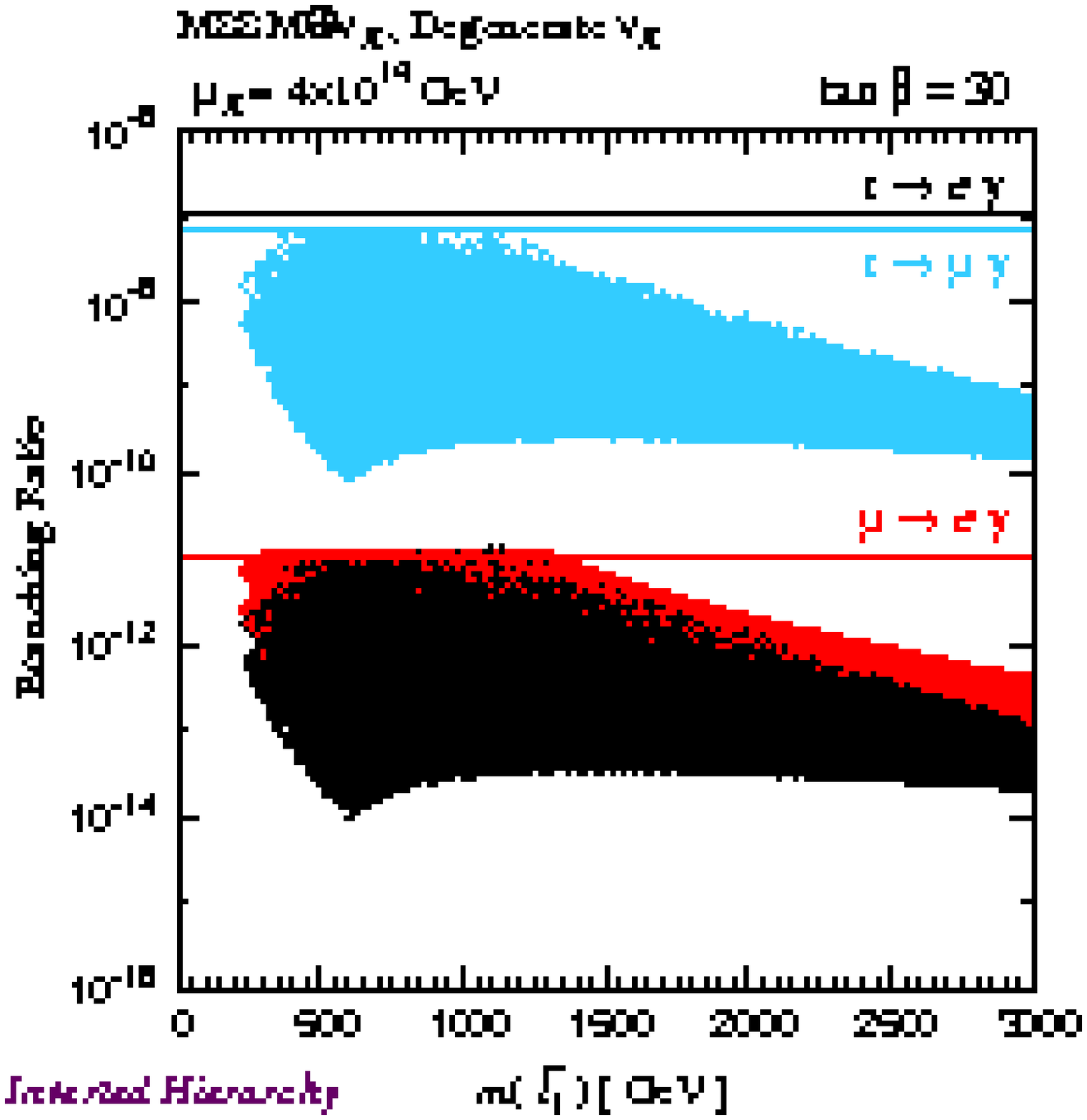} &
\includegraphics[scale=.3]{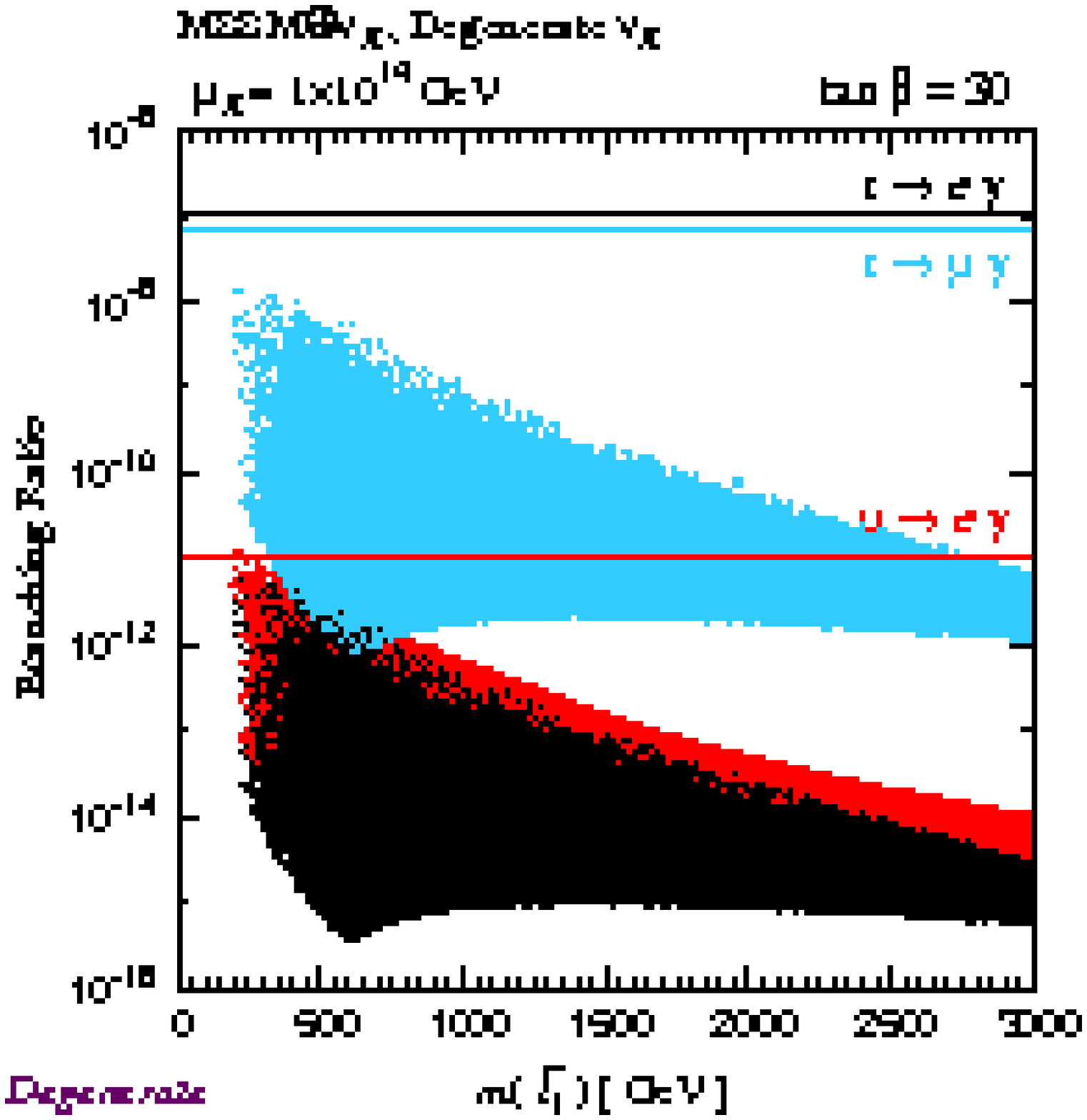} \\
(a) & (b) & (c) \\
\includegraphics[scale=.3]{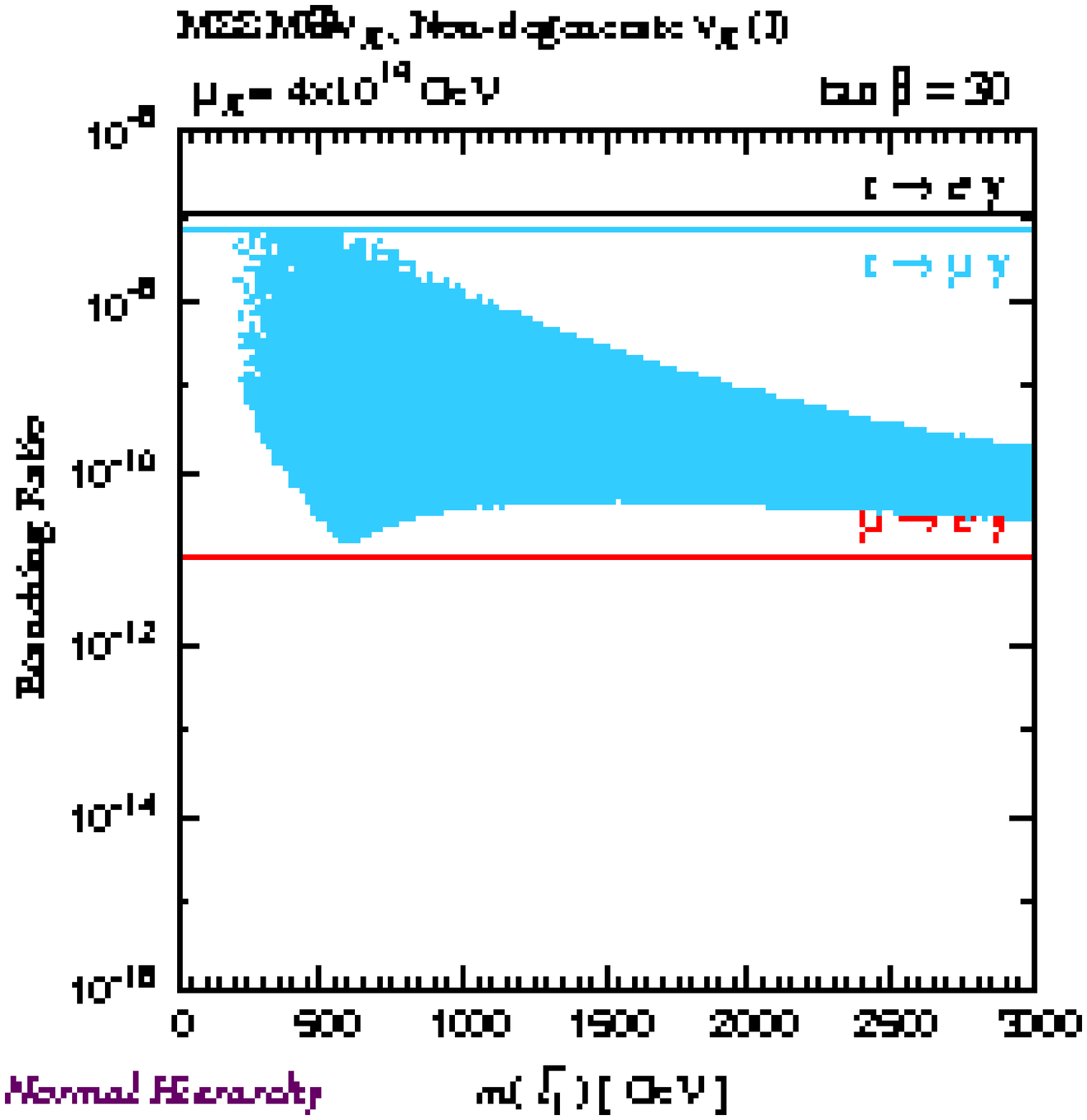} &
\includegraphics[scale=.3]{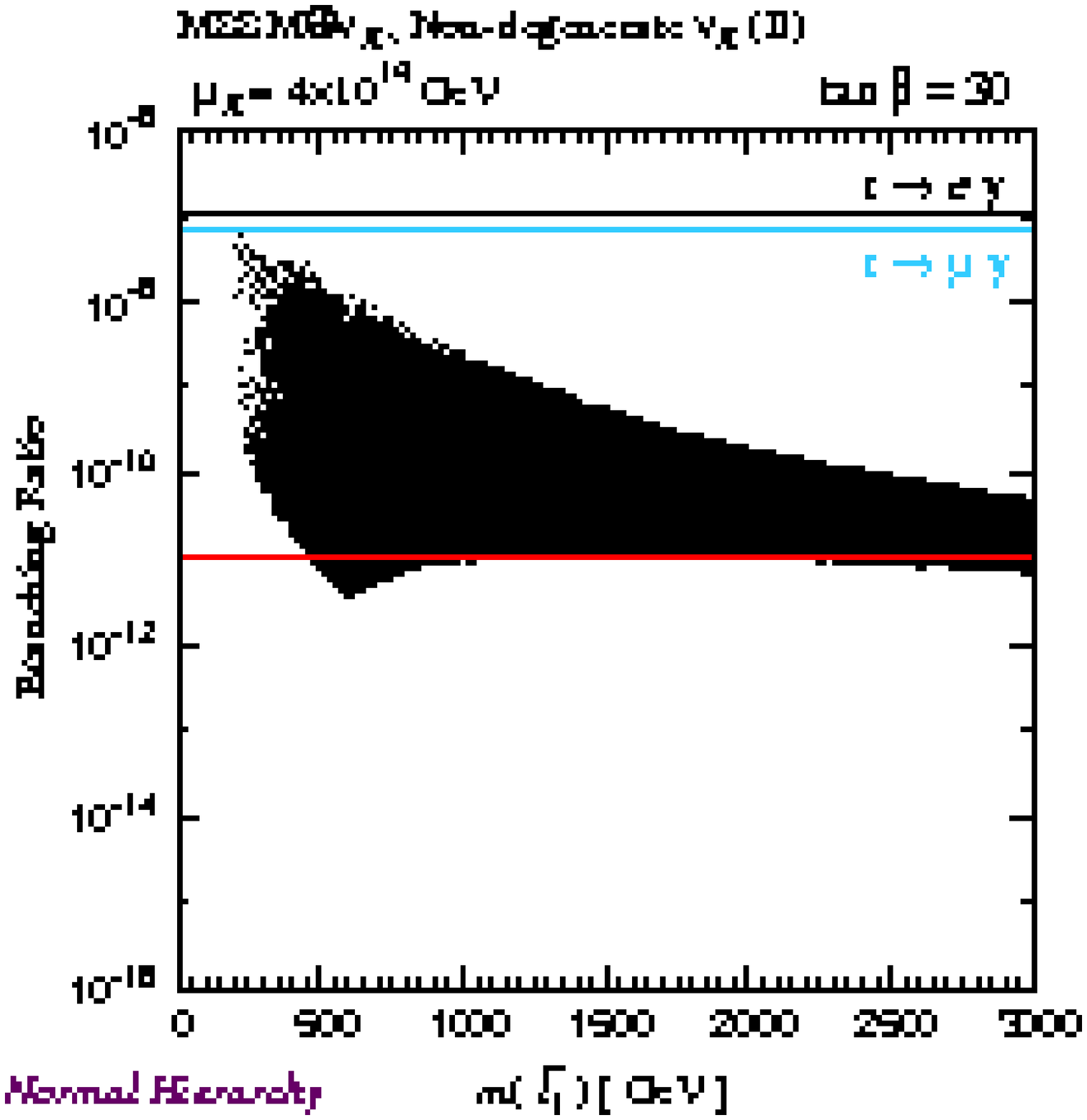} & \\
(d) & (e) & 
\end{tabular}
\caption{%
(Color online)
Branching ratios of lepton flavor violation processes $\mu\to e\gamma$
(grey/red), $\tau\to \mu\gamma$ (light-grey/light-blue), and
$\tau\to e\gamma$ (black) as functions of the lightest charged slepton
mass $m(\tilde{l}_1)$ for
MSSM with right-handed neutrinos.
Horizontal lines denote experimental upper limits.
In the plot (d), $\mu\to e\gamma$ and $\tau\to e\gamma$ are strongly
suppressed.
In the plot (e), $\mu\to e\gamma$ and $\tau\to \mu\gamma$ are strongly 
suppressed.
}
\label{fig:ml-LFV}
\end{figure}
\begin{figure}
\begin{tabular}{ccc}
\includegraphics[scale=.3]{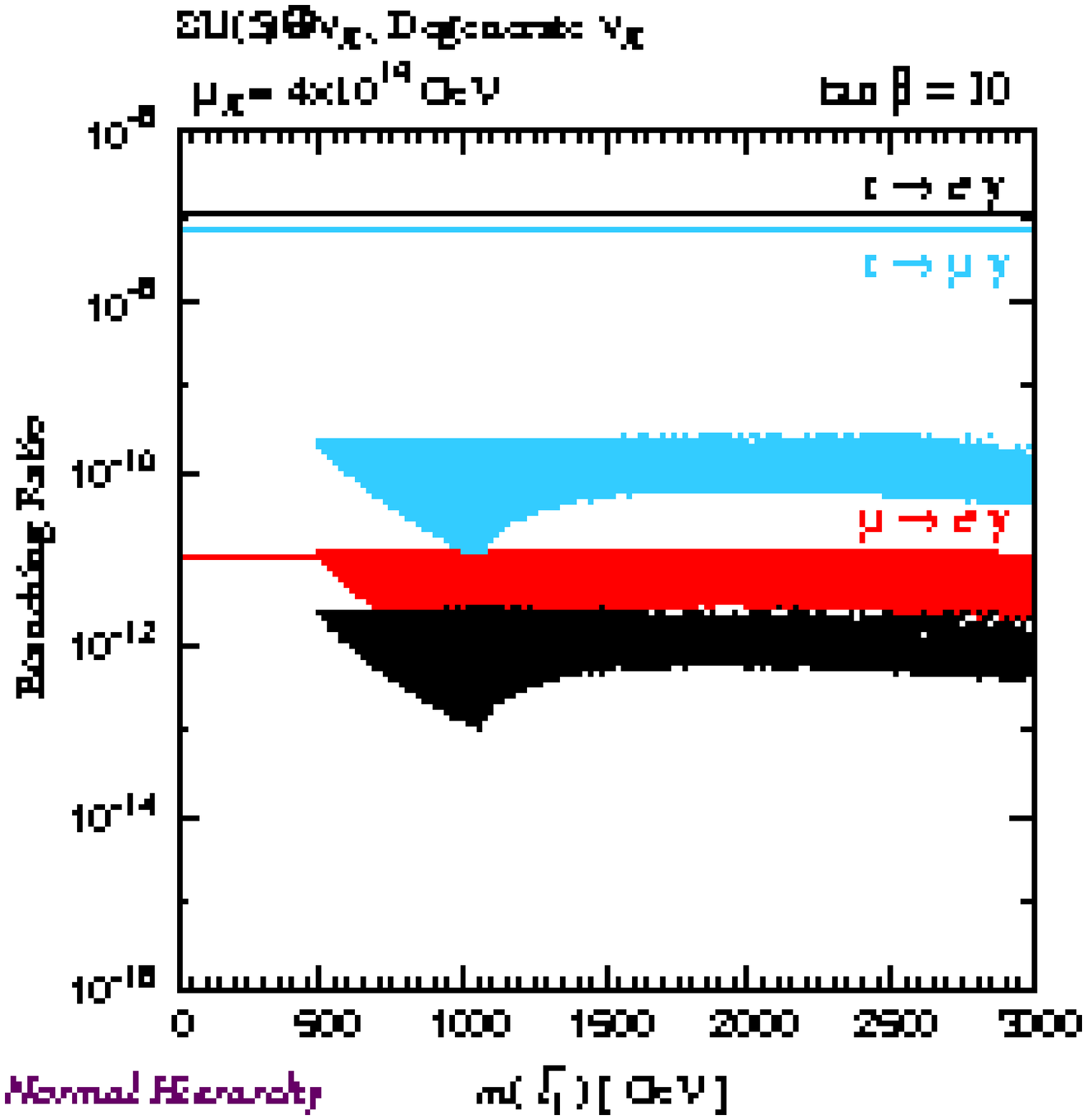} &
\includegraphics[scale=.3]{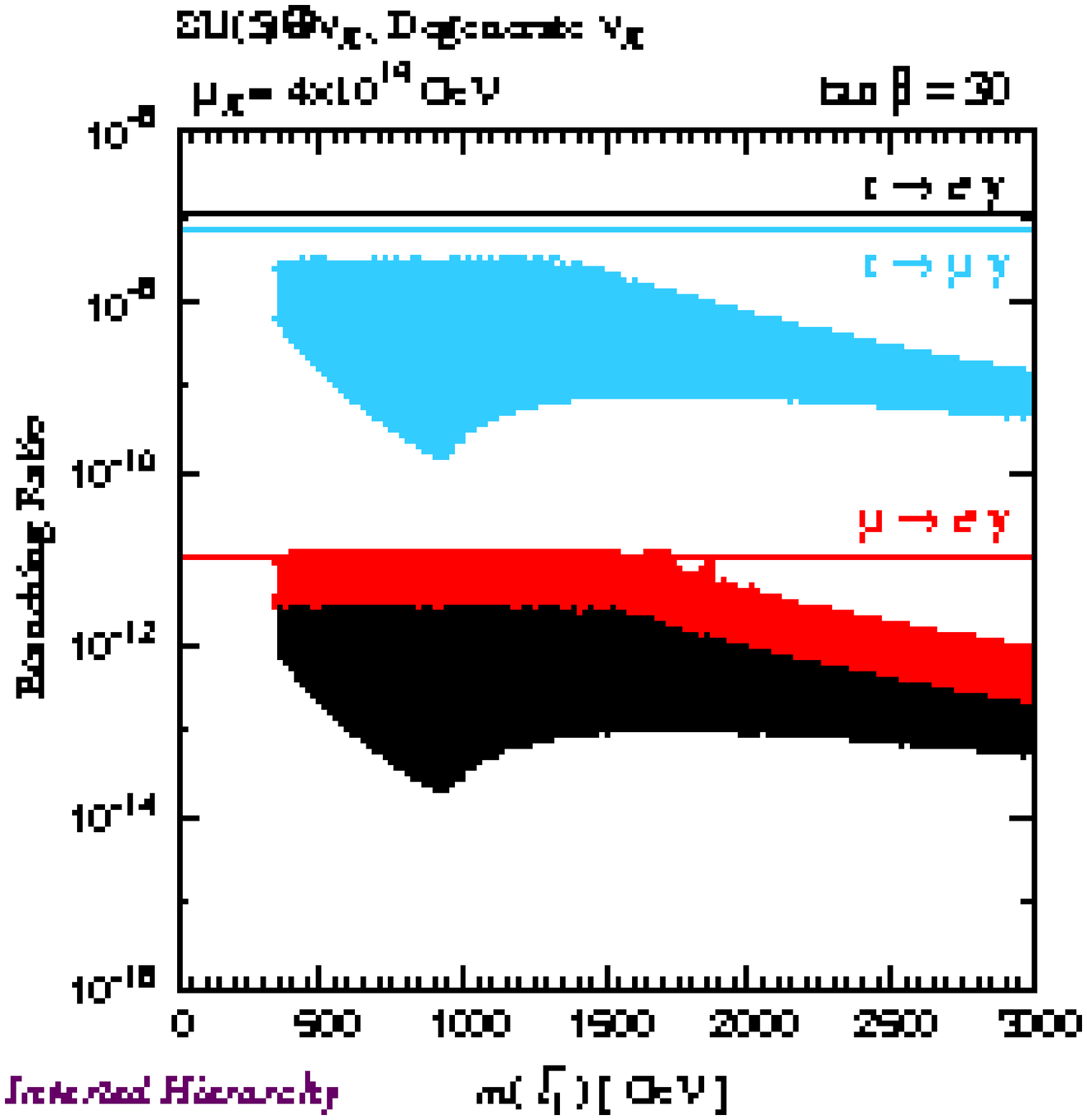} &
\includegraphics[scale=.3]{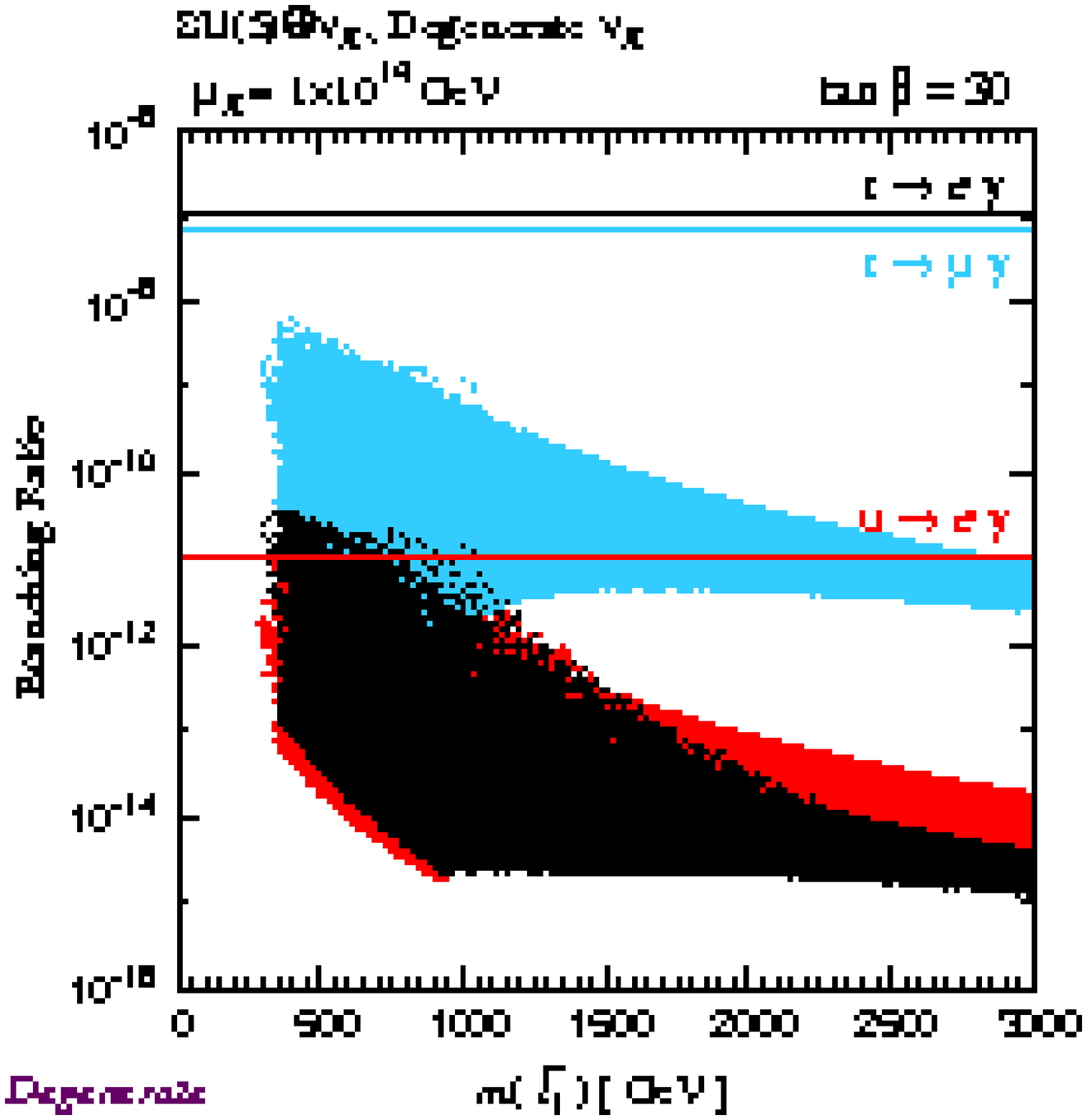} \\
(a) & (b) & (c) \\
\includegraphics[scale=.3]{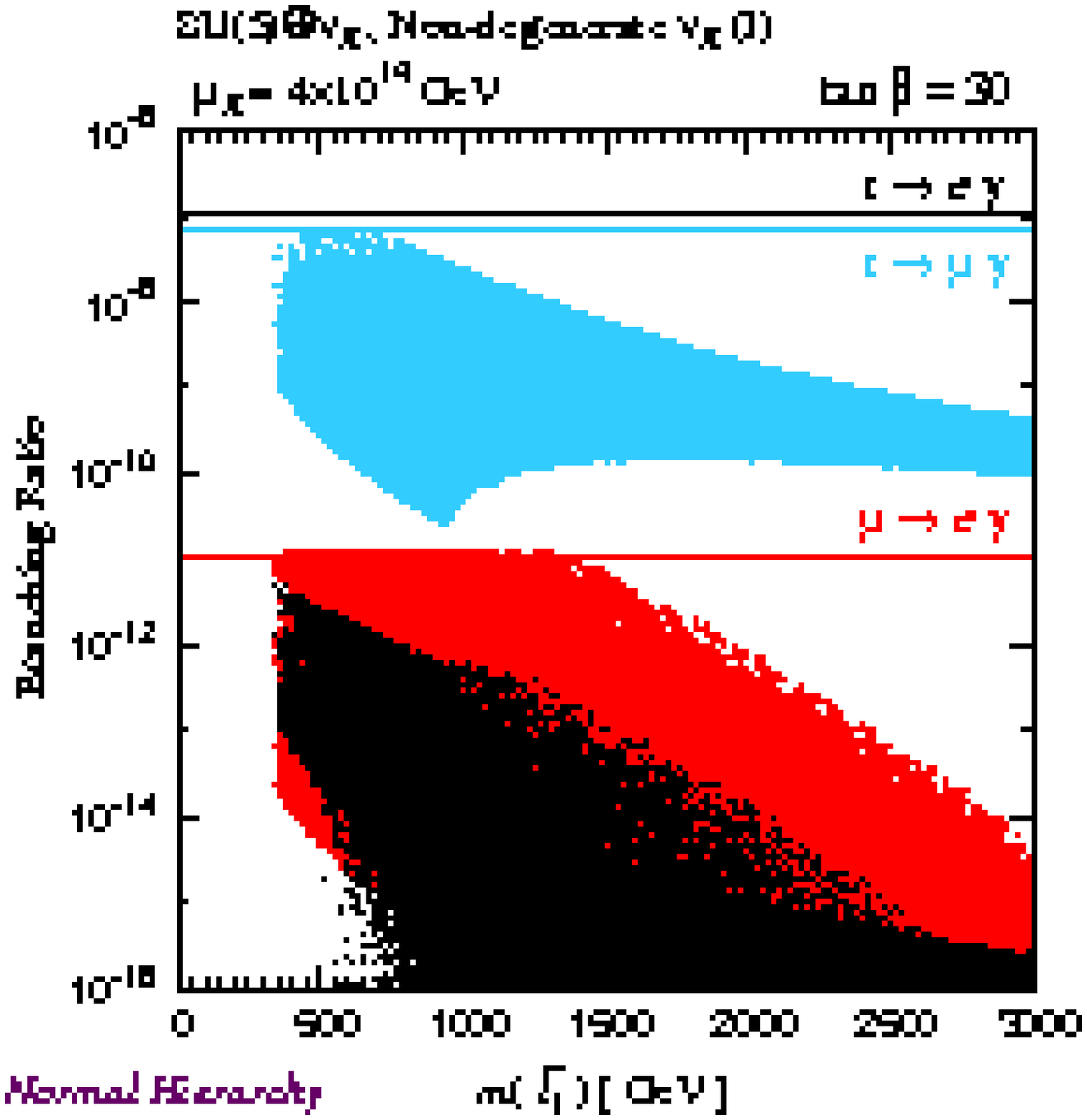} &
\includegraphics[scale=.3]{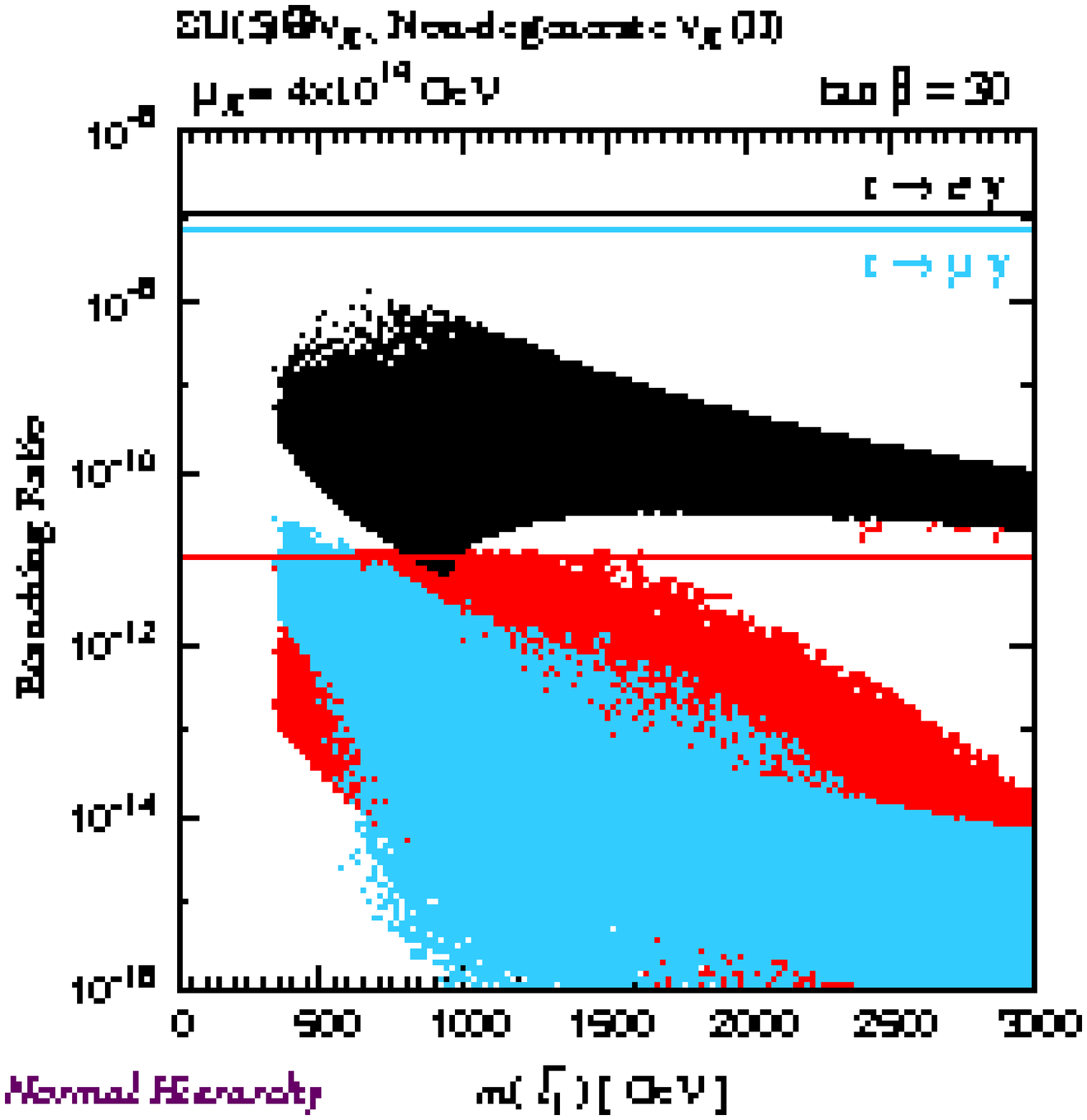} & \\
(d) & (e) &
\end{tabular}
\caption{%
(Color online)
Branching ratios of lepton flavor violation processes as functions of the lightest charged slepton
mass for SU(5) SUSY GUT with right-handed neutrinos.
Notations are the same as those in Fig.~\ref{fig:ml-LFV}.
}
\label{fig:ml-LFV-su5}  
\end{figure}

\begin{figure}[htbp]
\begin{tabular}{ccc}
\includegraphics[scale=.3]{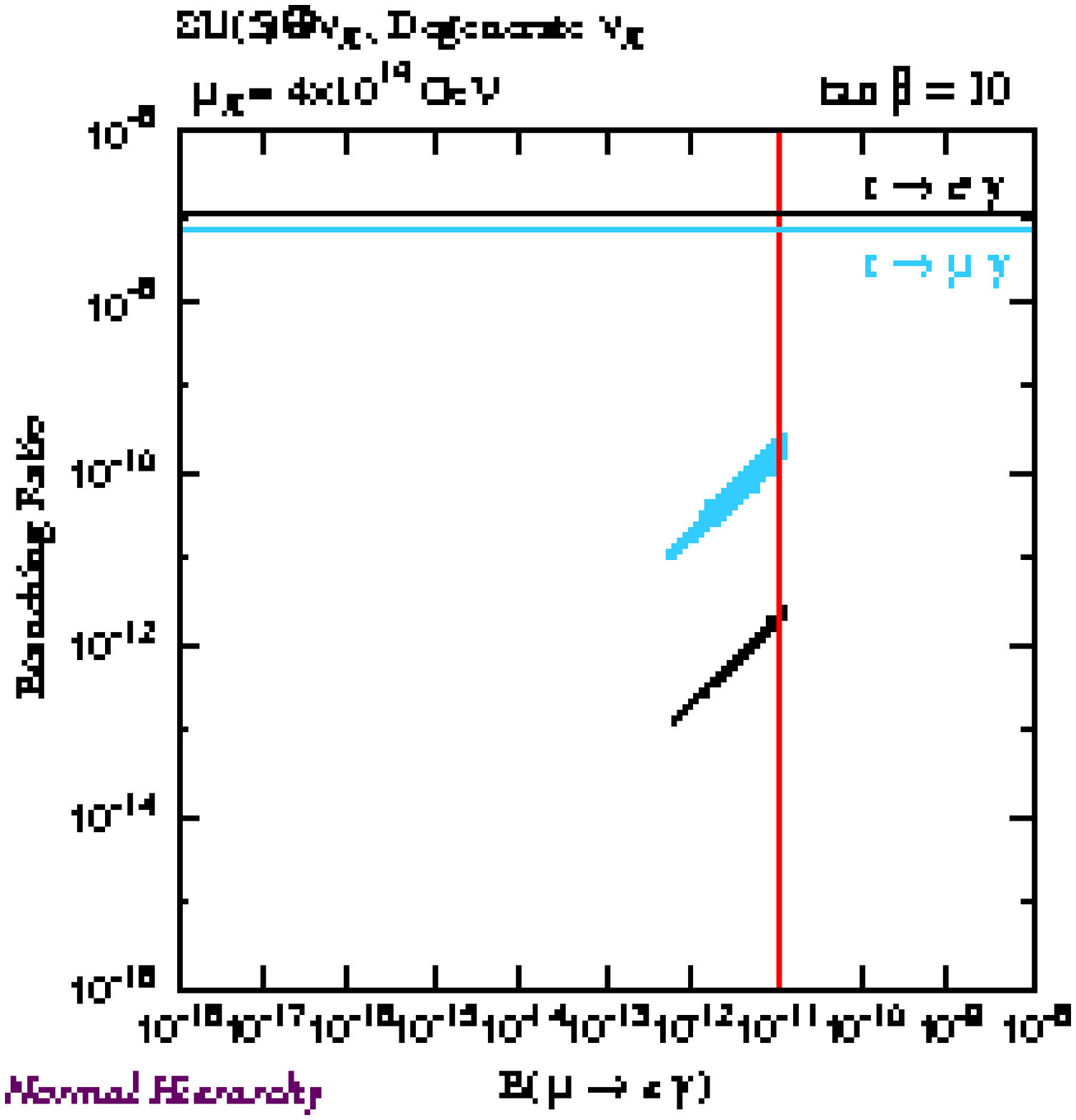} &
\includegraphics[scale=.3]{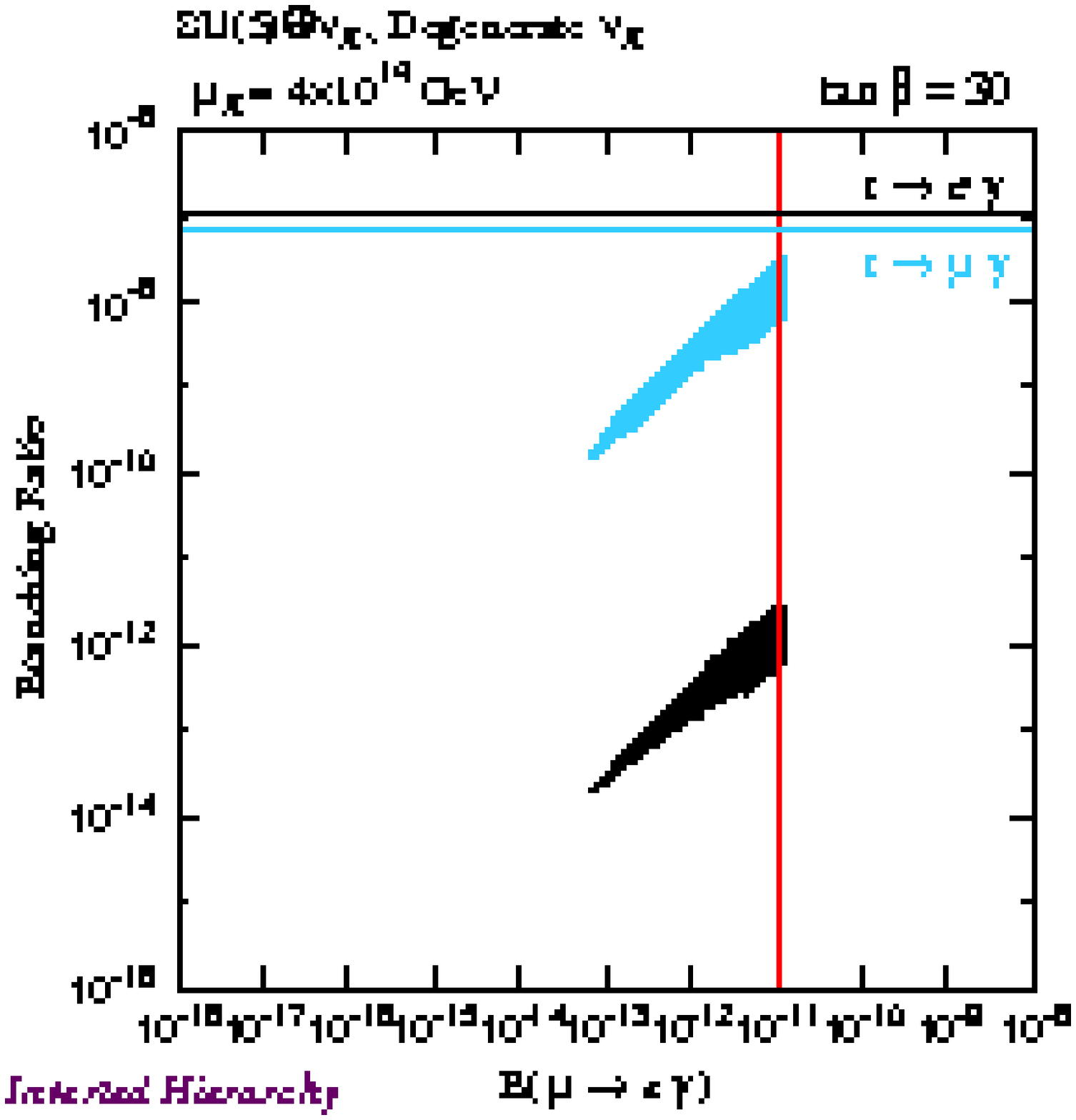} &
\includegraphics[scale=.3]{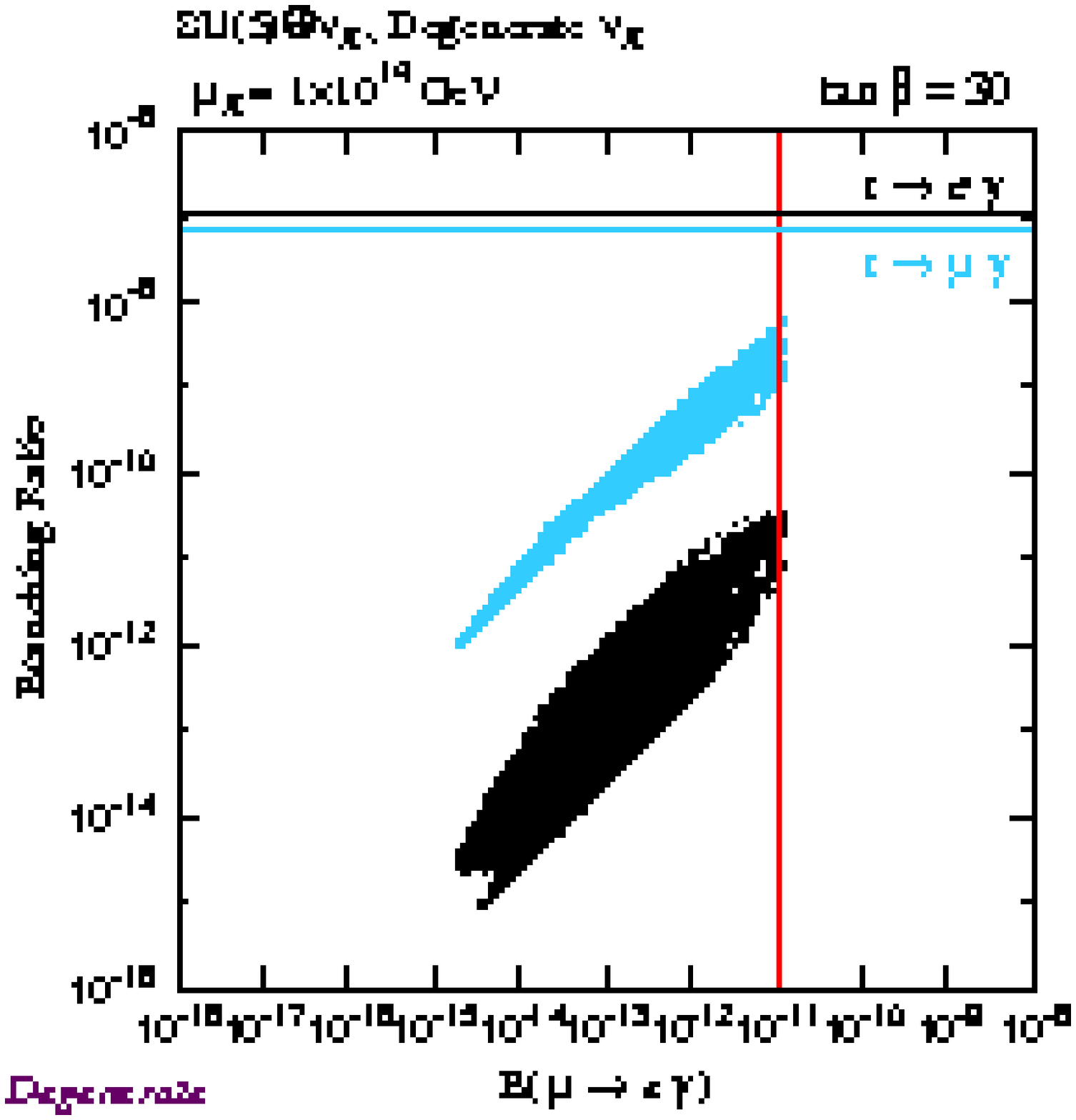} \\
(a) & (b) & (c) \\
\includegraphics[scale=.3]{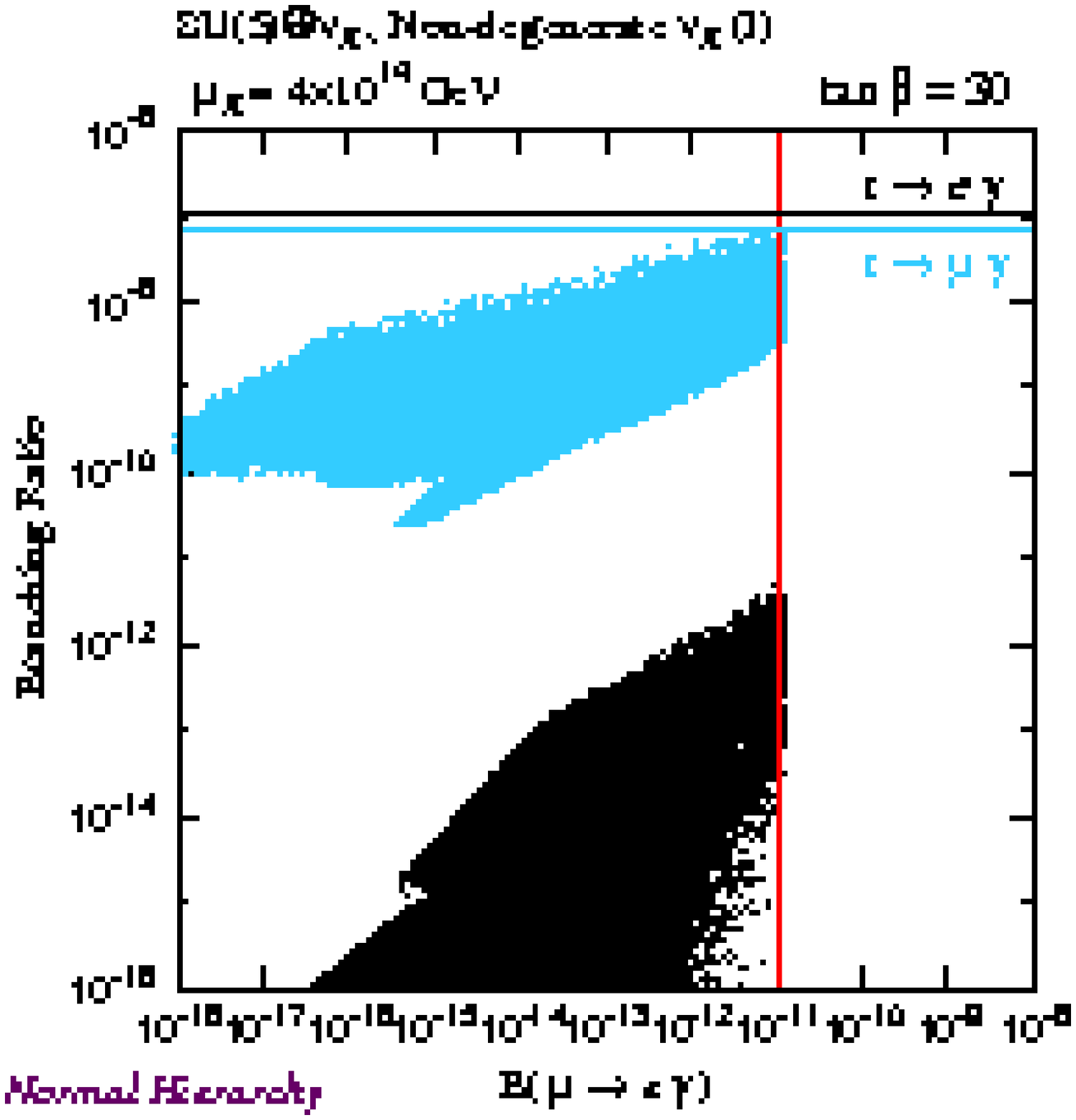} &
\includegraphics[scale=.3]{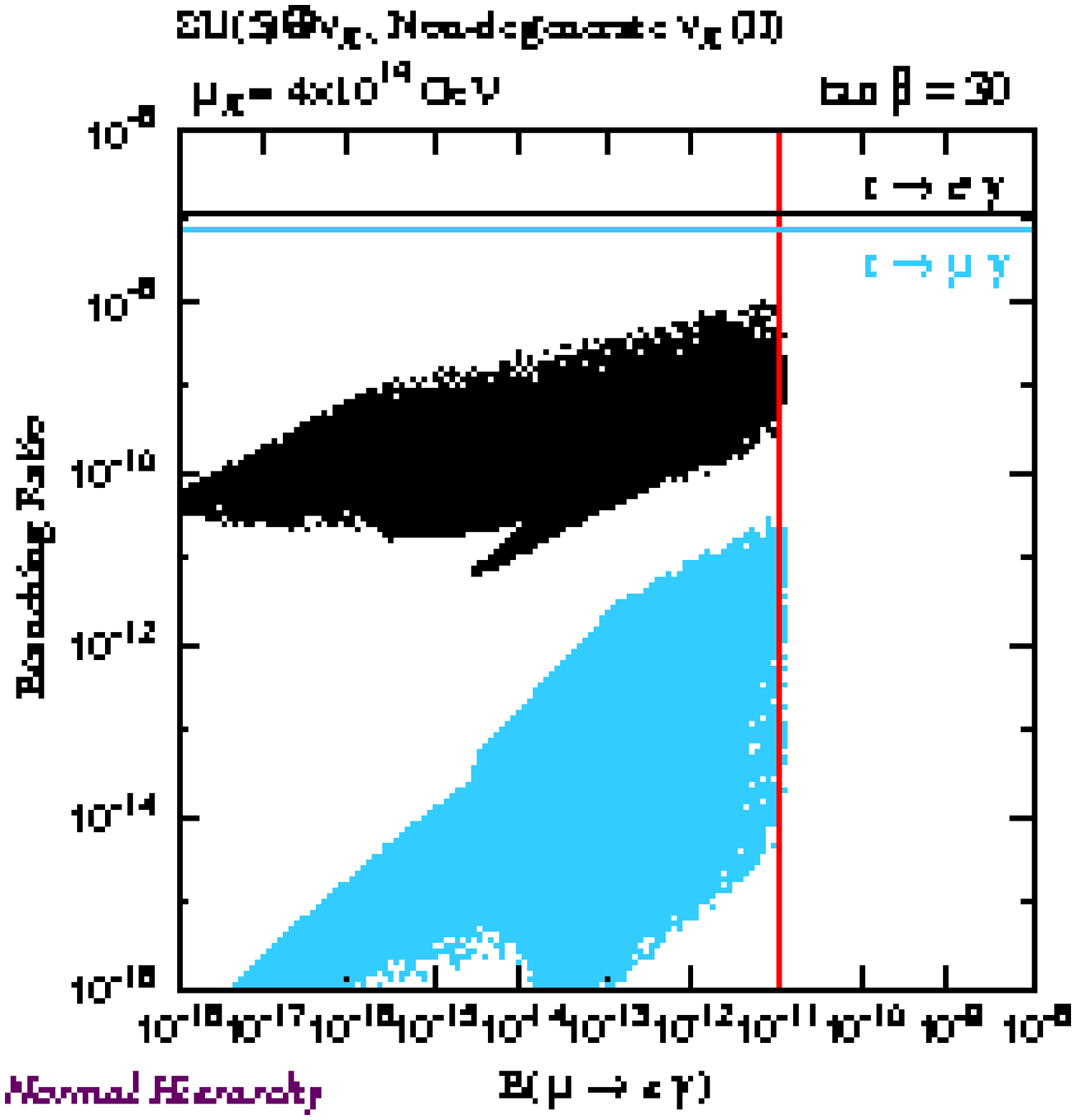} & \\
(d) & (e) &
\end{tabular}
\caption{%
(Color online)
Correlations between $\text{B}(\tau\to \mu(e)\gamma)$ and
$\text{B}(\mu\to e\gamma)$.
Light-grey (light-blue) and black dots denote $\tau\to \mu\gamma$ and
$\tau\to e\gamma$, respectively.
Experimental upper limits of the branching ratios are shown by horizontal
and vertical lines in each plot.
}
\label{fig:meg-LFV}
\end{figure}

There are lepton flavor mixings in the slepton sector of
the MSSM with right-handed neutrinos and the SU(5) SUSY GUT with
right-handed neutrinos.
It comes through the running between the right-handed neutrino mass
scale and the cut-off scale where the universal soft breaking mass terms
are generated.
On the other hand, no such slepton flavor mixings exist in the mSUGRA.

In Fig.~\ref{fig:ml-LFV} and  Fig.~\ref{fig:ml-LFV-su5}, the branching
ratios of $\mu\to e\gamma$, $\tau\to \mu\gamma$ and $\tau\to e\gamma$
are displayed as a function of the lightest charged slepton mass
$m(\tilde{l}_1)$ for the MSSM with right-handed neutrinos and the SU(5)
SUSY GUT with right-handed neutrinos, respectively.
For each model, we show the results for five cases of the neutrino
masses and Yukawa coupling matrix as explained in
Sec.~\ref{sec:inputparameters}.
The right-handed neutrino mass scale $\mu_R$ is taken as
$\mu_R=4\times10^{14}\text{GeV}$ for the normal and inverted hierarchy
cases, which corresponds to the neutrino Yukawa couplings of $O(1)$.
In the degenerate ($m_{\nu_1}=0.1\text{eV}$) case, we take
$\mu_R=1\times10^{14}\text{GeV}$ since the neutrino Yukawa coupling
blows up below the Planck scale for $\mu_R=4\times10^{14}\text{GeV}$.
It is known that branching ratios are enhanced by a factor of
$\tan^2\beta$ for large values of $\tan\beta$.
In the presented plots, we take $\tan\beta=30$ except for the case of
the degenerate $\nu_R$ with normal hierarchical neutrinos (D$\nu_R$-NH)
in the SU(5) SUSY GUT with right-handed neutrinos, where we show
the result for $\tan\beta=10$.
When we take $\tan\beta=30$ for D$\nu_R$-NH in the SU(5) SUSY GUT with
right-handed neutrinos, almost all the data points in the scanned
parameter space are excluded due to the $\text{B}(\mu\to e\gamma)$
constraint.

We can see that the $\mu\to e\gamma$ decay rate is enhanced in the
normal hierarchy with degenerate $\nu_R$ cases.
In fact, even for the slepton as heavy as 3TeV,
$\text{B}(\mu\to e\gamma)$ is close to (or above) the experimental
upper limit.
After applying the constraint from $\text{B}(\mu\to e\gamma)$, the
branching ratio of $\tau\to\mu\gamma$ can be $10^{-9}$ at most.
On the other hand, in the inverted hierarchy and degenerate cases (with
degenerate $\nu_R$), $\mu\to e\gamma$ and $\tau\to e\gamma$ are
relatively suppressed.
This behavior is understood in the following way.
From the neutrino Yukawa coupling matrix (\ref{eq:yNdegenerate}) and the
PMNS matrix (\ref{eq:VPMNS}) with $s_{13}=0$, the off-diagonal elements
of $y_N^\dagger y_N$ are written as
\begin{equation}
  (y_N^\dagger y_N)_{12} =
  \frac{\hat{M}_N}{\langle h_2\rangle^2}
  c_{\odot} s_{\odot}c_{\text{atm}}
  \frac{m_{\nu_2}^2-m_{\nu_1}^2}{m_{\nu_2}+m_{\nu_1}}
\;,
\qquad
  (y_N^\dagger y_N)_{13} = -
  \frac{\hat{M}_N}{\langle h_2\rangle^2}
  c_{\odot} s_{\odot} s_{\text{atm}}
  \frac{m_{\nu_2}^2-m_{\nu_1}^2}{m_{\nu_2}+m_{\nu_1}}
\;.
\end{equation}
Therefore the $1-2$ and $1-3$ slepton mixings are suppressed for a
larger value of $m_{\nu_2}+m_{\nu_1}$ when $m_{\nu_2}^2-m_{\nu_1}^2$ and
$\hat{M}_N=\mu_R$ are fixed.

In the non-degenerate $\nu_R$ (I) case, $\text{B}(\mu\to e\gamma)$ is
suppressed compared to the degenerate $\nu_R$ cases, so that the
constraint is weakened.
In particular, there is an approximate electron-number conservation in
the MSSM with right-handed neutrinos with the Yukawa coupling matrix of
the structure Eq.~(\ref{eq:Y-non-degenerate-I}), which leads to the
suppression of both $\mu\to e\gamma$ and $\tau\to e\gamma$.
The branching ratio of $\tau\to \mu\gamma$ can be as large as the
current experimental upper limit.
In the SU(5) SUSY GUT with right-handed neutrinos, the electron-number
conservation is broken by GUT interactions.
As a result, $\text{B}(\mu\to e\gamma)$ can be also as large as the
current experimental upper limit.
In the non-degenerate $\nu_R$ (II) cases, the role of $e$ and $\mu$ are
interchanged due to the Yukawa structure
Eq.~(\ref{eq:Y-non-degenerate-II}).

Correlations between $\text{B}(\tau\to \mu(e)\gamma)$ and
$\text{B}(\mu\to e\gamma)$ in
the SU(5) SUSY GUT with right-handed neutrinos are shown in
Fig.~\ref{fig:meg-LFV}.
Since the MEG experiment can measure $\text{B}(\mu\to e\gamma)$ down to
$10^{-13}$ and the Super $B$ factory can measure
$\text{B}(\tau\to \mu\gamma)$ and $\text{B}(\tau\to e\gamma)$ of
$10^{-9}$, it is possible to distinguish the structure of the slepton
flavor mixing if the slepton mass is less than 1TeV.

\subsubsection{Quark flavor signals}

\begin{figure}[htbp]
\begin{tabular}{cccc}
\includegraphics[scale=.23]{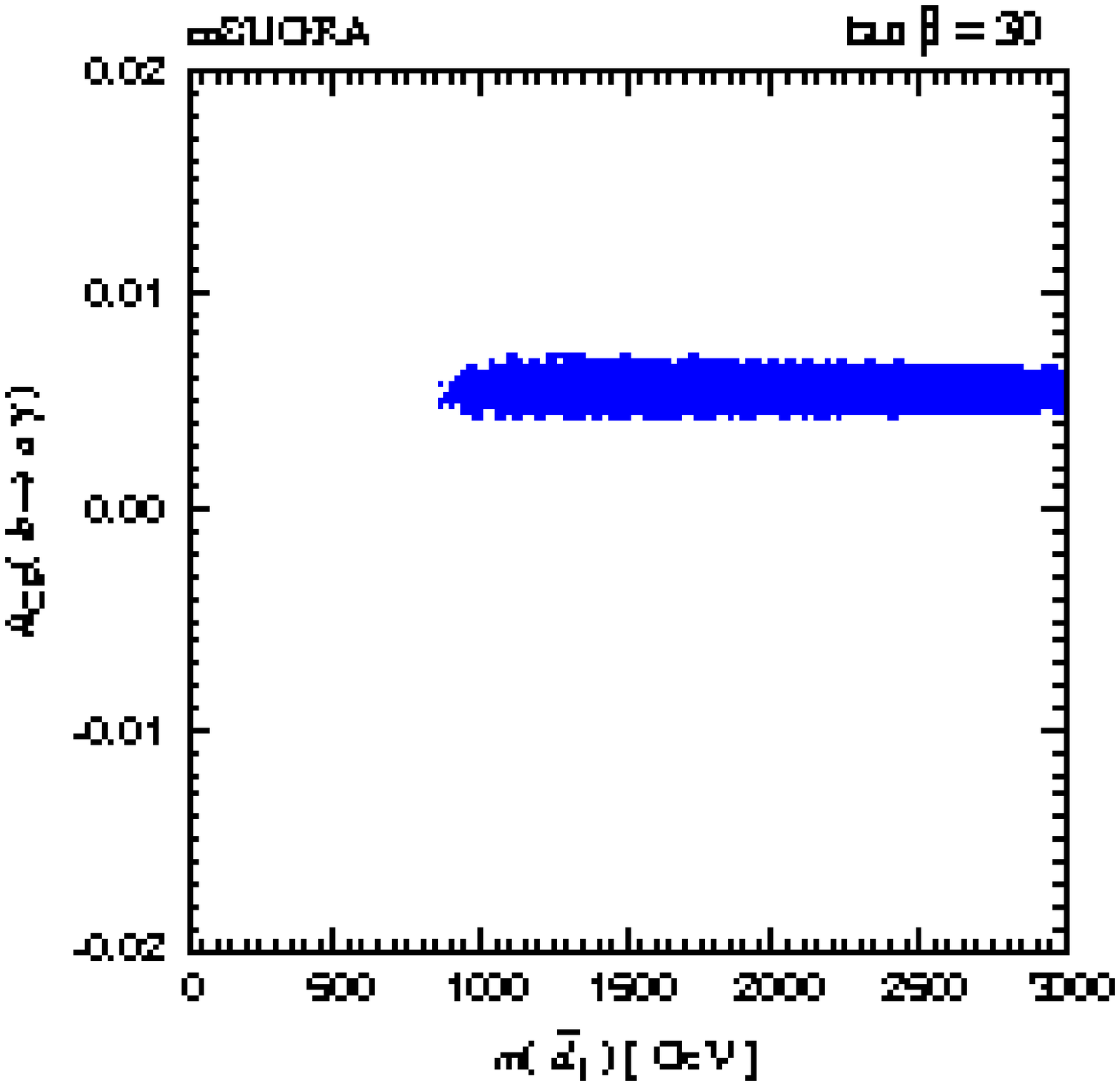} &
\includegraphics[scale=.23]{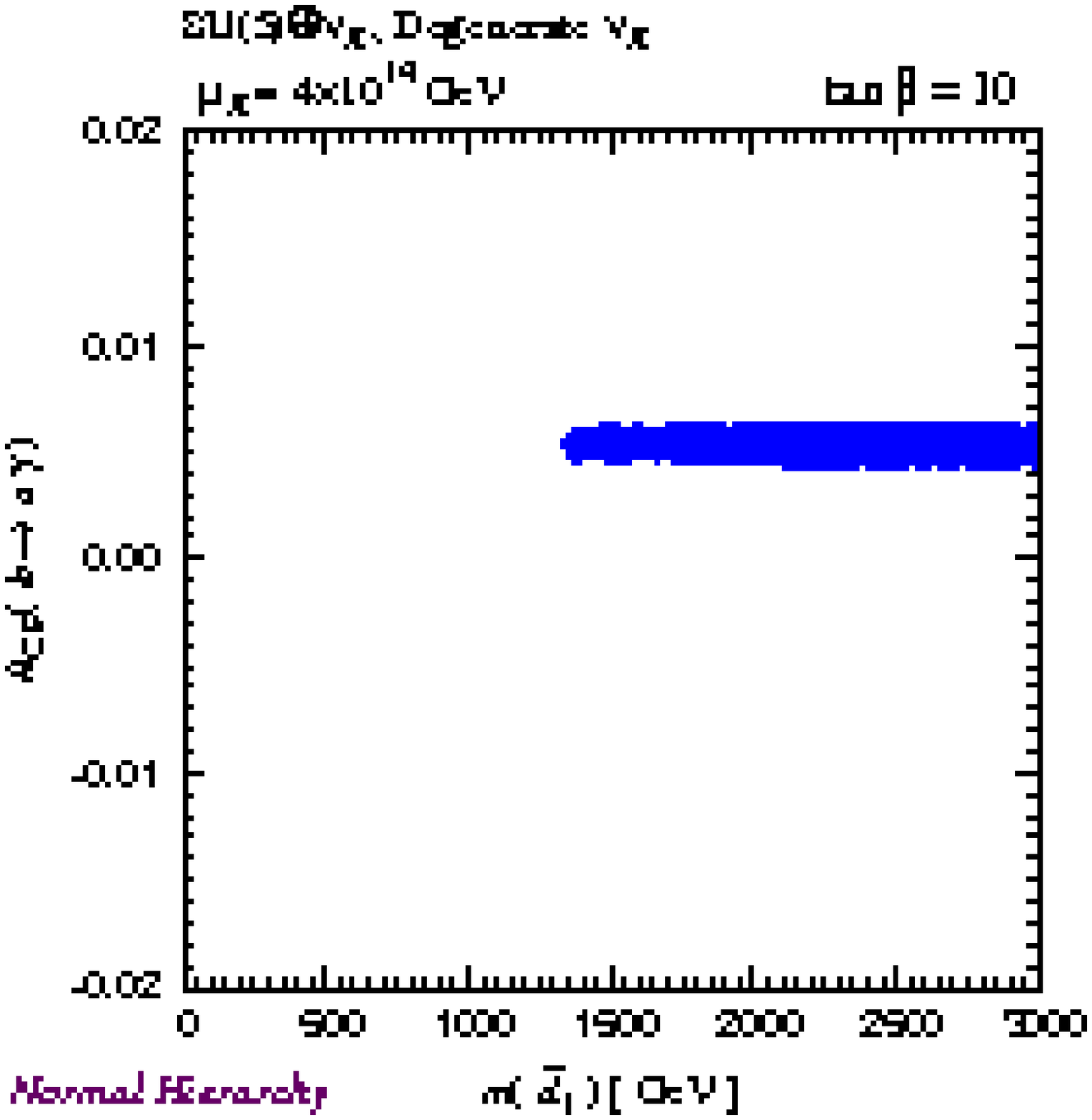} &
\includegraphics[scale=.23]{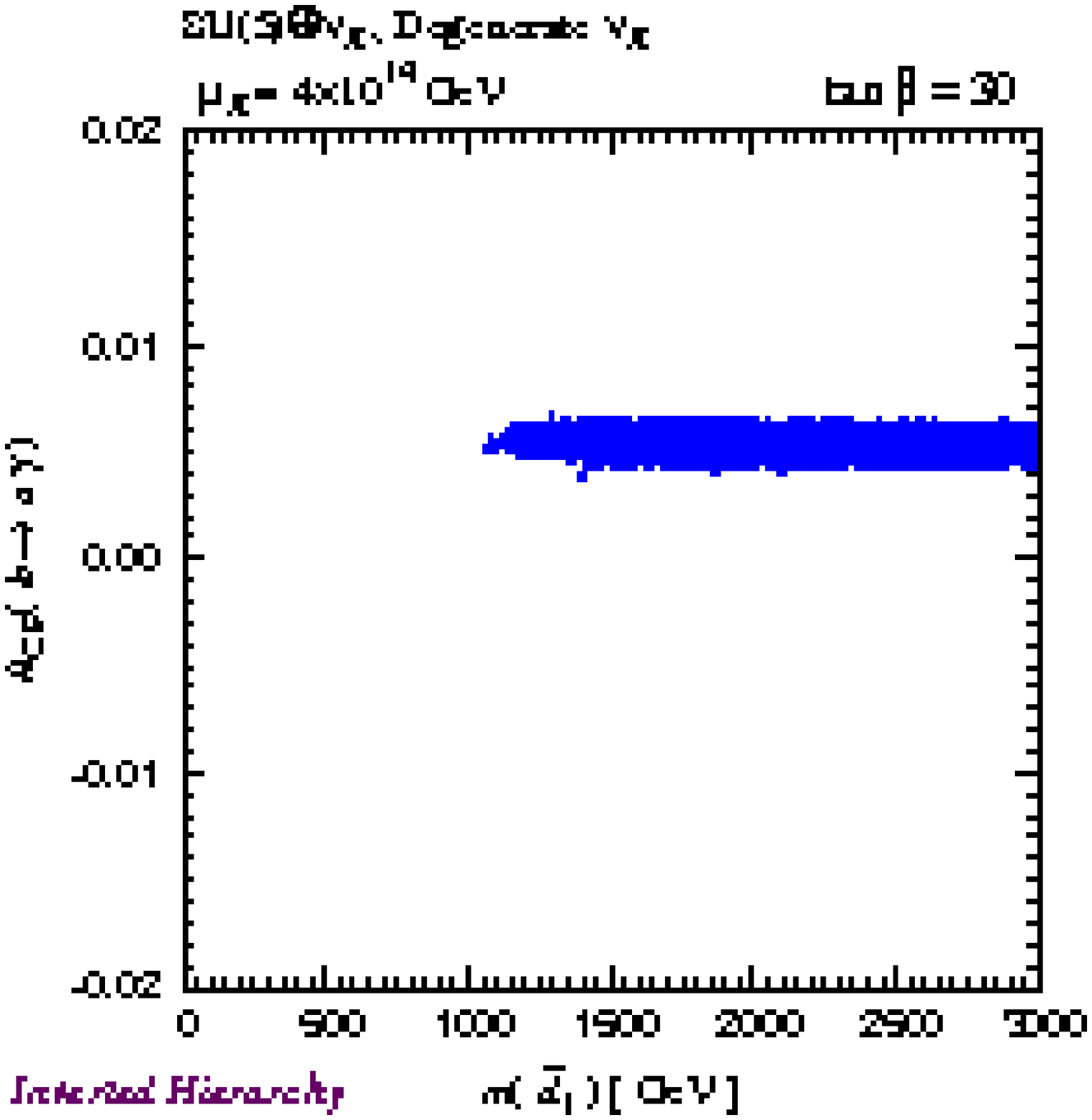} &
\includegraphics[scale=.23]{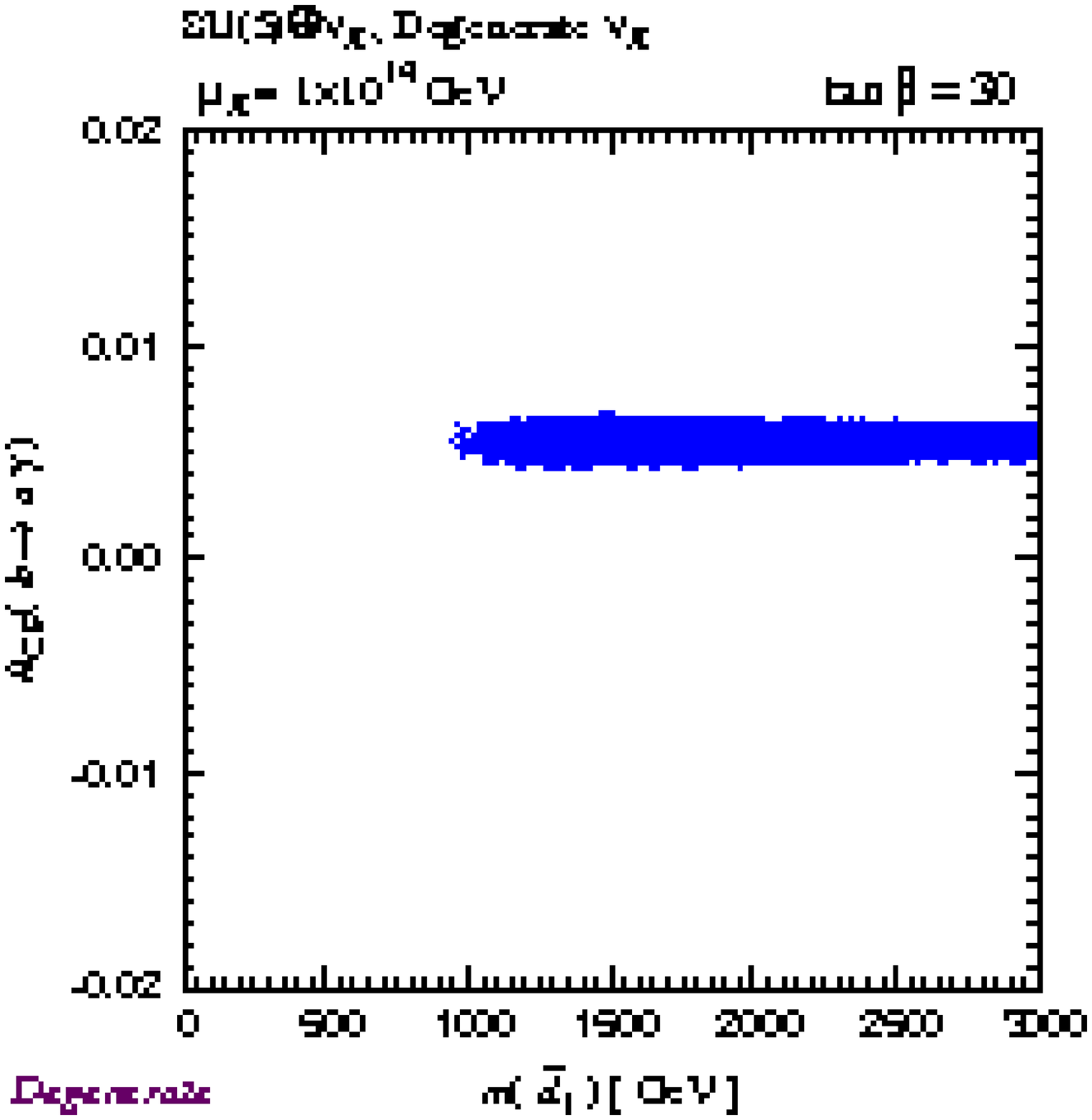}
\\
(a) & (b) & (c) & (d)
\\
\includegraphics[scale=.23]{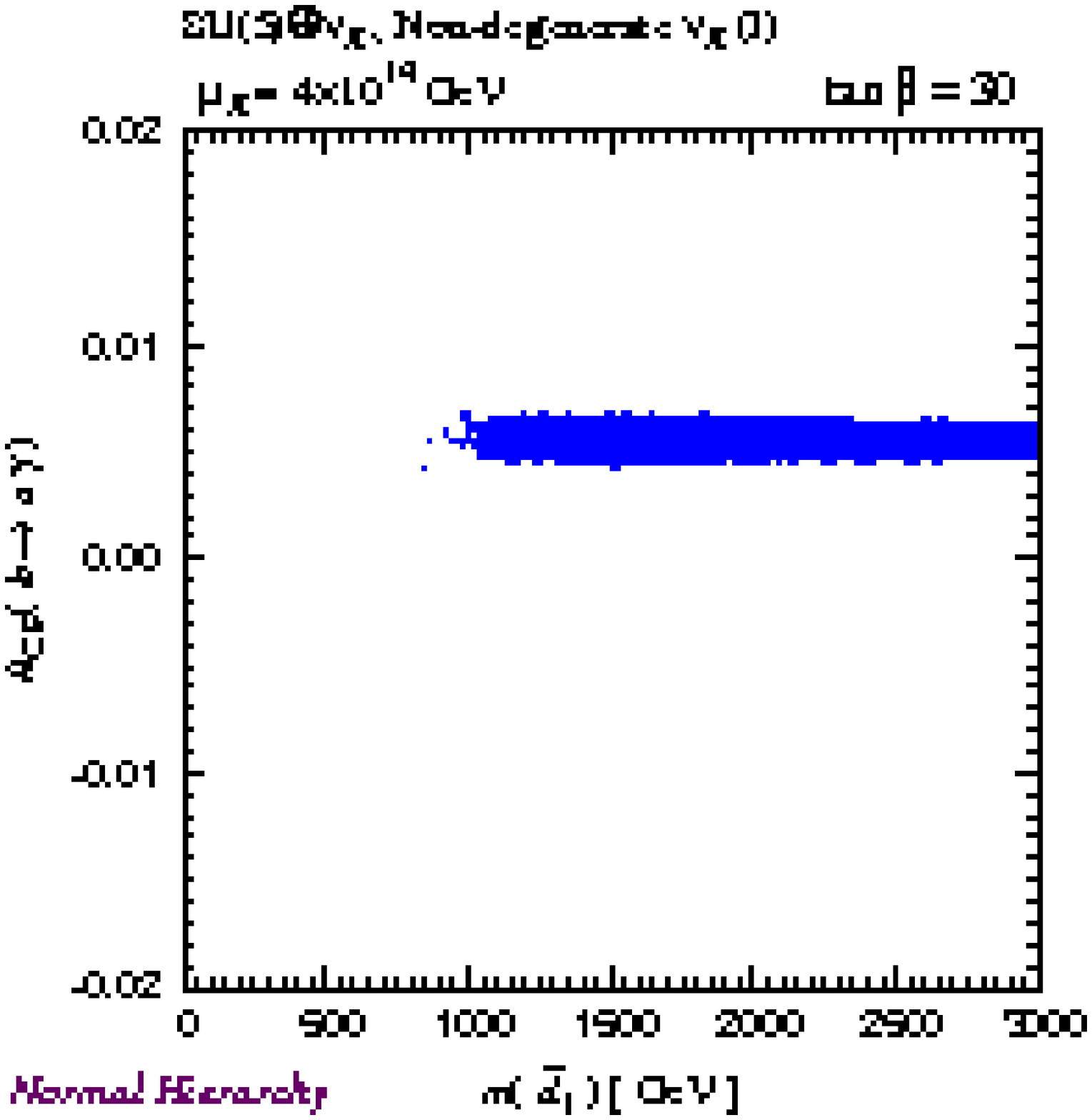} &
\includegraphics[scale=.23]{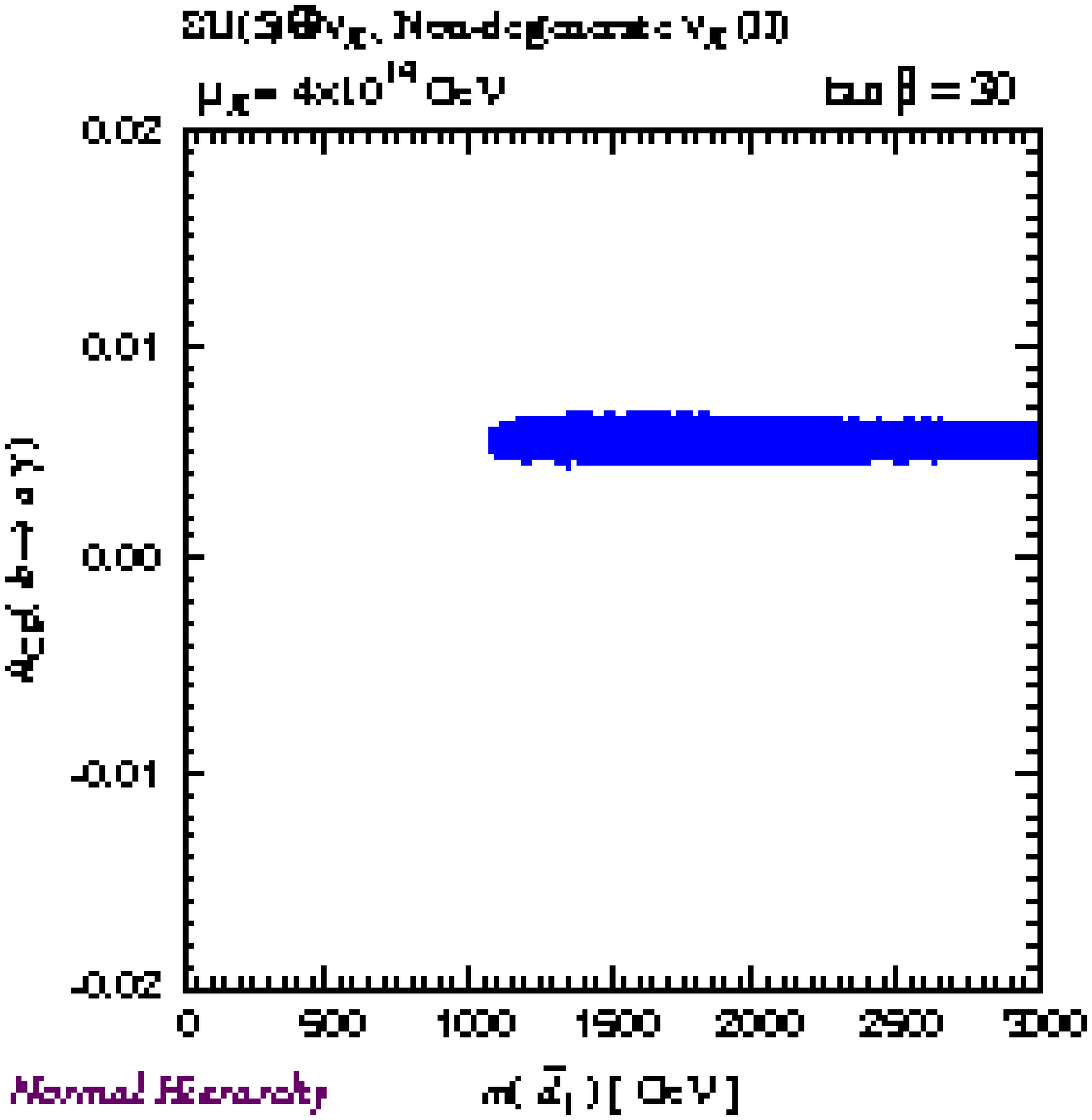} &
\includegraphics[scale=.23]{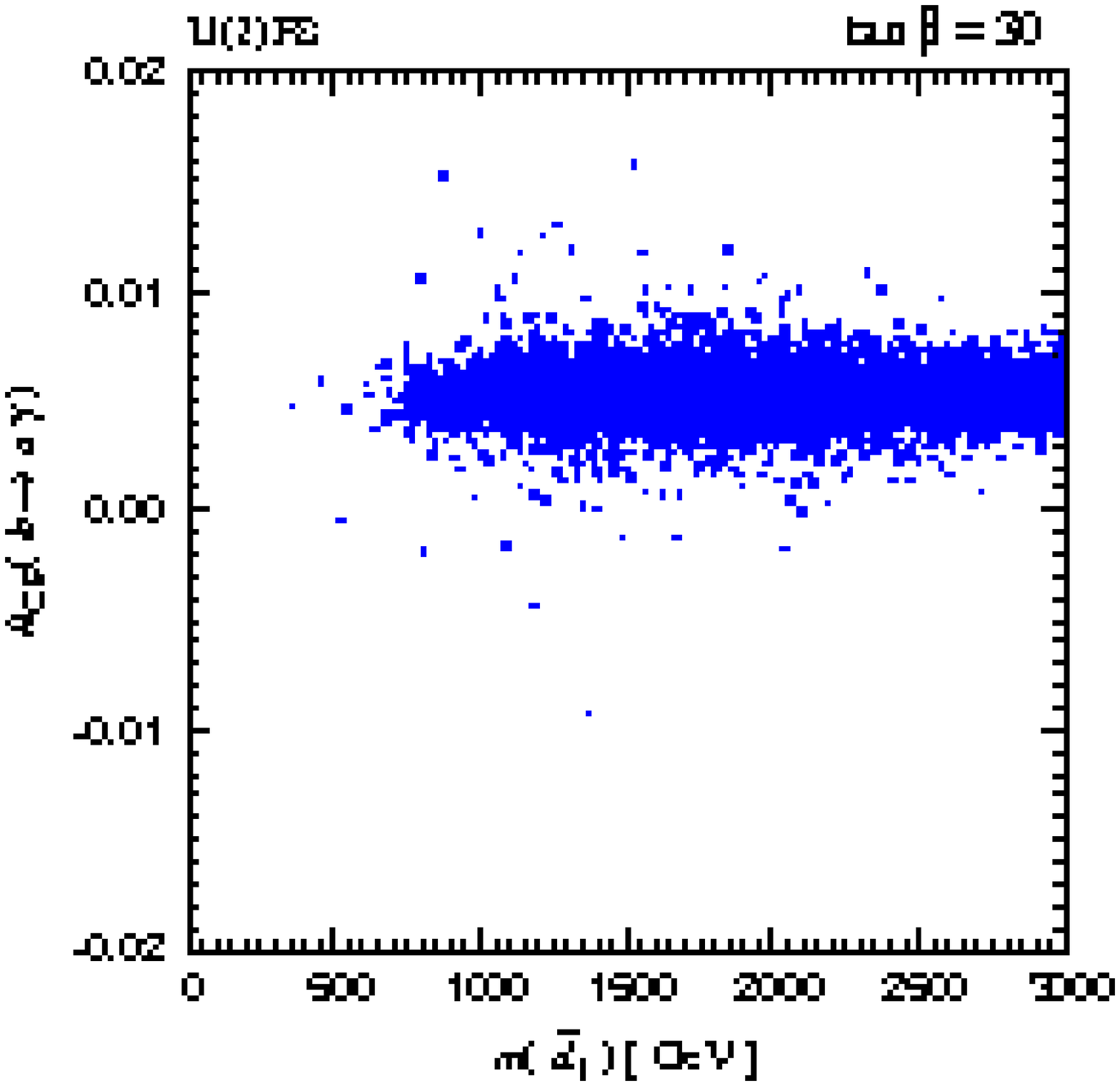} &
\\
(e) & (f) & (g) &
\end{tabular}
\caption{%
The direct CP asymmetry in $b\to s\gamma$ as functions of the lightest
down-type squark mass $m(\tilde{d}_1)$ for (a) mSUGRA,
(b)--(f) five cases of the SU(5) SUSY GUT with right-handed
neutrinos and (g) U(2) model.
}
\label{fig:md-ACP-bsg}
\end{figure}

\begin{figure}[htbp]
\begin{tabular}{cccc}
\includegraphics[scale=.23]{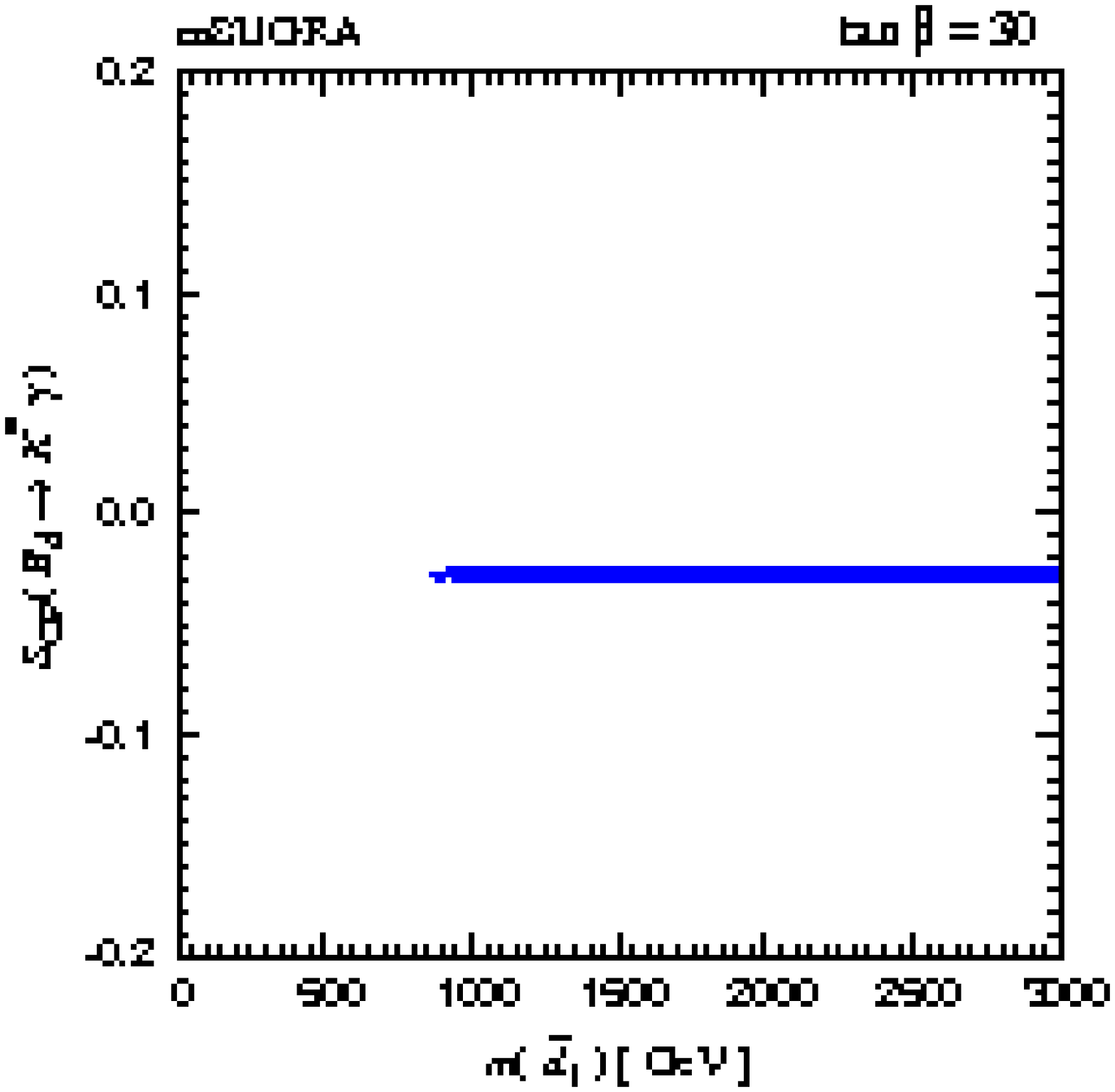} &
\includegraphics[scale=.23]{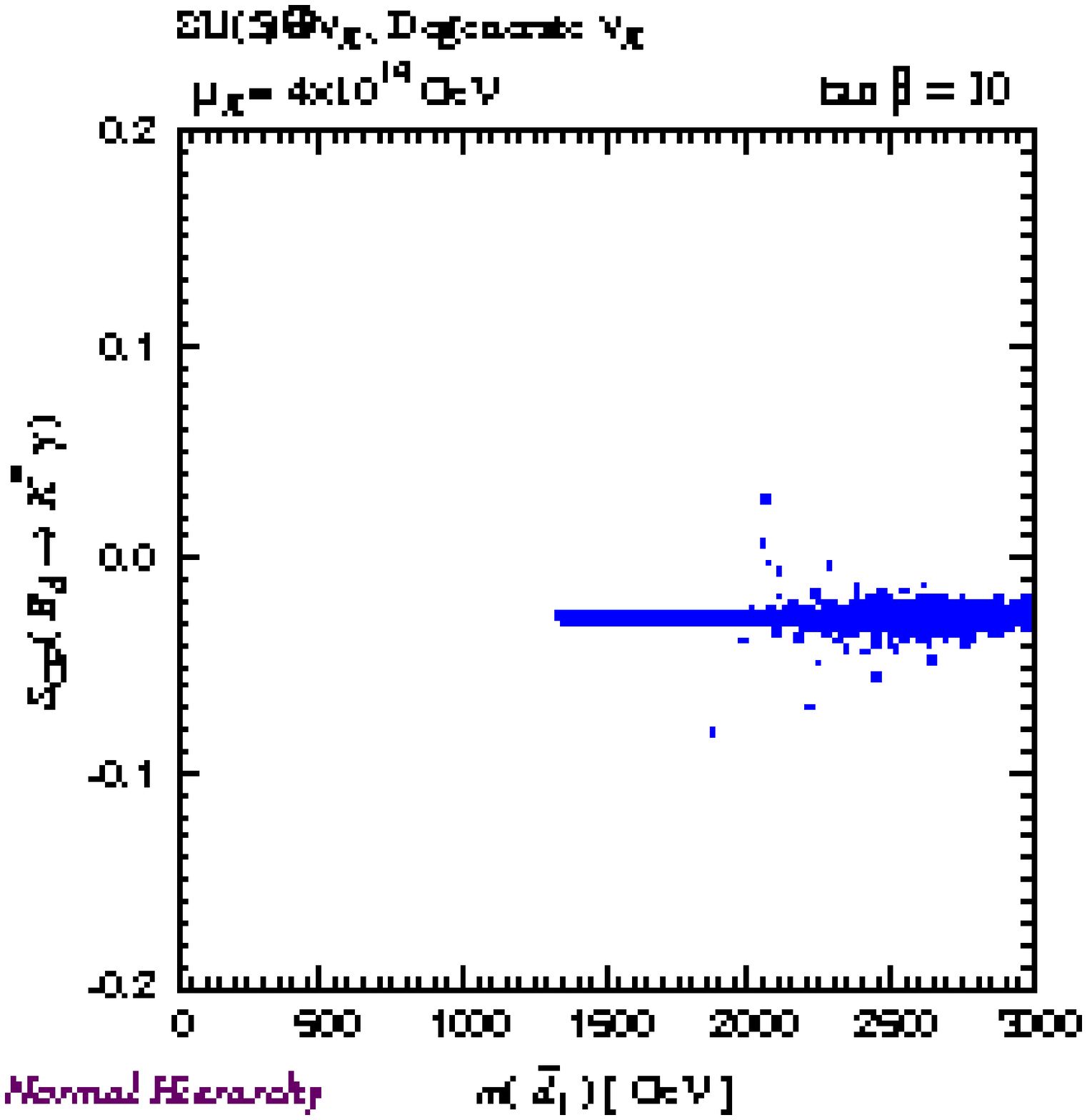} &
\includegraphics[scale=.23]{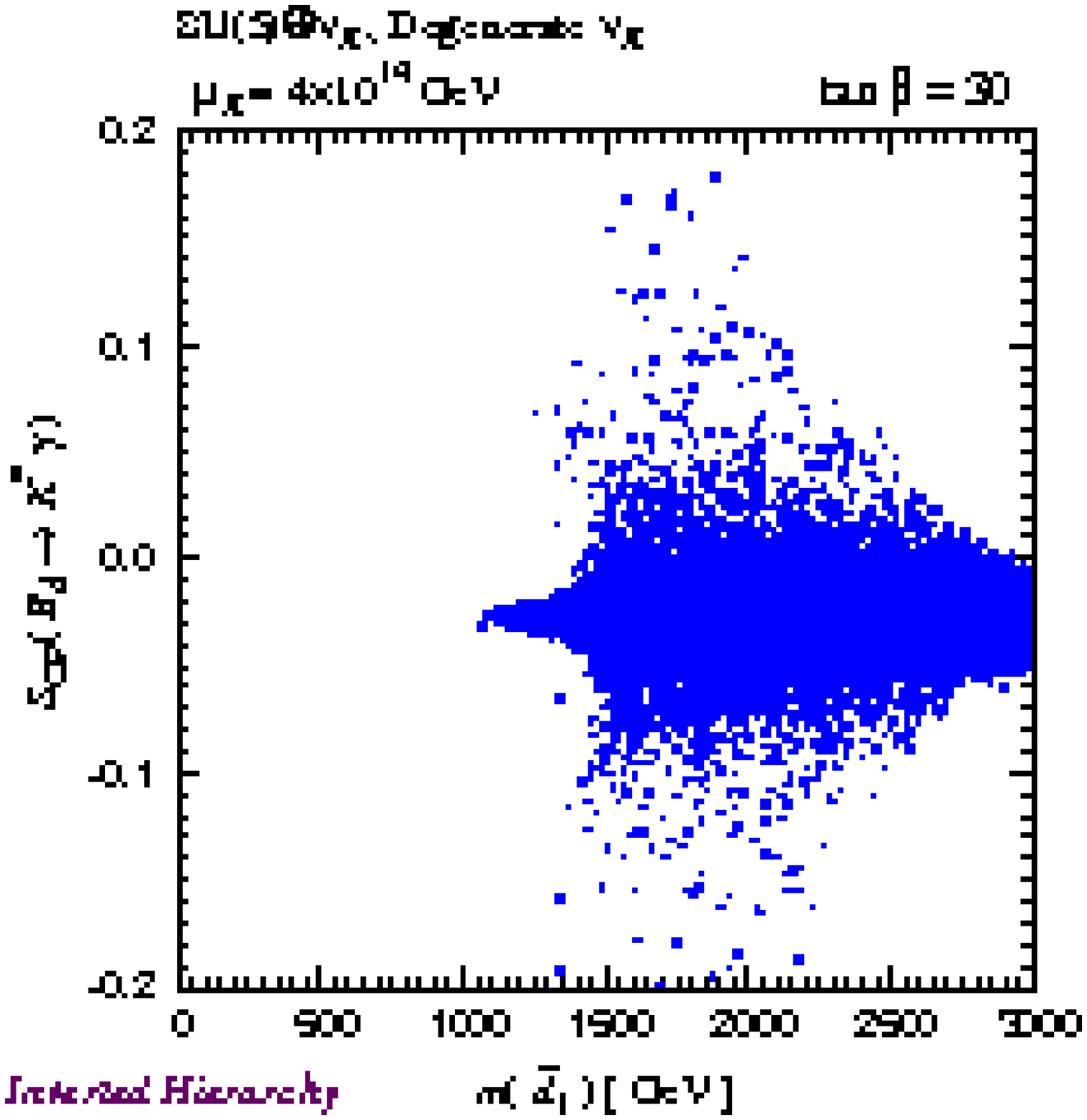} &
\includegraphics[scale=.23]{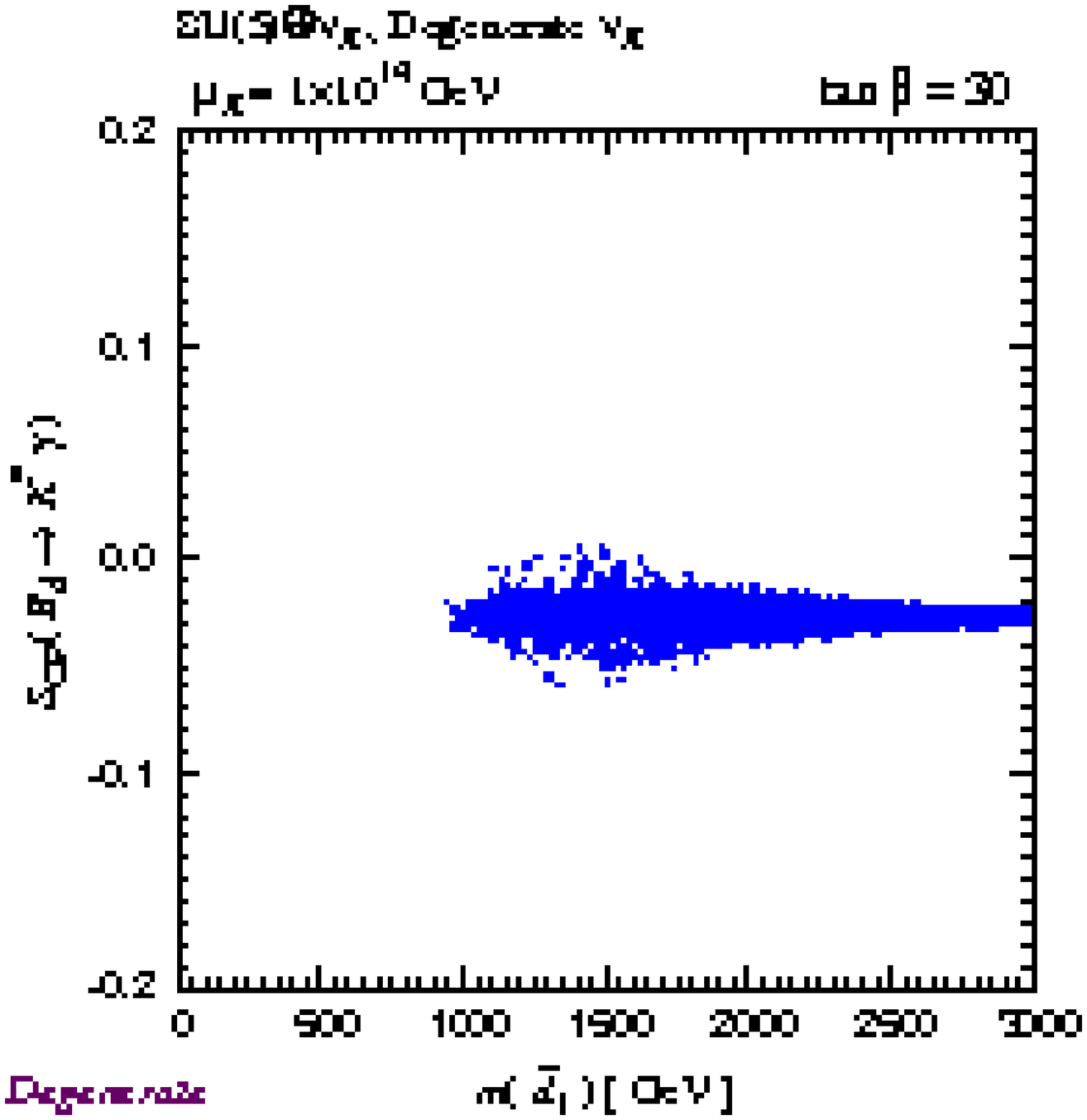}
\\
(a) & (b) & (c) & (d)
\\
\includegraphics[scale=.23]{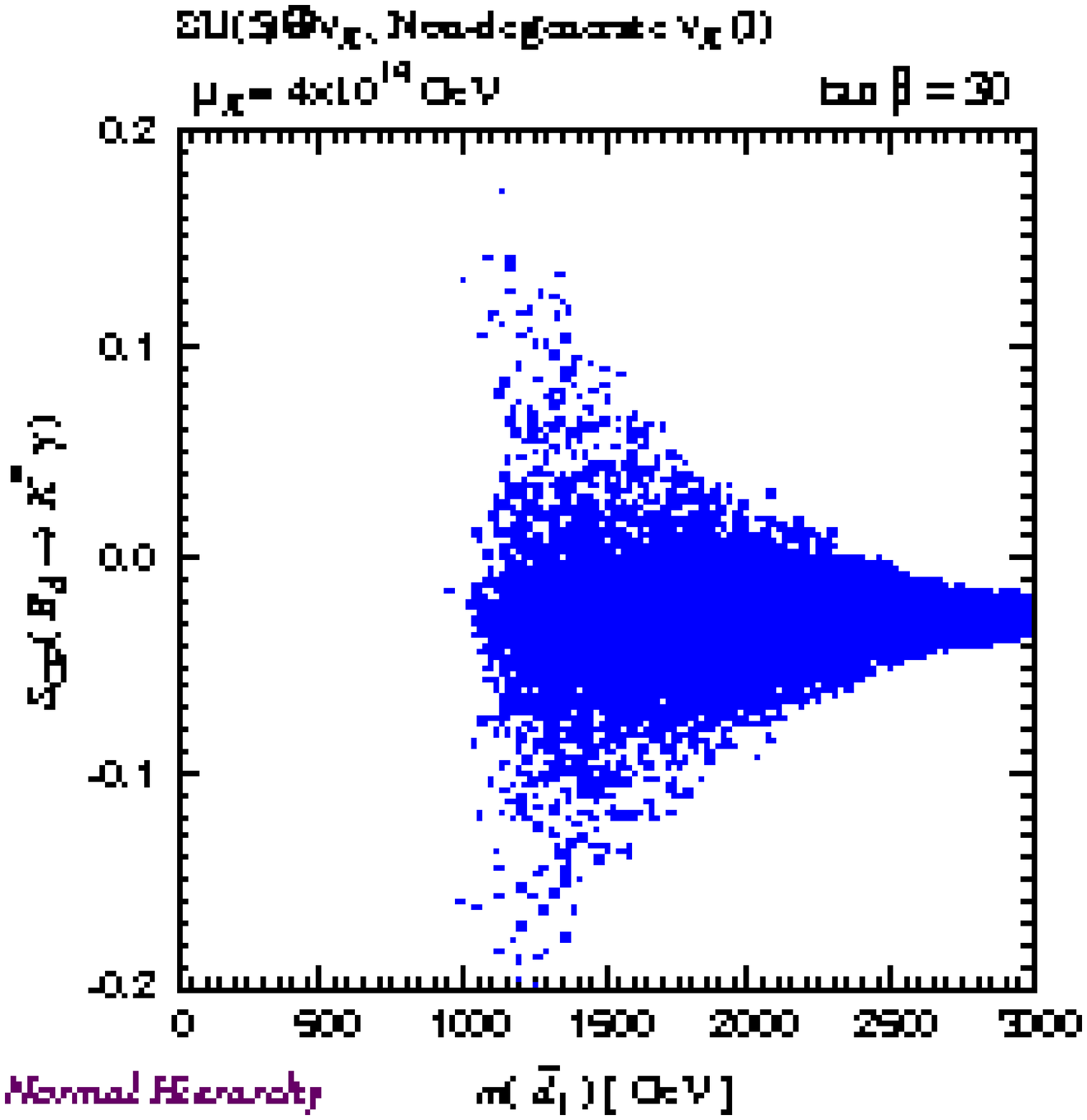} &
\includegraphics[scale=.23]{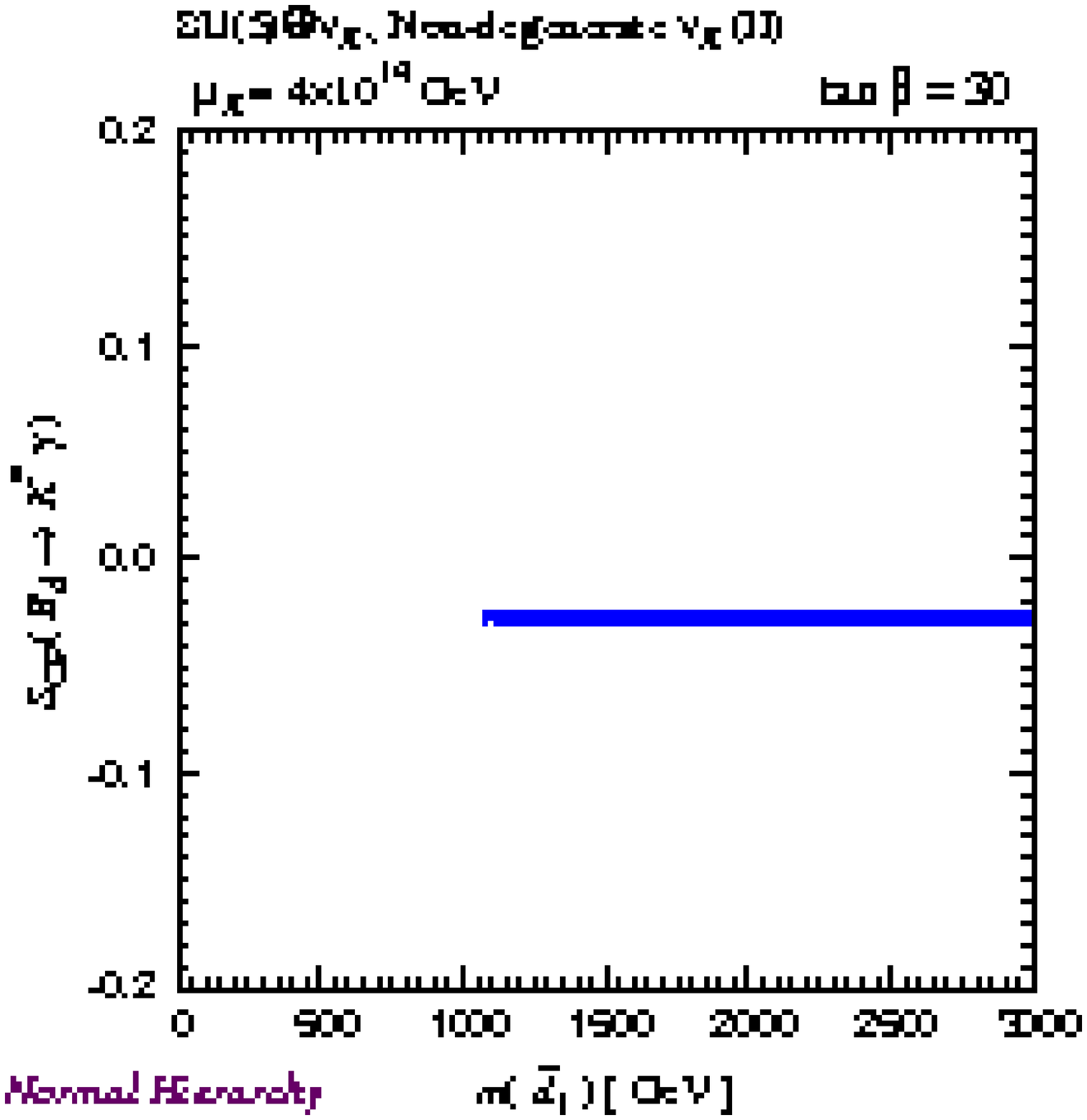} &
\includegraphics[scale=.23]{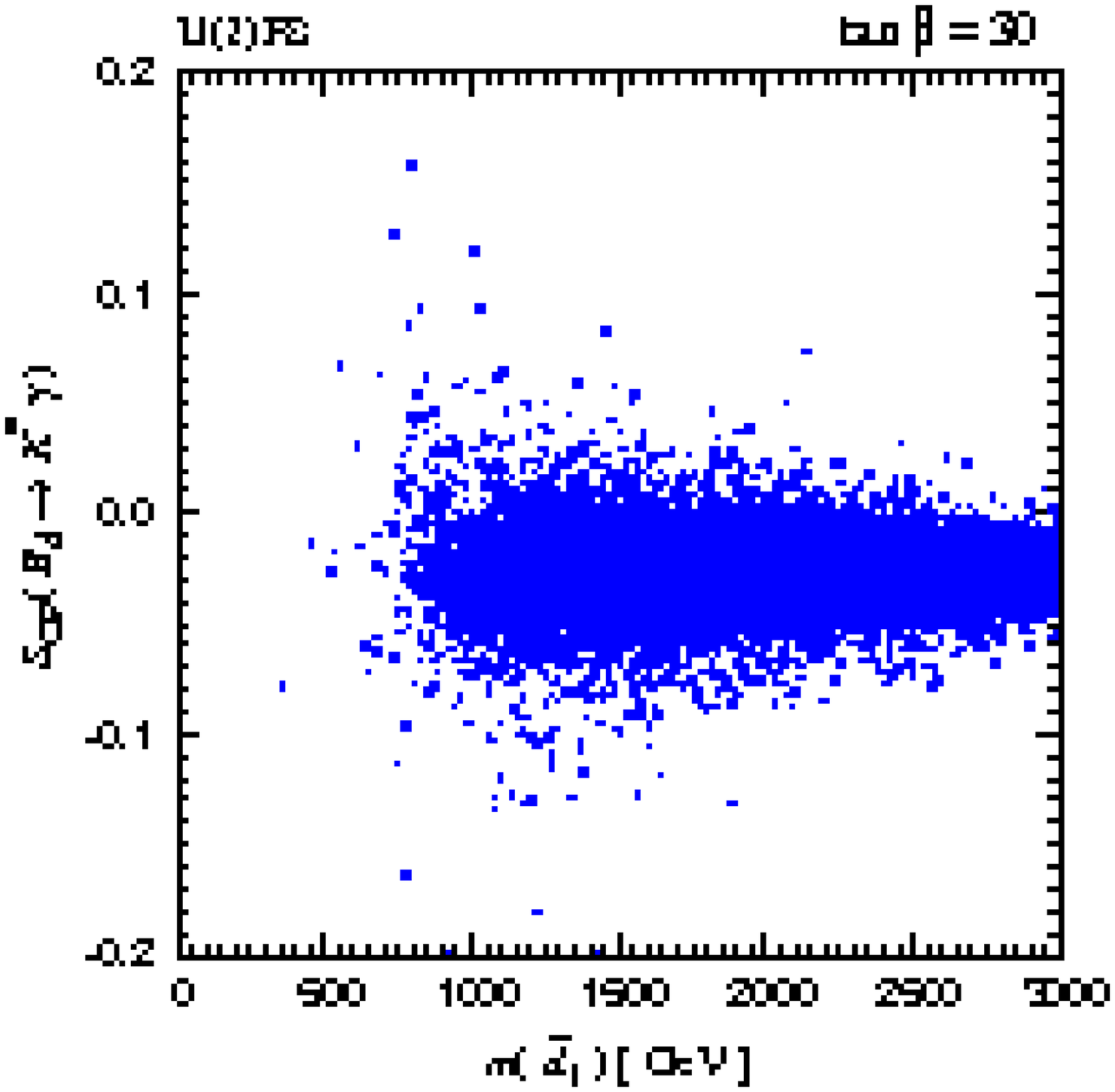} &
\\
(e) & (f) & (g) &
\end{tabular}
\caption{%
The mixing-induced CP asymmetry in $B_d\to K^*\gamma$ as functions of
$m(\tilde{d}_1)$ for the same parameter sets as those for
Fig.~\ref{fig:md-ACP-bsg}.
}
\label{fig:md-SCP-bsg}
\end{figure}

\begin{figure}[htbp]
\begin{tabular}{cccc}
\includegraphics[scale=.23]{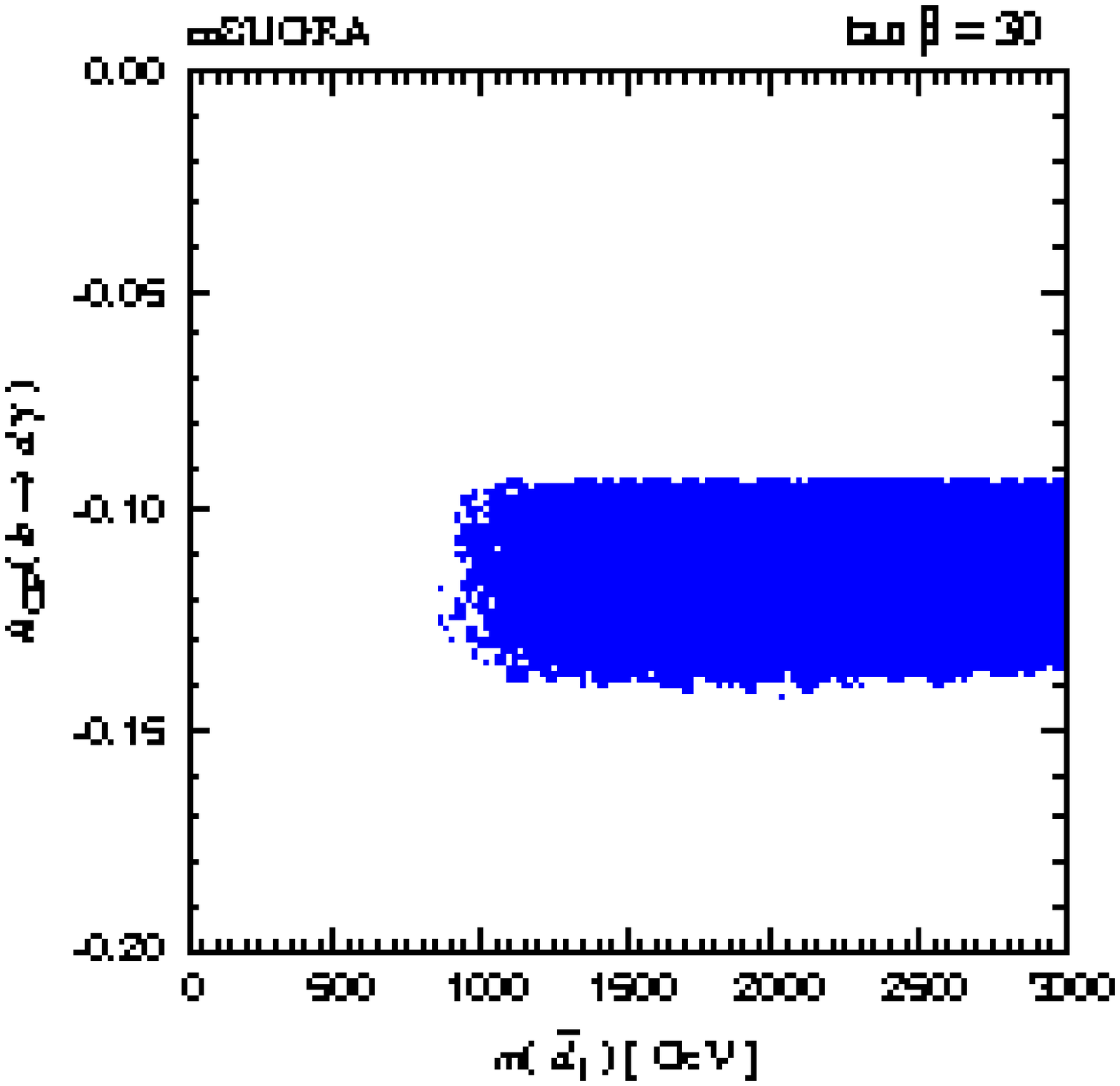} &
\includegraphics[scale=.23]{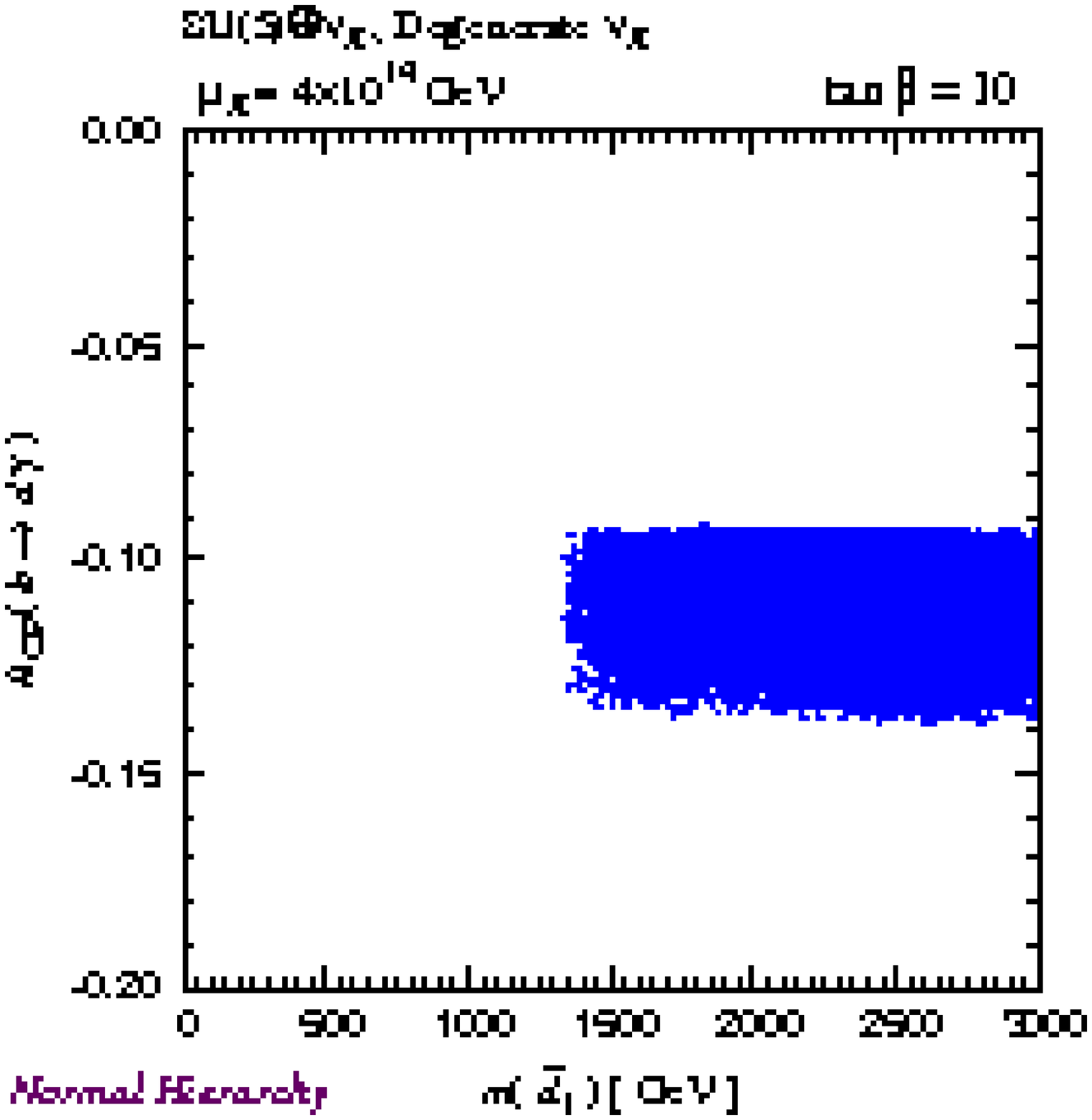} &
\includegraphics[scale=.23]{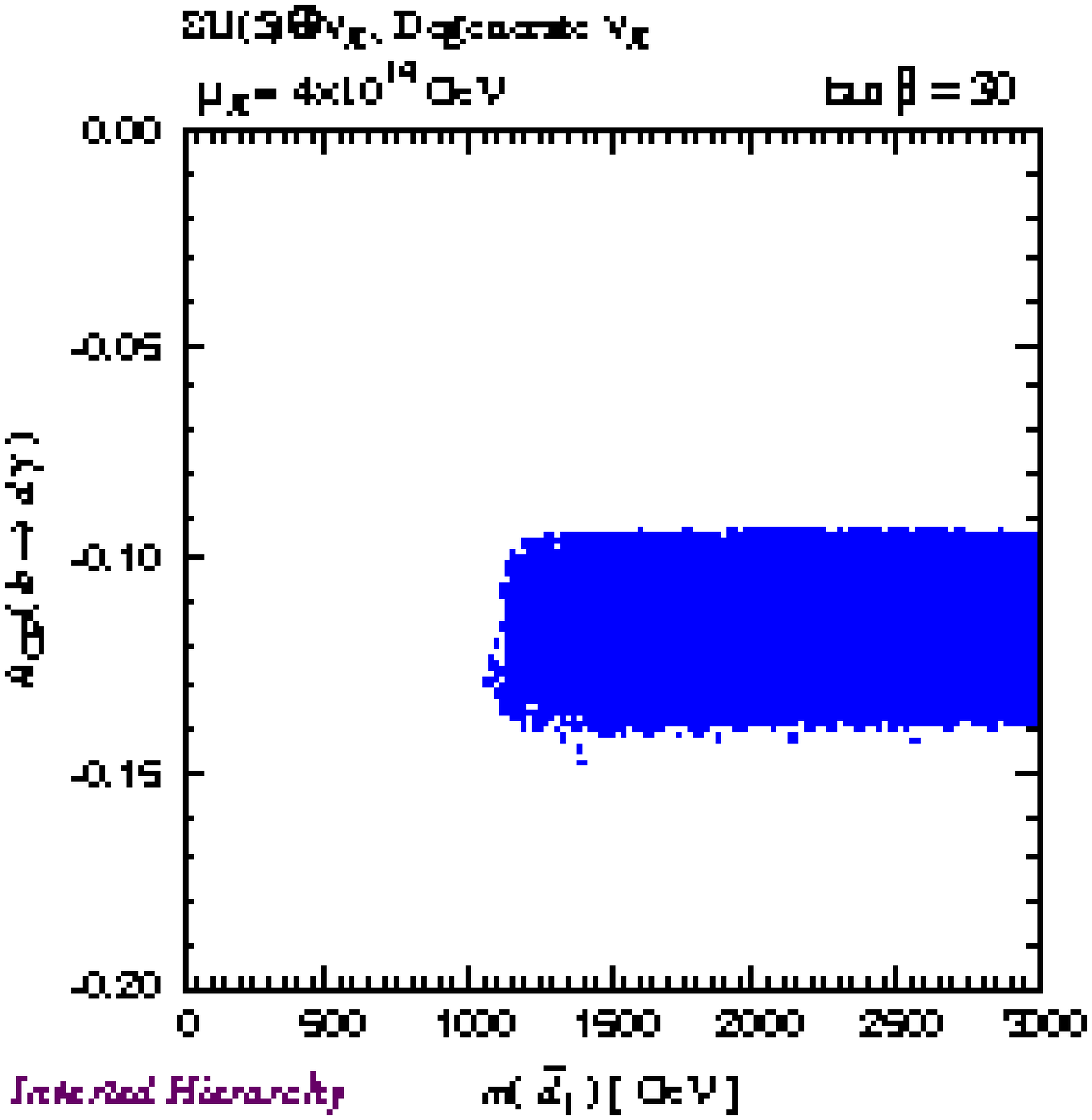} &
\includegraphics[scale=.23]{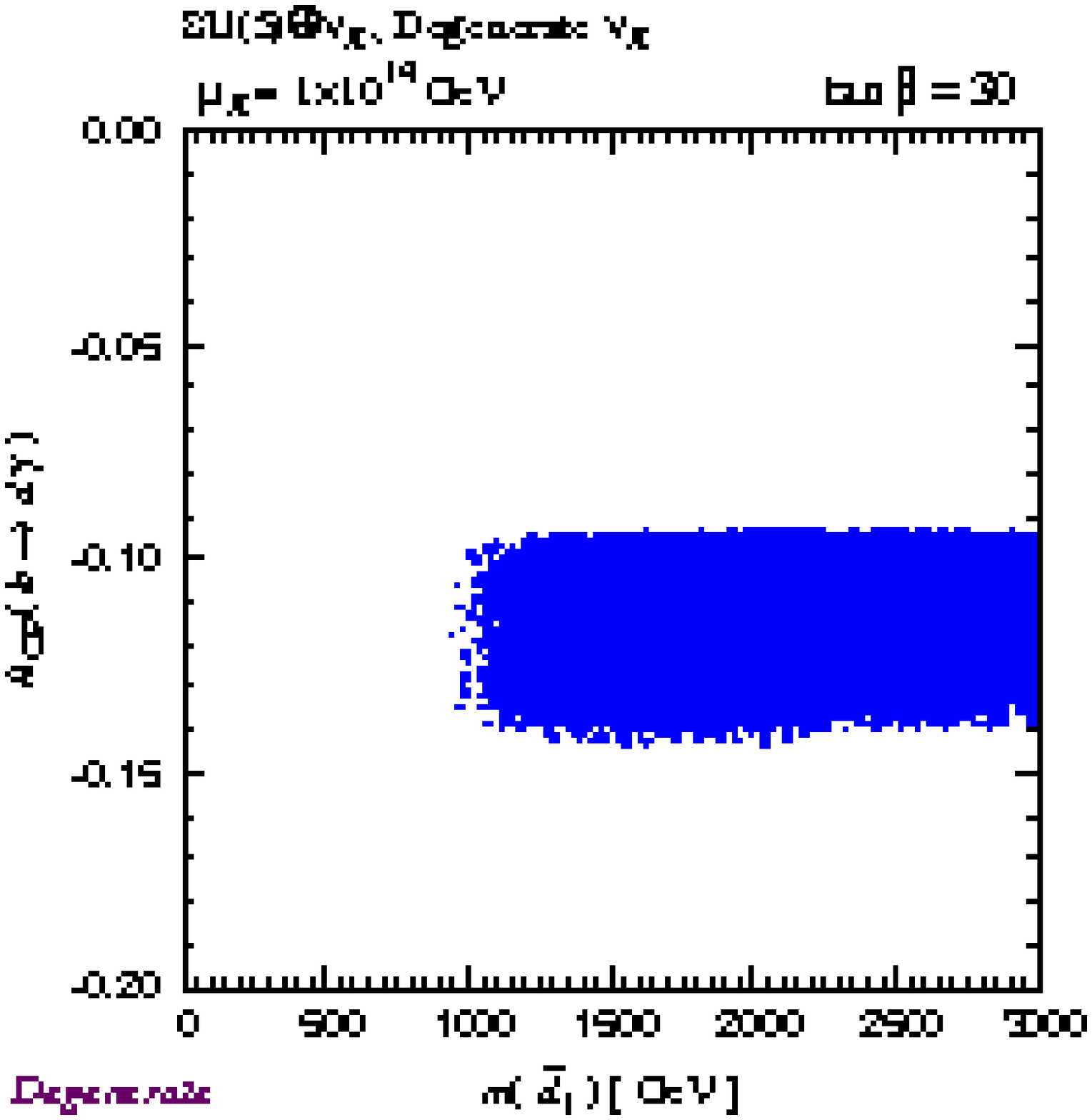}
\\
(a) & (b) & (c) & (d)
\\
\includegraphics[scale=.23]{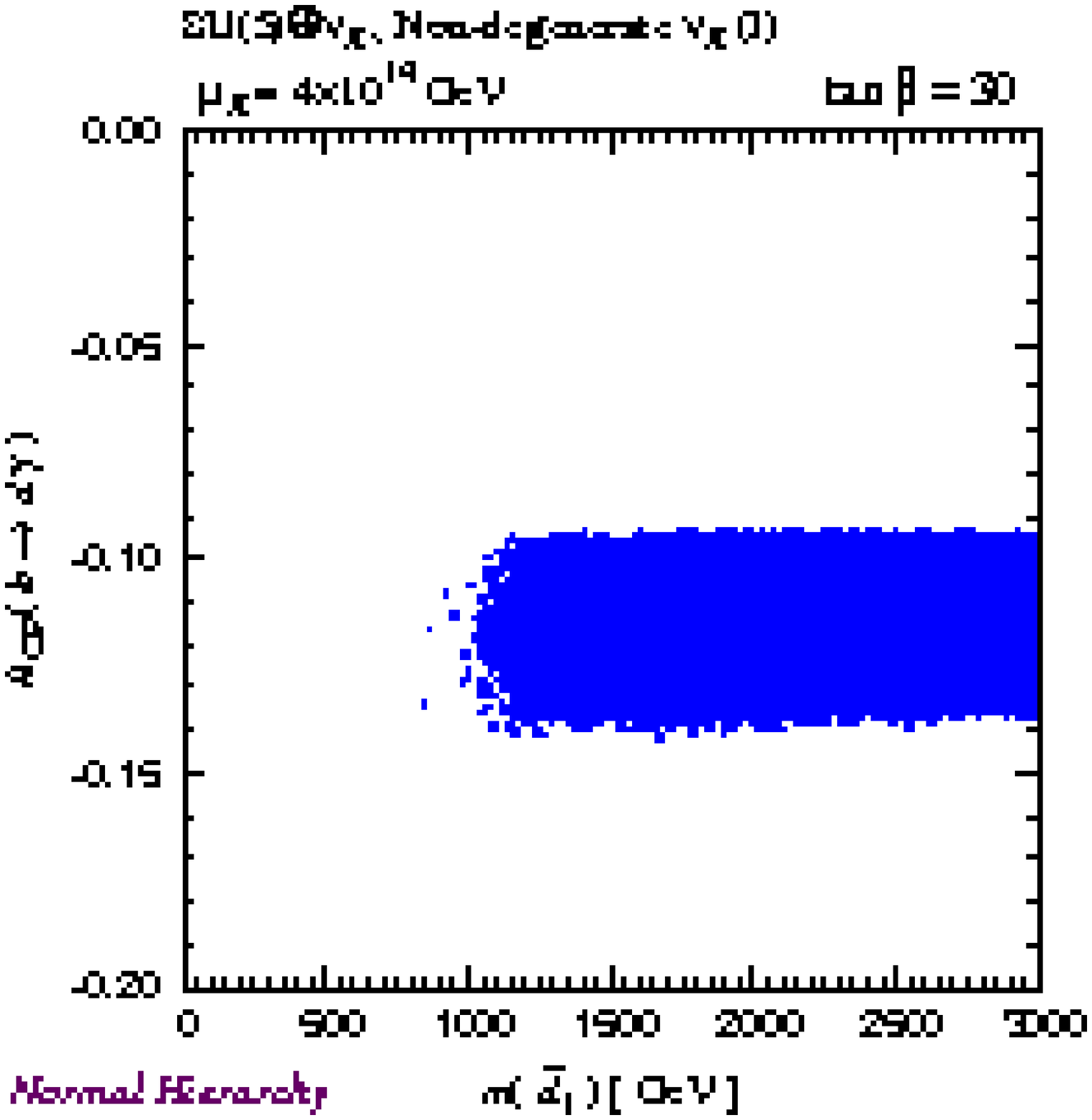} &
\includegraphics[scale=.23]{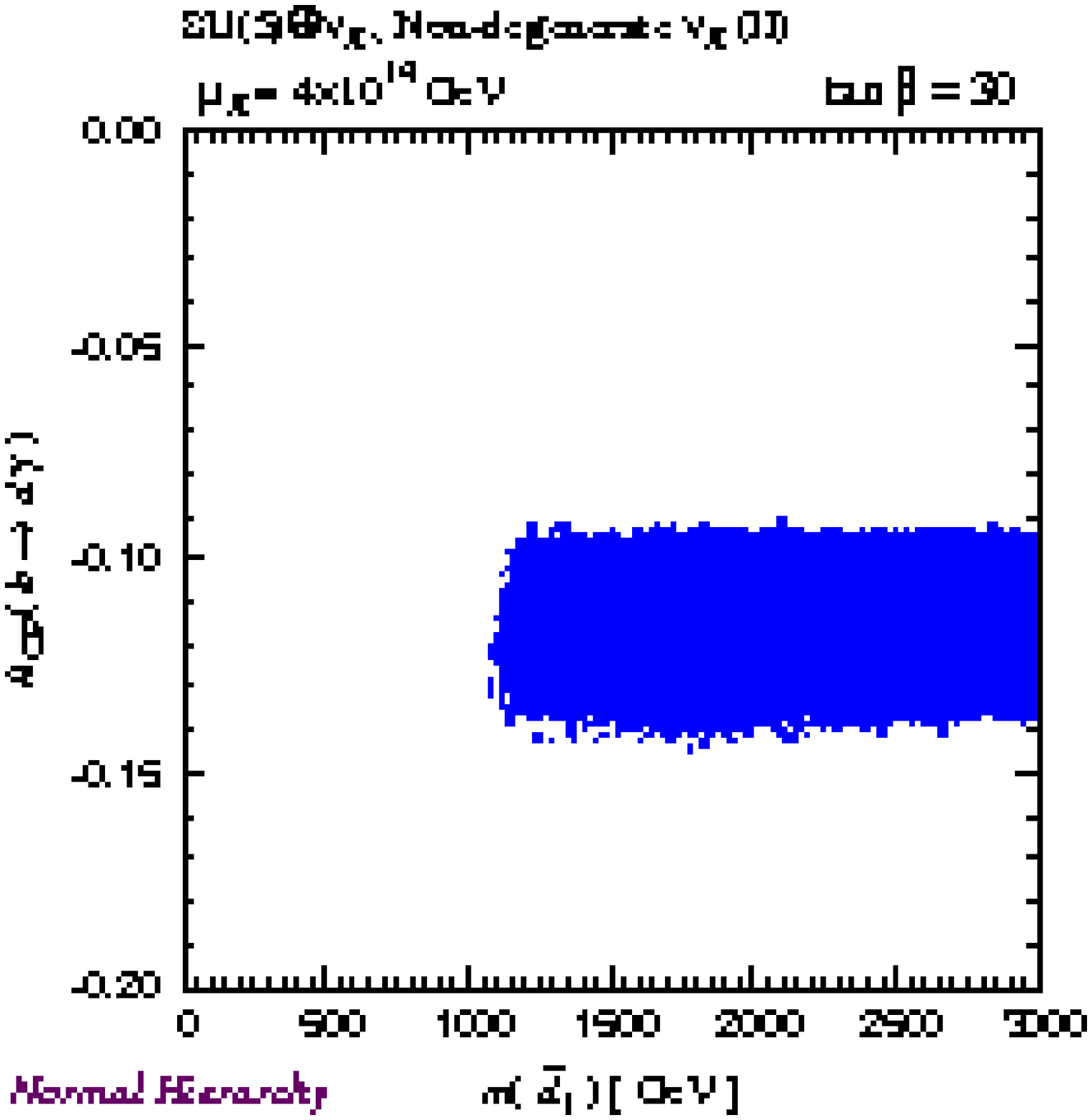} &
\includegraphics[scale=.23]{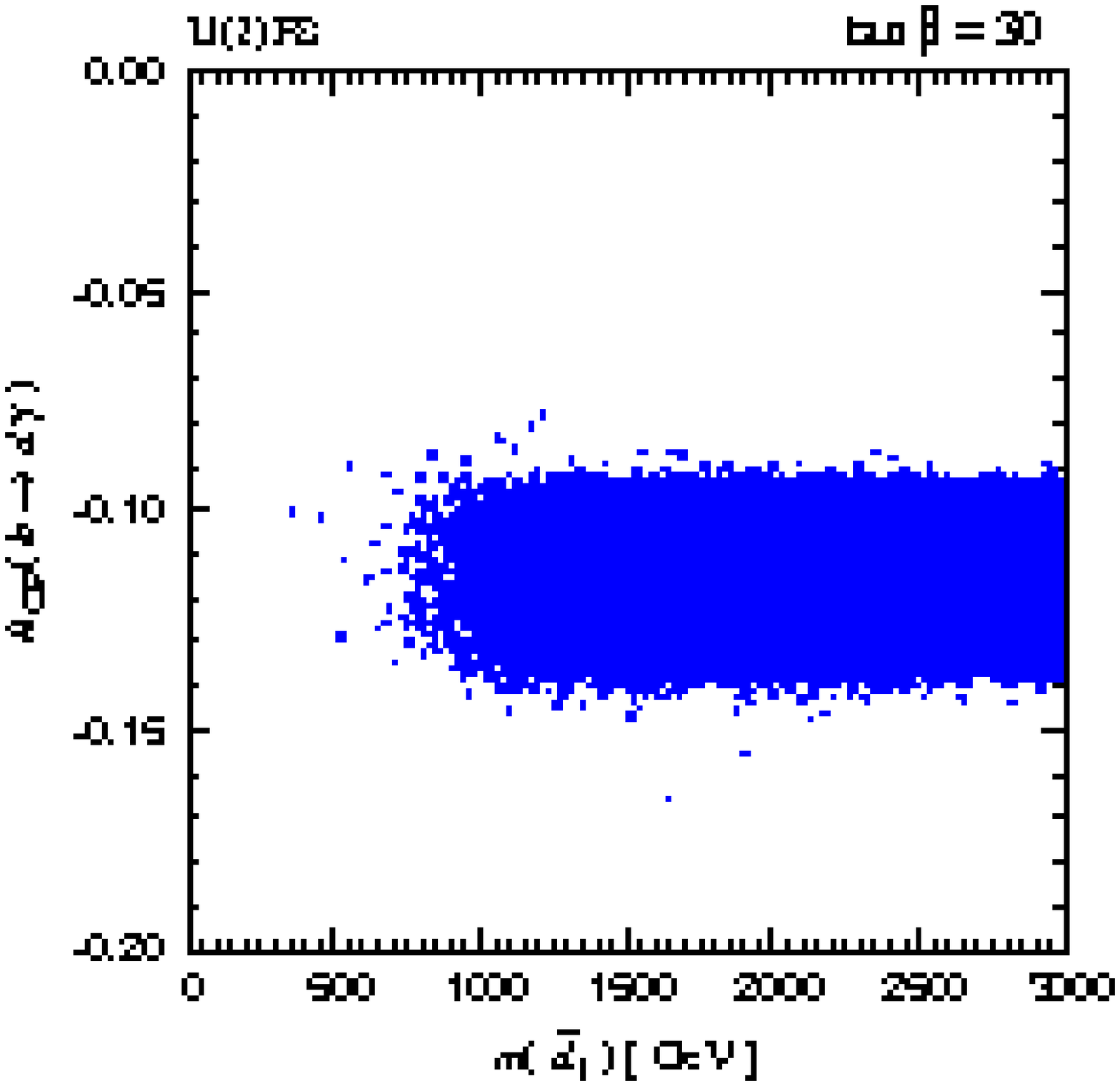} &
\\
(e) & (f) & (g) &
\end{tabular}
\caption{%
The direct CP asymmetry in $b\to d\gamma$ as functions of
$m(\tilde{d}_1)$ for the same parameter sets as those for
Fig.~\ref{fig:md-ACP-bsg}.
}
\label{fig:md-ACP-bdg}
\end{figure}

\begin{figure}[htbp]
\begin{tabular}{cccc}
\includegraphics[scale=.23]{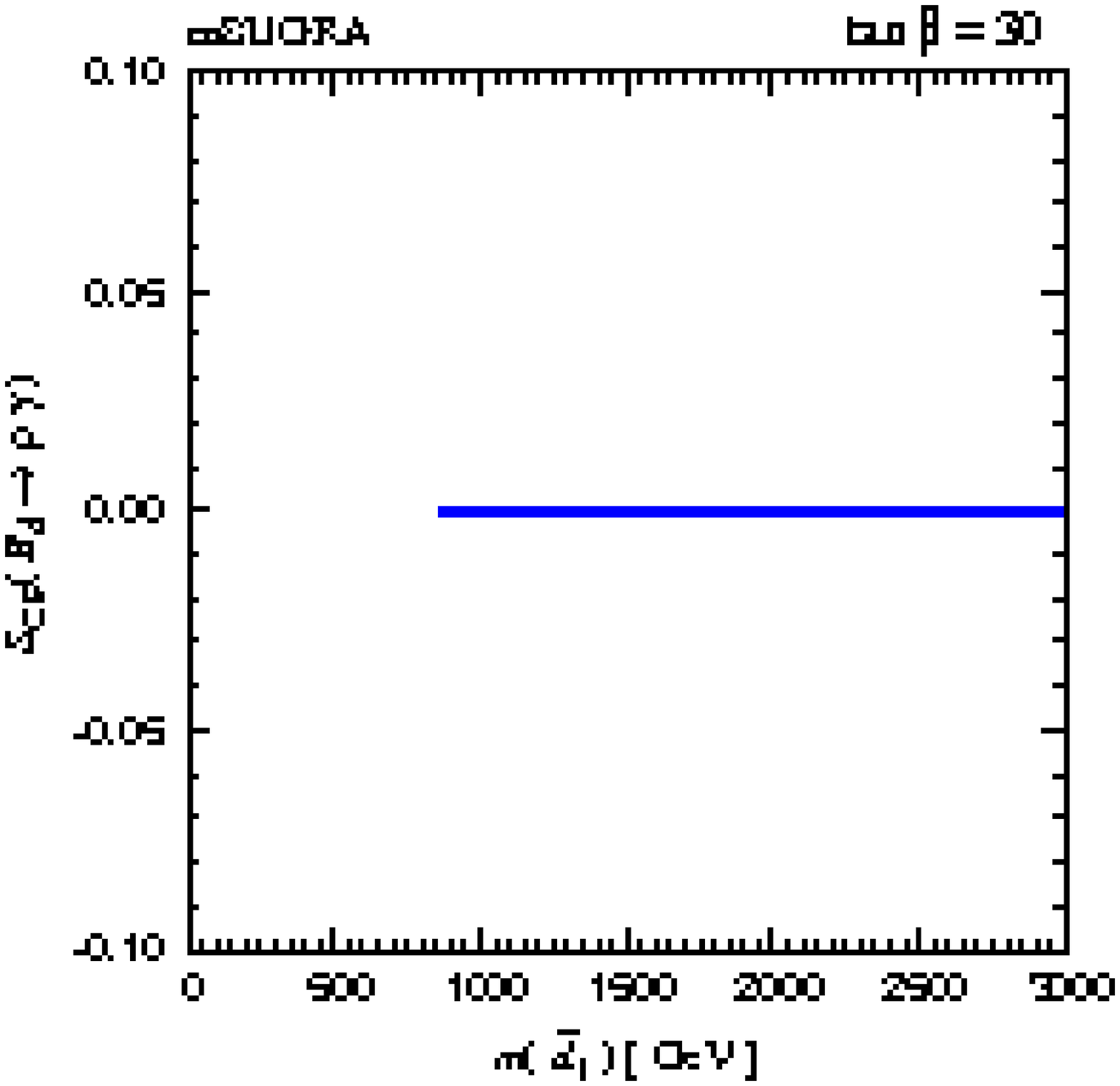} &
\includegraphics[scale=.23]{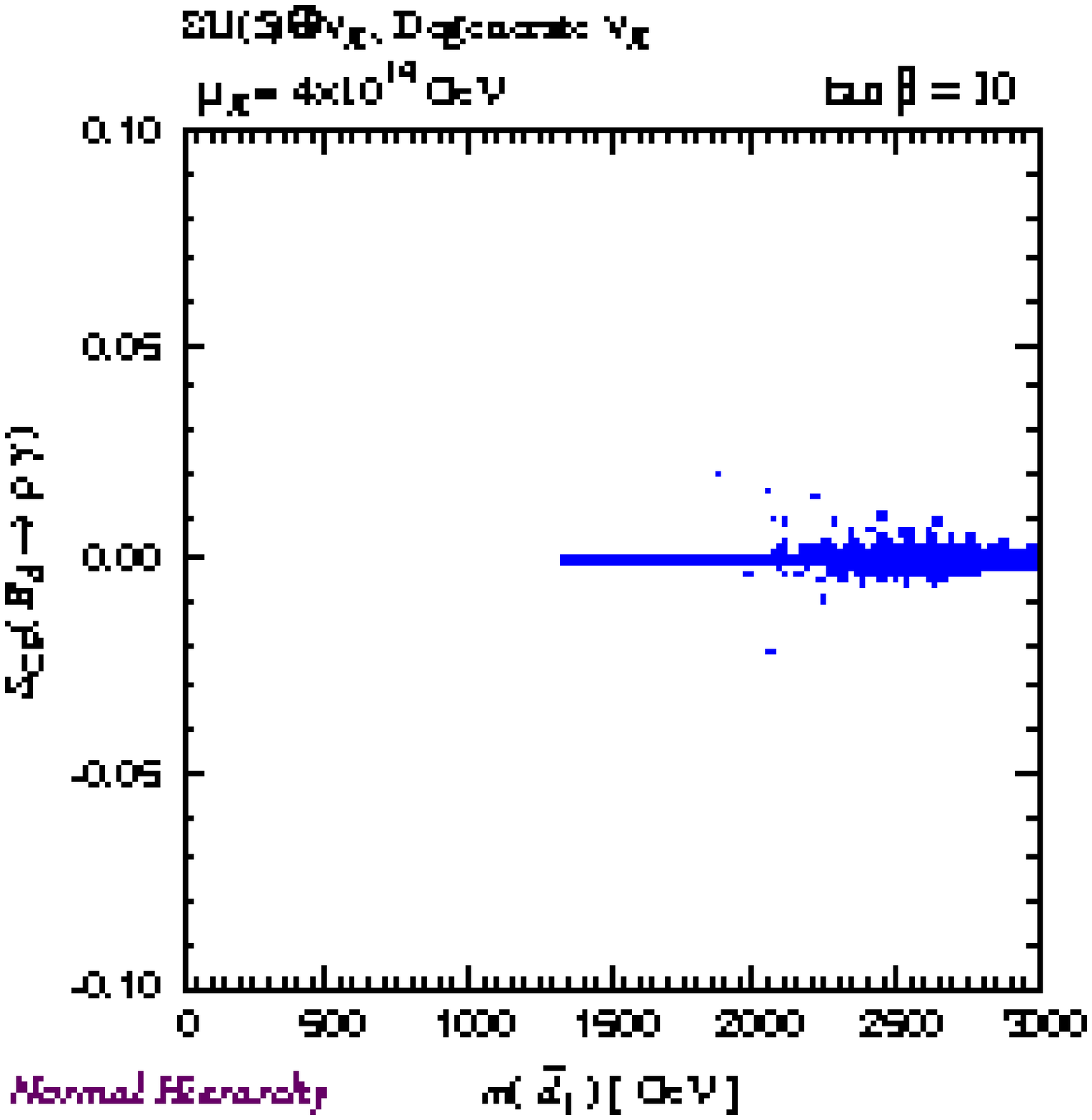} &
\includegraphics[scale=.23]{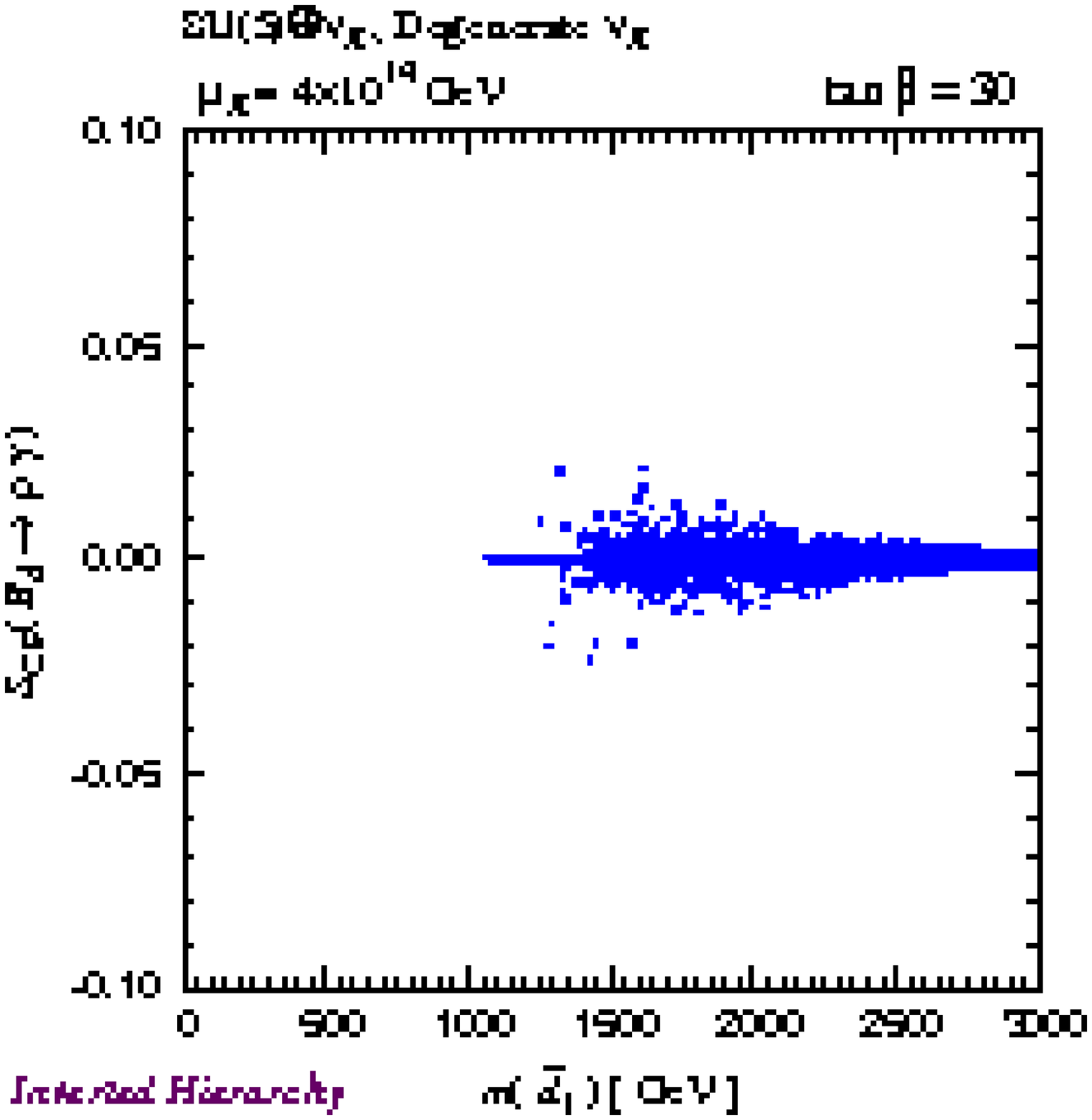} &
\includegraphics[scale=.23]{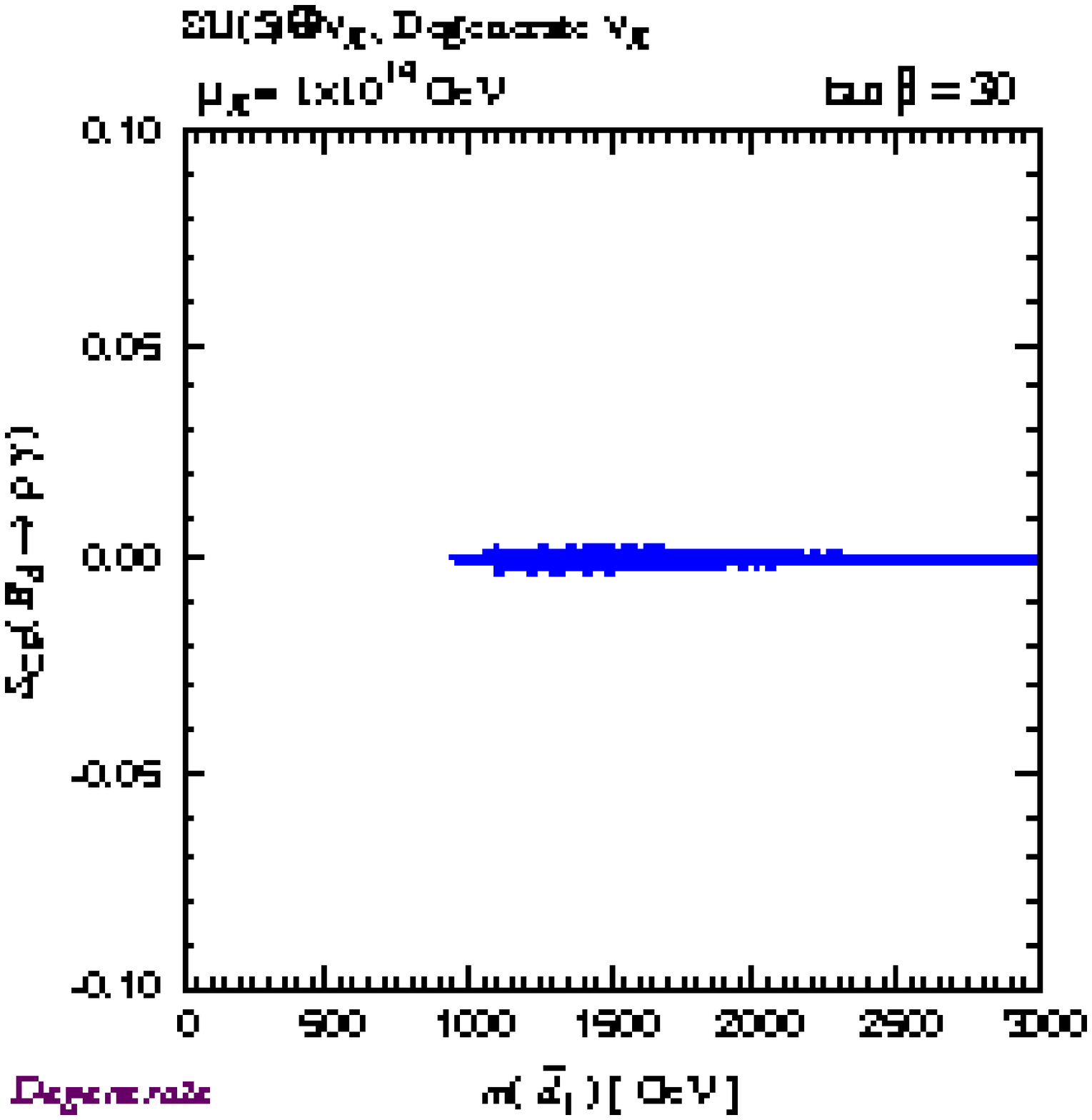}
\\
(a) & (b) & (c) & (d)
\\
\includegraphics[scale=.23]{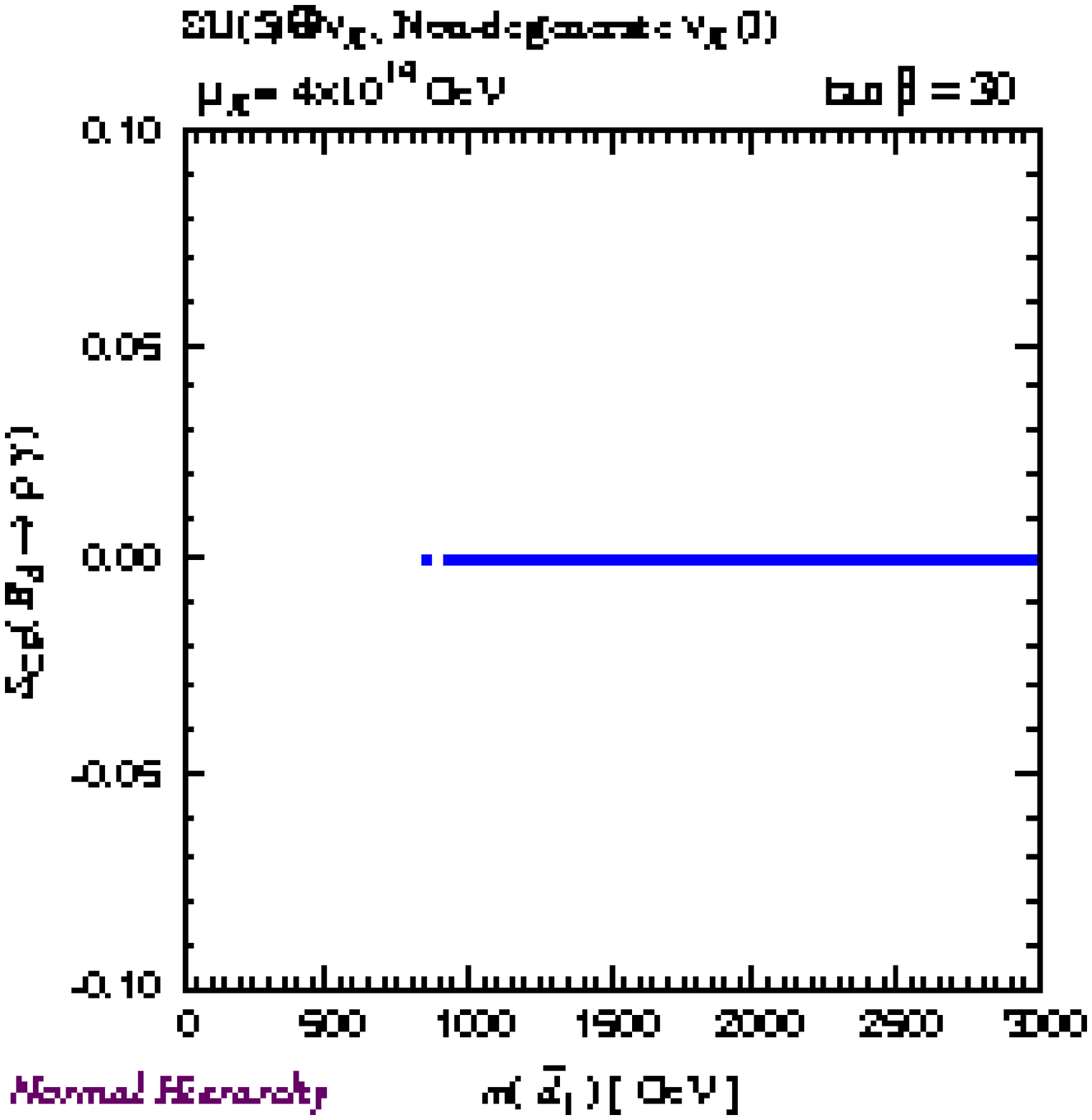} &
\includegraphics[scale=.23]{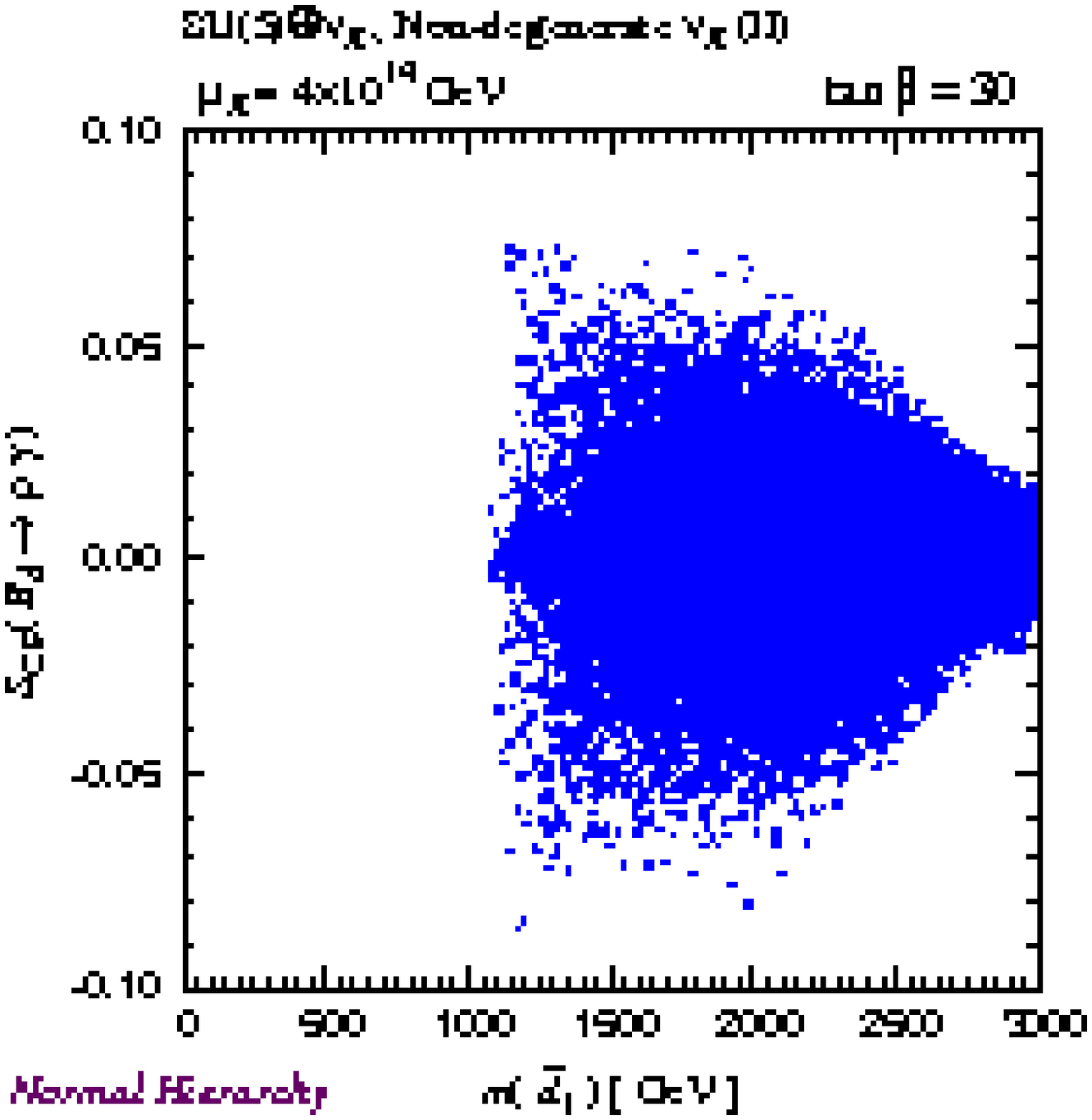} &
\includegraphics[scale=.23]{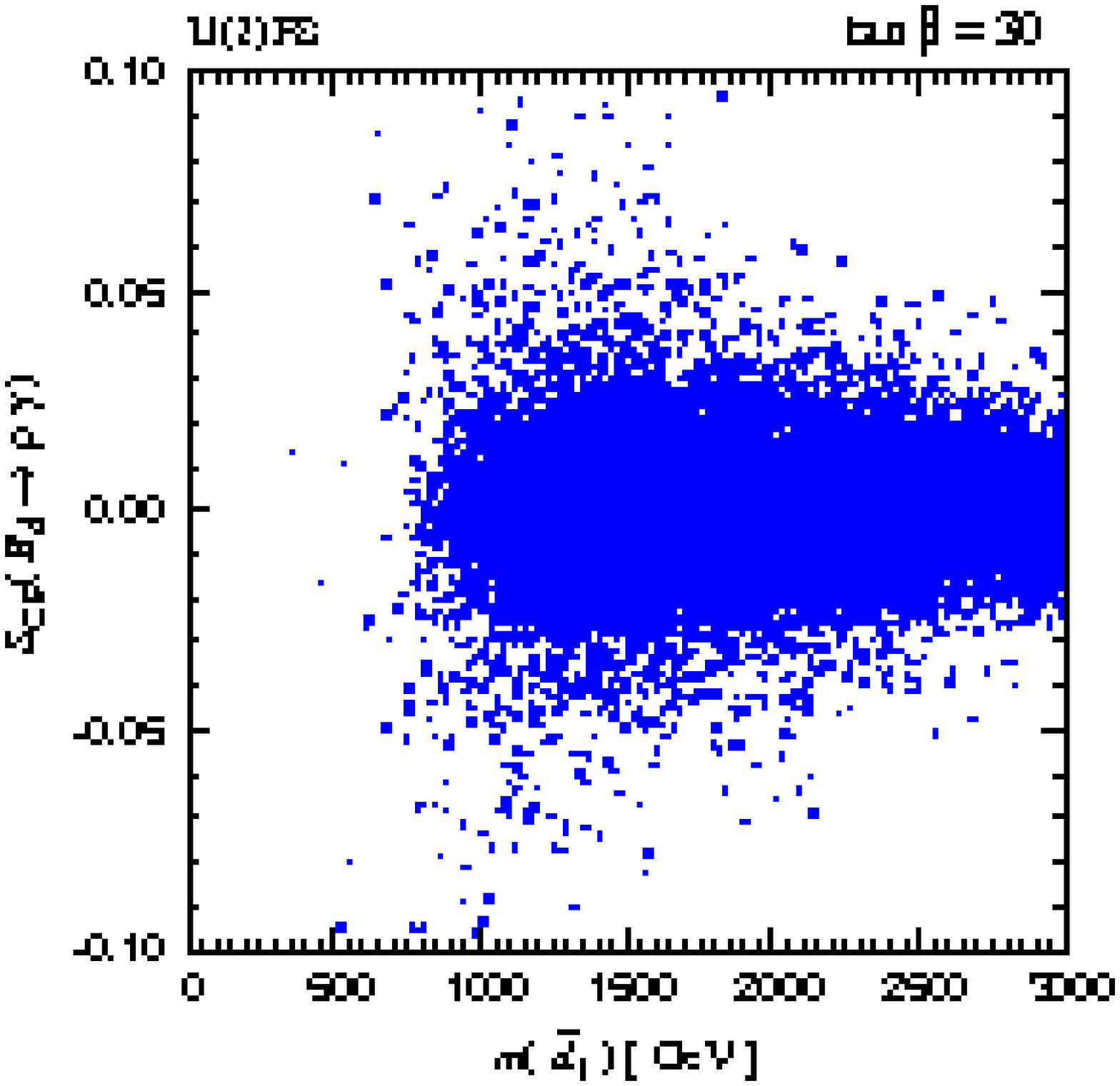} &
\\
(e) & (f) & (g) &
\end{tabular}
\caption{%
The mixing-induced CP asymmetry in $B_d\to \rho\gamma$ as functions of
$m(\tilde{d}_1)$ for the same parameter sets as those for
Fig.~\ref{fig:md-ACP-bsg}.
}
\label{fig:md-SCP-bdg}
\end{figure}

\begin{figure}
\begin{tabular}{cccc}
\includegraphics[scale=.23]{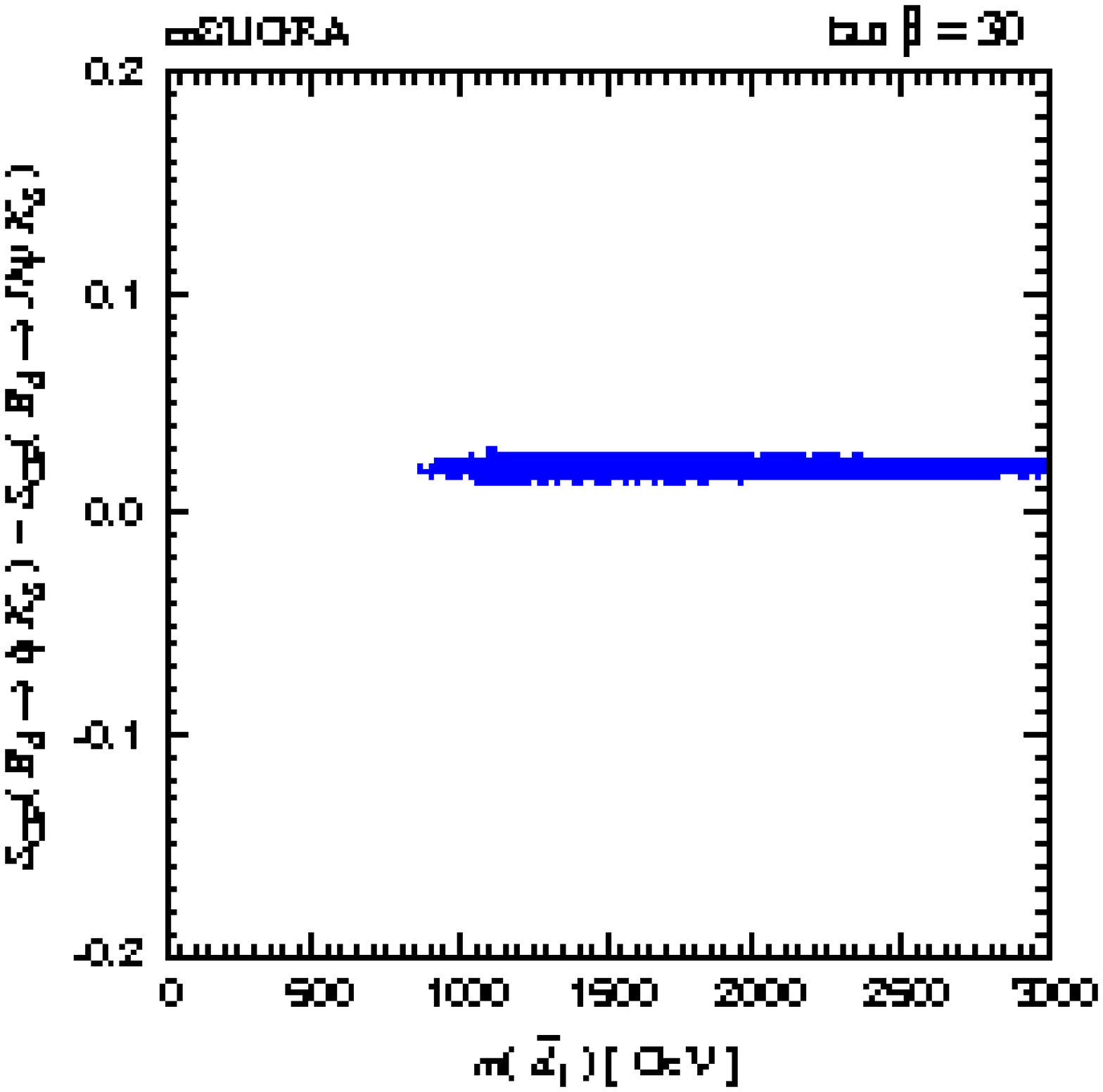} &
\includegraphics[scale=.23]{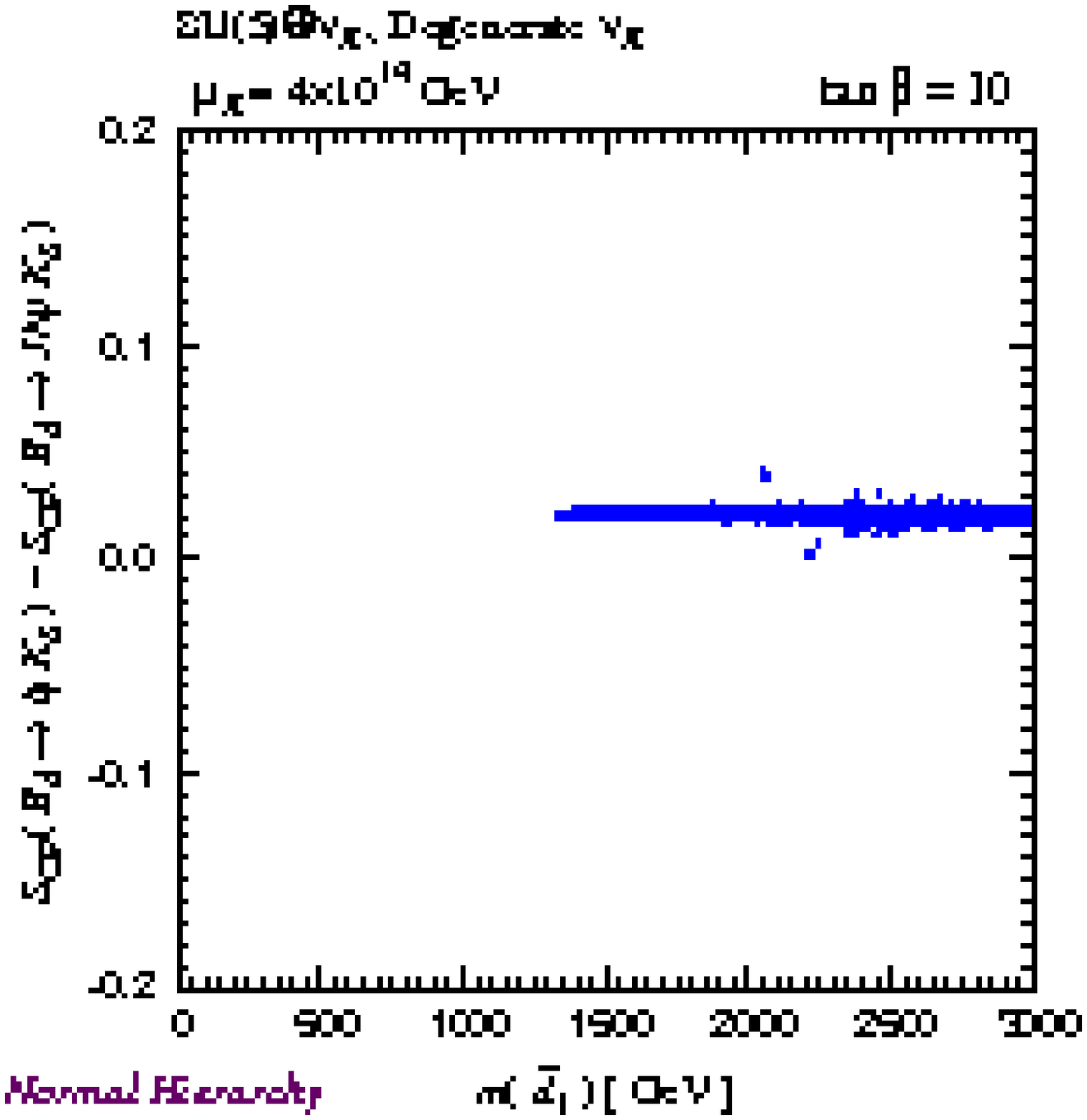} &
\includegraphics[scale=.23]{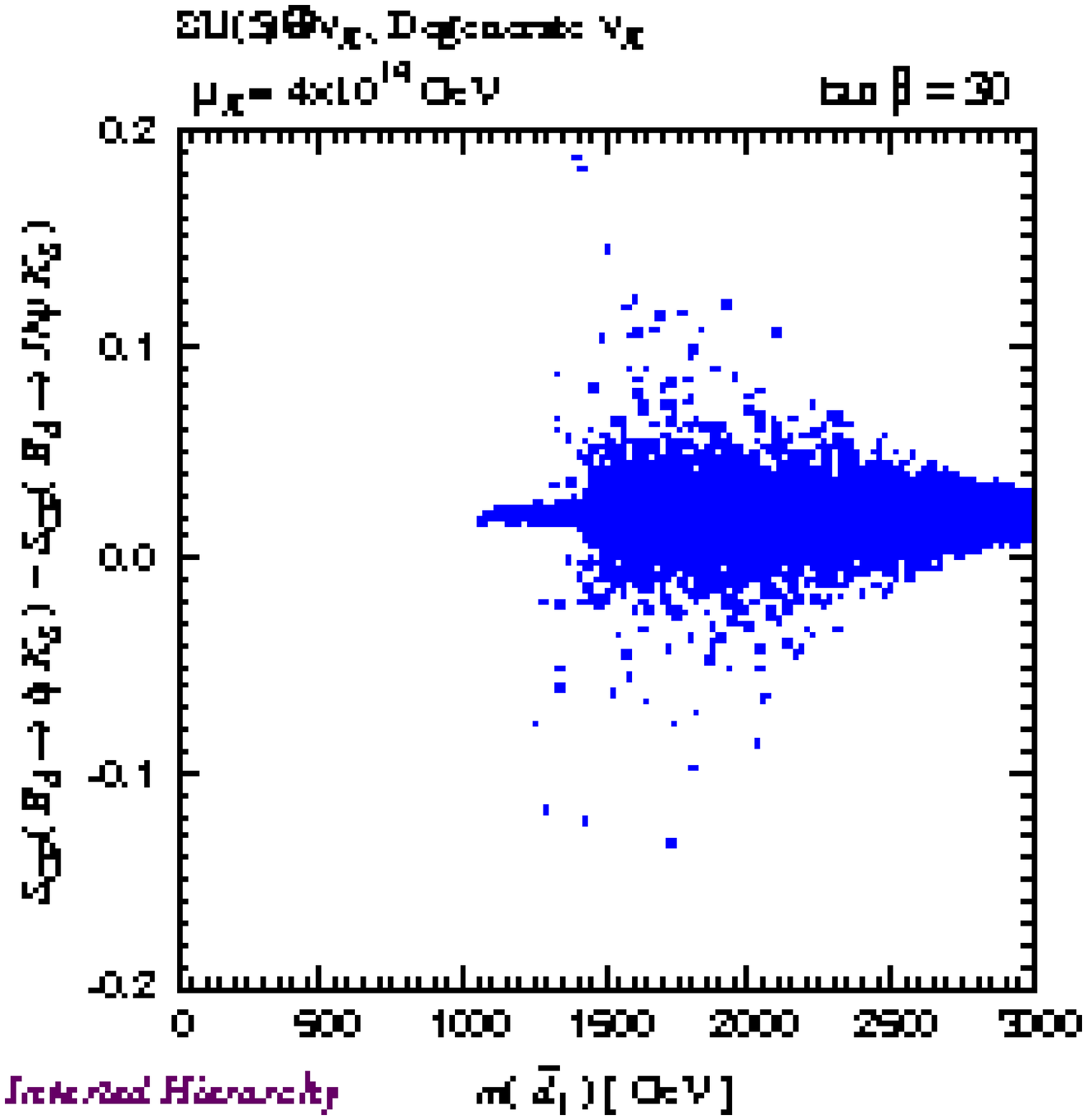} &
\includegraphics[scale=.23]{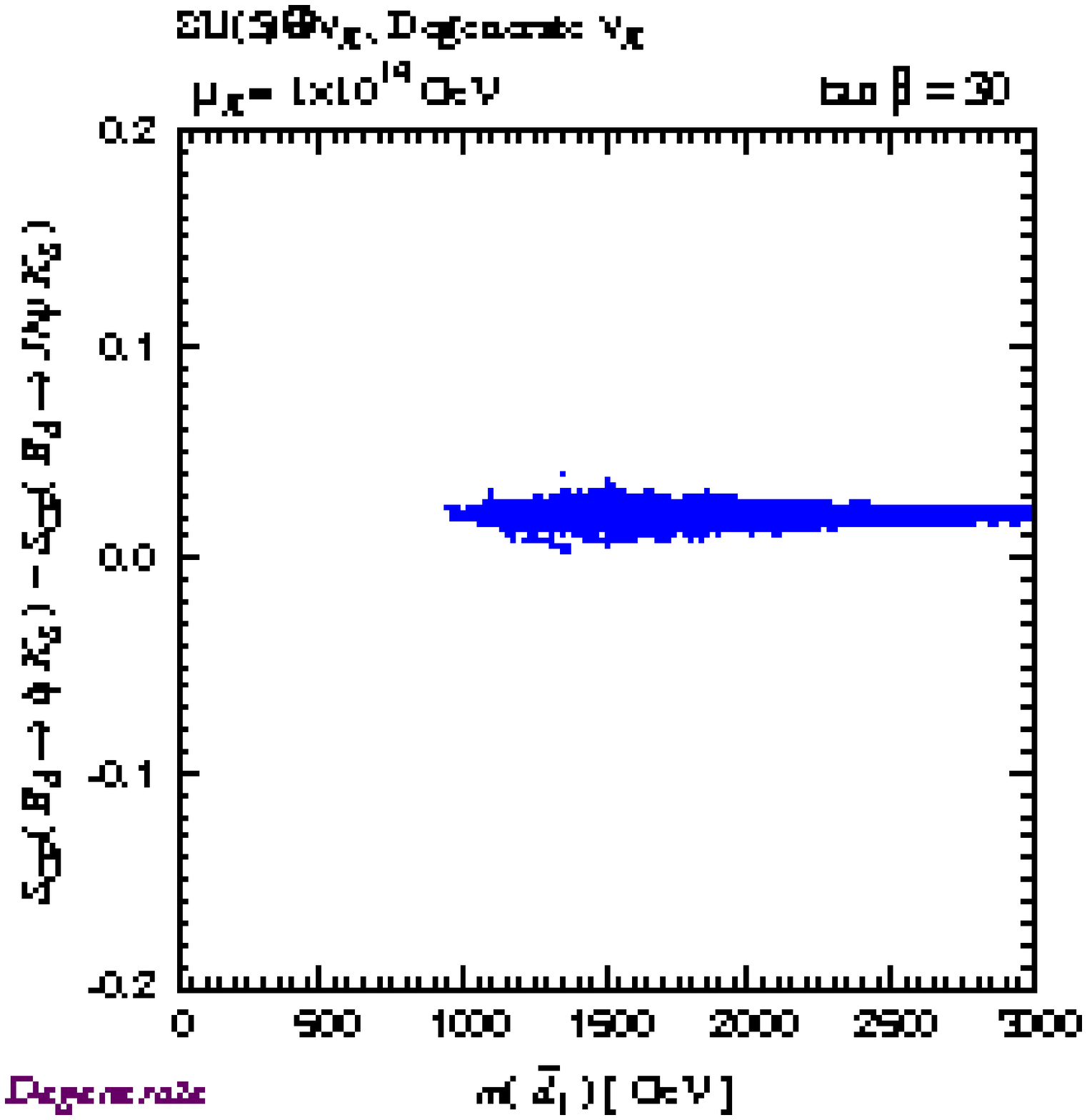}
\\
(a) & (b) & (c) & (d)
\\
\includegraphics[scale=.23]{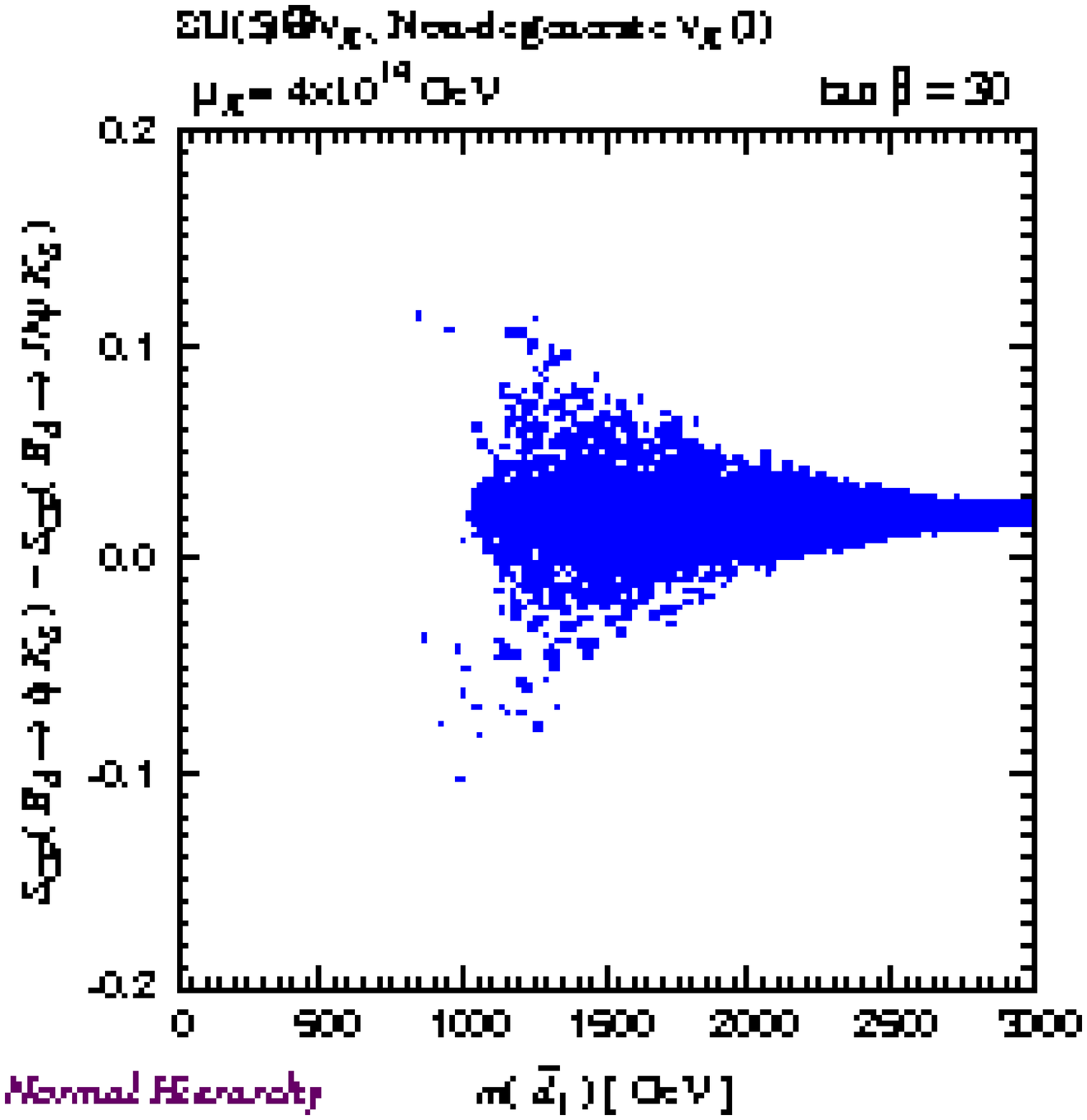} &
\includegraphics[scale=.23]{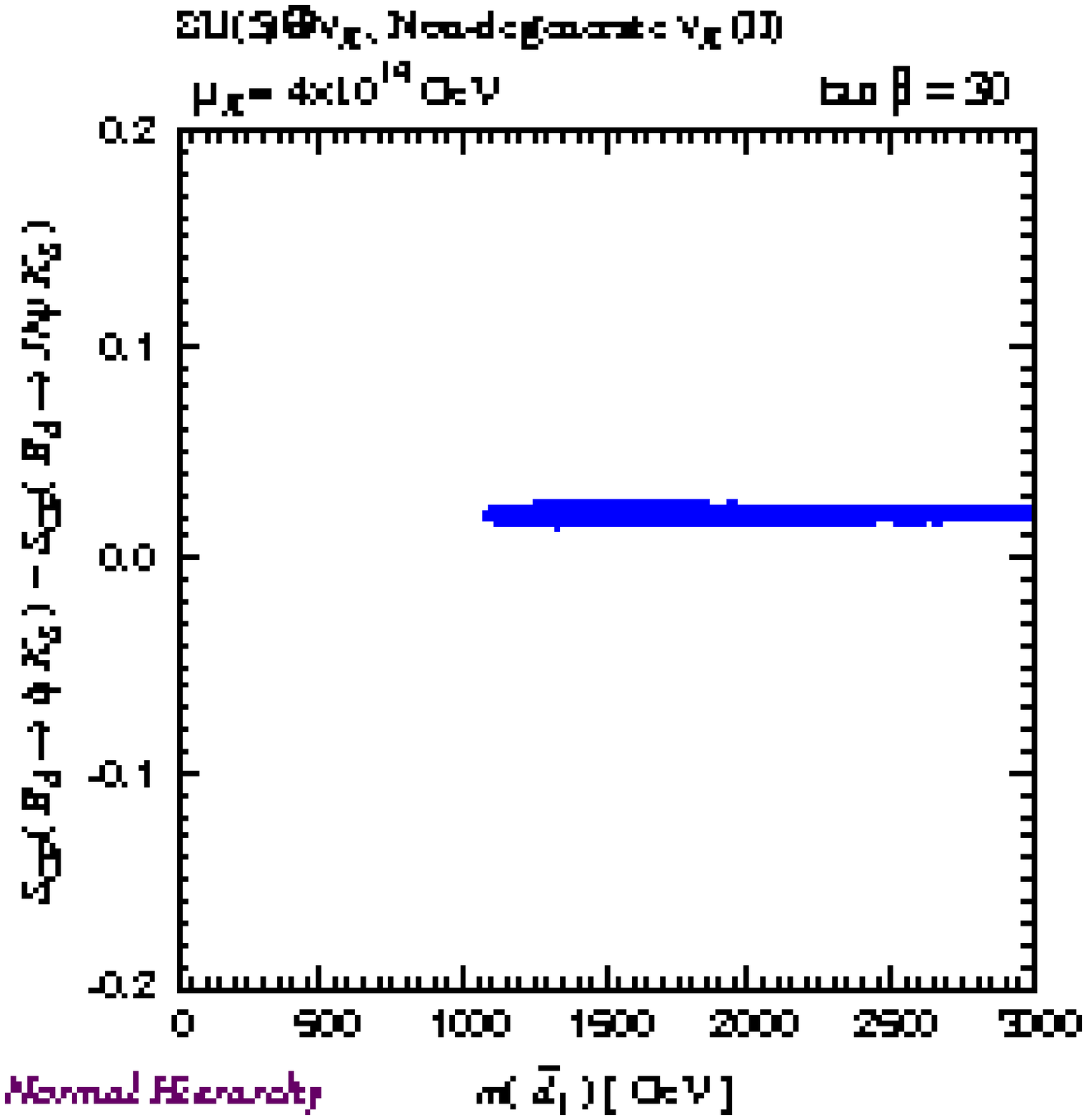} &
\includegraphics[scale=.23]{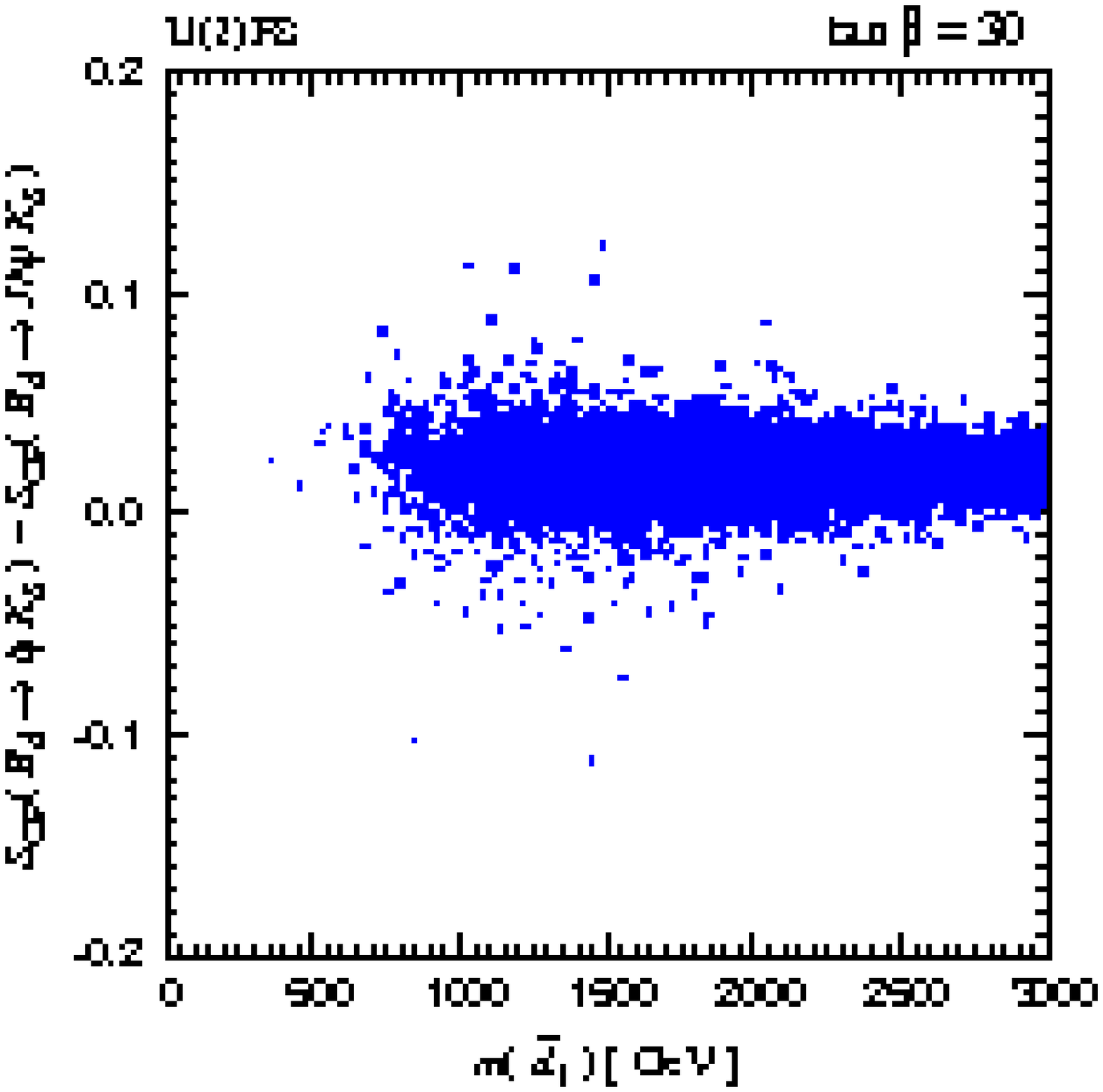} & \\
(e) & (f) & (g) &
\end{tabular}
\caption{%
The difference between mixing-induced CP asymmetries in the
$B_d\to \phi K_S$ and $B_d\to J/\psi K_S$ modes as functions of
$m(\tilde{d}_1)$ for the same parameter sets as those in
Fig.~\ref{fig:md-ACP-bsg}.
}
\label{fig:md-BdphiK}
\end{figure}

\begin{figure}[htbp]
\begin{tabular}{cccc}
\includegraphics[scale=.23]{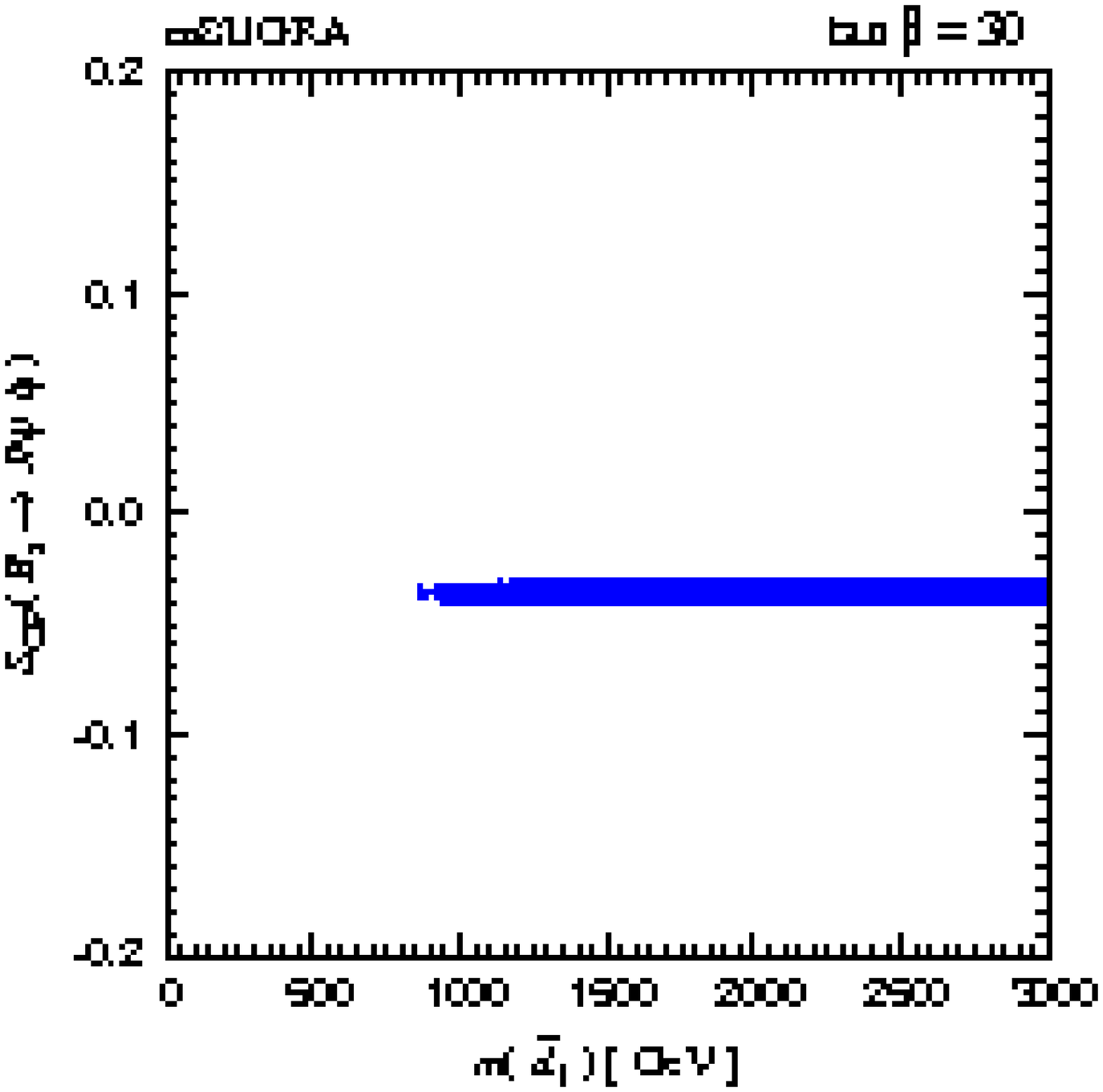} &
\includegraphics[scale=.23]{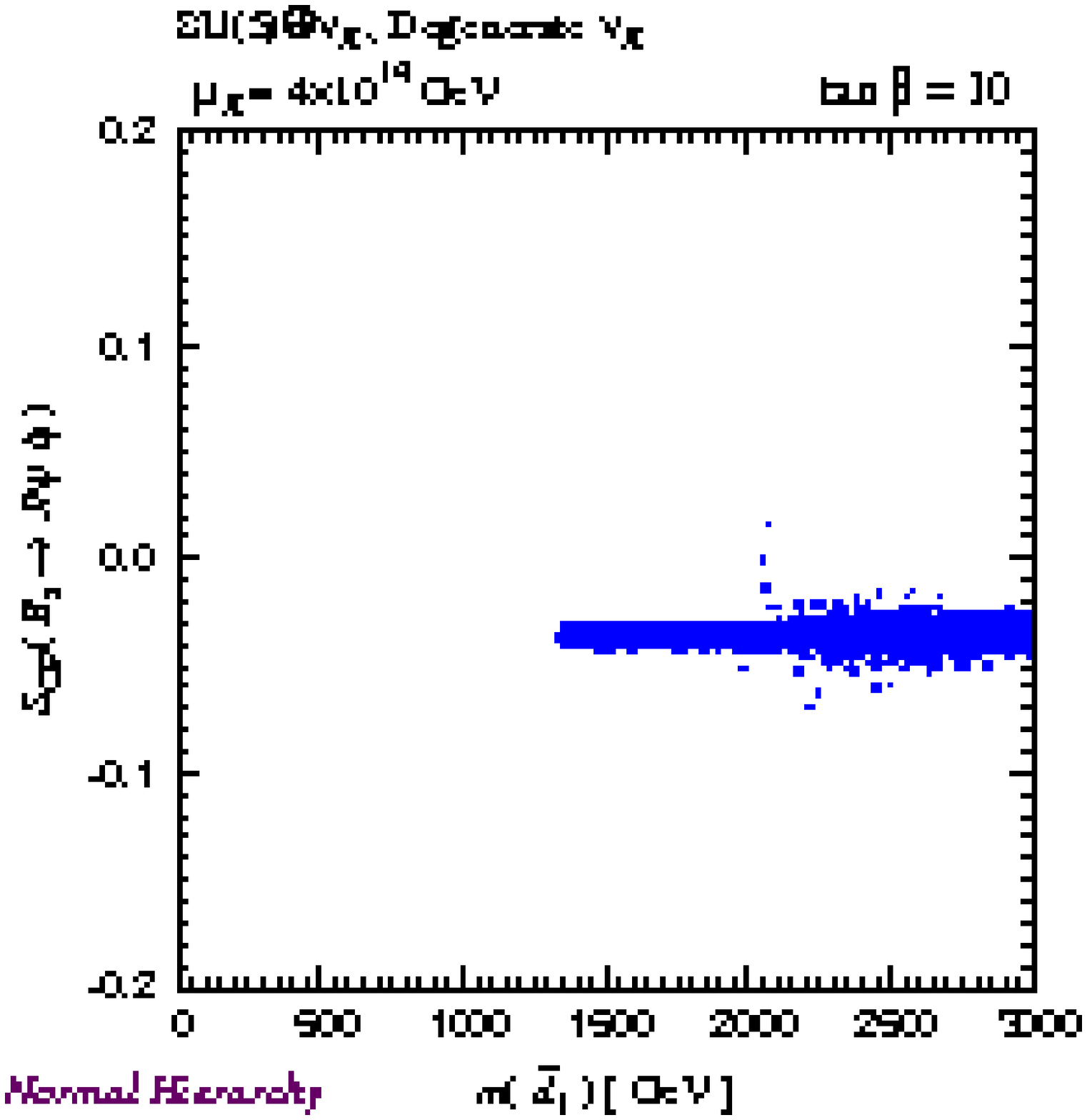} &
\includegraphics[scale=.23]{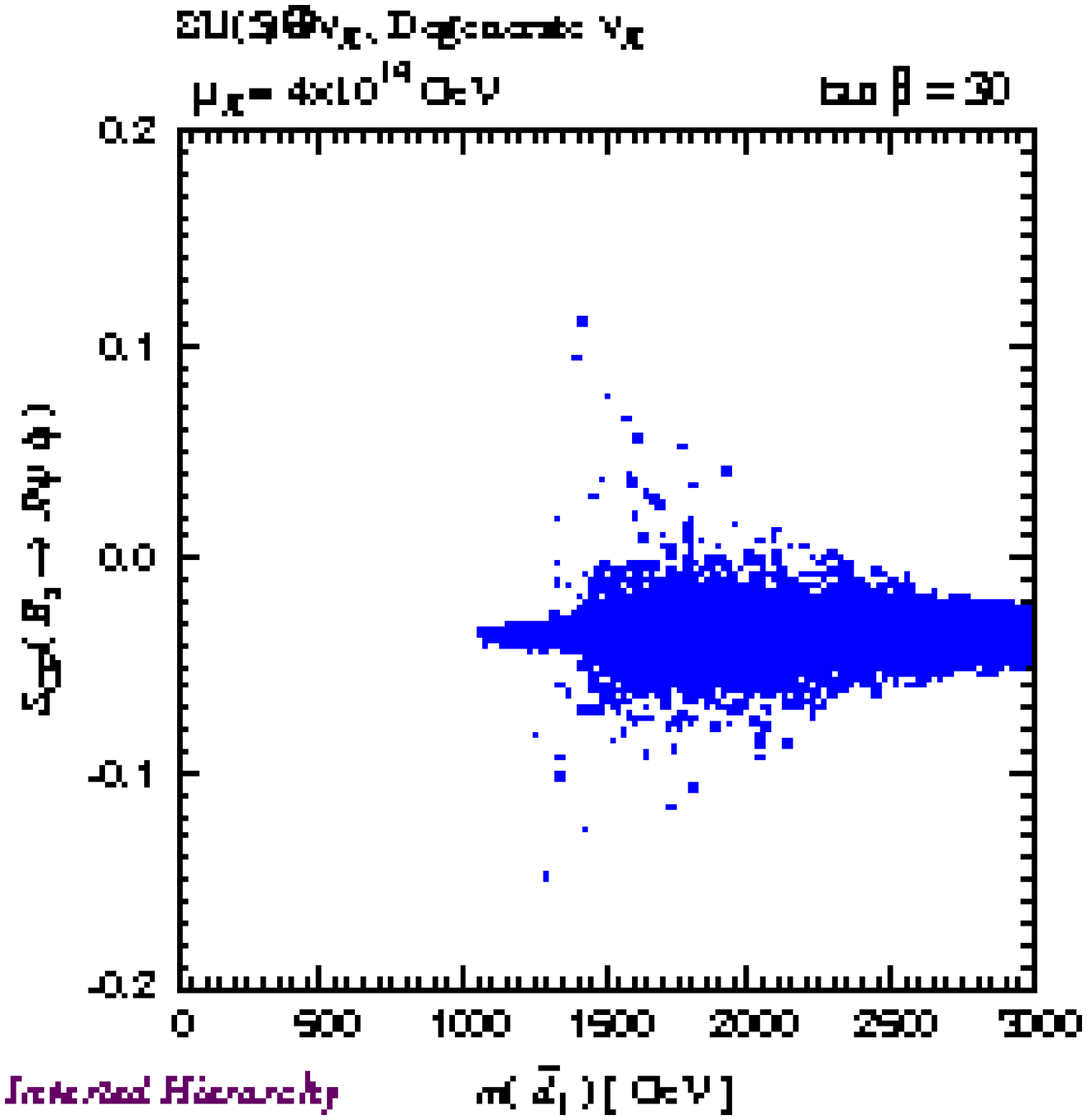} &
\includegraphics[scale=.23]{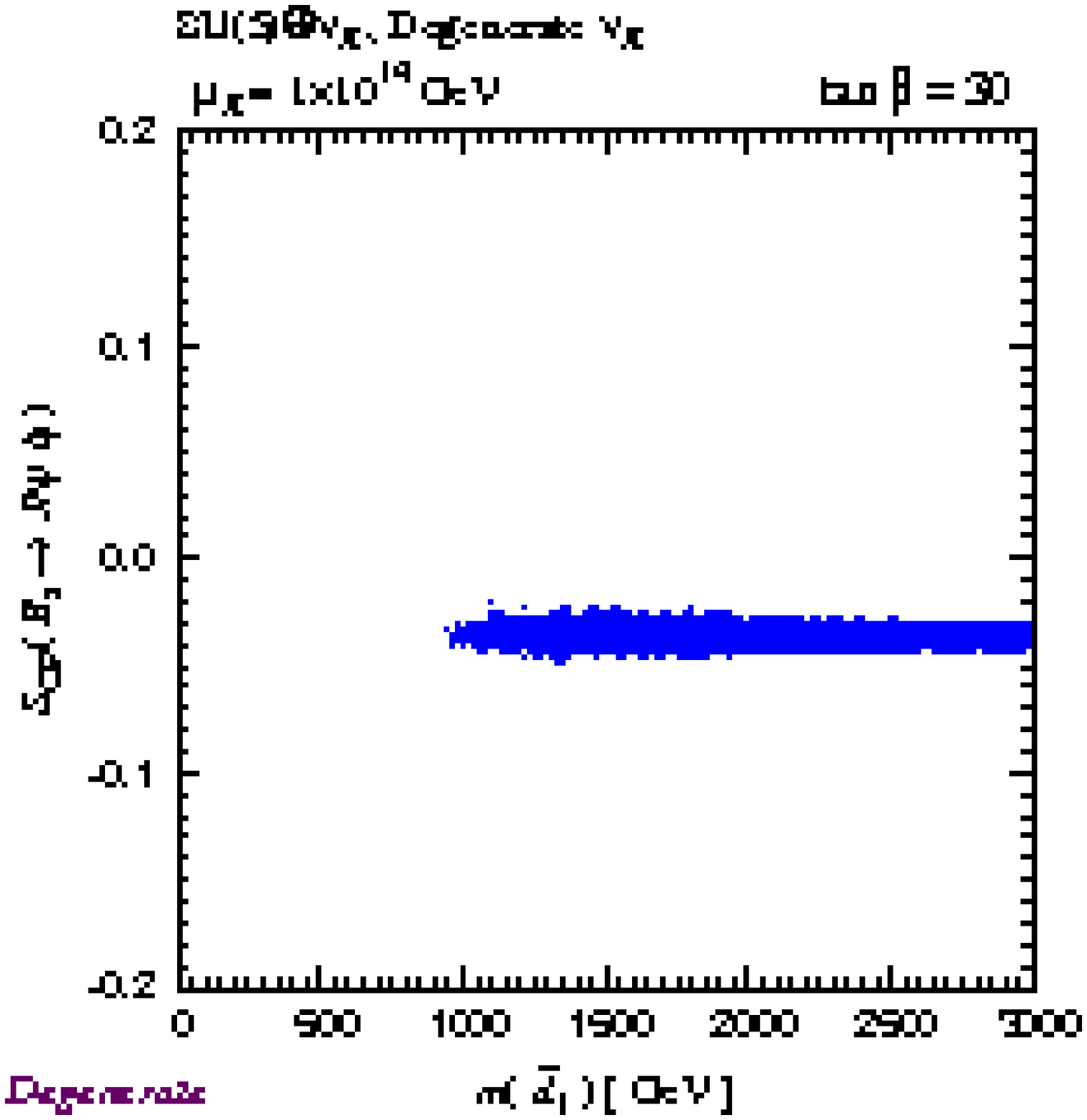}
\\
(a) & (b) & (c) & (d)
\\
\includegraphics[scale=.23]{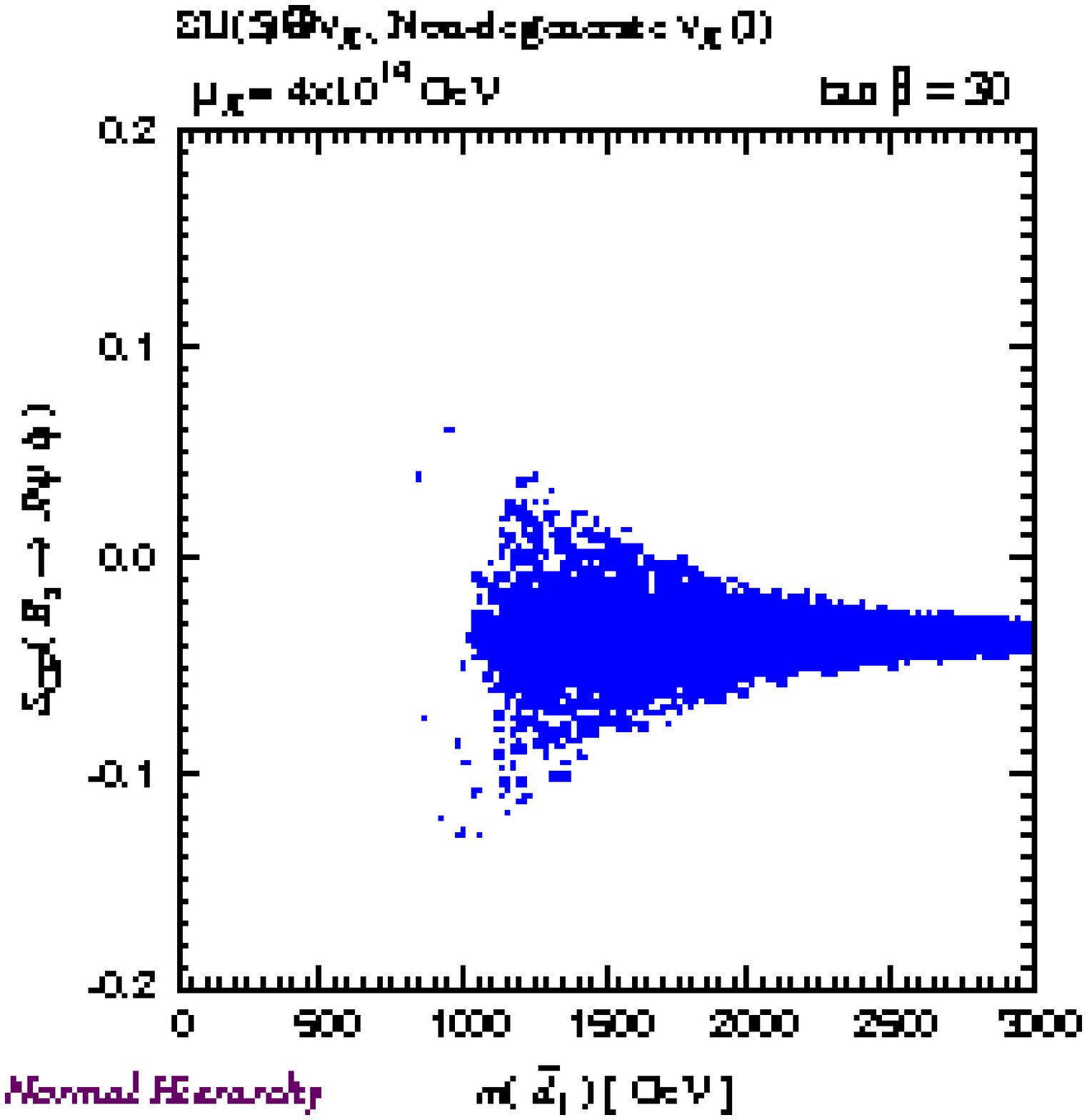} &
\includegraphics[scale=.23]{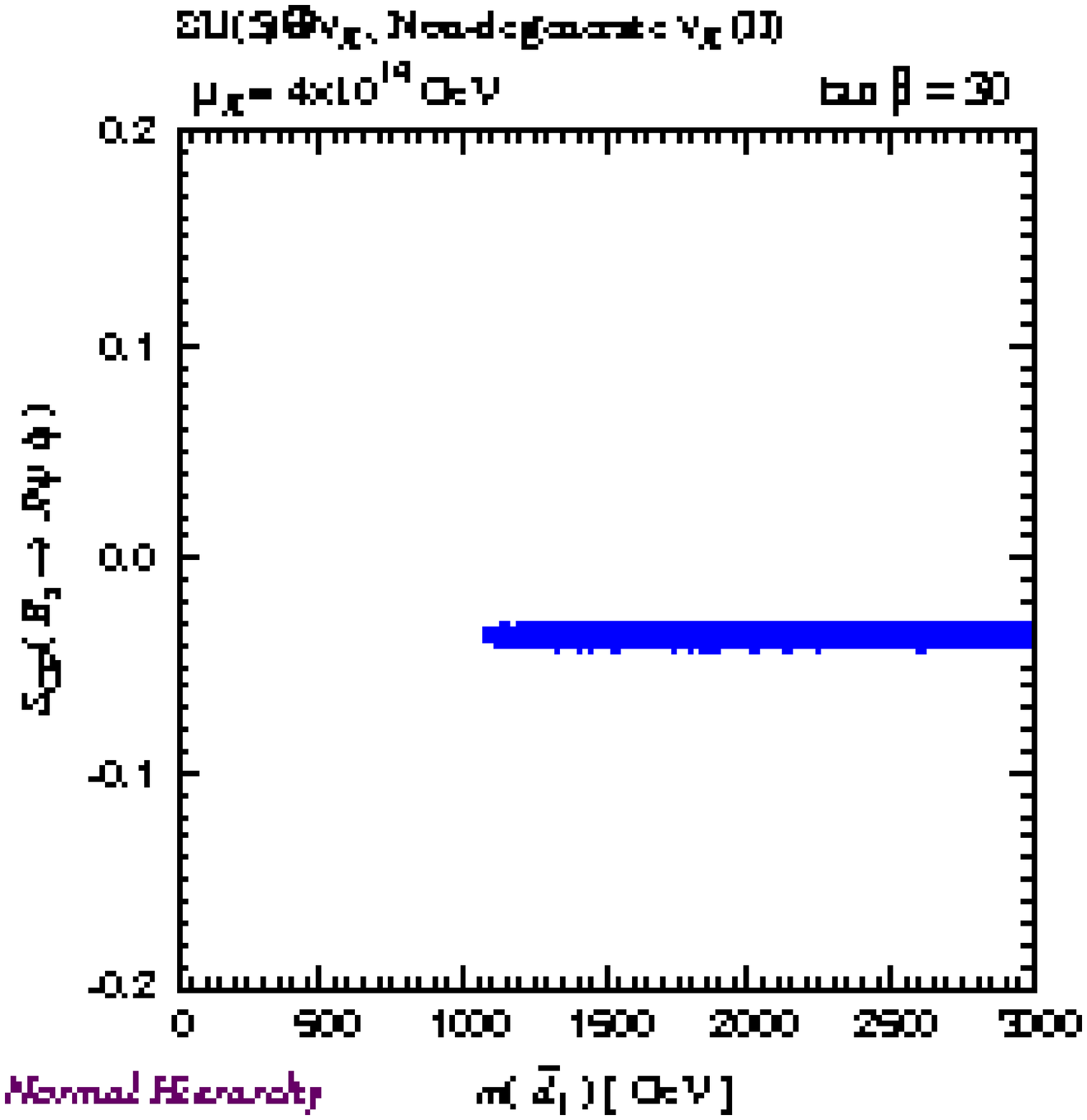} &
\includegraphics[scale=.23]{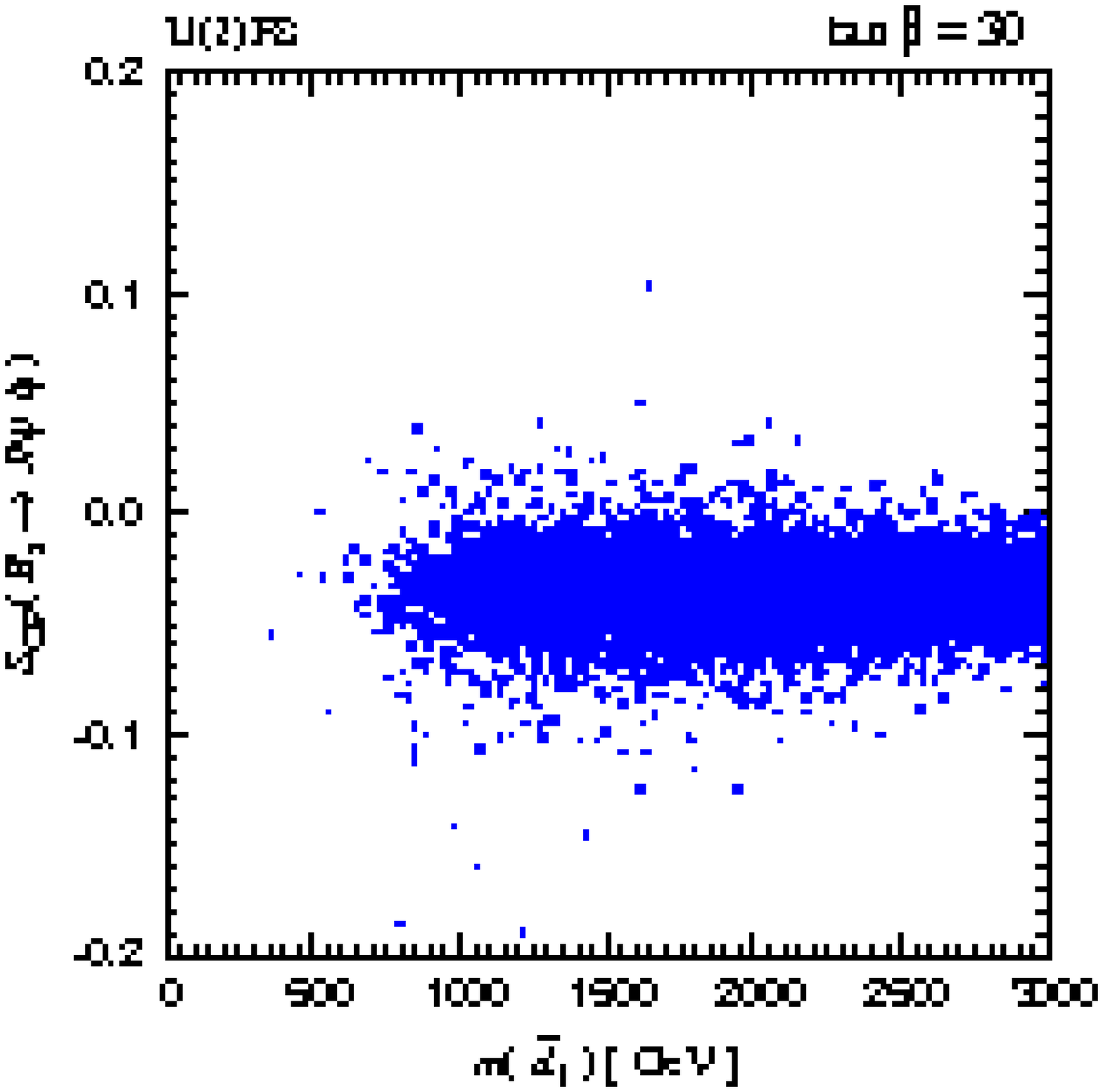} & \\
(e) & (f) & (g) &
\end{tabular}
\caption{%
Predicted value of the mixing-induced CP asymmetry in 
$B_s\to J/\psi \phi$ as a function
of $m(\tilde{d}_1)$ for the same parameter sets as those for
Fig.~\ref{fig:md-ACP-bsg}.
}
\label{fig:md-BsJphi}
\end{figure}

\begin{figure}
\begin{tabular}{cccc}
\includegraphics[scale=.23]{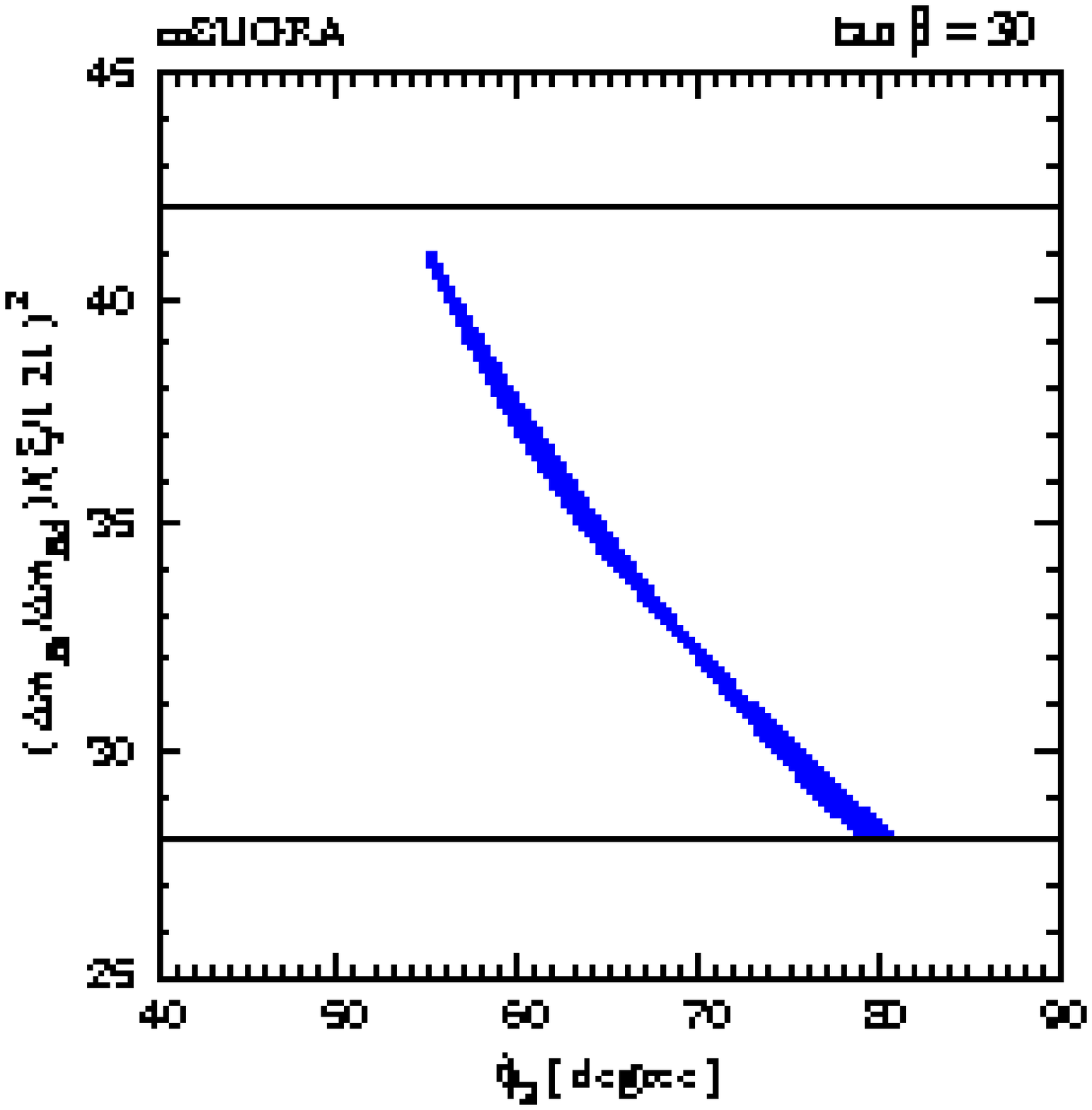} &
\includegraphics[scale=.23]{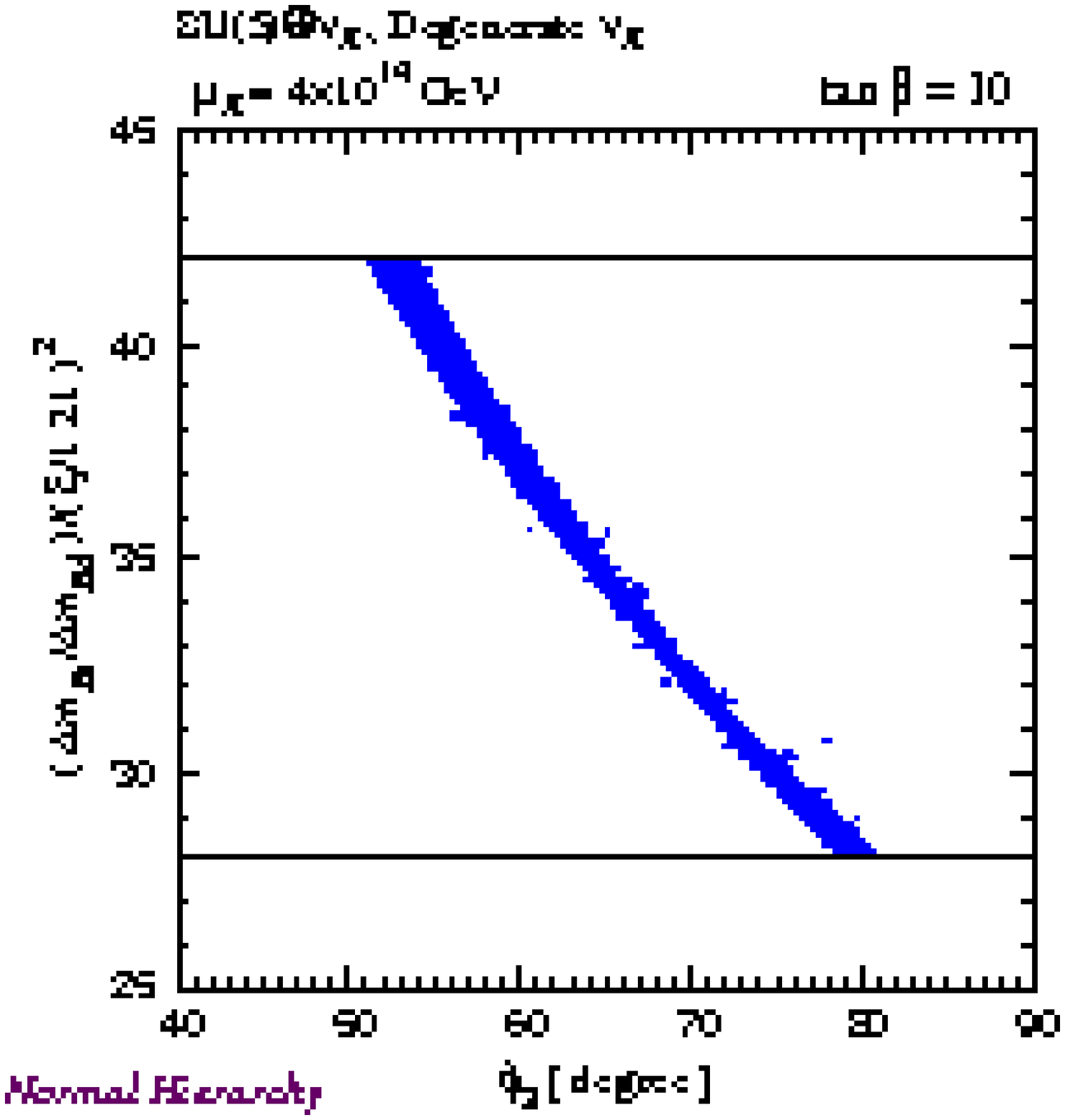} &
\includegraphics[scale=.23]{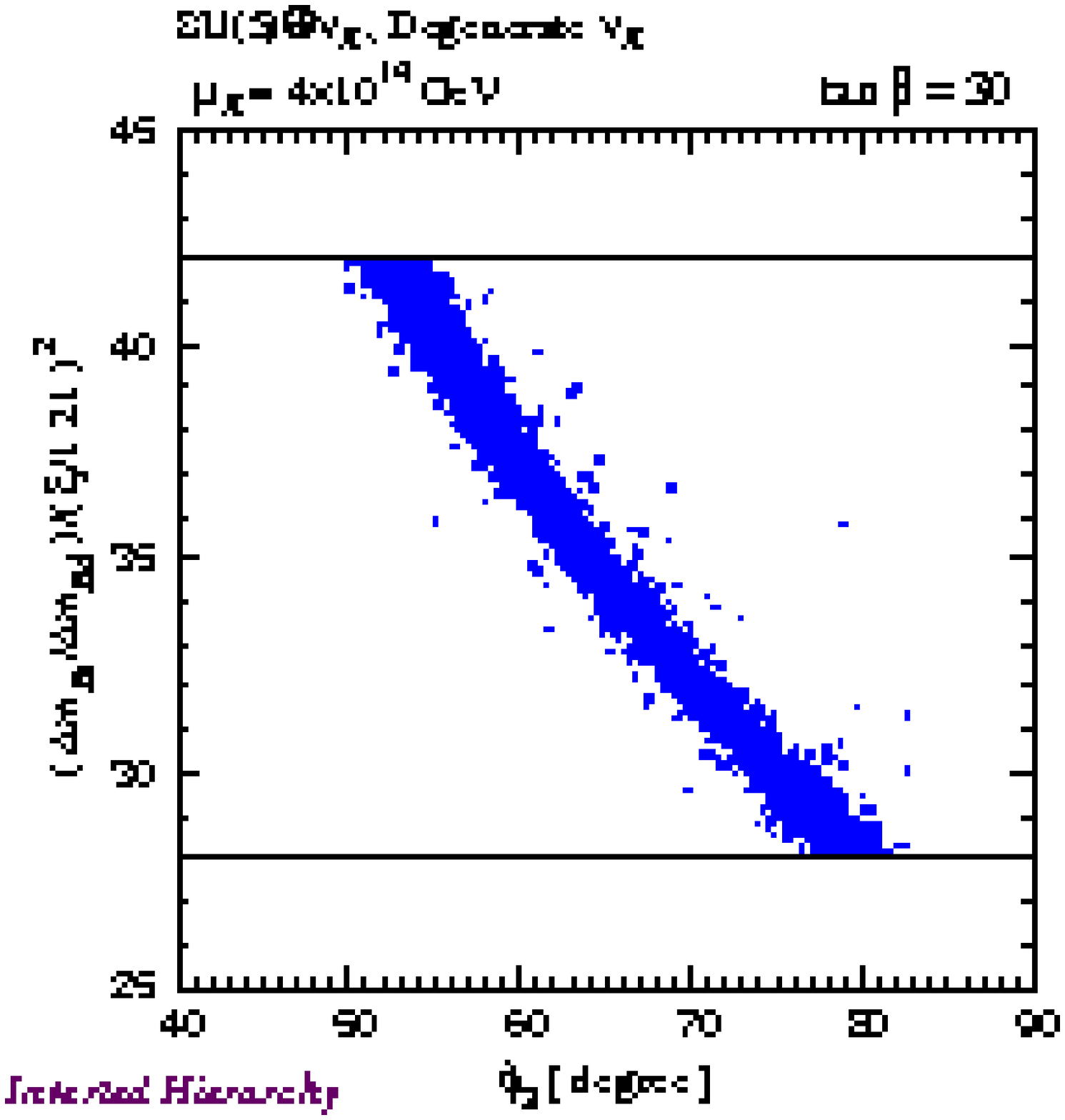} &
\includegraphics[scale=.23]{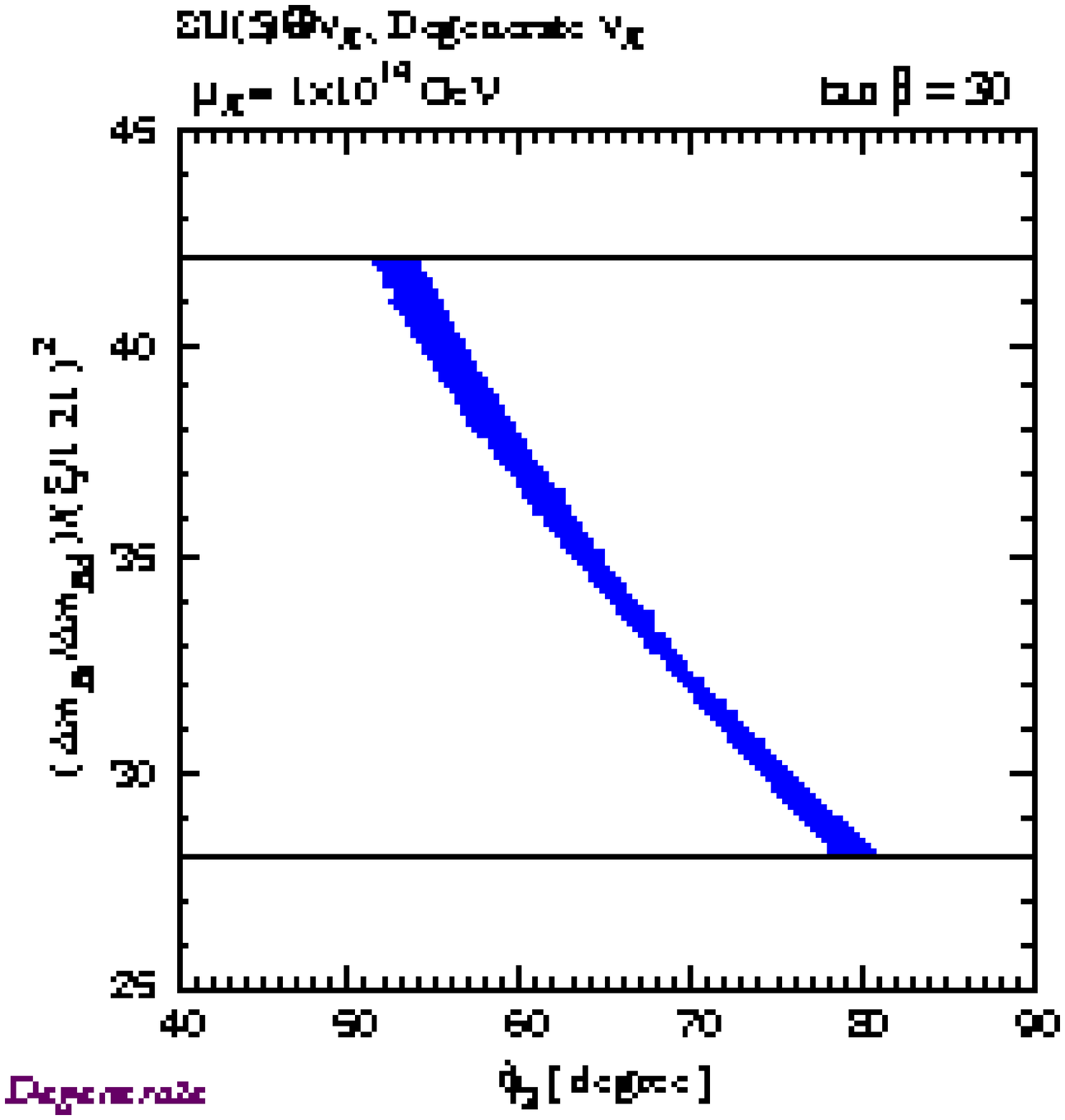}
\\
(a) & (b) & (c) & (d)
\\
\includegraphics[scale=.23]{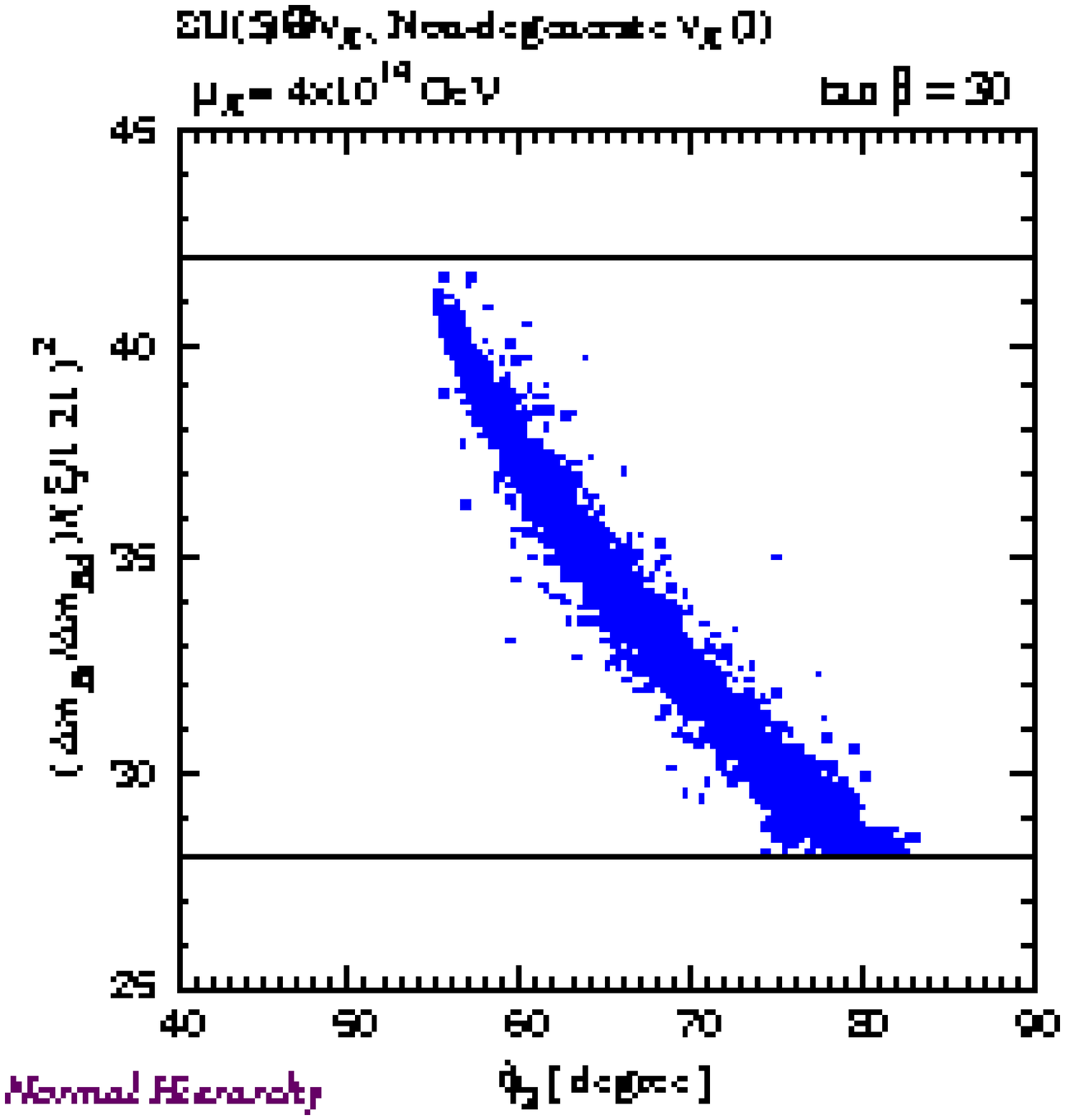} &
\includegraphics[scale=.23]{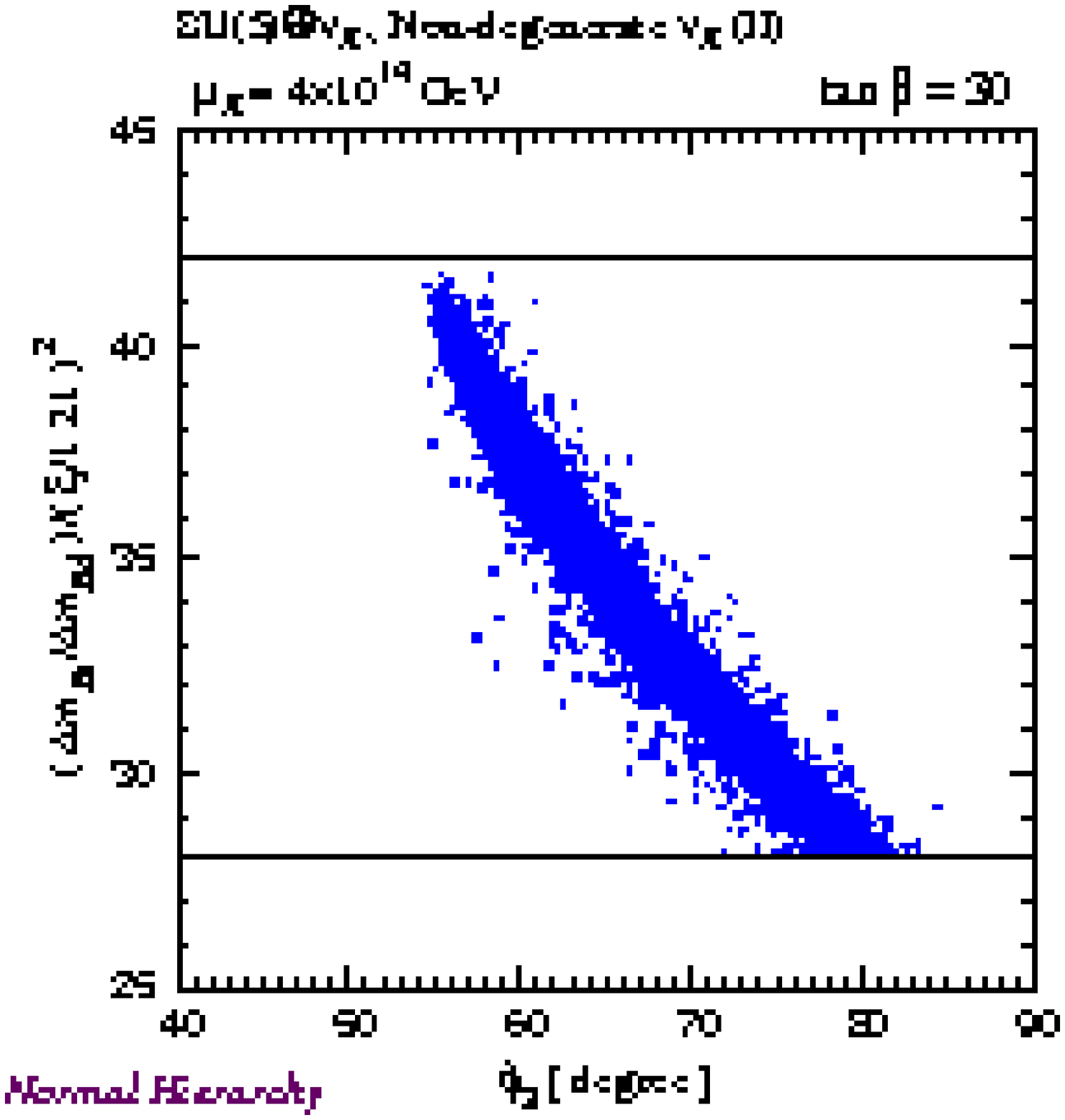} &
\includegraphics[scale=.23]{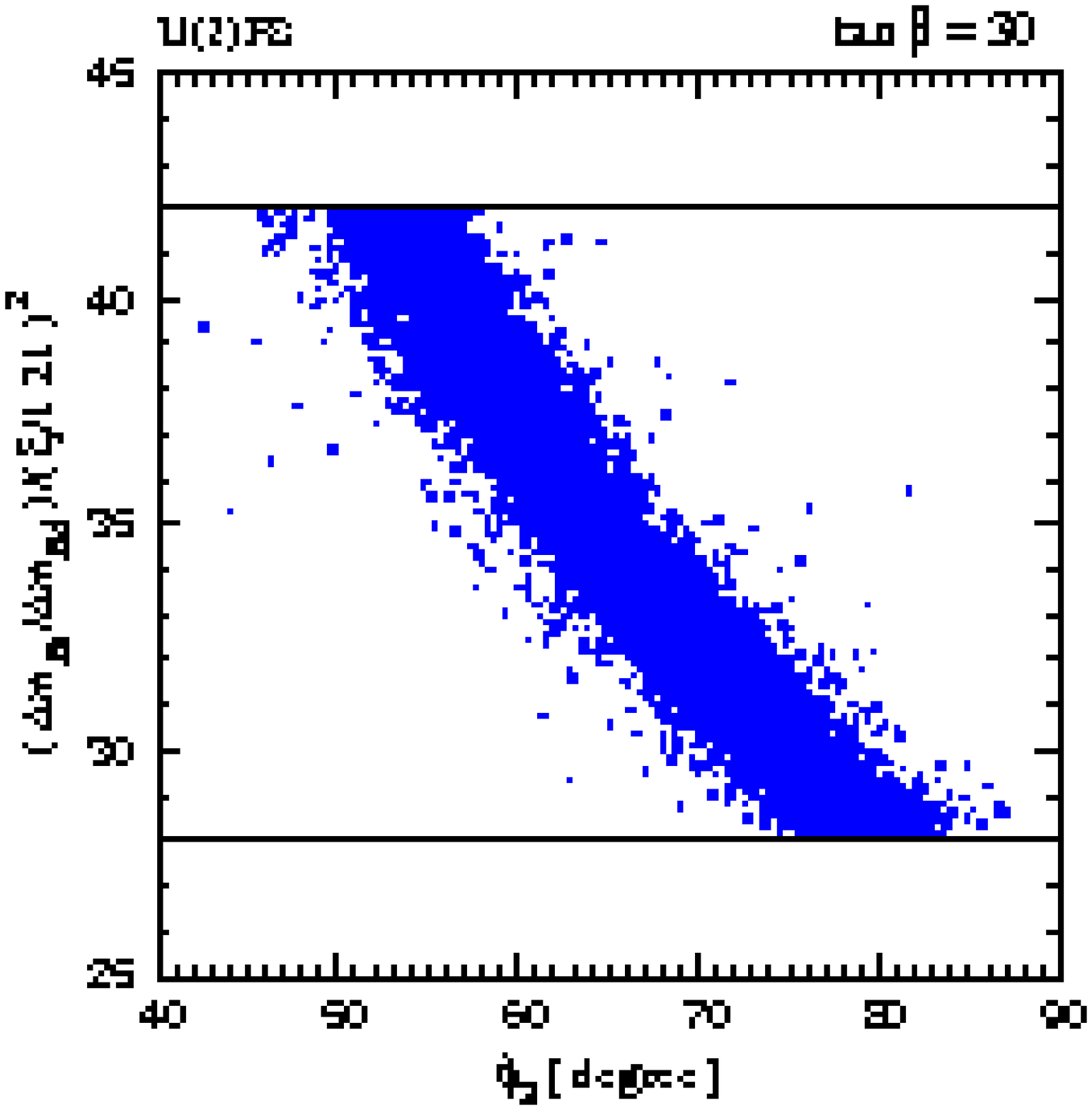} &
\\
(e) & (f) & (g) &
\end{tabular}
\caption{%
Correlation between $\phi_3$ and $\Delta m_{B_s}/\Delta m_{B_d}$ for the
same parameter sets as those in Fig.~\ref{fig:md-ACP-bsg}.
}
\label{fig:UT}
\end{figure}

We show quark flavor signals in the mSUGRA, SU(5) SUSY GUT with
right-handed neutrinos and the U(2) flavor symmetry models.
In the MSSM with right-handed neutrinos, there is no new source of the
squark flavor mixing other than the CKM matrix.
The effect of the neutrino Yukawa couplings appears in the squark sector
only through the renormalization of the Higgs fields.
Consequently the flavor structure of the squarks is essentially the same
as the mSUGRA case.
In fact, we have checked that the plots of the quark flavor signals look
similar to those in the mSUGRA, except that the allowed SUSY parameter
region is largely affected by the constraints from the LFV processes.
That is why we do not show the plots in the MSSM with
right-handed-neutrinos here.
Quark flavor signals in the SU(5) SUSY GUT with right-handed neutrinos
are affected by the existence of the neutrino Yukawa coupling through
the GUT running between $\mu_P$ and $\mu_G$.
Large $b-s$ and $b-d$ mixings in the right-handed down-type squark
sector are induced by the large $2-3$ and $1-3$ mixing in the neutrino
Yukawa coupling in the non-degenerate $\nu_R$ (I) and (II) cases,
respectively.
In the degenerate $\nu_R$ cases, the parameter region excluded by the
$\mu\to e\gamma$ constraint depends on the low energy neutrino mass
spectrum.
In particluar, the region with sizable squark mixings is excluded due to
the strong $\mu\to e\gamma$ constraint in the normal hierarchy case.
On the other hand, region with large $2-3$ squark mixing remains in the
inverted hierarchy case.
The U(2) model has large $2-3$ mixings in both right-handed and
left-handed squark sector at the cut-off scale.

Here we show our results on the following observables.
\begin{itemize}
\item
The direct CP asymmetry in $b\to s\gamma$ decay
(Fig.~\ref{fig:md-ACP-bsg}), which is sensitive to the effect of new CP
violating phase in $b\to s\gamma$ decay amplitude.
\item
The mixing-induced CP asymmetry in $B_d\to K^*\gamma$
(Fig.~\ref{fig:md-SCP-bsg}).
This asymmetry is enhanced by the $b\to s\gamma$ decay amplitude
with the chirality opposite to the SM one.
\item 
The direct CP asymmetry in $b\to d\gamma$ decay
(Fig.~\ref{fig:md-ACP-bdg}), which is sensitive to the effect of new CP
violating phase in $b\to d\gamma$ decay amplitude.
\item
The mixing-induced CP asymmetry in $B_d\to \rho\gamma$
(Fig.~\ref{fig:md-SCP-bdg}), which is enhanced by the $b\to d\gamma$
decay amplitude with the chirality opposite to the SM one.
\item 
The mixing-induced CP asymmetry in $B_d\to \phi K_S$ decay
(Fig.~\ref{fig:md-BdphiK}).
The difference between this quantity and the mixing-induced CP asymmetry
in $B_d\to J/\psi K_S$, $\Delta S_{\text{CP}}(B_d\to \phi K_S) \equiv
S_{\text{CP}}(B_d\to \phi K_S) -  S_{\text{CP}}(B_d\to J/\psi K_S)$, is
sensitive to the new CP violating phase in $b\to s\bar{s}s$ decay
amplitude.
\item 
The mixing-induced CP asymmetry in $B_s\to J/\psi \phi$ decay
(Fig.~\ref{fig:md-BsJphi}), which is affected by the new CP violating
phase in $B_s-\bar{B}_s$ mixing matrix element.
\end{itemize}

From Figs.~\ref{fig:md-ACP-bsg}, \ref{fig:md-SCP-bsg},
\ref{fig:md-ACP-bdg}, \ref{fig:md-SCP-bdg}, \ref{fig:md-BdphiK} and
\ref{fig:md-BsJphi}, we can draw the following conclusions.
For the mSUGRA case, we do not see significant deviations in any of
above observables.
In the cases of the degenerate $\nu_R$ with normal hierarchical (light)
neutrinos (D$\nu_R$-NH) and the degenerate $\nu_R$ with degenerate
neutrinos (D$\nu_R$-D) for the SU(5) SUSY GUT with right-handed
neutrinos, the parameter region is strongly constrained by the
$\text{B}(\mu\to e\gamma)$ as already discussed.
There are some points in which deviations are apparent in
$S_{\text{CP}}(B_d\to K^*\gamma)$, $S_{\text{CP}}(B_d\to \rho\gamma)$,
$\Delta S_{\text{CP}}(B_d\to \phi K_S)$ and
$S_{\text{CP}}(B_s\to J/\psi\phi)$.
These points could be distinguished by future measurements such as LHCb,
in which the precision in the determination of the phase 
of $B_s-\bar{B}_s$ mixing matrix element 
is expected to be 0.01 radian level \cite{ref:Nakada:SUSY2010s}.
In the degenerate $\nu_R$ with inverted hierarchical neutrinos
(D$\nu_R$-IH) and the non-degenerate $\nu_R$ (I) with normal
hierarchical neutrinos (ND$\nu_R$(I)-NH) cases of the SU(5) SUSY GUT with
right-handed neutrinos, the SUSY contributions to mixing-induced CP
asymmetries in $B_s\to J/\psi\phi$, $B_d\to K^*\gamma$, and
$B_d\to \phi K_S$ can be significant.
On the other hand, in the non-degenerate $\nu_R$ (II) with normal
hierarchical neutrinos (ND$\nu_R$(II)-NH) case of SU(5) SUSY GUT
with right-handed neutrinos, there is a significant SUSY contribution to
$b\to d\gamma$ decay amplitude, so that
$S_{\text{CP}}(B_d\to \rho\gamma)$ can be as large as $\pm 0.1$.
Large SUSY contributions can be found for almost all modes we analyze
in the U(2) model.
Only the direct CP asymmetry in $b\to d\gamma$ does not show any
significant deviation from the SM.

The correlation between $\phi_3$ and $\Delta m_{B_s}/\Delta m_{B_d}$
are shown in Fig.~\ref{fig:UT}.
$\Delta m_{B_s}/\Delta m_{B_d}$ is sensitive to the new physics
contributions to the $B_d-\bar{B}_d$ and $B_s-\bar{B}_s$ mixing matrix
elements unless the contributions cancel in the ratio.
For the mSUGRA case, the deviation is negligible and the plot in this
plane is the same as in the SM.
The lower limit of $\phi_3$ is determined by the constraint from
$\varepsilon_K$.
In the D$\nu_R$-NH and D$\nu_R$-D cases of SU(5) SUSY GUT with
right-handed neutrinos, the deviation in the correlation is not so
significant.
In the D$\nu_R$-IH and the non-degenerate $\nu_R$ cases of SU(5) SUSY
GUT with right-handed neutrinos, as well as the U(2) model, some
deviations appear in the correlation plots.
In the D$\nu_R$-IH and ND$\nu_R$(I)-NH cases the deviation comes from
the SUSY contribution to the $B_s-\bar{B}_s$ mixing matrix element,
while $B_d-\bar{B}_d$ receive sizable SUSY correction in
ND$\nu_R$(II)-NH.
In the U(2) model SUSY contributions show up in both matrix elements.
In order to identify the deviation in the correlation in future, it is
required that the evaluation of $\xi$ parameter by the lattice QCD
calculation is significantly improved and that the $\phi_3$ is precisely
measured from tree-level dominant processes.

\begin{figure}[htbp]
\begin{tabular}{cccc}
\includegraphics[scale=.23]{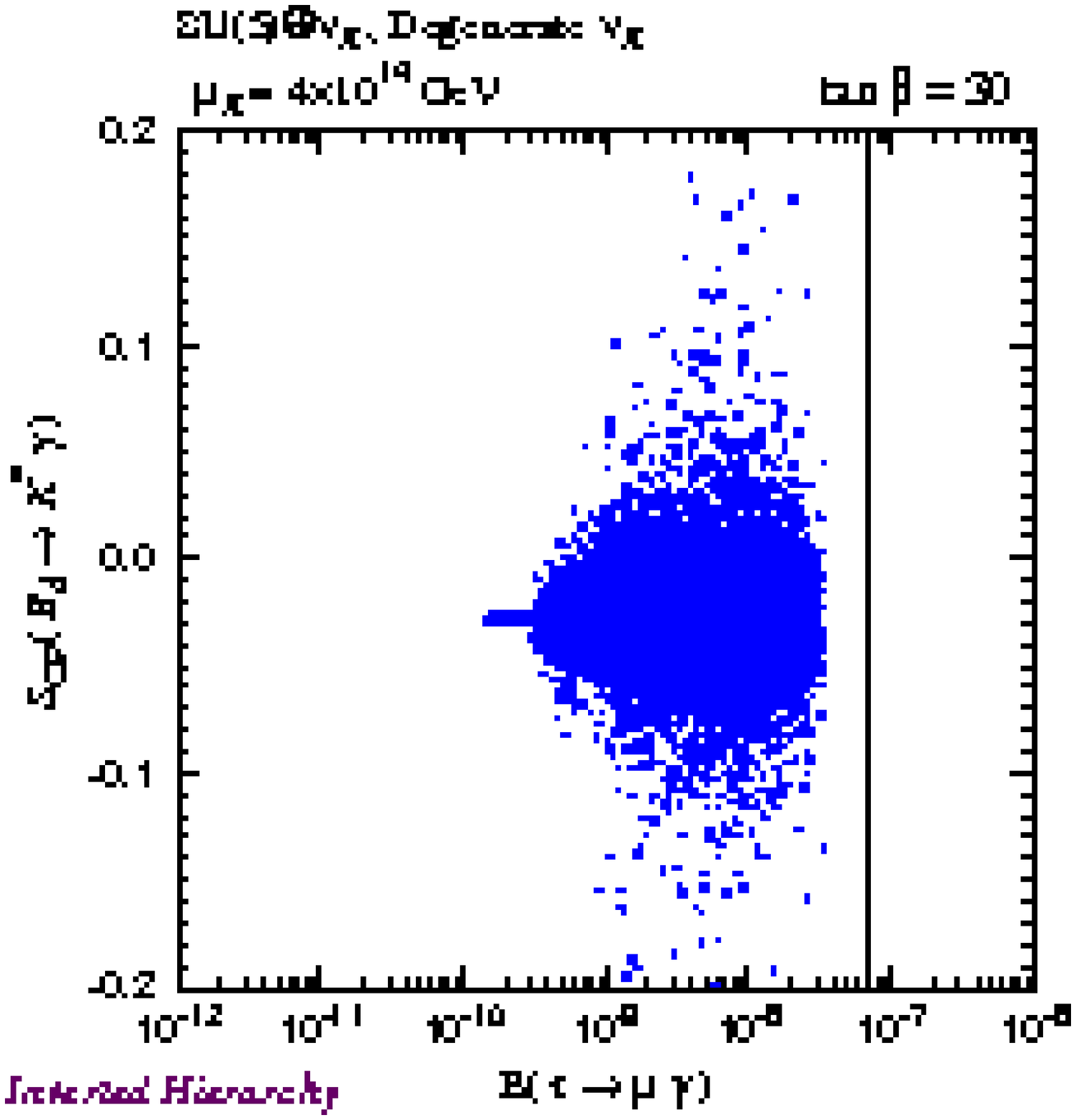} &
\includegraphics[scale=.23]{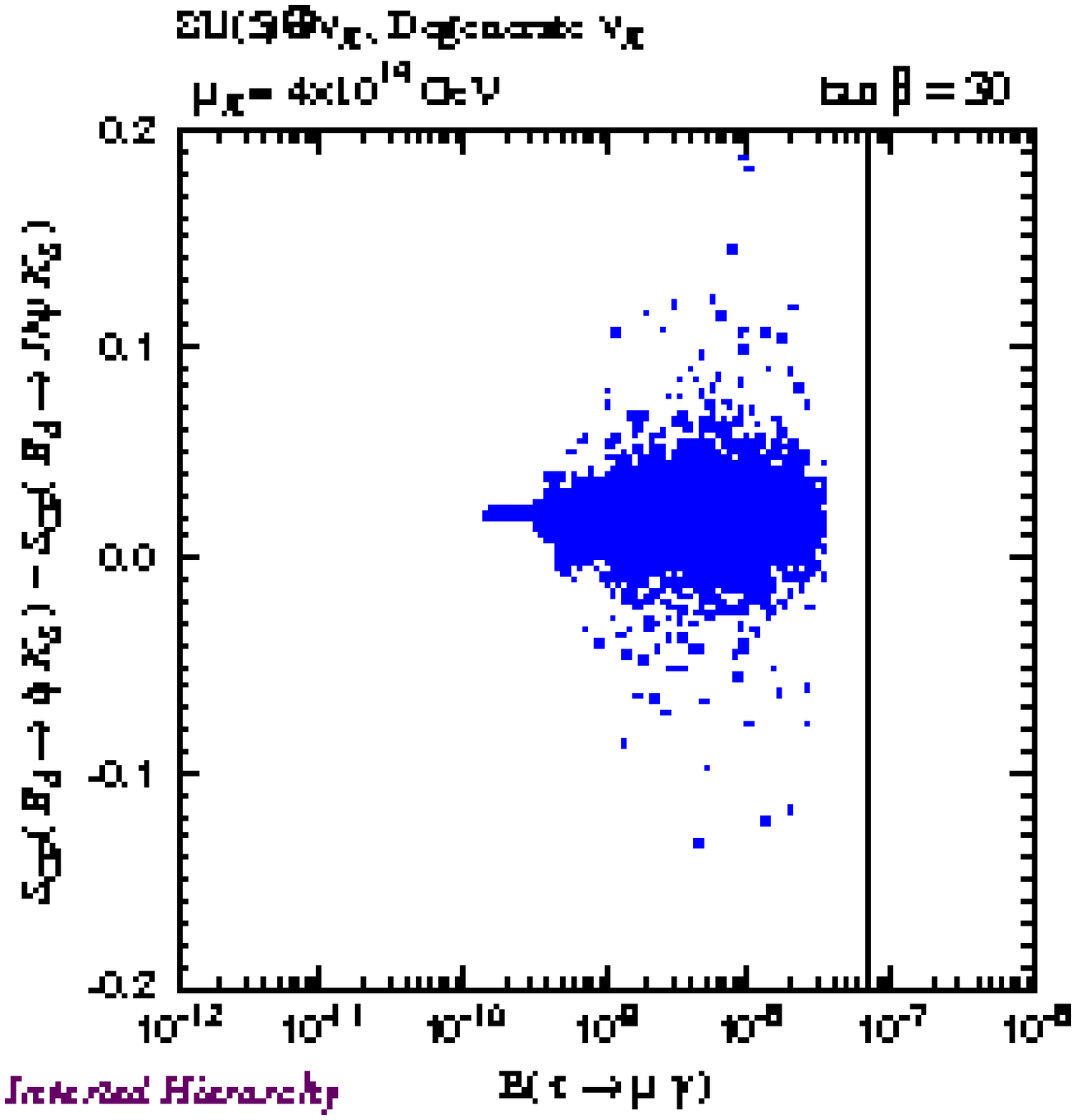} &
\includegraphics[scale=.23]{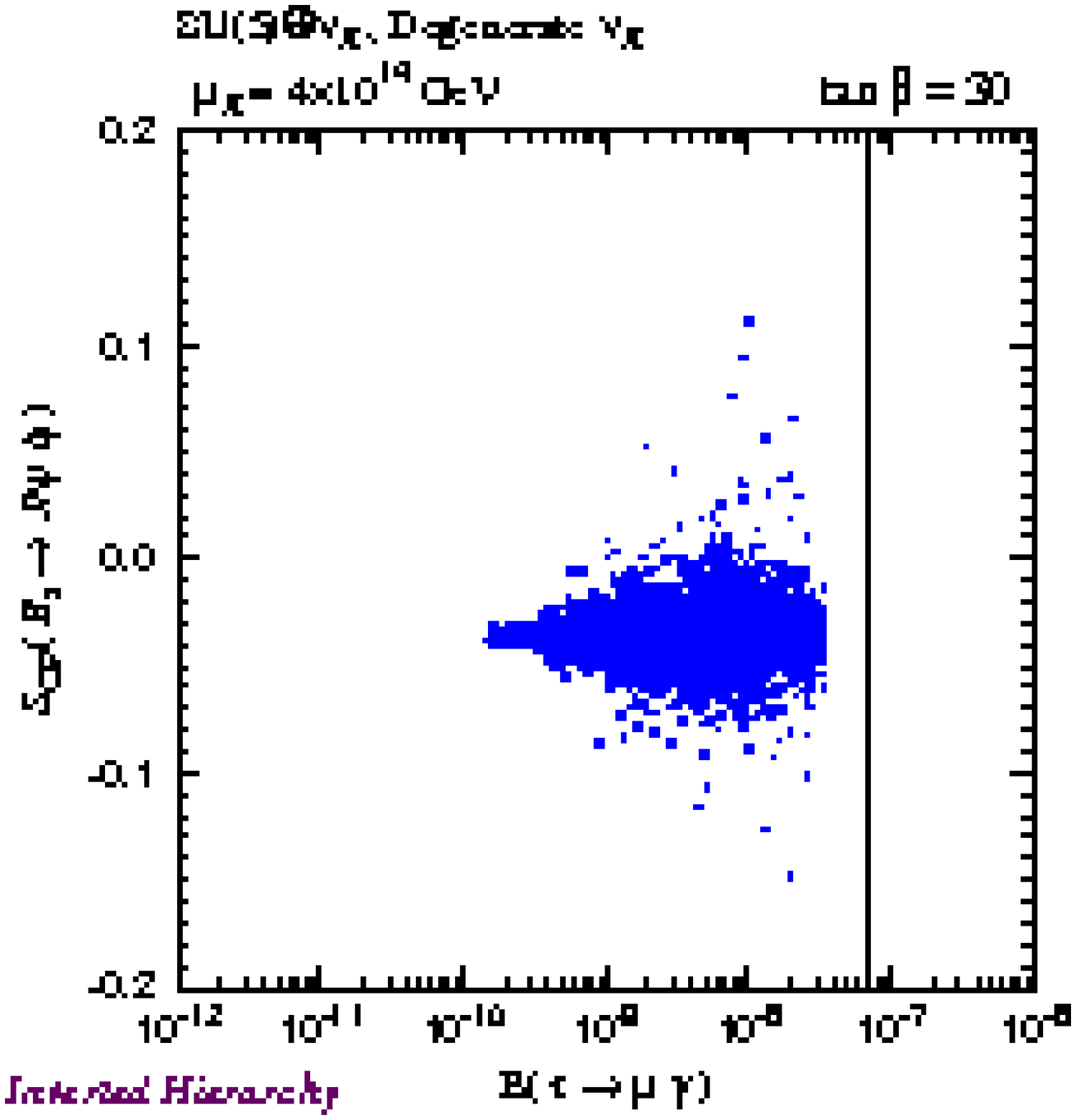} &
\includegraphics[scale=.23]{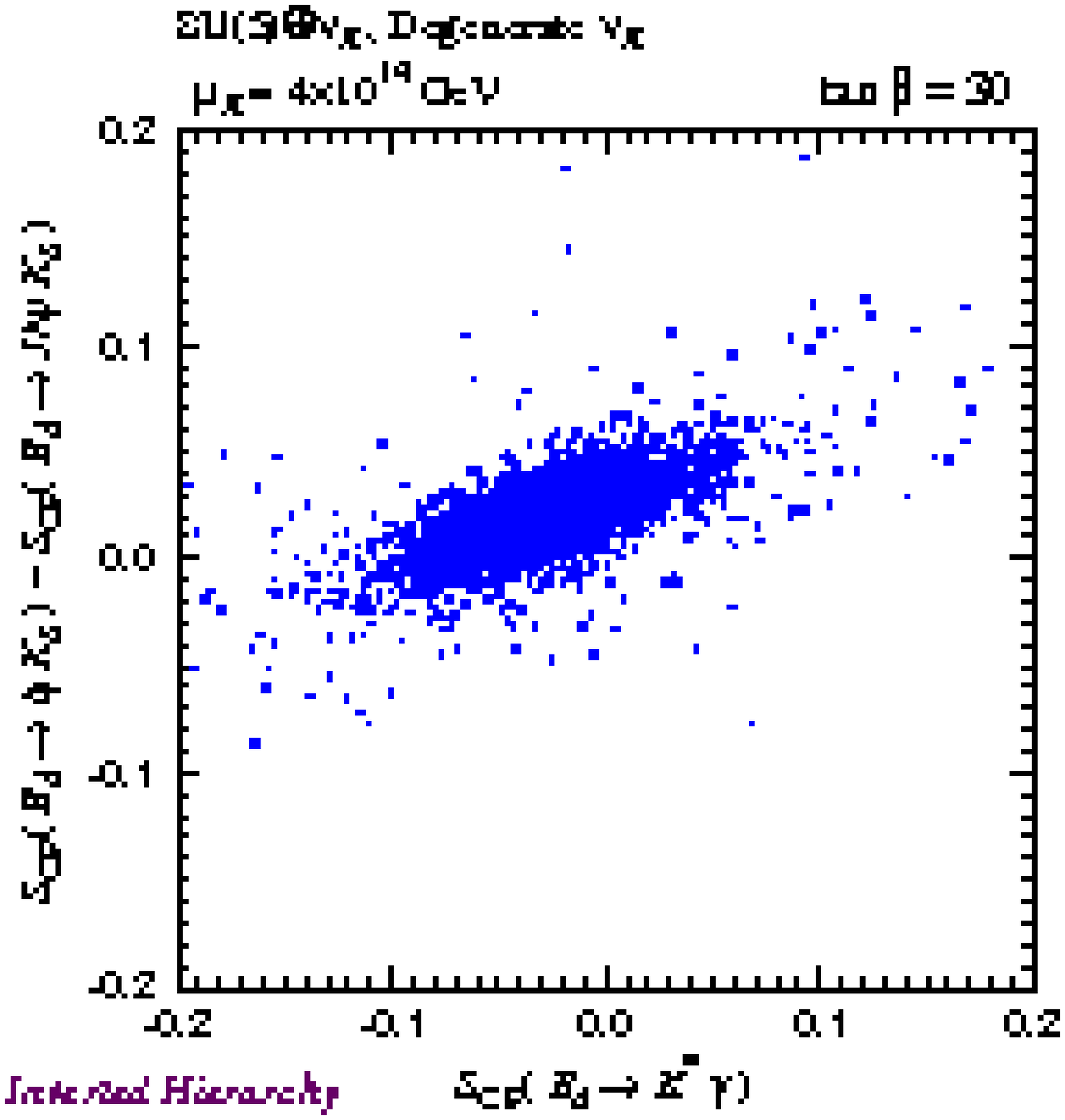} \\
(a) & (b) & (c) & (d) \\
\includegraphics[scale=.23]{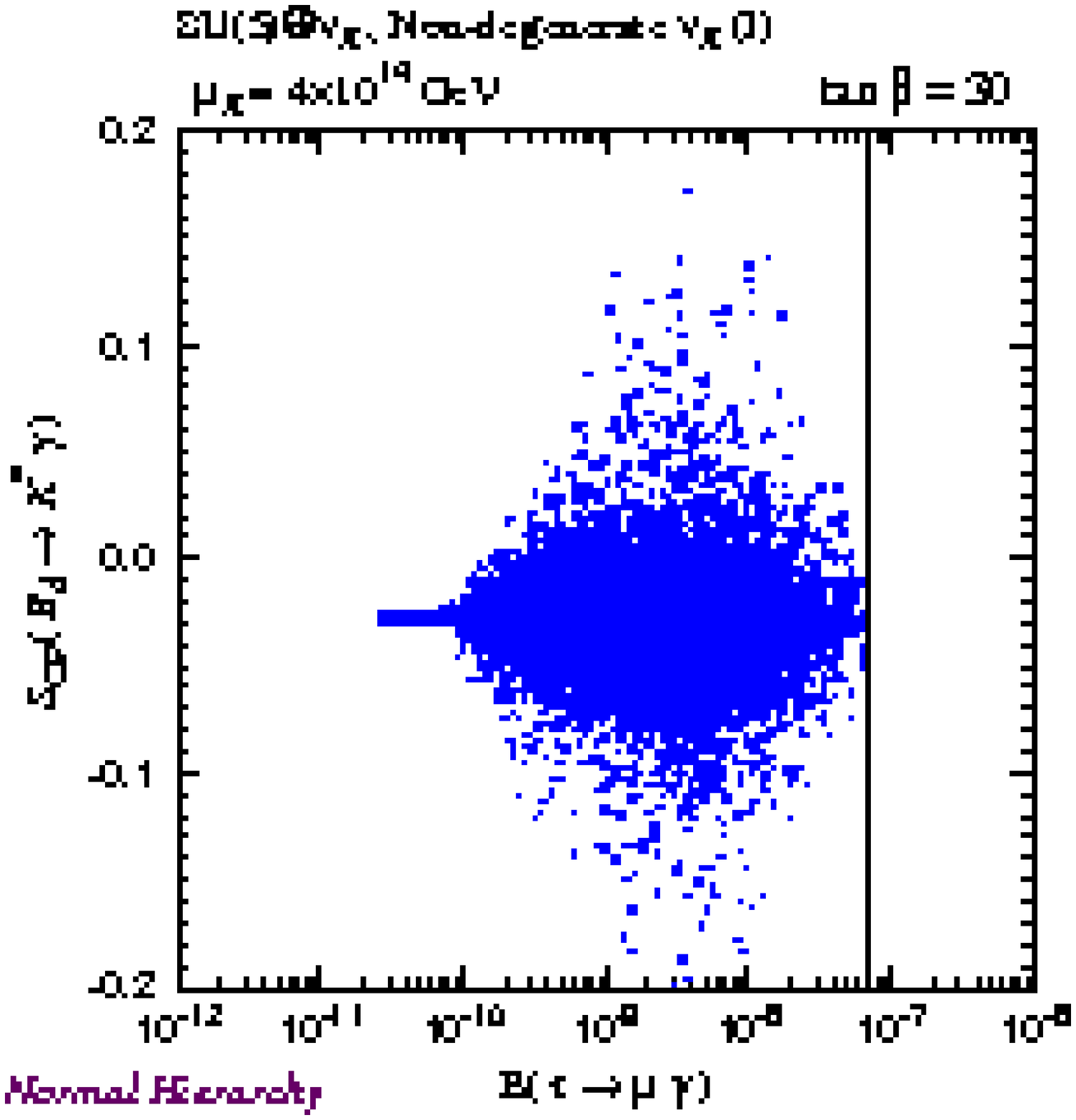} &
\includegraphics[scale=.23]{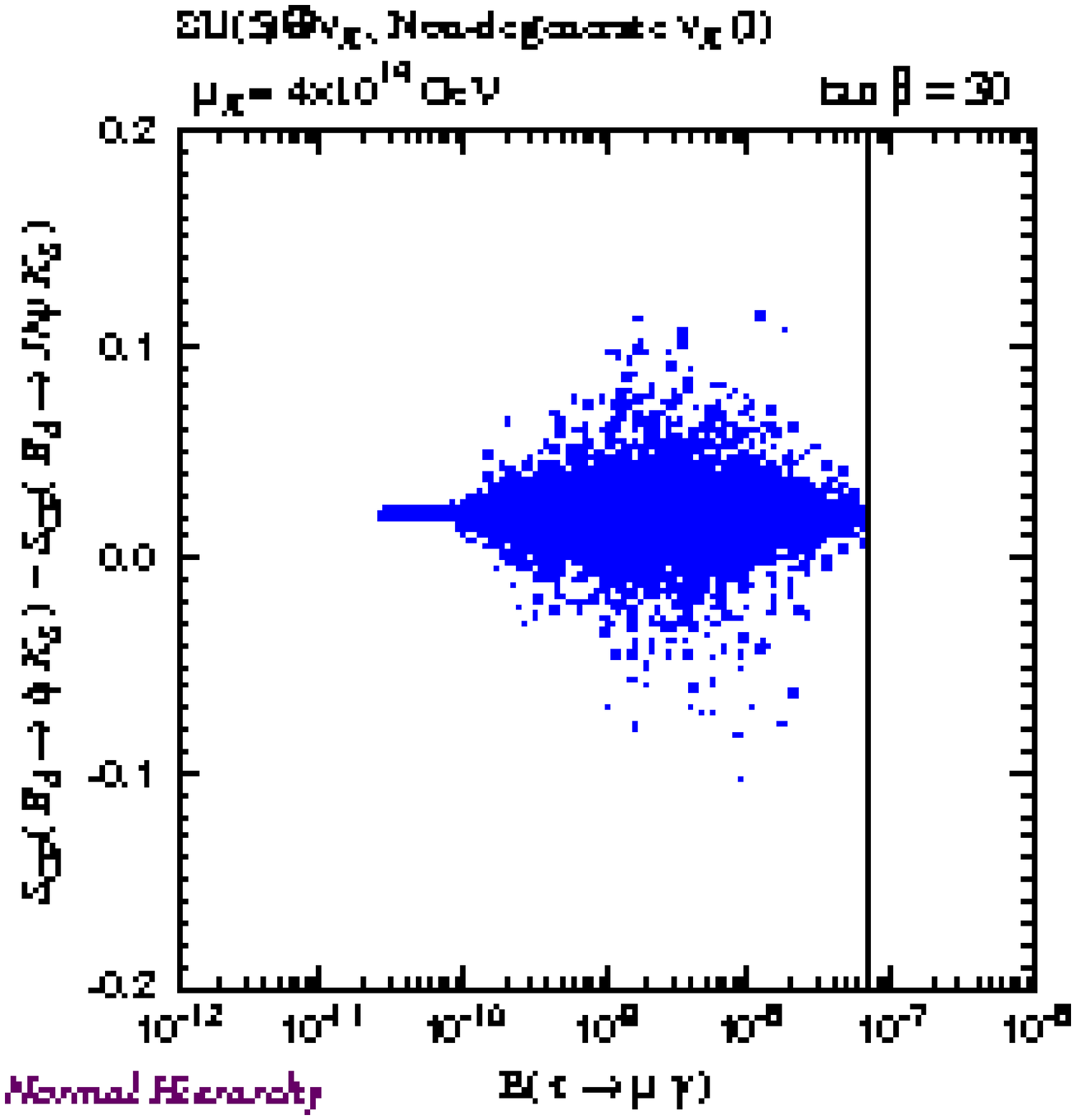} &
\includegraphics[scale=.23]{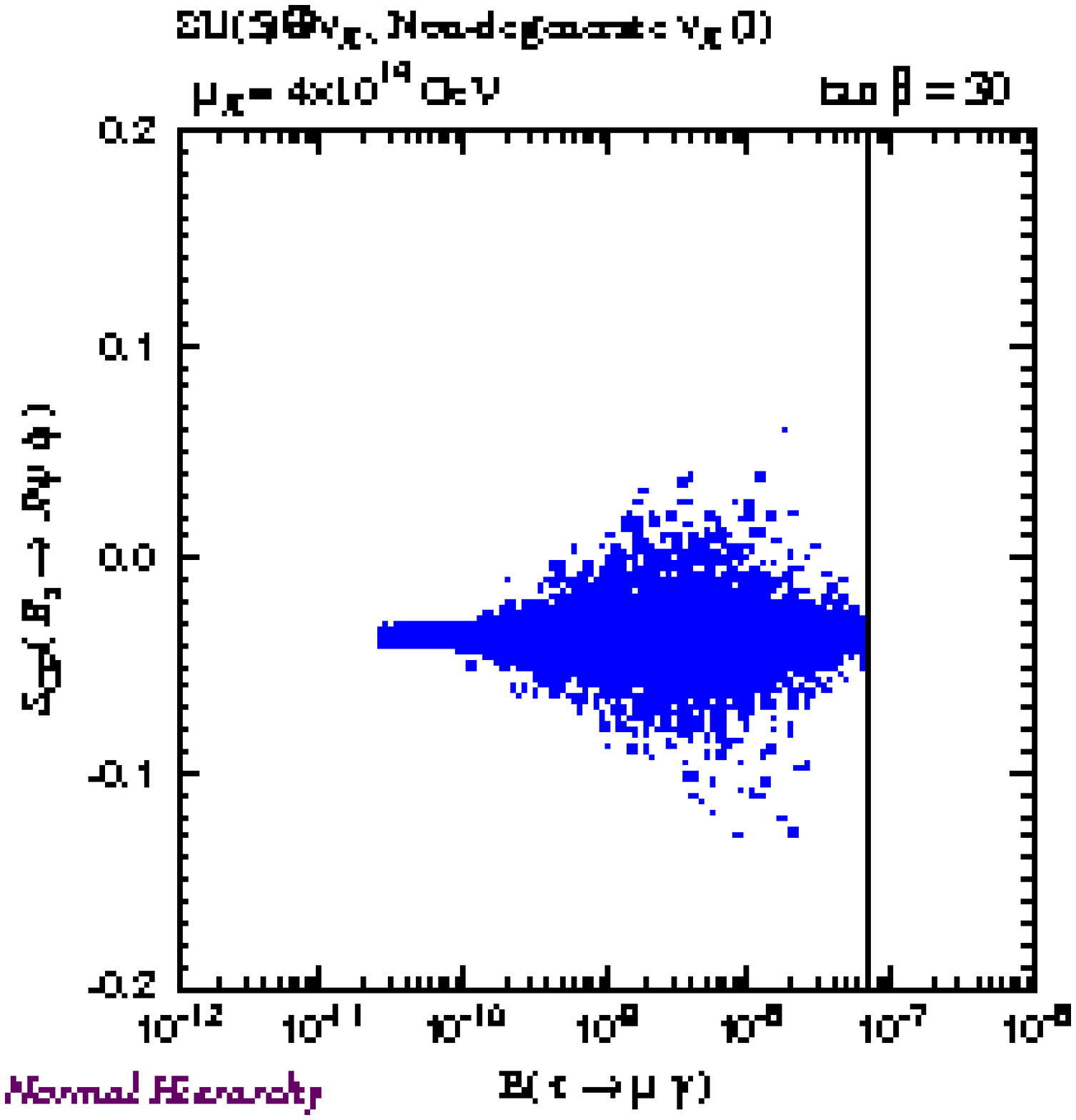} &
\includegraphics[scale=.23]{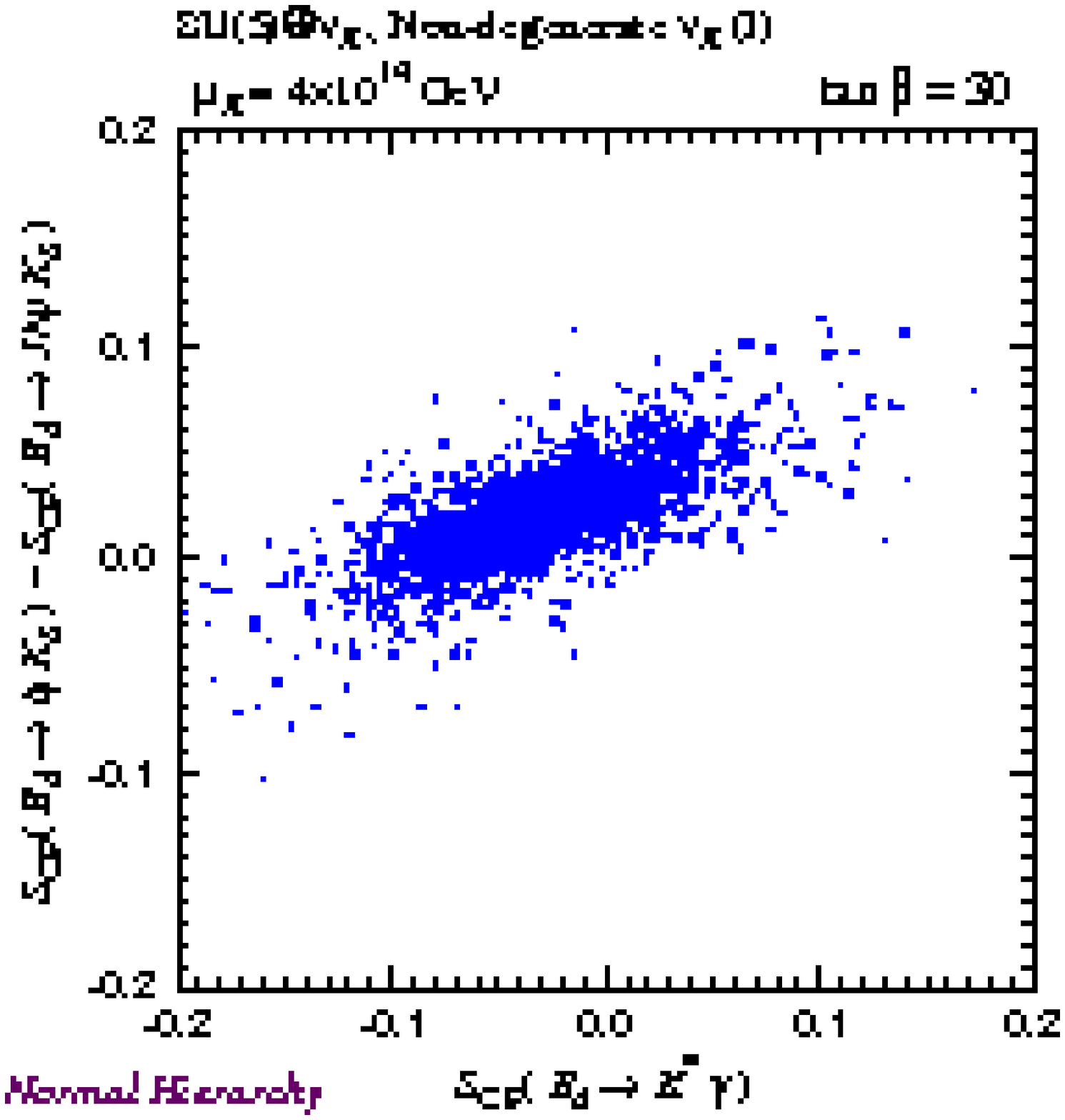} \\
(e) & (f) & (g) & (h)
\end{tabular}
\caption{%
Correlations among $b\to s$ observables and
$\text{B}(\tau\to\mu\gamma)$.
}
\label{fig:correlations}
\end{figure}

In Fig.~\ref{fig:correlations}, we show the correlations among
$S_{\text{CP}}(B_d\to K^*\gamma)$,
$\Delta S_{\text{CP}}(B_d\to\phi K_S)$,
$S_{\text{CP}}(B_s\to J/\psi\phi)$, and $\text{B}(\tau\to\mu\gamma)$ for
D$\nu_R$-IH and ND$\nu_R$(I)-NH cases of the SU(5) SUSY GUT with
right-handed neutrinos, where these quantities are significantly
affected.
We can see that large deviations in $b\to s$ transitions occur in the
region with $\text{B}(\tau\to\mu\gamma)\gtrsim 10^{-9}$.
Also there is a positive correlation between
$S_{\text{CP}}(B_d\to K^*\gamma)$ and
$\Delta S_{\text{CP}}(B_d\to\phi K_S)$.

We also calculate the branching ratio and the forward-backward asymmetry
of $b\to sl^+l^-$, which are sensitive to the amplitudes from photon-
and $Z$-penguin and box diagrams.
In all the cases we consider here, we find the deviations are negligible.

\subsubsection{EDM constraints}

\begin{figure}[htbp]
\begin{tabular}{cccc}
\includegraphics[scale=.23]{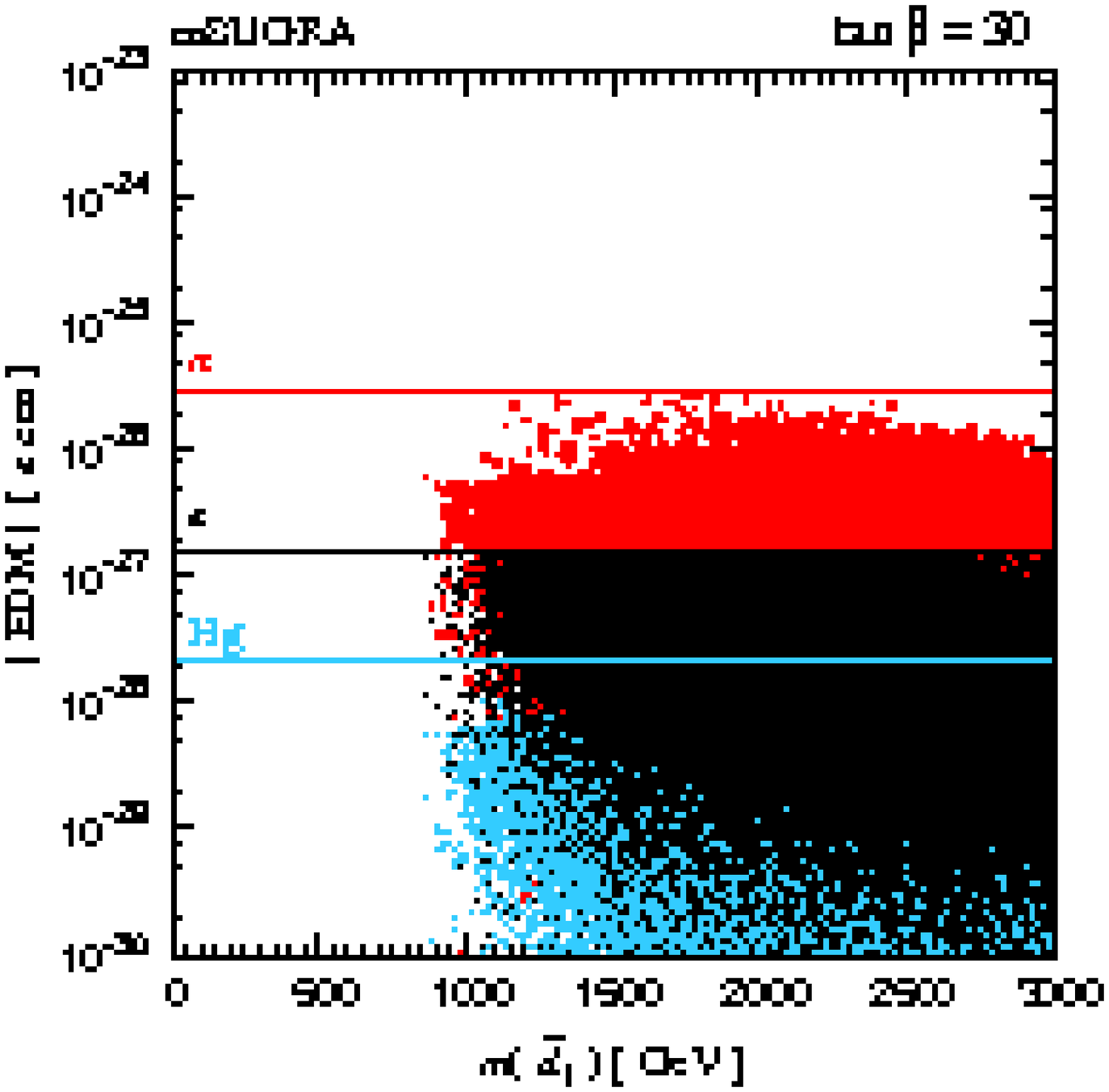} &
\includegraphics[scale=.23]{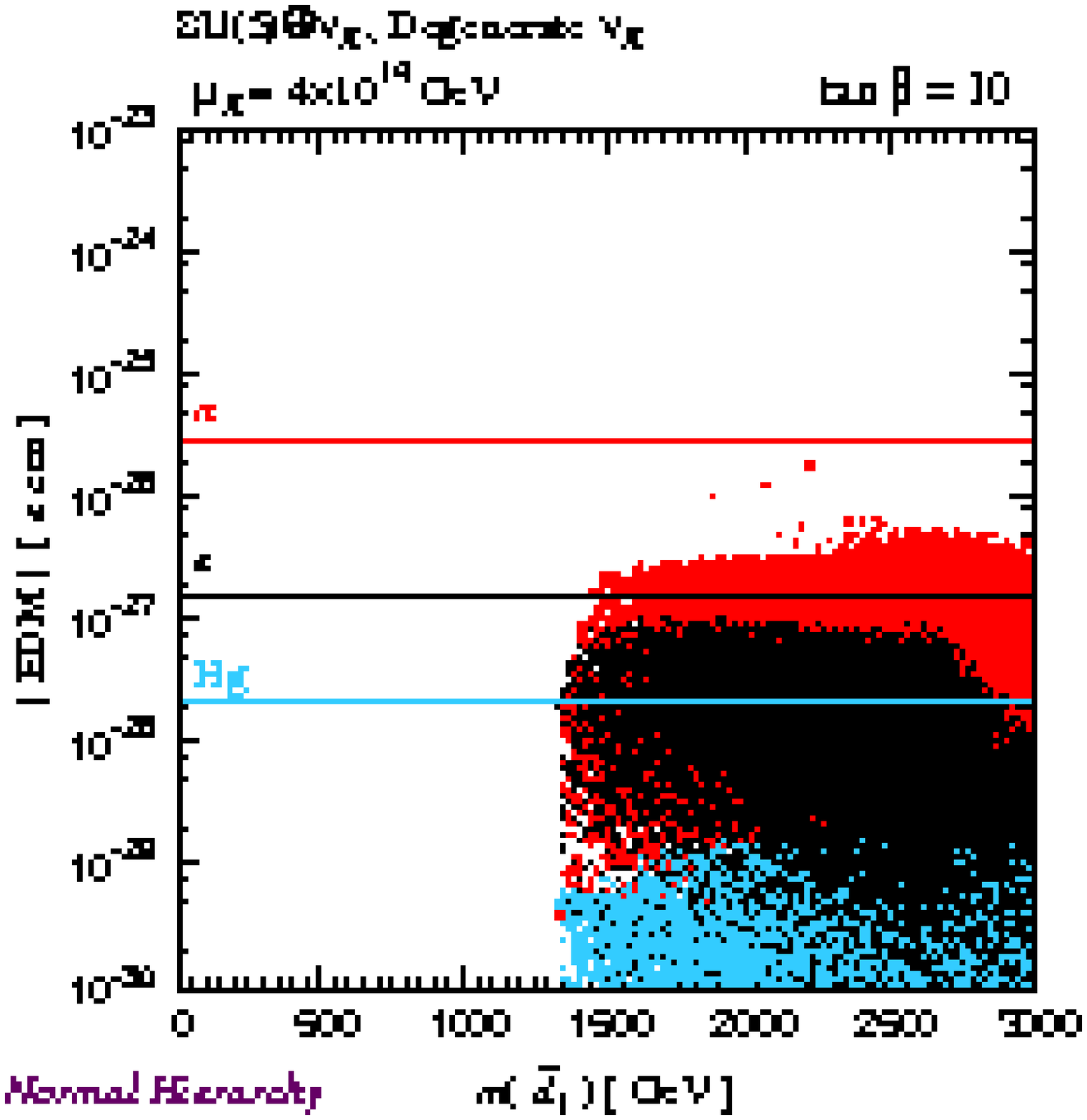} &
\includegraphics[scale=.23]{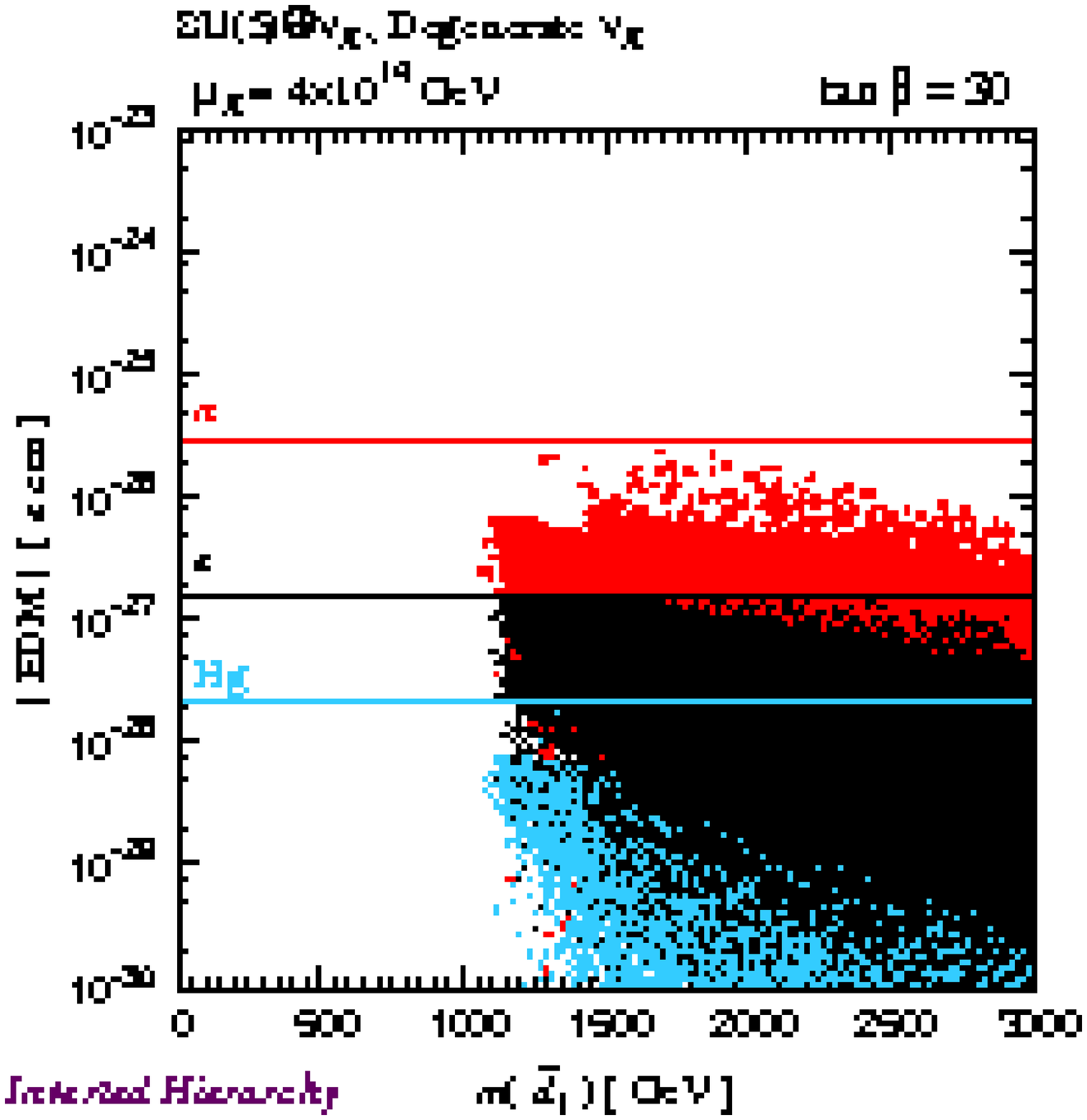} &
\includegraphics[scale=.23]{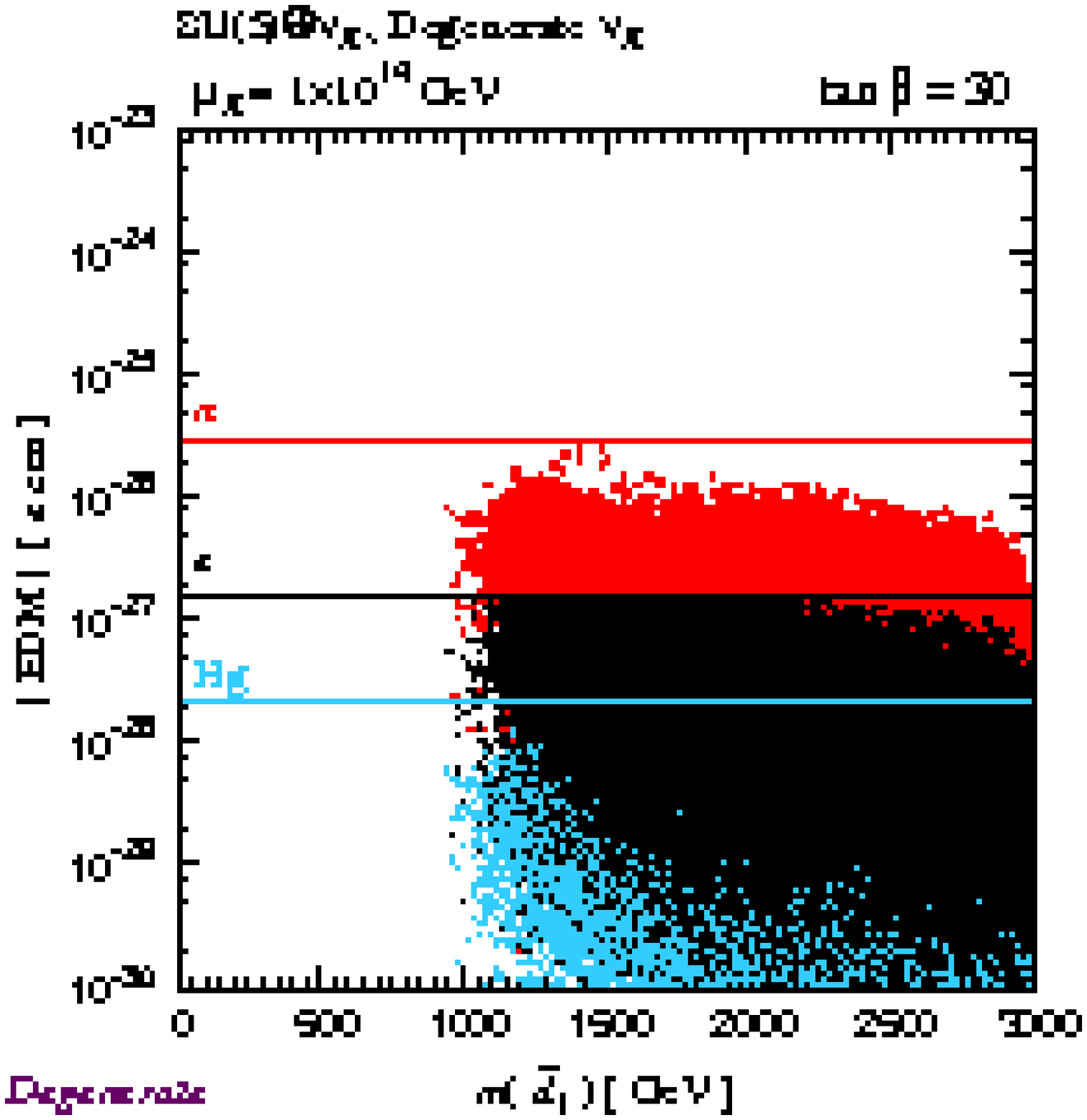}
\\
(a) & (b) & (c) & (d)
\\
\includegraphics[scale=.23]{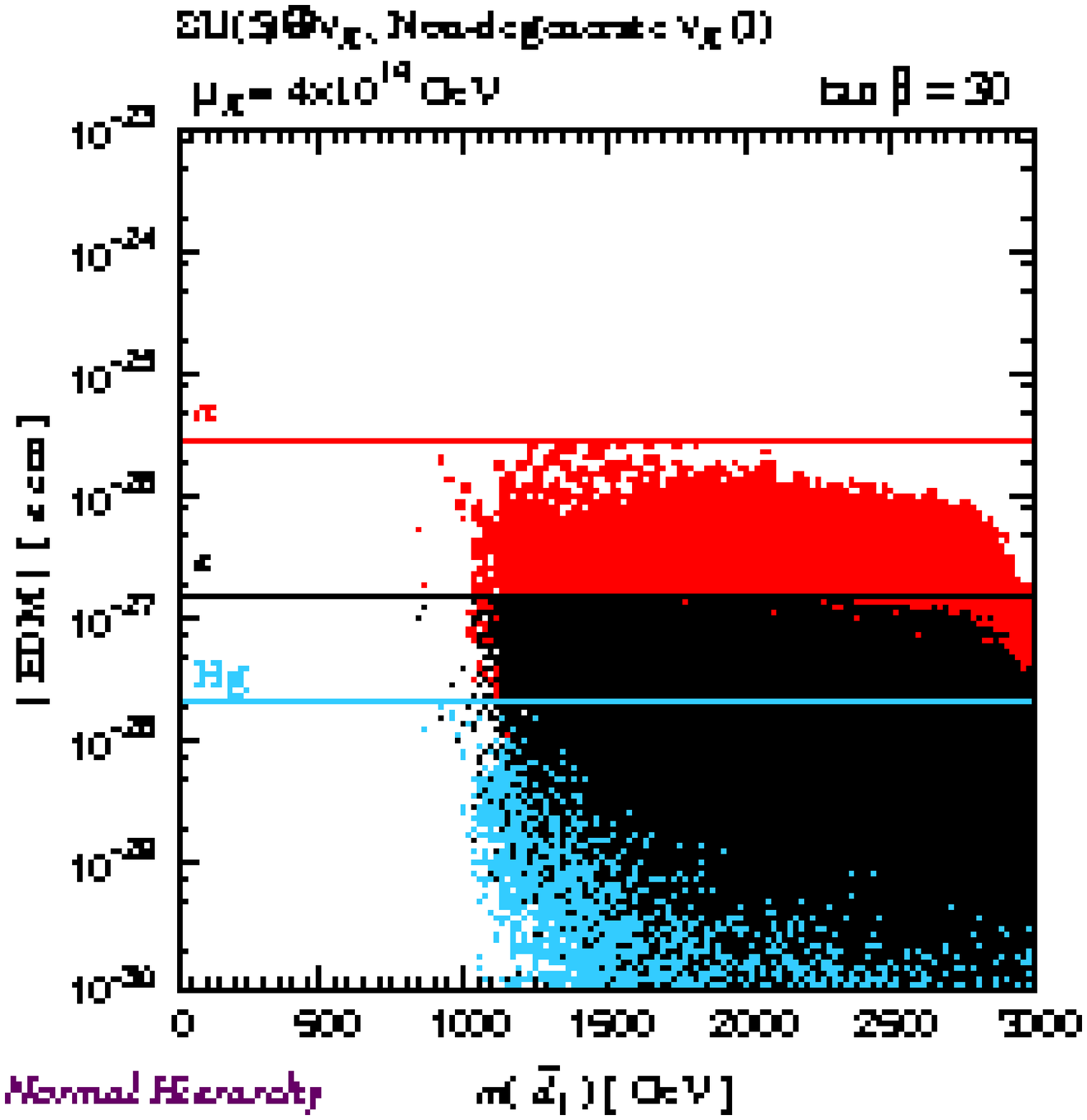} &
\includegraphics[scale=.23]{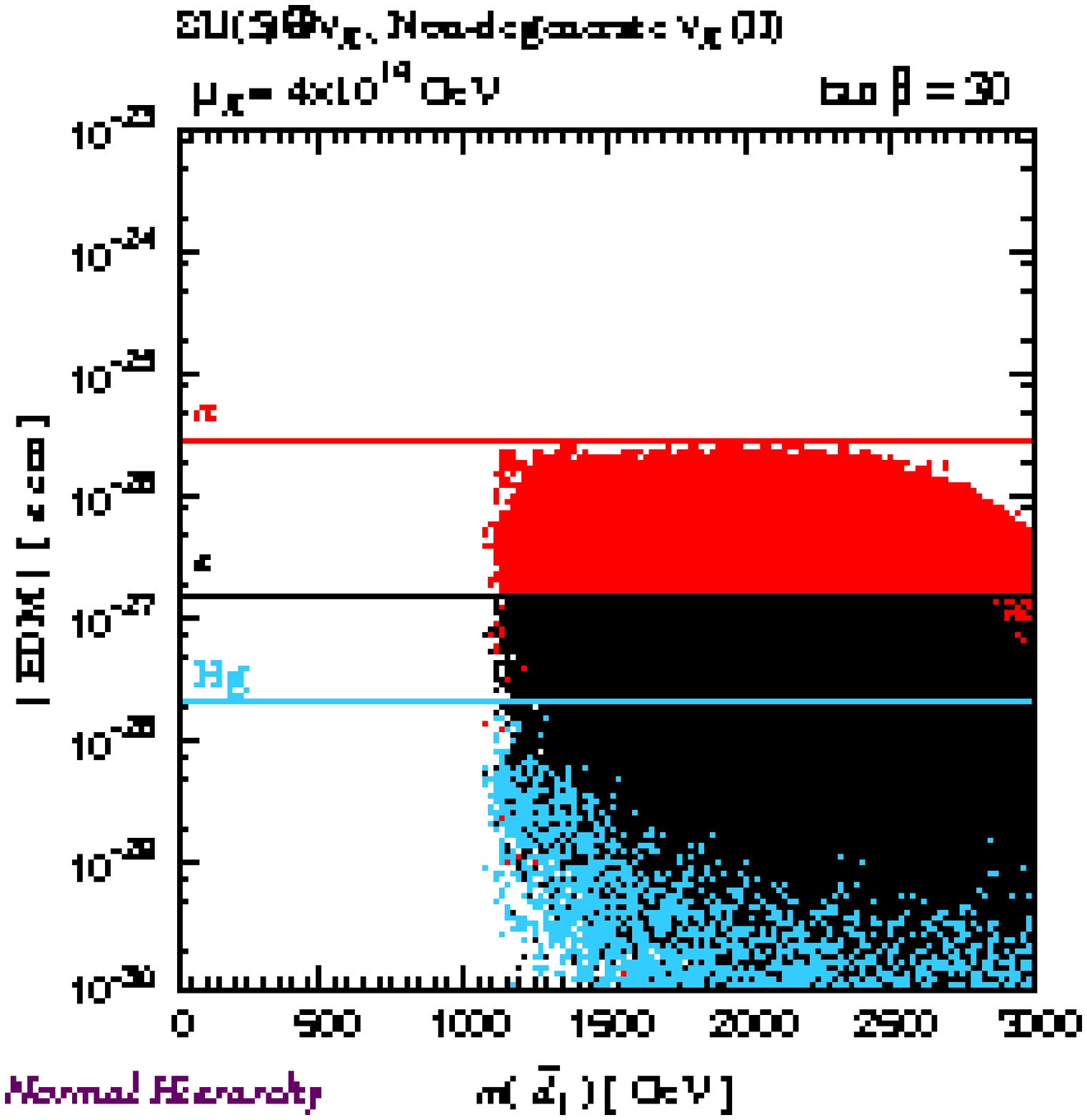} &
\includegraphics[scale=.23]{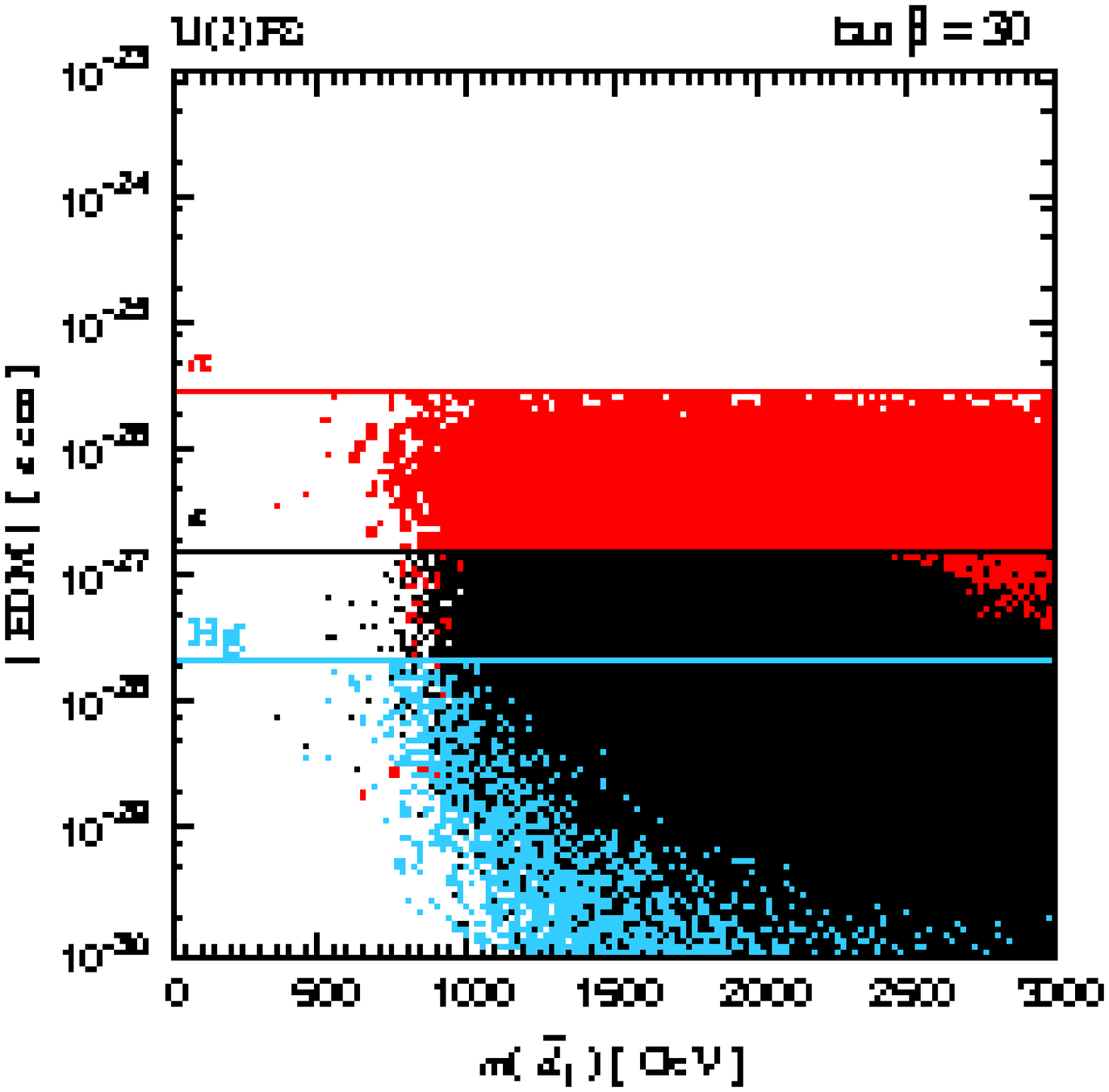} &
\\
(e) & (f) & (g) &
\end{tabular}
\caption{%
(Color online)
Electric dipole moments of the neutron (grey/red), $^{199}\mathrm{Hg}$
(light-grey/light-blue) and the electron (black) as functions of
$m(\tilde{d}_1)$ for the same parameter sets as those in
Fig.~\ref{fig:md-ACP-bsg}.
Horizontal lines show the experimental upper limits.
The neutron EDM is calculated by the NDA formula.
}
\label{fig:edm-msd1}
\end{figure}

\begin{figure}[htbp]
\begin{tabular}{cccc}
\includegraphics[scale=.23]{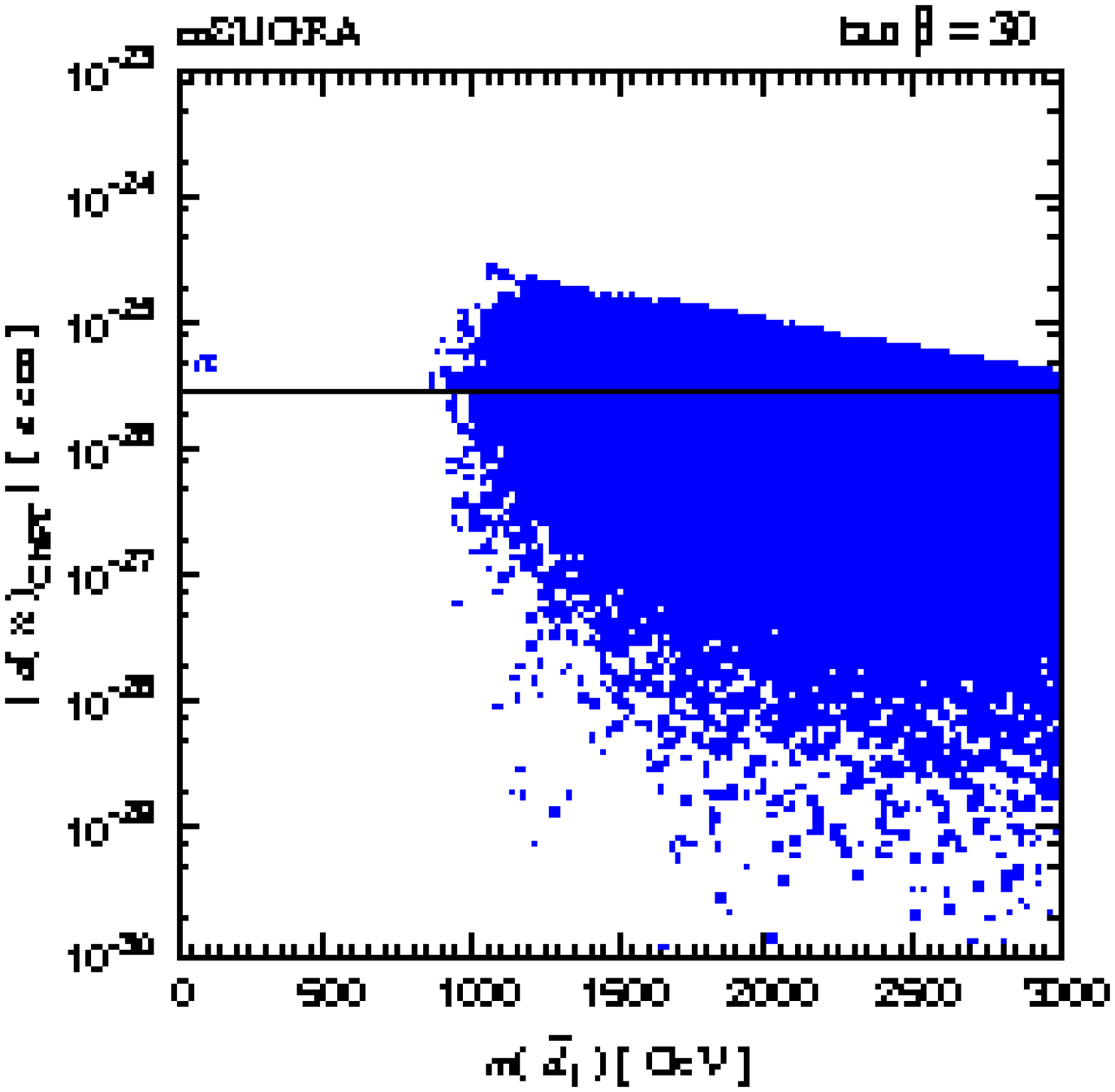} &
\includegraphics[scale=.23]{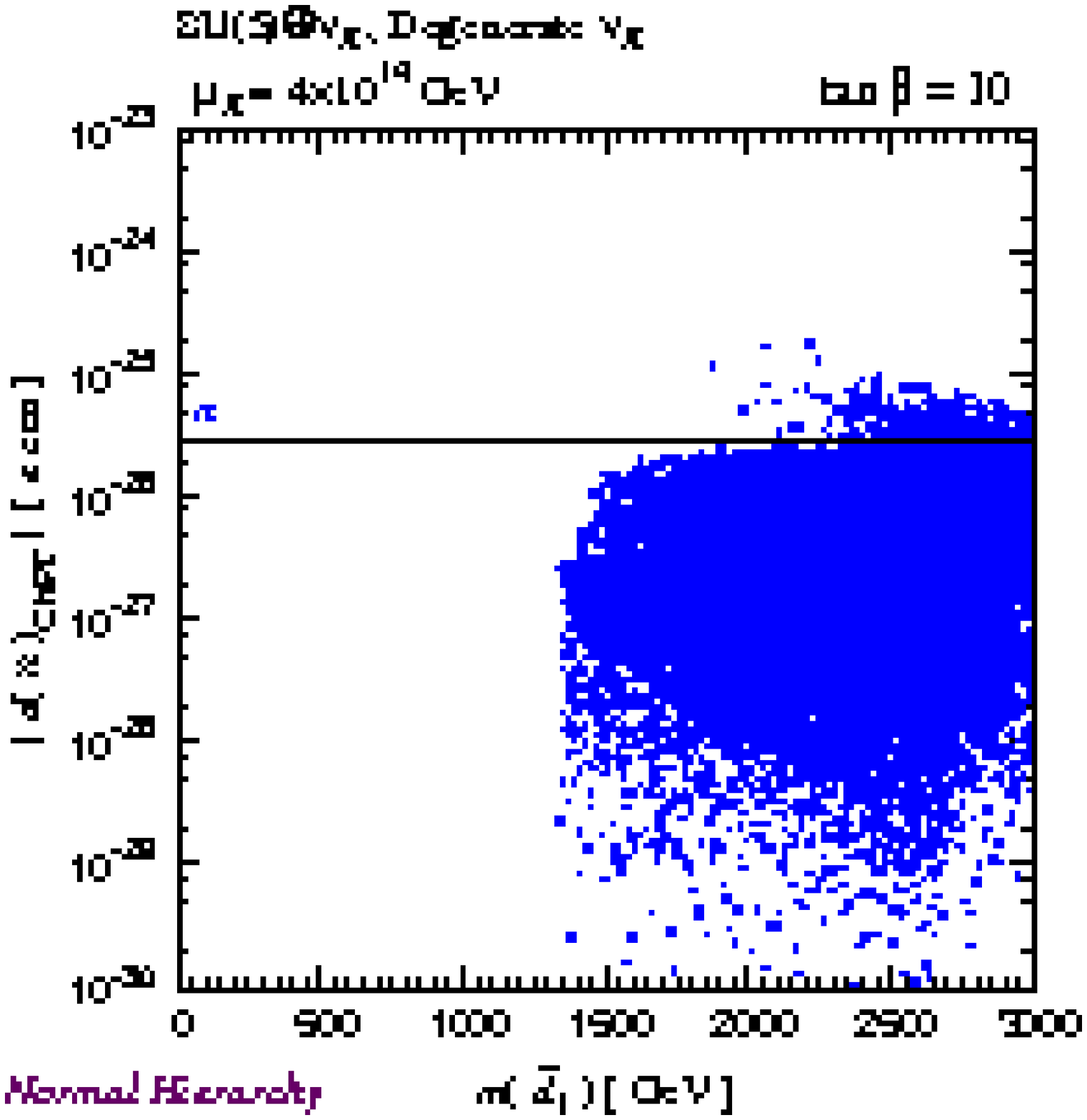} &
\includegraphics[scale=.23]{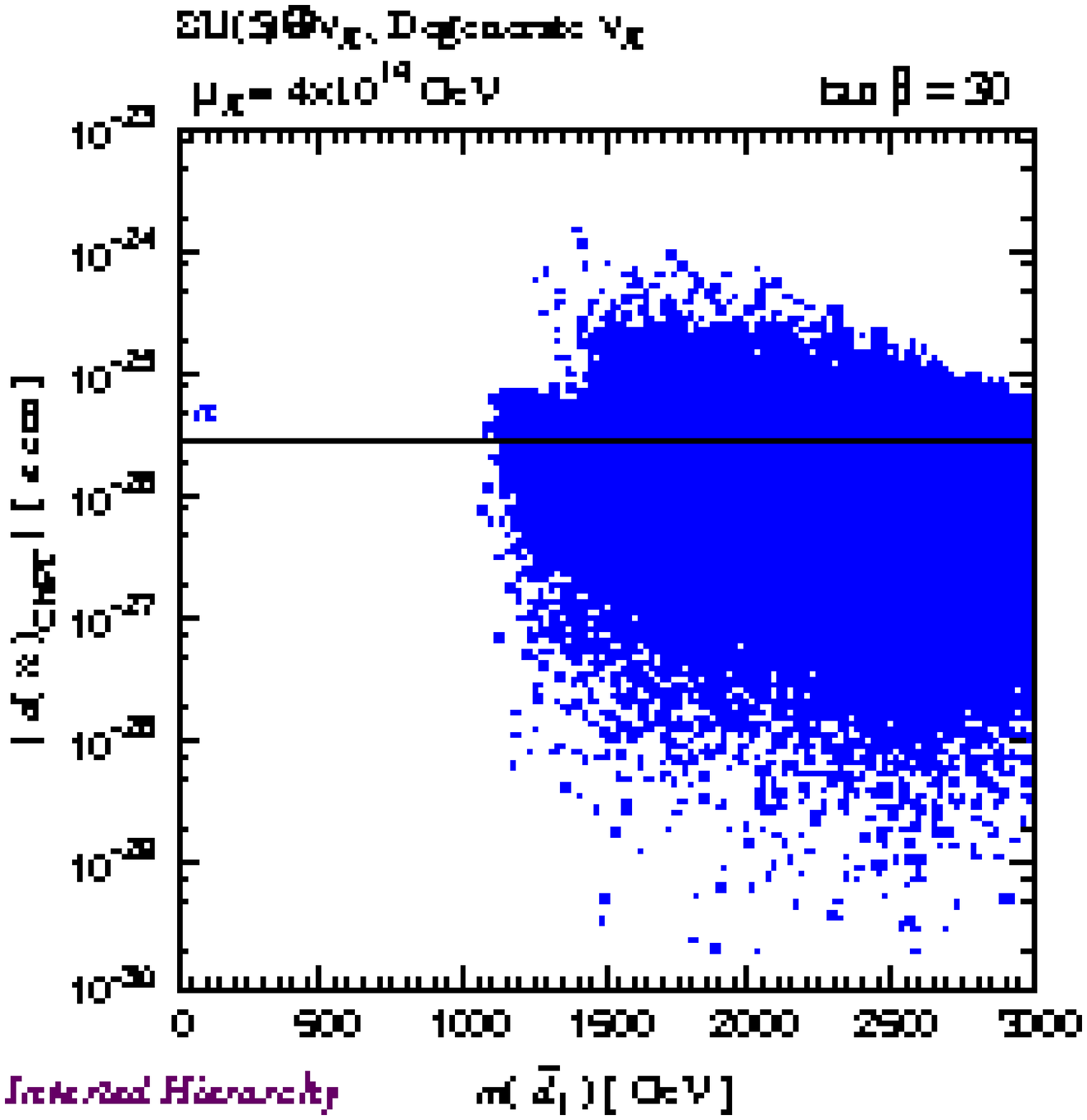} &
\includegraphics[scale=.23]{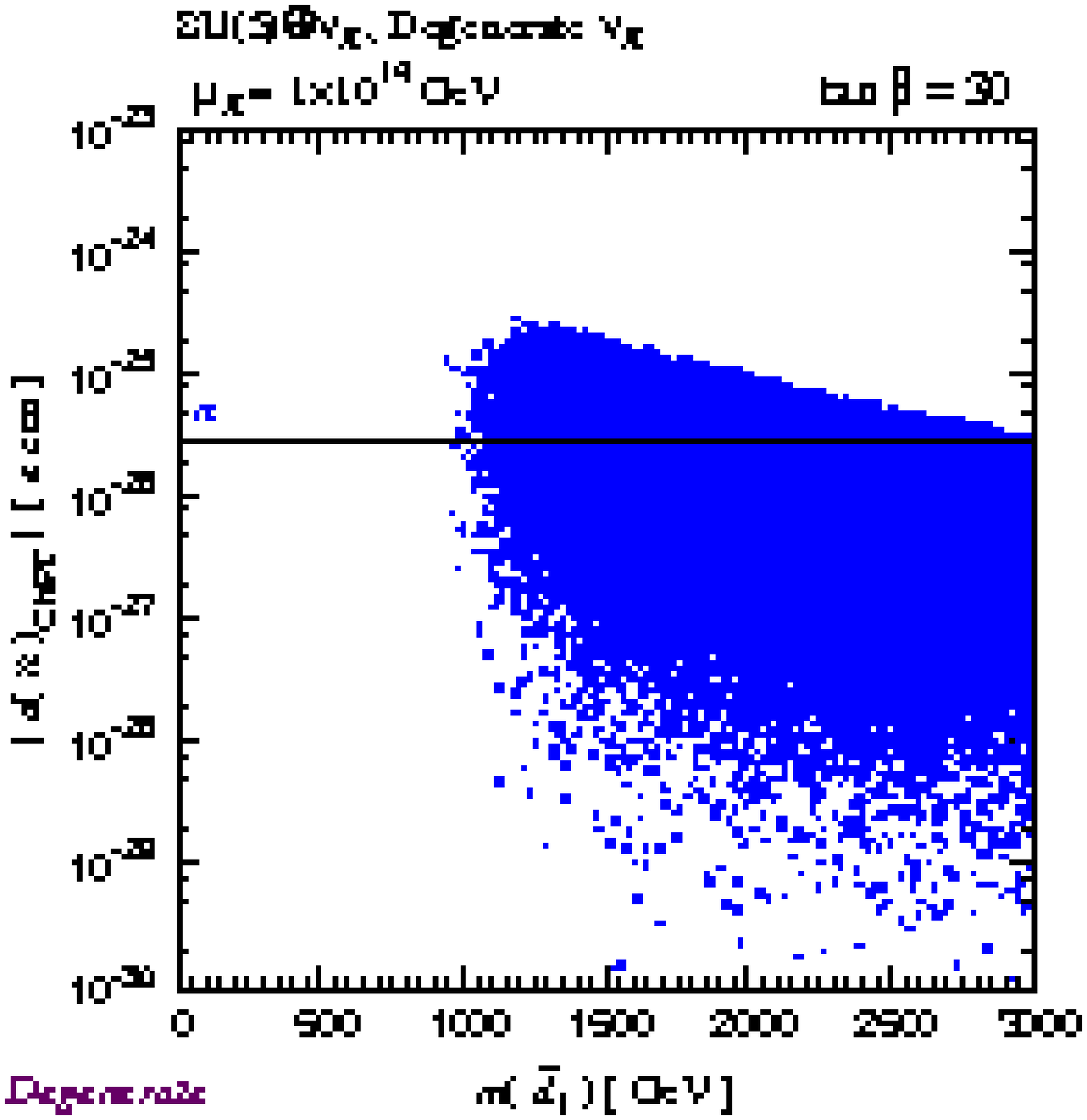}
\\
(a) & (b) & (c) & (d)
\\
\includegraphics[scale=.23]{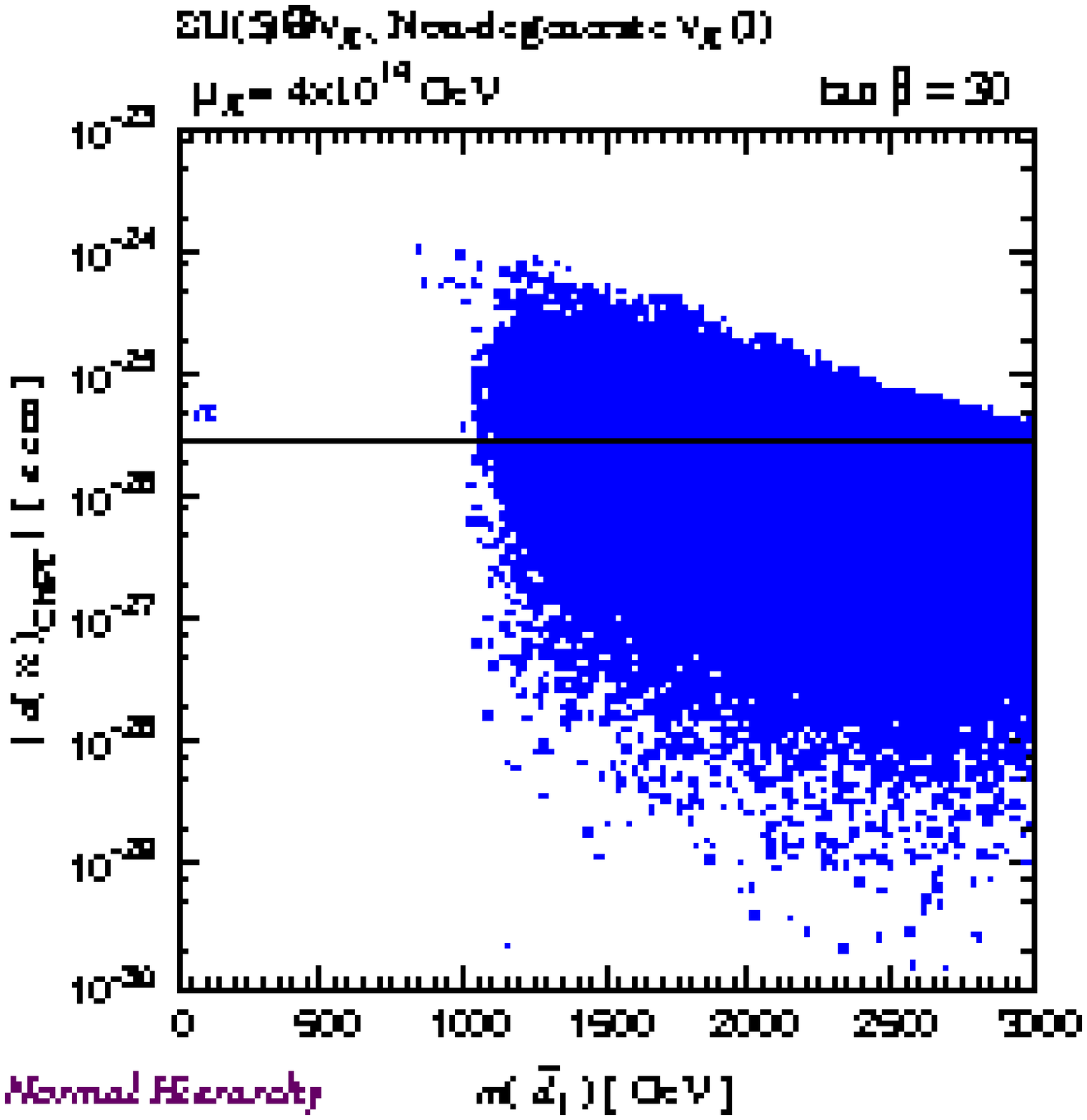} &
\includegraphics[scale=.23]{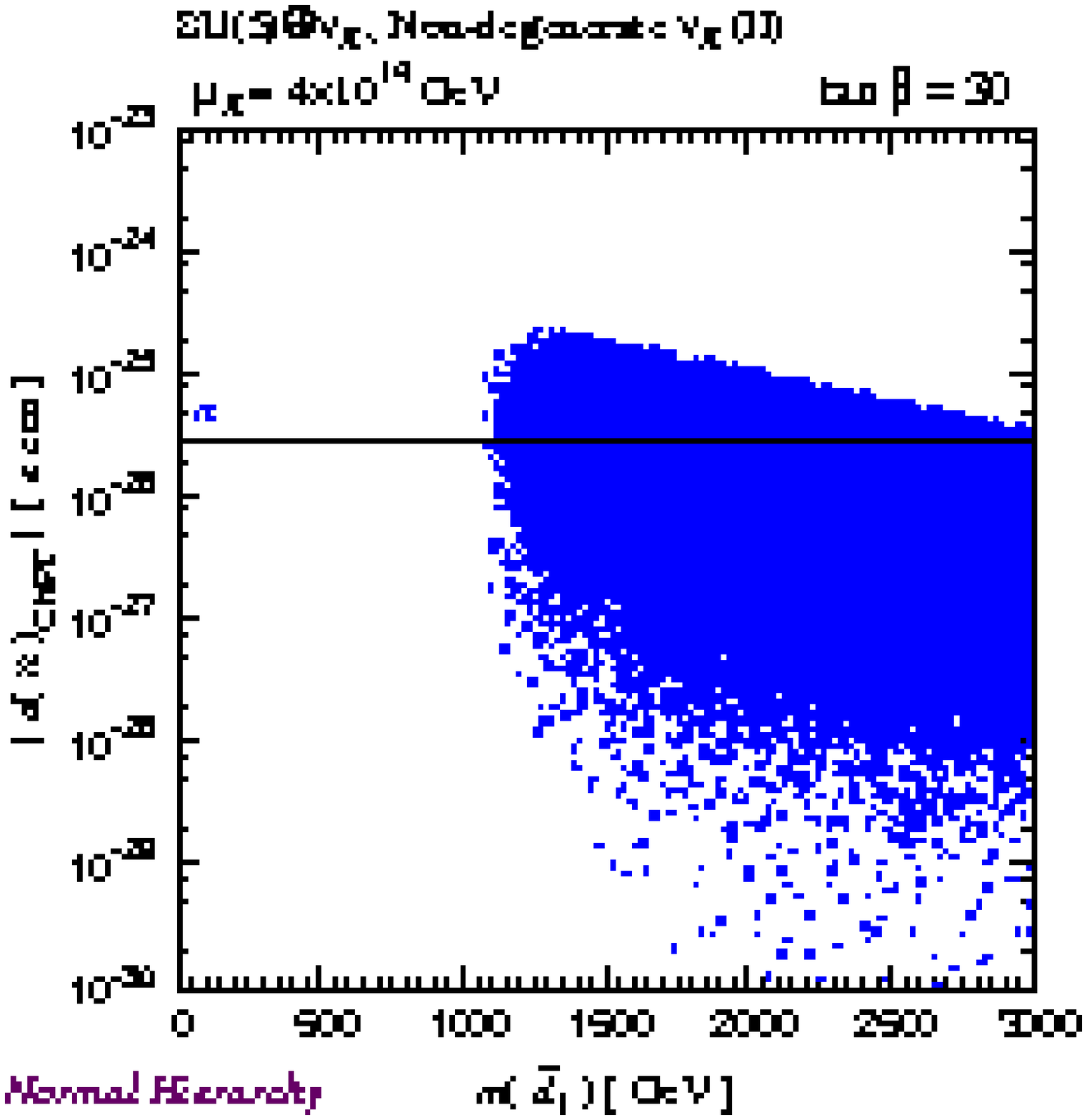} &
\includegraphics[scale=.23]{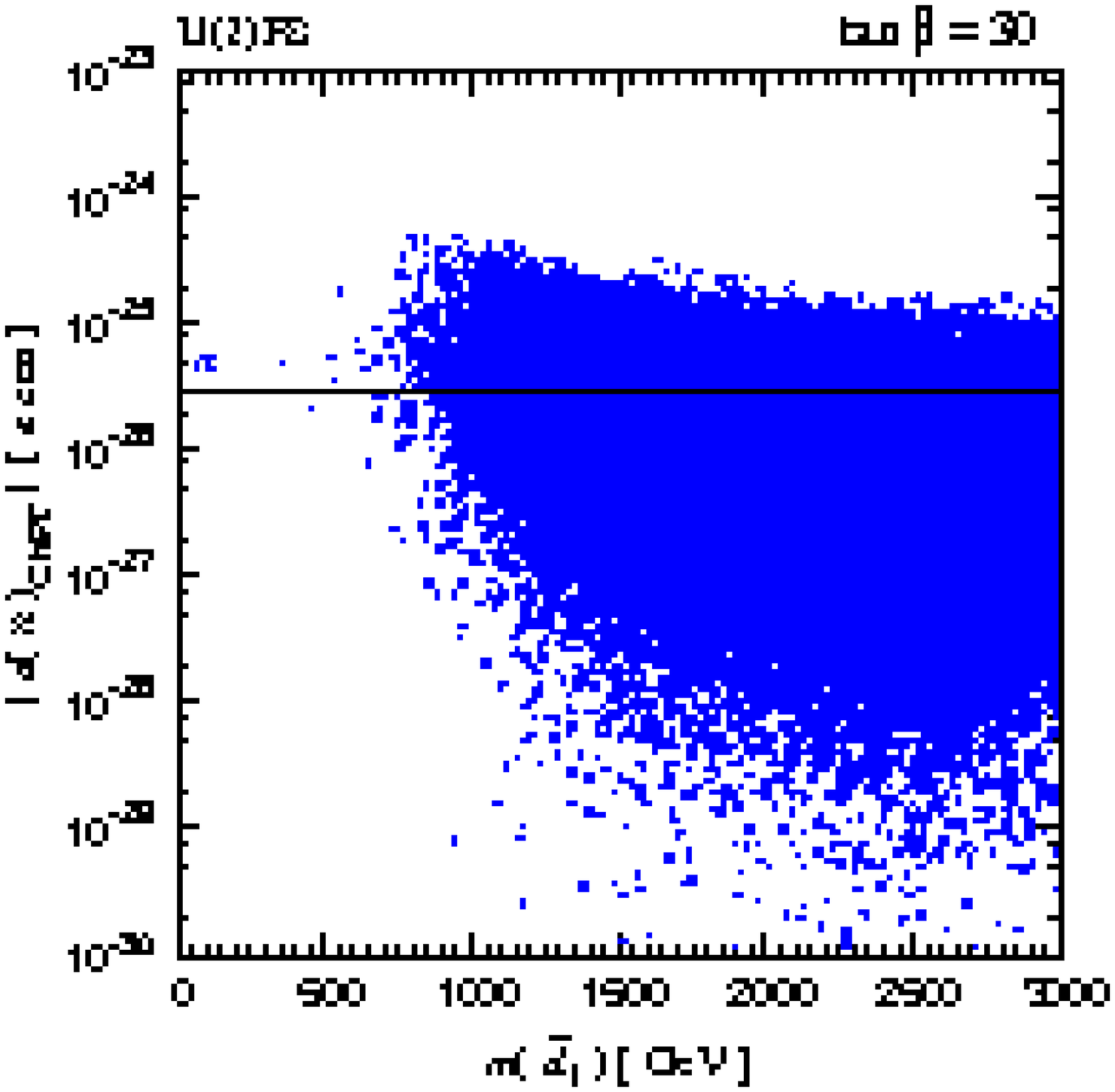} & \\
(e) & (f) & (g) &
\end{tabular}
\caption{%
The neutron EDM calculated by the ChPT formula for the same
parameter sets as those in Fig.~\ref{fig:edm-msd1}.
}
\label{fig:edmnXP-msd1}
\end{figure}

\begin{figure}[htbp]
\begin{tabular}{cccc}
\includegraphics[scale=.23]{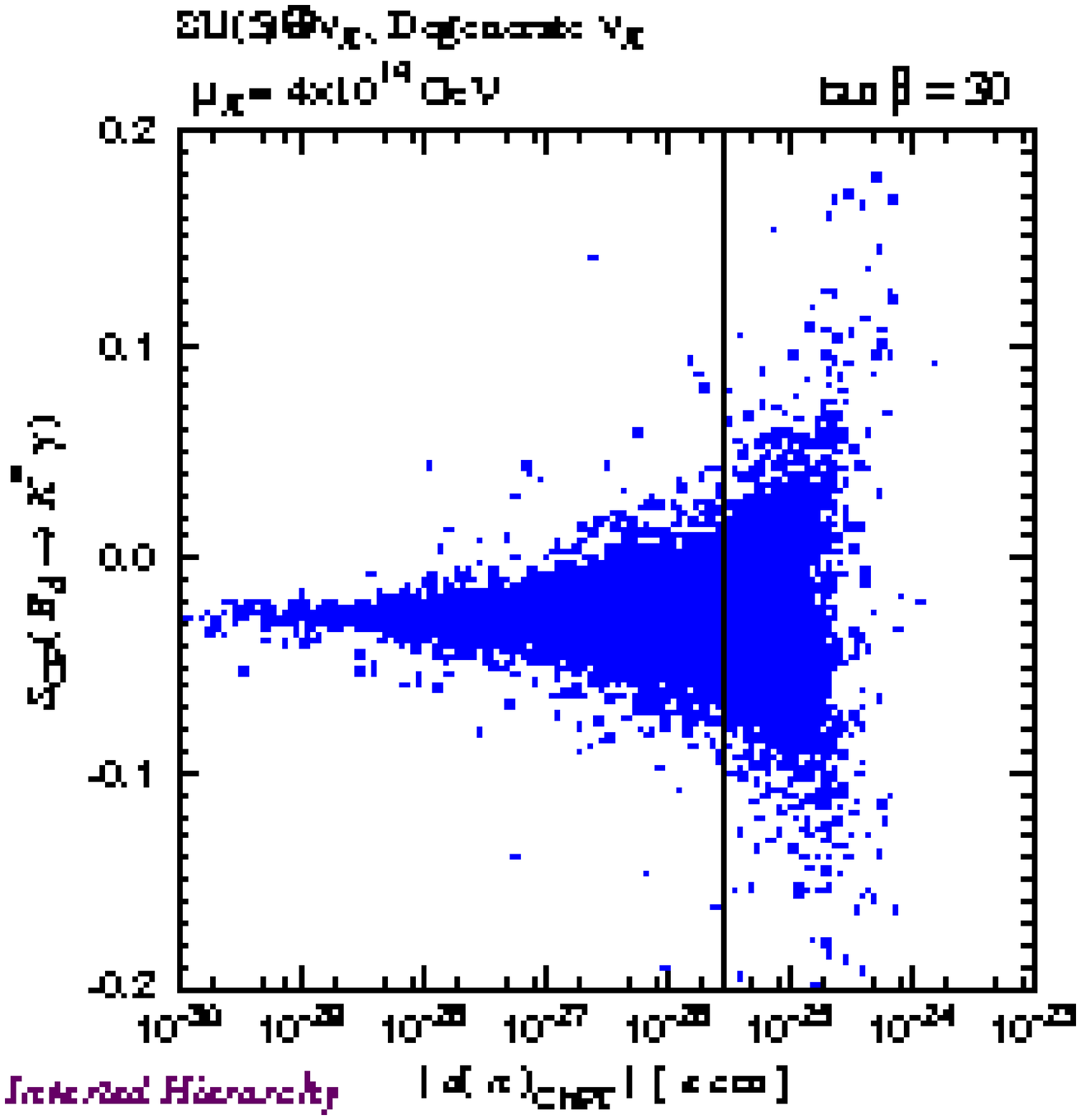} &
\includegraphics[scale=.23]{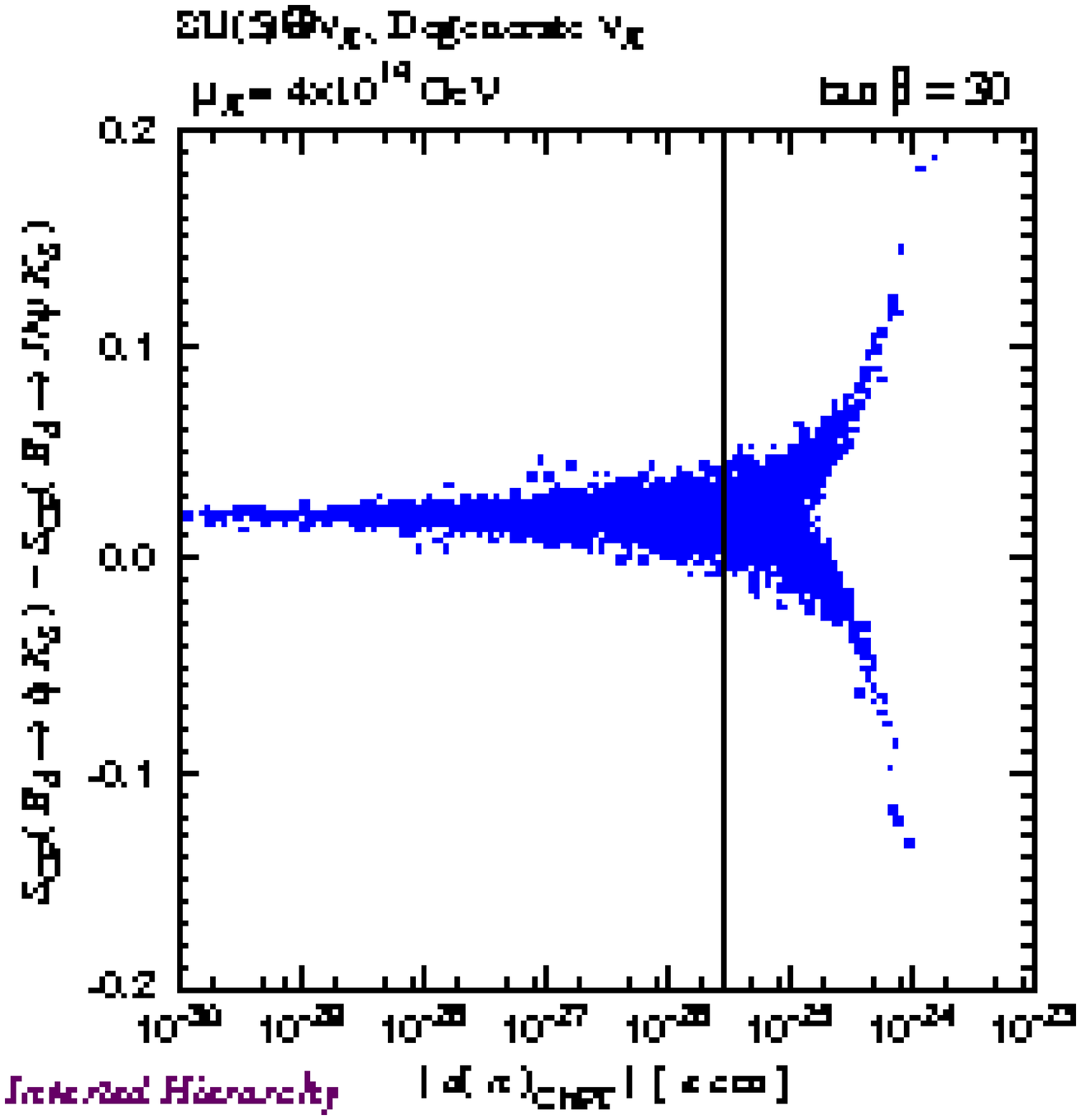} &
\includegraphics[scale=.23]{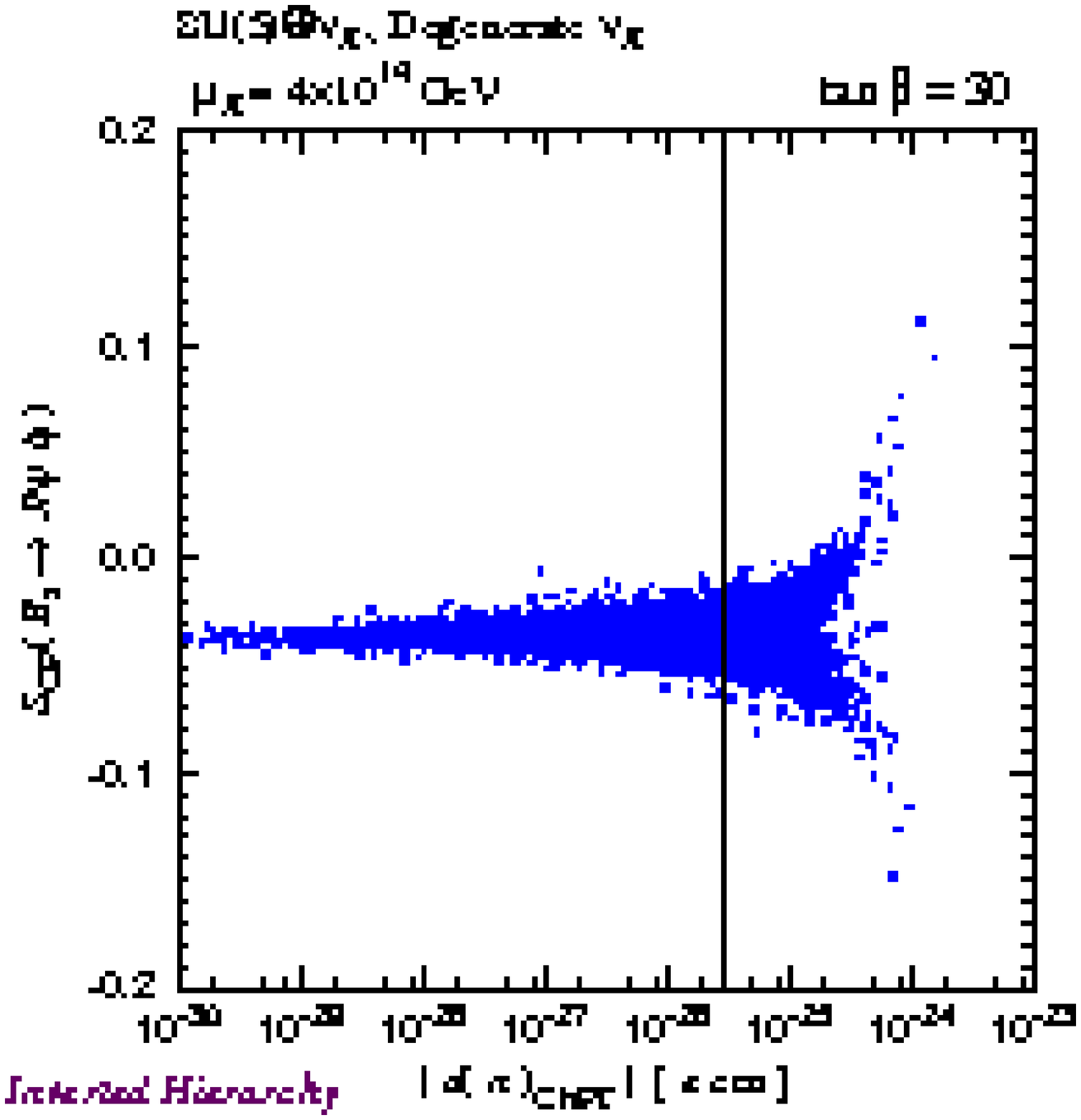} &
\includegraphics[scale=.23]{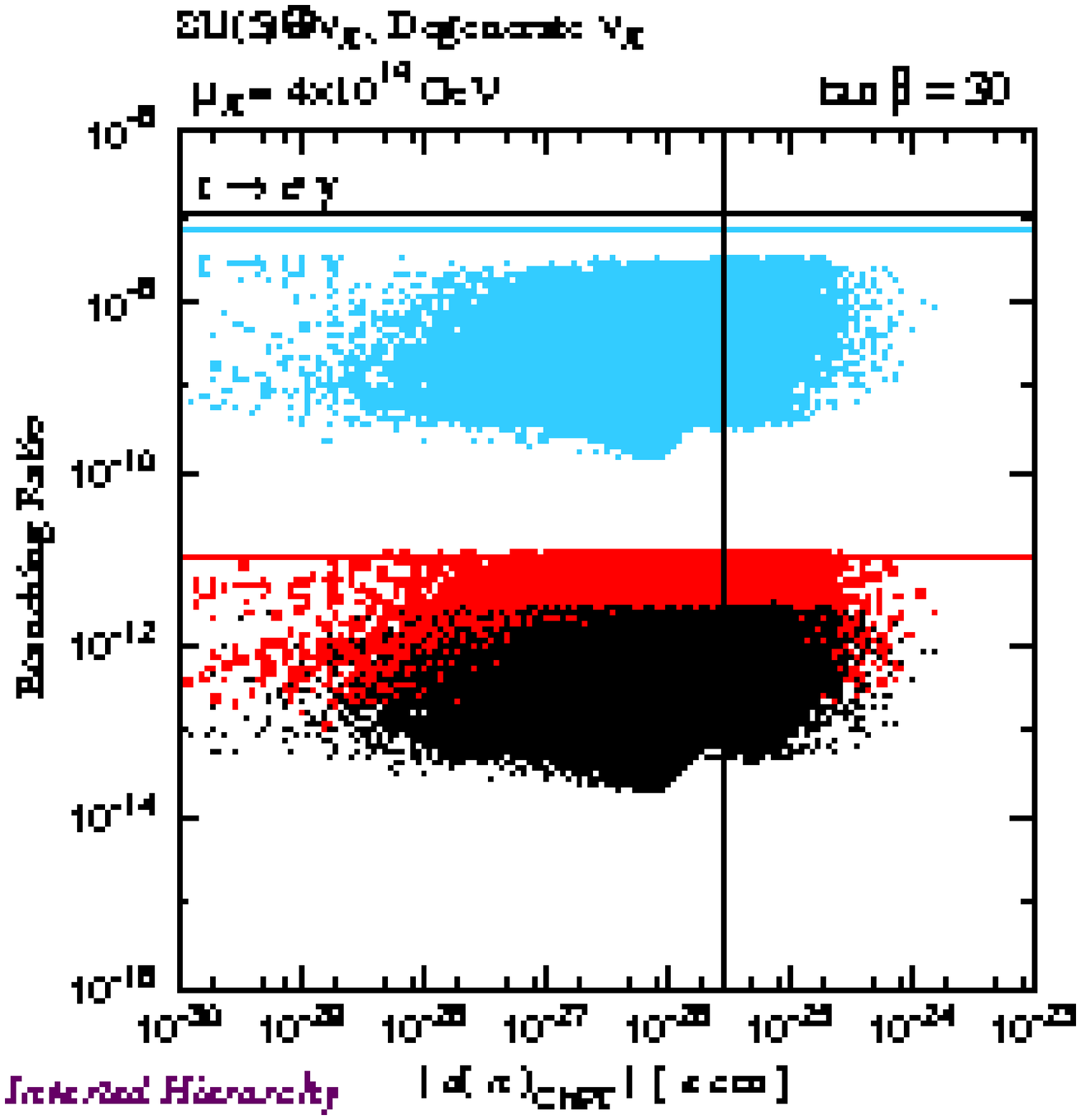} \\
(a) & (b) & (c) & (d) \\
\includegraphics[scale=.23]{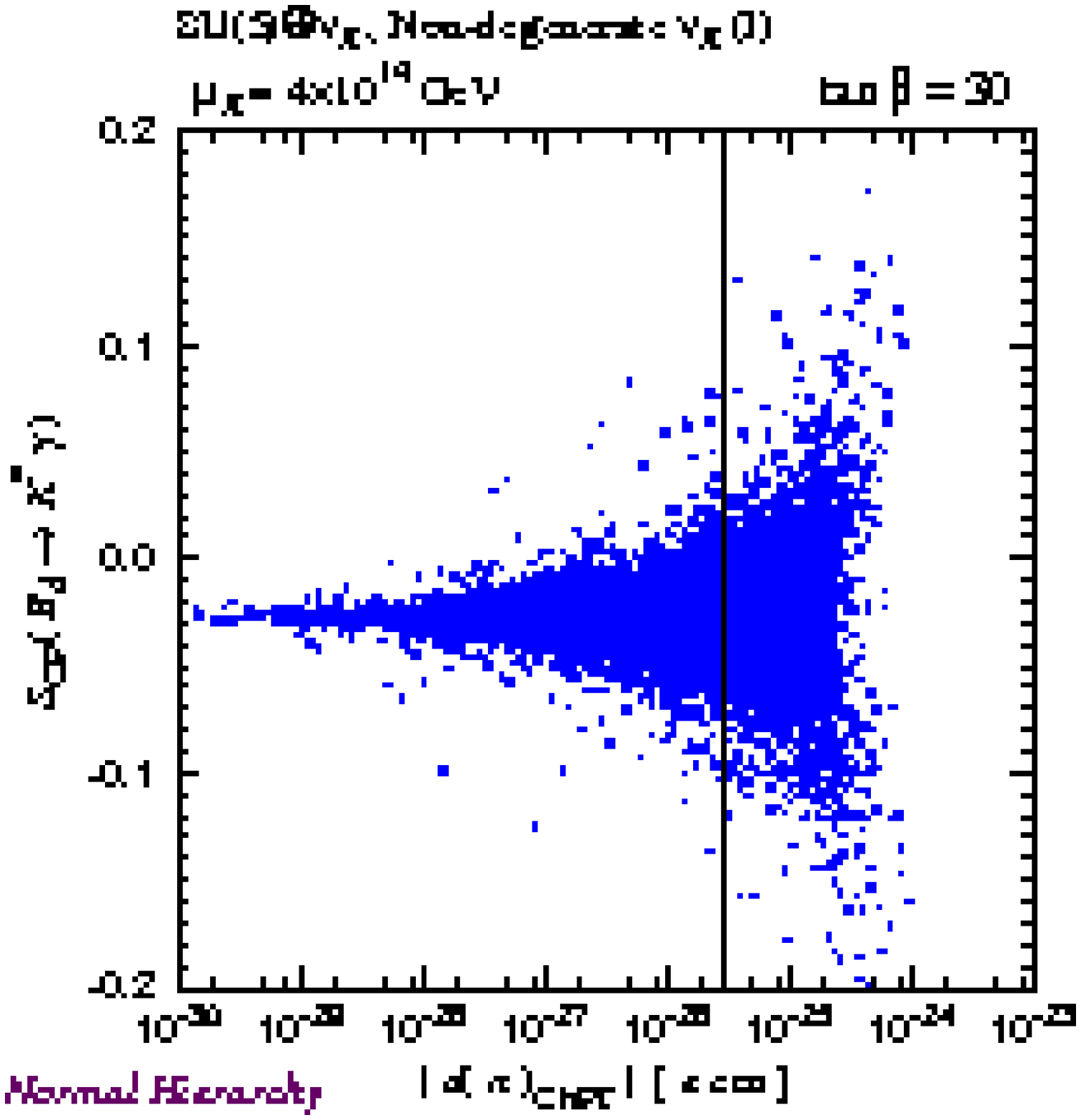} &
\includegraphics[scale=.23]{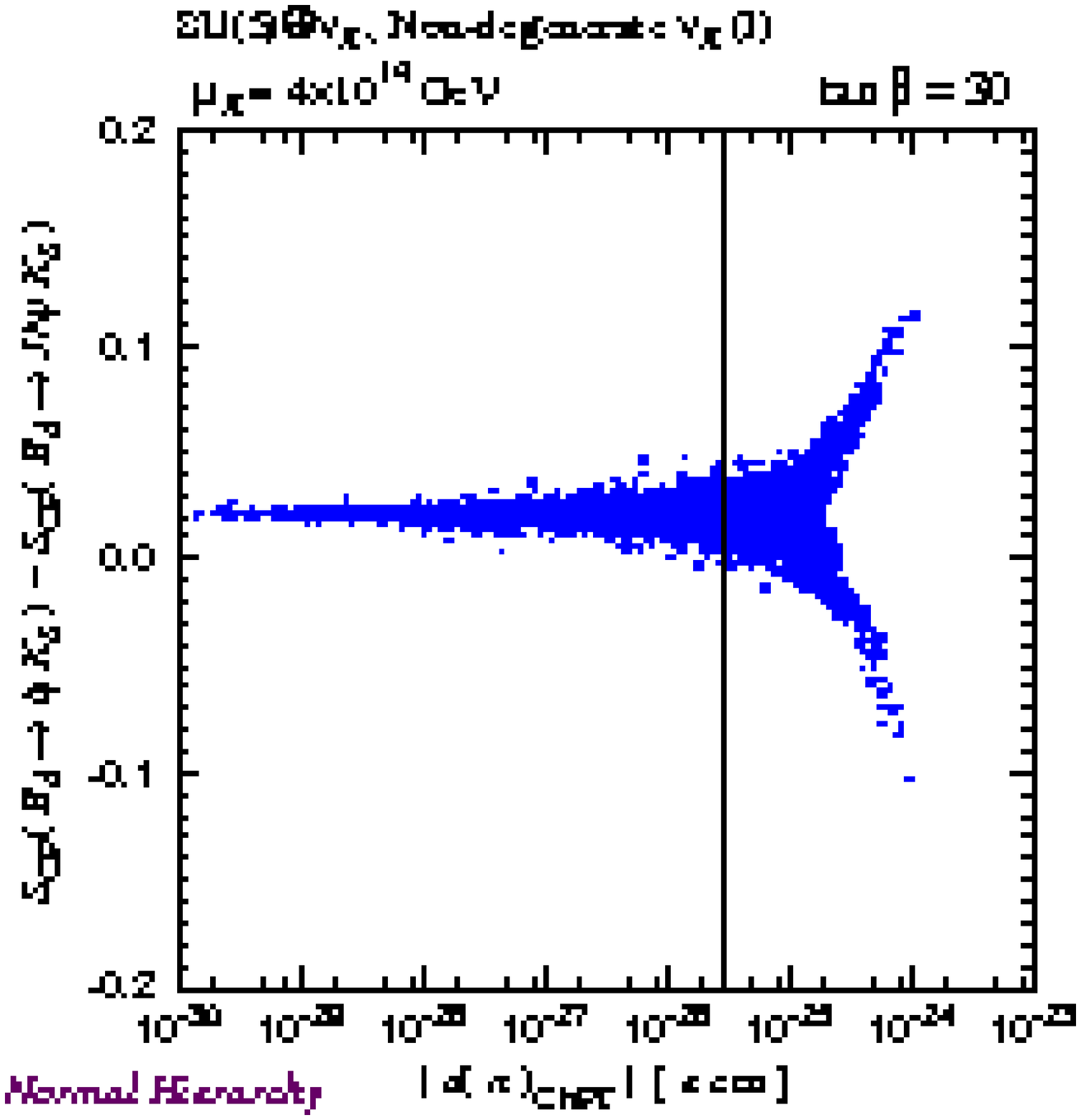} &
\includegraphics[scale=.23]{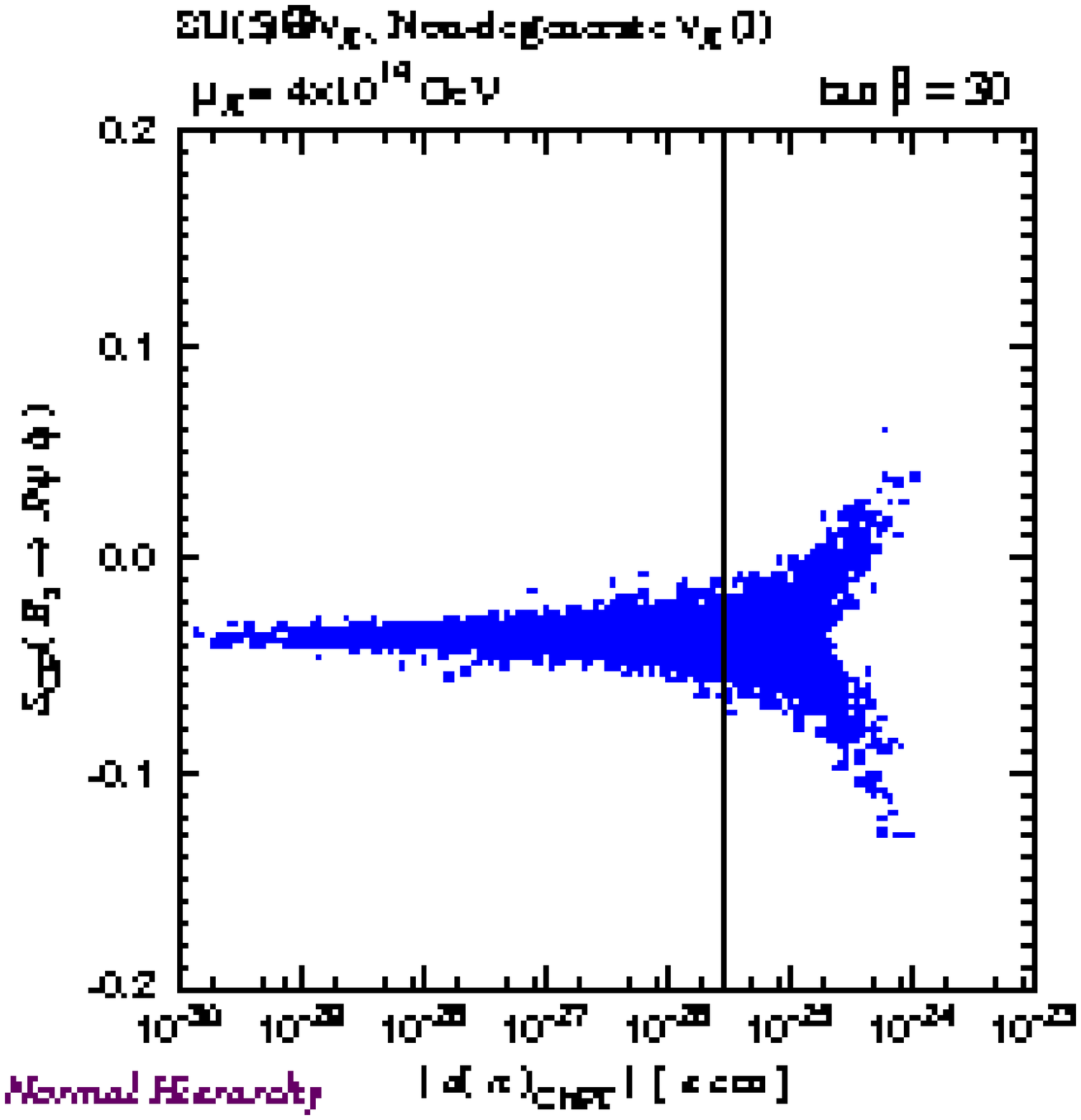} &
\includegraphics[scale=.23]{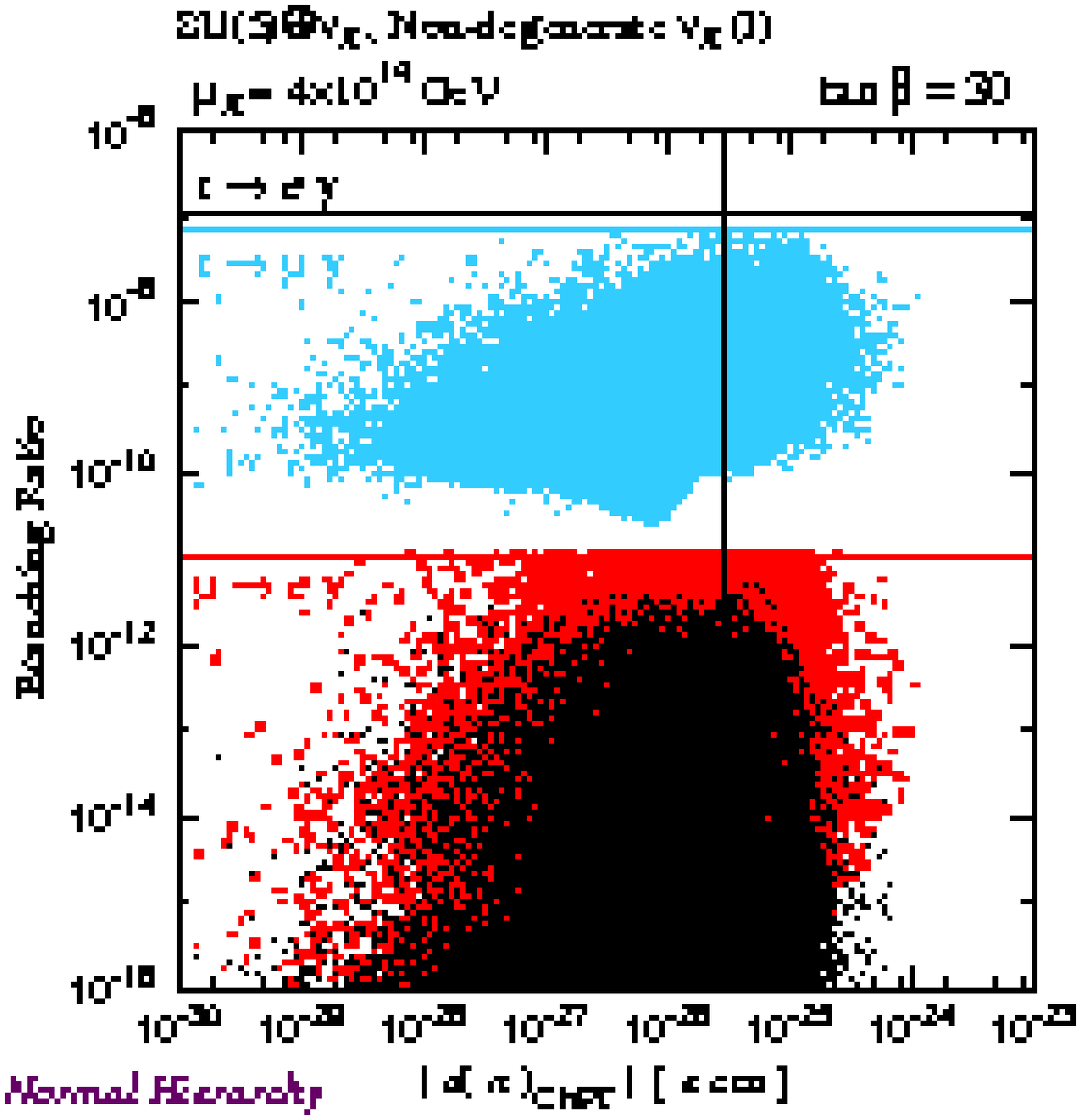} \\
(e) & (f) & (g) & (h)
\end{tabular}
\caption{%
(Color online)
The CP asymmetries in $b\to s$ decays and the LFV branching ratios as
functions of the neutron EDM given by the chiral perturbation formula in
SU(5) SUSY GUT with right-handed neutrinos.
(a)--(d) and (e)--(h) are plots in the D$\nu_R$-IH and the
ND$\nu_R$(I)-NH cases, respectively.
}
\label{fig:edmnXP-su5nr-nd1}
\end{figure}

\begin{figure}[htbp]
\begin{tabular}{ccc}
\includegraphics[scale=.3]{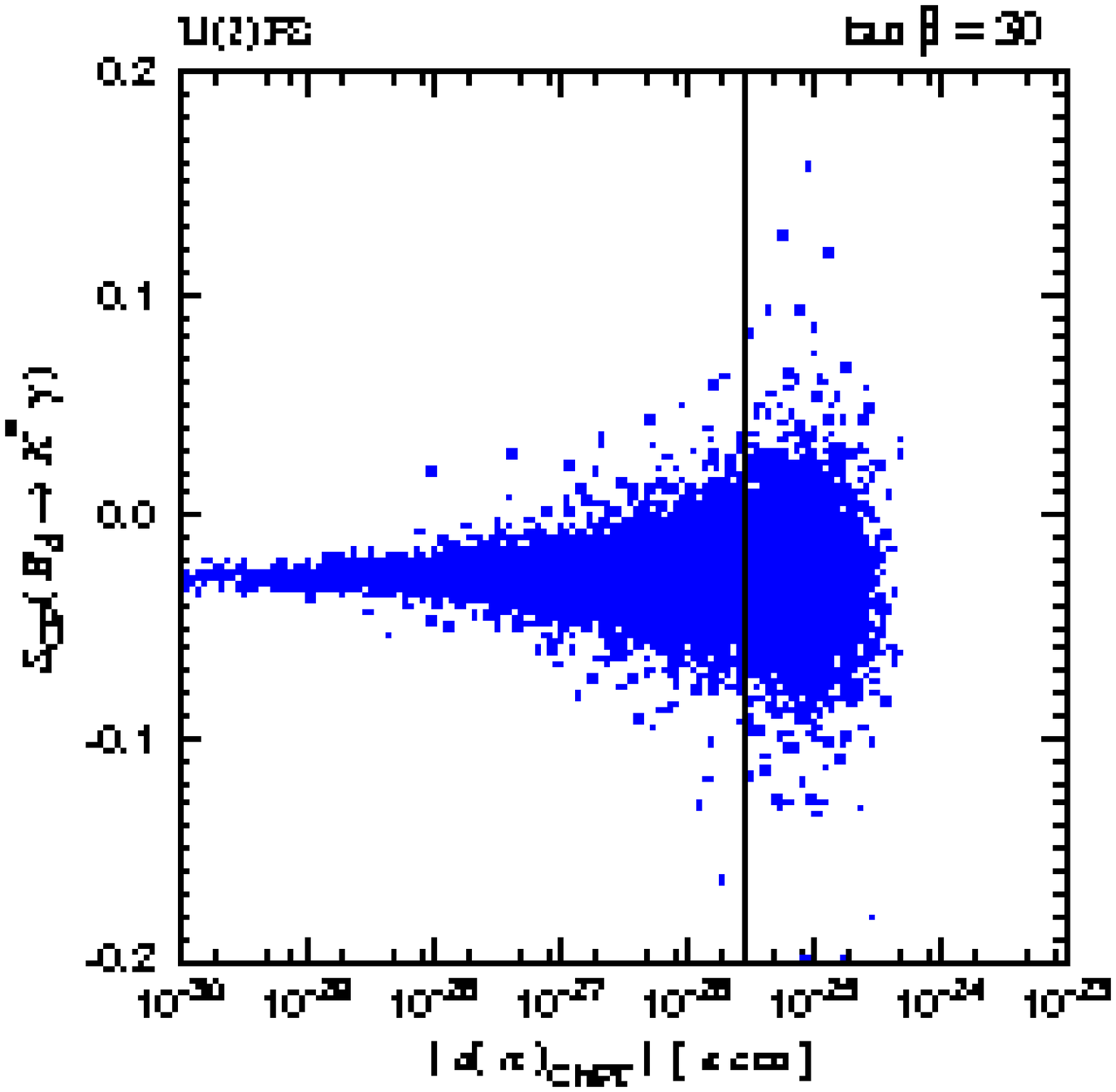} &
\includegraphics[scale=.3]{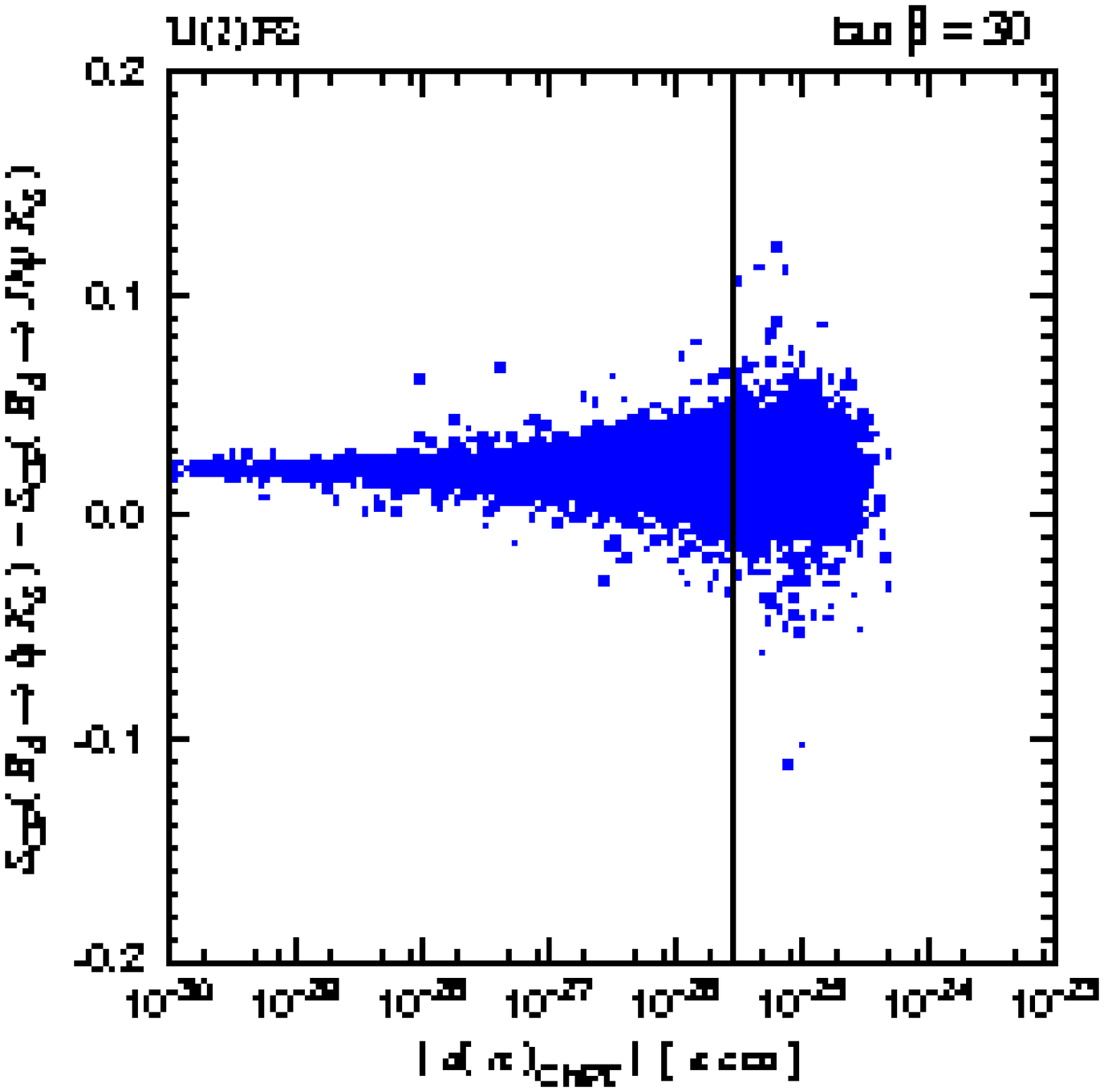} &
\includegraphics[scale=.3]{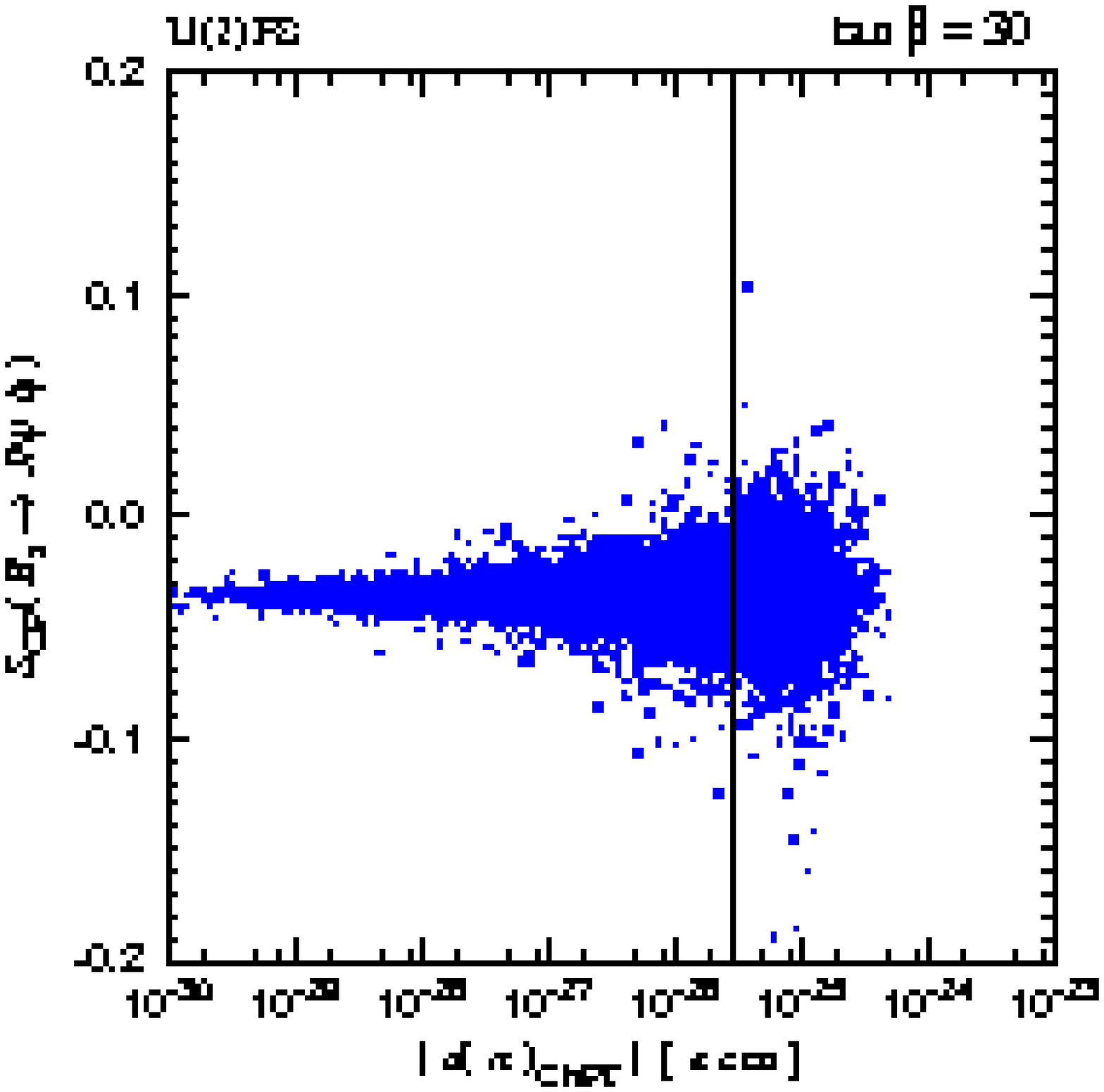} \\
(a) & (b) & (c)
\end{tabular}
\caption{%
The CP asymmetries in $b\to s$ decays as functions of the neutron EDM
given by the chiral perturbation formula in the U(2) flavor symmetry
model.
}
\label{fig:edmnXP-u2}
\end{figure}

We show the EDMs of the neutron, $^{199}\mathrm{Hg}$ and the electron
as functions of the lightest down-type squark mass in
Fig.~\ref{fig:edm-msd1}.
Here we use the NDA formula for the neutron EDM.
Primary source of these EDMs is the phase of the $A_0$, since we fix
the phase of the higgsino mass parameter as $\phi_\mu=0$ in the present
analysis.
The EDMs for $\phi_\mu=O(1)$ are larger than those for $\phi_\mu=0$ by
one or two orders of magnitude and easily exceed the experimental upper
limits in large portions of the parameter space.
In the present case, we can see that the upper limit of the electron EDM
mainly constrain the parameter space, while the constraints from other
two EDMs are slightly weaker.

Let us discuss how the possible quark and lepton flavor signals
change if we use the formula for the neutron EDM based on the chiral
perturbation theory, as mentioned in Sec.~\ref{sec:EDM}.
The main difference between the NDA and ChPT formulae is the treatment
of the strange quark (chromo-)EDM.
In ChPT the contribution from the strange quark is taken into account,
while it is simply neglected in NDA.
In the ND$\nu_R$(I)-NH and D$\nu_R$-IH of the SU(5) SUSY GUT with
right-handed neutrinos and the U(2) cases, the $2-3$ generation mixings
and CP violating phases exist in both left- and right-handed down-type
squark mass matrices, which enhance the chromo-EDM of the strange quark.
Therefore, the SUSY contributions to the CP asymmetries in $b\to s$
decays correlate to the chromo-EDM of the strange quark.

We show the plots of the neutron EDM calculated by the ChPT formula
\cite{ref:EDM-3} in Fig.~\ref{fig:edmnXP-msd1}.
It can be seen that the ChPT formula leads to typically $1-2$ orders of
magnitude larger value of the neutron EDM than the NDA formula does, in
the $m(\tilde{d}_1)\sim 1\text{TeV}$ region.
Therefore, a larger portion of the parameter space with new CP violating
phases is excluded if we adopt the ChPT formula for the evaluation of
the neutron EDM, and possible deviations in the CP violation observables
are also affected.

In Fig.~\ref{fig:edmnXP-su5nr-nd1}, we show the correlations between
the neutron EDM calculated by the ChPT formula
and the CP asymmetries
$S_{\text{CP}}(B_d\to K^*\gamma)$, 
$\Delta S_{\text{CP}}(B_d\to \phi K_S)$ and
$S_{\text{CP}}(B_s\to J/\psi\phi)$.
The correlations between the neutron EDM and the LFV decay branching
ratios are also shown.
We show the correlation plots for the U(2) model in
Fig.~\ref{fig:edmnXP-u2}.
In particular for the ND$\nu_R$(I)-NH and D$\nu_R$-IH of the SU(5) SUSY
GUT with right-handed neutrinos, the parameter region with large
deviations of $\Delta S_{\text{CP}}(B_d\to \phi K_S)$ and
$S_{\text{CP}}(B_s\to J/\psi\phi)$ is excluded if we adopt the ChPT
formula for the evaluation of the neutron EDM.
On the other hand, large deviations remain for other cases.

\subsection{Summary of results and experimental prospects}

There are good experimental prospects for future improvements in the
observables considered above.
From recent study of Super $B$ Factories \cite{Browder:2007gg}, the
precision of determination
for $50-75\text{ab}^{-1}$ is
0.02--0.03 for $S_{\text{CP}}(B_d\to K^*\gamma)$,
0.08--0.12 for $S_{\text{CP}}(B_d\to \rho\gamma)$ and
0.02--0.03 for $S_{\text{CP}}(B_d\to \phi K_S)$ for mixing-induced CP
asymmetries.
For the direct CP asymmetries of the radiative $B$ decays, the expected
sensitivity reach 0.004 for $A_{\text{CP}}(b\to s\gamma)$ and 0.01 for
$A_{\text{CP}}(b\to (s+d)\gamma)$.
The CP asymmetry of $B_s\to J/\psi\phi$ mode is determined up to 0.01
from LHCb with 10fb$^{-1}$ \cite{ref:Nakada:SUSY2010s}.
The precision of the $\phi_3$ determination is expected at 2.4$^\circ$
for LHCb at 10fb$^{-1}$ \cite{ref:Nakada:SUSY2010s}, and further
improvement is expected at Super $B$ Factory.
In order to extract new physics effect from the correlation between
$\Delta m_{B_s}/\Delta m_{B_d}$ and $\phi_3$ we need to improve the
determination of $\xi$ factor up to a percent level.
The $\mu\to e\gamma$ branching ratio will be searched for at the level of
$10^{-13}$ level at the MEG experiment.
Current upper bounds of $\text{B}(\tau\to\mu\gamma)$ and
$\text{B}(\tau\to e\gamma)$ are $6.8\times 10^{-8}$ and
$1.1\times 10^{-7}$, respectively, at the $B$ factory experiments, and
future improvement by $1-2$ orders of magnitude is expected at Super $B$
factory.

Comparing with these prospects, we can determine the significance of the
deviations observed in Figs.~\ref{fig:ml-LFV}--\ref{fig:UT}.
Our results of lepton and quark flavor signals are summarized in
Table~\ref{tab:resultssummary}.
We list various quark flavor signals in $b-s$ and $b-d$ transition for
the mSUGRA, three cases of MSSM with right-handed neutrinos, and SU(5)
SUSY GUT with right-handed neutrinos, and the U(2) flavor symmetry
model.
The $\mu$ and $\tau$ LFV processes are also included for the cases
except for the U(2) model.
The observable with a mark $\surd$ indicates that a large deviation is
possible.
The mark $\bullet$ means that there are some points that the deviation
could be identified with future improvements of experimental
measurements and/or theoretical understanding of uncertainty.
From the table, we can see that significant flavor signals are expected
in the lepton sector for the MSSM with right-handed neutrinos and the
SU(5) SUSY GUT with right-handed neutrinos.
These lepton flavour violation signals depend on the texture of the
neutrino Yukawa coupling matrix, {\it i.e.} $\tau\to\mu\gamma$ can be
large in D$\nu_R$-IH, D$\nu_R$-D and ND$\nu_R$(I)-NH cases and $\tau\to
e\gamma$ can be large in the ND$\nu_R$(II)-NH cases while satisfying the
present experimental bound on $\mu\to e\gamma$.
In D$\nu_R$-NH cases, $\mu\to e\gamma$ is the most promising mode
among these three lepton flavour violation processes.
In the SU(5) SUSY GUT with right-handed neutrinos, 
in addition to the above texture dependent signals, 
$\mu\to e\gamma$ can be enhanced as large as the present experimental
bound due to GUT interactions even in the non-degenerate $\nu_R$ (I) and
(II) cases.
As for the quark flavor signals,
we can expect that significant CP violating asymmetries in 
$b\to s$ and $b\to d$ transitions in the SU(5) SUSY GUT 
with right-handed neutrinos and in the U(2) model. 
The pattern of the deviations from the SM predictions also depends on 
the texture of the neutrino Yukawa coupling matrix in 
the SU(5) SUSY GUT with right-handed neutrinos.
Examining the pattern of deviations from the SM in the quark 
and lepton flavor signals, we can gain insights on the flavor structure
in the SUSY models.

\begin{table}[htbp]
\includegraphics[angle=90,scale=.7]{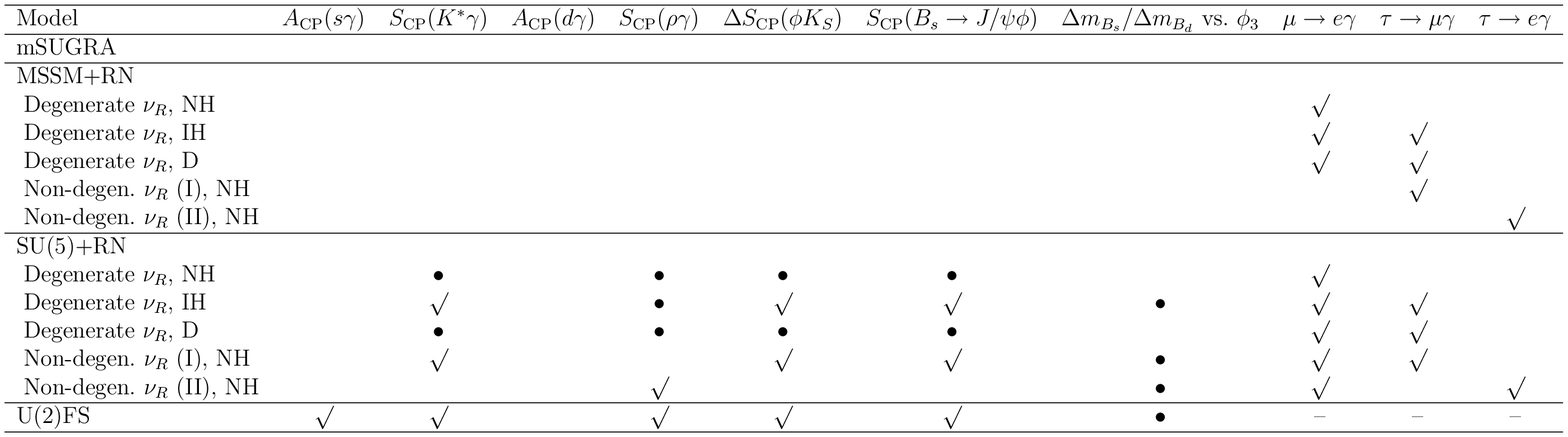}
\caption{%
Summary of expected flavor signals for each model.
``NH'', ``IH'', and ``D'' denote normal hierarchy, inverted
hierarchy and degenerate, respectively, for the low energy neutrino
spectrum.
The observable with a mark $\surd$ indicates that a large deviation is
possible.
The mark $\bullet$ means that there are some points that the deviation
could be identified with future improvements of experimental
measurements and/or theoretical understanding of uncertainty.
We do not consider LFV processes for the U(2)FS model (--).
}
\label{tab:resultssummary}
\end{table}

In addition to experimental progress, it is important to reduce
theoretical uncertainties to identify the deviations.
In particular, theoretical issue to predict
mixing-induced CP asymmetries in $B_d\to K^*\gamma$
\cite{ref:SCPKstargamma} and $B_d\to\phi K_S$ \cite{ref:SCPphiK} modes
within the SM need to be clarified because the deviation we expect is
up to 10\% level.

Notice that the significant flavor signals in the models with
right-handed neutrinos appear in the case with sufficiently large
neutrino Yukawa couplings, which corresponds to the right-handed
neutrino mass scale $\mu_R=O(10^{14})\text{GeV}$.
When we take a smaller value of $\mu_R$, all the flavor signals are
suppressed.
As already mentioned previously, the effects of the neutrino Yukawa
couplings are negligibly small for $\mu_R\ll 10^{12}\text{GeV}$.

In this paper we do not include the heavy Higgs exchange contributions
to various FCNC and LFV processes.
These contributions are known to play an important role for particular
cases of SUSY parameter sets due to large corrections to Yukawa
coupling constants through SUSY loop diagrams \cite{ref:SUSYYukawacorr}.
The relevant parameter set corresponds to large values of $\tan\beta$
and relatively small values of heavy Higgs boson masses with large
values of $\mu$.
The Higgs exchange contribution induces drastic effects in processes
like $B_s\to\mu^+\mu^-$ and $b\to sl^+l^-$ \cite{ref:Higgs2B} especially
for a large value of $\tan\beta$ ($=50-60$) because of a high power
dependence of $\tan\beta$.
In some restricted parts of our analysis, we may have additional flavor
signals due to the Higgs exchange effects.

\section{Conclusions}
\label{sec:conclusions}

We have analyzed quark flavor signals associated with $b\to s$ and
$b\to d$ transitions and lepton flavor violations in various cases of
supersymmetric models.
Extensive study is carried out in terms of observables for
representative SUSY models.
Our result is summarized in Table~\ref{tab:resultssummary}.
We have improved computational methods and updated phenomenological
constraints from our works in previous publications.
The most important effect is the inclusion of the constraint from the
$B_s-\bar{B}_s$ mixing from recent Tevatron experiments.
The maximum deviation for various $b\to s$ transition processes turn out
to be 10\% level, compared to the previous results where the deviation
at the level of 50\% was possible.
In this work, we also present predictions of tau lepton flavor
processes.
Under the constraint of $\mu\to e\gamma$, the tau LFV processes are
promising to look for new physics effects, which are also related to
$b\to s$ and $b\to d$ transition processes in SUSY GUT models.
The pattern of deviation from the SM prediction provides us with an
important clue on physics determining the structure of the SUSY breaking
sector, and a future $B$ factory plays a central role in such
investigation along with on-going flavor experiments such as MEG and
LHCb.

\begin{acknowledgments}
The work of T. G. and Y. O. is supported in part by the Grant-in-Aid for
Science Research, Ministry of Education, Culture, Sports, Science and
Technology, Japan, No. 16081211.
The work of Y. O. is supported in part by the Grant-in-Aid for Science
Research, Ministry of Education, Culture, Sports, Science and 
Technology, No. 17540286.
The work of T. S. is supported in part by the INFN under the program 
``Fisica Astroparticellare'', and by the Italian MIUR
(Internazionalizzazione Program).
The numerical calculations were carried out in part on Altix3700 BX2 at
YITP in Kyoto University.
\end{acknowledgments}


\begin{thebibliography}{99}
\bibitem{ref:CKM}
  N.~Cabibbo,
  Phys.\ Rev.\ Lett.\  {\bf 10}, 531 (1963);
  M.~Kobayashi and T.~Maskawa,
  Prog.\ Theor.\ Phys.\  {\bf 49}, 652 (1973).

\bibitem{ref:sol_atm_nu}
  Y.~Ashie {\it et al.}  [Super-Kamiokande Collaboration],
  Phys.\ Rev.\  D {\bf 71}, 112005 (2005)
  [arXiv:hep-ex/0501064];
  J.~Hosaka {\it et al.}  [Super-Kamkiokande Collaboration],
  Phys.\ Rev.\  D {\bf 73}, 112001 (2006)
  [arXiv:hep-ex/0508053];
  B.~Aharmim {\it et al.}  [SNO Collaboration],
  Phys.\ Rev.\  C {\bf 75}, 045502 (2007)
  [arXiv:nucl-ex/0610020];
  M.~Altmann {\it et al.}  [GNO Collaboration],
  Phys.\ Lett.\  B {\bf 616}, 174 (2005)
  [arXiv:hep-ex/0504037];
  W.~Hampel {\it et al.}  [GALLEX Collaboration],
  Phys.\ Lett.\  B {\bf 447}, 127 (1999);
  J.~N.~Abdurashitov {\it et al.}  [SAGE Collaboration],
  J.\ Exp.\ Theor.\ Phys.\  {\bf 95}, 181 (2002)
  [Zh.\ Eksp.\ Teor.\ Fiz.\  {\bf 122}, 211 (2002)]
  [arXiv:astro-ph/0204245].

\bibitem{ref:K2K}
  S.~Yamamoto {\it et al.}  [K2K Collaboration],
  Phys.\ Rev.\ Lett.\  {\bf 96}, 181801 (2006)
  [arXiv:hep-ex/0603004];
  M.~H.~Ahn {\it et al.}  [K2K Collaboration],
  Phys.\ Rev.\  D {\bf 74}, 072003 (2006)
  [arXiv:hep-ex/0606032].

\bibitem{ref:KamLAND}
  T.~Araki {\it et al.}  [KamLAND Collaboration],
  Phys.\ Rev.\ Lett.\  {\bf 94}, 081801 (2005)
  [arXiv:hep-ex/0406035].

\bibitem{ref:reactor_neutrino}
  M.~Apollonio {\it et al.}  [CHOOZ Collaboration],
  Eur.\ Phys.\ J.\  C {\bf 27}, 331 (2003)
  [arXiv:hep-ex/0301017];
  A.~Piepke  [Palo Verde Collaboration],
  Prog.\ Part.\ Nucl.\ Phys.\  {\bf 48}, 113 (2002).

\bibitem{ref:LHC}
ATLAS Collaboration, Technical Design Report, CERN/LHCC/99-15 (1999);
A.~Ball, M.~Della Negra, A.~Petrilli and L.~Foa  [CMS Collaboration],
  J.\ Phys.\ G {\bf 34}, 995 (2007).

\bibitem{ref:SUSY}
For reviews, see
  H.~P.~Nilles,
  Phys.\ Rept.\  {\bf 110}, 1 (1984);
  H.~E.~Haber and G.~L.~Kane,
  Phys.\ Rept.\  {\bf 117}, 75 (1985);
  S.~P.~Martin,
  arXiv:hep-ph/9709356.
  D.~J.~H.~Chung, L.~L.~Everett, G.~L.~Kane, S.~F.~King, J.~D.~Lykken
  and L.~T.~Wang,
  Phys.\ Rept.\  {\bf 407}, 1 (2005)
  [arXiv:hep-ph/0312378].

\bibitem{ref:SUSY_flavor_problem}
  J.~R.~Ellis and D.~V.~Nanopoulos,
  Phys.\ Lett.\  B {\bf 110}, 44 (1982).

\bibitem{Ellis:1982tk}
  J.~R.~Ellis, S.~Ferrara and D.~V.~Nanopoulos,
  Phys.\ Lett.\  B {\bf 114}, 231 (1982).


\bibitem{ref:MEG}
  L.~M.~Barkov {\it et al.}, PSI Proposal R-99-05 (1999);
  S.~Ritt  [MEG Collaboration],
  Nucl.\ Phys.\ Proc.\ Suppl.\  {\bf 162}, 279 (2006).

\bibitem{ref:LHCb}
  S.~Amato {\it et al.}  [LHCb Collaboration],
  CERN-LHCC-98-04, CERN-LHCC-P-4 (1998).

\bibitem{ref:Nakada:SUSY2010s}
  T.~Nakada, talk given at
  {\it Conference on Supersymmetry in 2010's},
  June 20--22, 2007, Hokkaido University, Sapporo, Japan.

\bibitem{ref:SuperKEKB}
  A.~G.~Akeroyd {\it et al.}  [SuperKEKB Physics Working Group],
  arXiv:hep-ex/0406071;
  S.~Hashimoto {\it et al.},
  KEK-REPORT-2004-4, Jun 2004.

\bibitem{ref:SuperBFrascati}
  M.~Bona {\it et al.},
  arXiv:0709.0451 [hep-ex].

\bibitem{Hewett:2004tv}
  J.~L.~Hewett {\it et al.},
  arXiv:hep-ph/0503261.

\bibitem{ref:massinsertion}
  F.~Gabbiani and A.~Masiero,
  Nucl.\ Phys.\  B {\bf 322}, 235 (1989);
  I.~I.~Y.~Bigi and F.~Gabbiani,
  Nucl.\ Phys.\  B {\bf 352}, 309 (1991).
  J.~S.~Hagelin, S.~Kelley and T.~Tanaka,
  Nucl.\ Phys.\  B {\bf 415}, 293 (1994).

\bibitem{ref:massinsertion-flavour}
  F.~Gabbiani, E.~Gabrielli, A.~Masiero and L.~Silvestrini,
  Nucl.\ Phys.\  B {\bf 477}, 321 (1996)
  [arXiv:hep-ph/9604387];
  M.~Ciuchini {\it et al.},
  JHEP {\bf 9810}, 008 (1998)
  [arXiv:hep-ph/9808328];
  T.~Besmer, C.~Greub and T.~Hurth,
  Nucl.\ Phys.\  B {\bf 609}, 359 (2001)
  [arXiv:hep-ph/0105292];
  E.~Lunghi and D.~Wyler,
  Phys.\ Lett.\  B {\bf 521}, 320 (2001)
  [arXiv:hep-ph/0109149].
  D.~Becirevic {\it et al.},
  Nucl.\ Phys.\  B {\bf 634}, 105 (2002)
  [arXiv:hep-ph/0112303];
  S.~Khalil and E.~Kou,
  Phys.\ Rev.\  D {\bf 67}, 055009 (2003)
  [arXiv:hep-ph/0212023];
  Phys.\ Rev.\ Lett.\  {\bf 91}, 241602 (2003)
  [arXiv:hep-ph/0303214];
  G.~L.~Kane, P.~Ko, H.~b.~Wang, C.~Kolda, J.~h.~Park and L.~T.~Wang,
  Phys.\ Rev.\  D {\bf 70}, 035015 (2004)
  [arXiv:hep-ph/0212092];
  Phys.\ Rev.\ Lett.\  {\bf 90}, 141803 (2003)
  [arXiv:hep-ph/0304239];
  M.~Ciuchini, E.~Franco, A.~Masiero and L.~Silvestrini,
  Phys.\ Rev.\  D {\bf 67}, 075016 (2003)
  [Erratum-ibid.\  D {\bf 68}, 079901 (2003)]
  [arXiv:hep-ph/0212397];
  K.~Agashe and C.~D.~Carone,
  Phys.\ Rev.\  D {\bf 68}, 035017 (2003)
  [arXiv:hep-ph/0304229].
  R.~Harnik, D.~T.~Larson, H.~Murayama and A.~Pierce,
  Phys.\ Rev.\  D {\bf 69}, 094024 (2004)
  [arXiv:hep-ph/0212180];
  J.~Foster, K.~i.~Okumura and L.~Roszkowski,
  JHEP {\bf 0508}, 094 (2005)
  [arXiv:hep-ph/0506146];

\bibitem{ref:massinsertion-bsbs}
  J.~Foster, K.~i.~Okumura and L.~Roszkowski,
  Phys.\ Lett.\  B {\bf 641}, 452 (2006)
  [arXiv:hep-ph/0604121];
  M.~Ciuchini and L.~Silvestrini,
  Phys.\ Rev.\ Lett.\  {\bf 97}, 021803 (2006)
  [arXiv:hep-ph/0603114];
  M.~Endo and S.~Mishima,
  Phys.\ Lett.\  B {\bf 640}, 205 (2006)
  [arXiv:hep-ph/0603251].


\bibitem{Bouquet:1984pp}
  A.~Bouquet, J.~Kaplan and C.~A.~Savoy,
  Phys.\ Lett.\  B {\bf 148}, 69 (1984).
  A.~Bouquet, J.~Kaplan and C.~A.~Savoy,
  Nucl.\ Phys.\  B {\bf 262}, 299 (1985).

\bibitem{Hall:1985dx}
  L.~J.~Hall, V.~A.~Kostelecky and S.~Raby,
  Nucl.\ Phys.\  B {\bf 267}, 415 (1986).

\bibitem{Barbieri:1994pv}
  R.~Barbieri and L.~J.~Hall,
  Phys.\ Lett.\  B {\bf 338}, 212 (1994)
  [arXiv:hep-ph/9408406].

\bibitem{Bertolini:1991if}
S.~Bertolini, F.~Borzumati, A.~Masiero, and G.~Ridolfi,
Nucl.\ Phys.\ B {\bf 353}, 591 (1991);
T.~Goto, T.~Nihei, and Y.~Okada,
Phys.\ Rev.\ D {\bf 53}, 5233 (1996)
[Erratum-ibid.\ D {\bf 54}, 5904 (1996)];
T.~Goto, Y.~Okada, and Y.~Shimizu,
arXiv:hep-ph/9908499;
A.~Bartl, T.~Gajdosik, E.~Lunghi, A.~Masiero, W.~Porod,
H.~Stremnitzer, and O.~Vives,
Phys.\ Rev.\ D {\bf 64}, 076009 (2001).

\bibitem{Baek:2000sj}
  S.~Baek, T.~Goto, Y.~Okada and K.-i.~Okumura,
  Phys.\ Rev.\  D {\bf 63}, 051701 (2001)
  [arXiv:hep-ph/0002141];
  {\it ibid.} {\bf 64}, 095001 (2001)
  [arXiv:hep-ph/0104146].

\bibitem{Moroi:2000mr}
  T.~Moroi,
  JHEP {\bf 0003}, 019 (2000)
  [arXiv:hep-ph/0002208];
  T.~Moroi,
  Phys.\ Lett.\  B {\bf 493}, 366 (2000)
  [arXiv:hep-ph/0007328].

\bibitem{ref:flavoursignalSUSYGUT}
  R.~Barbieri, L.~J.~Hall and A.~Strumia,
  Nucl.\ Phys.\  B {\bf 445}, 219 (1995)
  [arXiv:hep-ph/9501334];
  {\it ibid.} {\bf 449}, 437 (1995)
  [arXiv:hep-ph/9504373];
  A.~Masiero, M.~Piai, A.~Romanino and L.~Silvestrini,
  Phys.\ Rev.\  D {\bf 64}, 075005 (2001)
  [arXiv:hep-ph/0104101];
  N.~Akama, Y.~Kiyo, S.~Komine and T.~Moroi,
  Phys.\ Rev.\  D {\bf 64}, 095012 (2001)
  [arXiv:hep-ph/0104263];
  J.~Hisano and Y.~Shimizu,
  Phys.\ Lett.\  B {\bf 565}, 183 (2003)
  [arXiv:hep-ph/0303071];
  {\it ibid.} {\bf 581}, 224 (2004)
  [arXiv:hep-ph/0308255];
  D.~Chang, A.~Masiero and H.~Murayama,
  Phys.\ Rev.\  D {\bf 67}, 075013 (2003)
  [arXiv:hep-ph/0205111];
  M.~Ciuchini, A.~Masiero, L.~Silvestrini, S.~K.~Vempati and O.~Vives,
  Phys.\ Rev.\ Lett.\  {\bf 92}, 071801 (2004)
  [arXiv:hep-ph/0307191];
  M.~Ciuchini, A.~Masiero, P.~Paradisi, L.~Silvestrini, S.~K.~Vempati
  and O.~Vives,
  Nucl.\ Phys.\  B {\bf 783}, 112 (2007)
  [arXiv:hep-ph/0702144].

\bibitem{Calibbi:2006nq}
  L.~Calibbi, A.~Faccia, A.~Masiero and S.~K.~Vempati,
  Phys.\ Rev.\  D {\bf 74}, 116002 (2006)
  [arXiv:hep-ph/0605139];

\bibitem{Goto:2002xt}
  T.~Goto, Y.~Okada, Y.~Shimizu, T.~Shindou and M.~Tanaka,
  Phys.\ Rev.\  D {\bf 66}, 035009 (2002)
  [arXiv:hep-ph/0204081].

\bibitem{Goto:2003iu}
   T.~Goto, Y.~Okada, Y.~Shimizu, T.~Shindou and M.~Tanaka,
   Phys.\ Rev.\  D {\bf 70}, 035012 (2004)
  [arXiv:hep-ph/0306093].

\bibitem{ref:BsBs}
P.~Ball and R.~Fleischer,
  Eur.\ Phys.\ J.\  C {\bf 48}, 413 (2006)
  [arXiv:hep-ph/0604249];
G.~Isidori and P.~Paradisi,
  Phys.\ Lett.\  B {\bf 639}, 499 (2006)
  [arXiv:hep-ph/0605012];
S.~Khalil,
  Phys.\ Rev.\  D {\bf 74}, 035005 (2006)
  [arXiv:hep-ph/0605021];
S.~Baek,
  JHEP {\bf 0609}, 077 (2006)
  [arXiv:hep-ph/0605182];
R.~Arnowitt, B.~Dutta, B.~Hu and S.~Oh,
  Phys.\ Lett.\  B {\bf 641}, 305 (2006)
  [arXiv:hep-ph/0606130];
B.~Dutta and Y.~Mimura,
  Phys.\ Rev.\ Lett.\  {\bf 97}, 241802 (2006)
  [arXiv:hep-ph/0607147];
S.~Nandi and J.~P.~Saha,
  Phys.\ Rev.\  D {\bf 74}, 095007 (2006)
  [arXiv:hep-ph/0608341].

\bibitem{ref:EDM-1}
  J.~Polchinski and M.~B.~Wise,
  Phys.\ Lett.\  B {\bf 125}, 393 (1983).
  Y.~Kizukuri and N.~Oshimo,
  Phys.\ Rev.\  D {\bf 45}, 1806 (1992);
  {\it ibid.} {\bf 46}, 3025 (1992);
  R.~Barbieri, A.~Romanino and A.~Strumia,
  Phys.\ Lett.\  B {\bf 369}, 283 (1996)
  [arXiv:hep-ph/9511305].
  T.~Falk, K.~A.~Olive and M.~Srednicki,
  Phys.\ Lett.\  B {\bf 354}, 99 (1995)
  [arXiv:hep-ph/9502401];
  T.~Falk and K.~A.~Olive,
  Phys.\ Lett.\  B {\bf 375}, 196 (1996)
  [arXiv:hep-ph/9602299];
  {\it ibid.} {\bf 439}, 71 (1998)
  [arXiv:hep-ph/9806236].

\bibitem{ref:EDM-2}
  T.~Ibrahim and P.~Nath,
  Phys.\ Lett.\  B {\bf 418}, 98 (1998)
  [arXiv:hep-ph/9707409];
  Phys.\ Rev.\  D {\bf 57}, 478 (1998)
  [arXiv:hep-ph/9708456];
  {\bf 58}, 019901(E) (1998);
  {\bf 60}, 079903(E) (1999);
  {\bf 60}, 119901(E) (1999);
  T.~Goto, Y.~Y.~Keum, T.~Nihei, Y.~Okada and Y.~Shimizu,
  Phys.\ Lett.\  B {\bf 460}, 333 (1999)
  [arXiv:hep-ph/9812369].

\bibitem{Chang:1998uc}
  D.~Chang, W.~Y.~Keung and A.~Pilaftsis,
  Phys.\ Rev.\ Lett.\  {\bf 82}, 900 (1999)
  [Erratum-ibid.\  {\bf 83}, 3972 (1999)]
  [arXiv:hep-ph/9811202].

\bibitem{Falk:1999tm}
  T.~Falk, K.~A.~Olive, M.~Pospelov and R.~Roiban,
  Nucl.\ Phys.\  B {\bf 560}, 3 (1999)
  [arXiv:hep-ph/9904393].

\bibitem{ref:EDM-3}
  J.~Hisano and Y.~Shimizu,
  Phys.\ Rev.\  D {\bf 70}, 093001 (2004)
  [arXiv:hep-ph/0406091];
  J.~Hisano, M.~Kakizaki, M.~Nagai and Y.~Shimizu,
  Phys.\ Lett.\  B {\bf 604}, 216 (2004)
  [arXiv:hep-ph/0407169].

\bibitem{ref:seesaw}  
P. Minkowski,
Phys.\ Lett.\  B {\bf  67} (1977) 421;
M. Gell-Mann, P. Ramond and R. Slansky,
{\it Proceedings of the Supergravity Stony Brook Workshop},
New York 1979,  eds. P. Van Nieuwenhuizen and D. Freedman;
T. Yanagida,
{\it Proceedinds of the Workshop on Unified Theories and Baryon Number
  in the Universe},  Tsukuba, Japan 1979, ed.s A. Sawada and A. Sugamoto;
R. N. Mohapatra and G. Senjanovic, Phys. Rev. Lett. {\bf 44} (1980) 912.

\bibitem{ref:PMNS}
B.~Pontecorvo,
  Sov.\ Phys.\ JETP {\bf 6}, 429 (1957)
  [Zh.\ Eksp.\ Teor.\ Fiz.\  {\bf 33}, 549 (1957)];
  Sov.\ Phys.\ JETP {\bf 7}, 172 (1958)
  [Zh.\ Eksp.\ Teor.\ Fiz.\  {\bf 34}, 247 (1957)];
  Sov.\ Phys.\ JETP {\bf 26}, 984 (1968)
  [Zh.\ Eksp.\ Teor.\ Fiz.\  {\bf 53}, 1717 (1967)].
  Z.~Maki, M.~Nakagawa and S.~Sakata,
  Prog.\ Theor.\ Phys.\  {\bf 28}, 870 (1962);

\bibitem{ref:BnuR}
Y.~Farzan,
  Phys.\ Rev.\  D {\bf 69}, 073009 (2004)
  [arXiv:hep-ph/0310055];
Y.~Farzan and M.~E.~Peskin,
  Phys.\ Rev.\  D {\bf 70}, 095001 (2004)
  [arXiv:hep-ph/0405214].
  Y.~Farzan,
  JHEP {\bf 0502}, 025 (2005)
  [arXiv:hep-ph/0411358].

\bibitem{ref:meg}
 F.~Borzumati and A.~Masiero,
  Phys.\ Rev.\ Lett.\  {\bf 57}, 961 (1986);
J.~Hisano, T.~Moroi, K.~Tobe and M.~Yamaguchi,
  Phys.\ Rev.\  D {\bf 53}, 2442 (1996)
  [arXiv:hep-ph/9510309];
S.~T.~Petcov, S.~Profumo, Y.~Takanishi and C.~E.~Yaguna,
  Nucl.\ Phys.\  B {\bf 676}, 453 (2004)
  [arXiv:hep-ph/0306195].

\bibitem{Ellis:1979hy}
  J.~R.~Ellis, M.~K.~Gaillard and D.~V.~Nanopoulos,
  Phys.\ Lett.\  B {\bf 88}, 320 (1979).

\bibitem{ref:U2-1}
  A.~Pomarol and D.~Tommasini,
  Nucl.\ Phys.\  B {\bf 466}, 3 (1996)
  [arXiv:hep-ph/9507462];
  R.~Barbieri, G.~R.~Dvali and L.~J.~Hall,
  Phys.\ Lett.\  B {\bf 377}, 76 (1996)
  [arXiv:hep-ph/9512388];
  R.~Barbieri and L.~J.~Hall,
  Nuovo Cim.\  A {\bf 110}, 1 (1997)
  [arXiv:hep-ph/9605224];
  R.~Barbieri, L.~Giusti, L.~J.~Hall and A.~Romanino,
  Nucl.\ Phys.\  B {\bf 550}, 32 (1999)
  [arXiv:hep-ph/9812239].

\bibitem{ref:U2-2}
  R.~Barbieri, L.~J.~Hall and A.~Romanino,
  Phys.\ Lett.\  B {\bf 401}, 47 (1997)
  [arXiv:hep-ph/9702315];
  R.~Barbieri, L.~J.~Hall, S.~Raby and A.~Romanino,
  Nucl.\ Phys.\  B {\bf 493}, 3 (1997)
  [arXiv:hep-ph/9610449].

\bibitem{ref:U2-neutrino}
  R.~Barbieri, P.~Creminelli and A.~Romanino,
  Nucl.\ Phys.\  B {\bf 559}, 17 (1999)
  [arXiv:hep-ph/9903460];
  T.~Bla\v{z}ek, S.~Raby and K.~Tobe,
  Phys.\ Rev.\  D {\bf 60}, 113001 (1999)
  [arXiv:hep-ph/9903340];
  {\it ibid.} {\bf 62}, 055001 (2000)
  [arXiv:hep-ph/9912482];
  A.~Aranda, C.~D.~Carone and R.~F.~Lebed,
  Phys.\ Lett.\  B {\bf 474}, 170 (2000)
  [arXiv:hep-ph/9910392];
  Phys.\ Rev.\  D {\bf 62}, 016009 (2000)
  [arXiv:hep-ph/0002044];
  M.~C.~Chen and K.~T.~Mahanthappa,
  Phys.\ Rev.\  D {\bf 62}, 113007 (2000)
  [arXiv:hep-ph/0005292];
  A.~Aranda, C.~D.~Carone and P.~Meade,
  Phys.\ Rev.\  D {\bf 65}, 013011 (2002)
  [arXiv:hep-ph/0109120];
  S.~Raby,
  Phys.\ Lett.\  B {\bf 561}, 119 (2003)
  [arXiv:hep-ph/0302027].

\bibitem{ref:DRbarRGE}
  S.~P.~Martin and M.~T.~Vaughn,
  Phys.\ Rev.\  D {\bf 50}, 2282 (1994)
  [arXiv:hep-ph/9311340].
  Y.~Yamada,
  Phys.\ Rev.\  D {\bf 50}, 3537 (1994)
  [arXiv:hep-ph/9401241].
  I.~Jack and D.~R.~T.~Jones,
  Phys.\ Lett.\  B {\bf 333}, 372 (1994)
  [arXiv:hep-ph/9405233].
 I.~Jack, D.~R.~T.~Jones, S.~P.~Martin, M.~T.~Vaughn and Y.~Yamada,
  Phys.\ Rev.\  D {\bf 50}, 5481 (1994)
  [arXiv:hep-ph/9407291].

\bibitem{Pierce:1996zz}
  D.~M.~Pierce, J.~A.~Bagger, K.~T.~Matchev and R.~j.~Zhang,
  Nucl.\ Phys.\  B {\bf 491}, 3 (1997)
  [arXiv:hep-ph/9606211].

\bibitem{ref:LatticeQCD}
  M.~Okamoto,
  PoS {\bf LAT2005}, 013 (2006)
  [arXiv:hep-lat/0510113];
  C.~R.~Allton {\it et al.},
  Phys.\ Lett.\  B {\bf 453}, 30 (1999)
  [arXiv:hep-lat/9806016];
  D.~Becirevic, V.~Gimenez, G.~Martinelli, M.~Papinutto and J.~Reyes,
  JHEP {\bf 0204}, 025 (2002)
  [arXiv:hep-lat/0110091].

\bibitem{ref:BtoVVangularanalysis}
  B.~Kayser, M.~Kuroda, R.~D.~Peccei and A.~I.~Sanda,
  Phys.\ Lett.\  B {\bf 237}, 508 (1990);
  I.~Dunietz, H.~R.~Quinn, A.~Snyder, W.~Toki and H.~J.~Lipkin,
  Phys.\ Rev.\  D {\bf 43}, 2193 (1991).


\bibitem{ref:CalcBphiK}
  N.~G.~Deshpande, X.~G.~He and J.~Trampetic,
  Phys.\ Lett.\  B {\bf 377}, 161 (1996)
  [arXiv:hep-ph/9509346];
  R.~Barbieri and A.~Strumia,
  Nucl.\ Phys.\  B {\bf 508}, 3 (1997)
  [arXiv:hep-ph/9704402];
  R.~Fleischer,
  Z.\ Phys.\  C {\bf 58}, 483 (1993);
  {\it ibid.} {\bf 62}, 81 (1994).

\bibitem{Kagan:1998bh}
  A.~L.~Kagan and M.~Neubert,
  Phys.\ Rev.\  D {\bf 58}, 094012 (1998)
  [arXiv:hep-ph/9803368].

\bibitem{ref:bsgamma-NLO}
  C.~Greub, T.~Hurth and D.~Wyler,
  Phys.\ Lett.\  B {\bf 380}, 385 (1996)
  [arXiv:hep-ph/9602281];
  Phys.\ Rev.\  D {\bf 54}, 3350 (1996)
  [arXiv:hep-ph/9603404];
  N.~Pott,
  Phys.\ Rev.\  D {\bf 54}, 938 (1996)
  [arXiv:hep-ph/9512252].

\bibitem{Atwood:1997zr}
  D.~Atwood, M.~Gronau and A.~Soni,
  Phys.\ Rev.\ Lett.\  {\bf 79}, 185 (1997)
  [arXiv:hep-ph/9704272].

\bibitem{Weinberg:1989dx}
  S.~Weinberg,
  Phys.\ Rev.\ Lett.\  {\bf 63}, 2333 (1989).

\bibitem{Kanemura:2005cq}
  S.~Kanemura, K.~Matsuda, T.~Ota, T.~Shindou, E.~Takasugi and
  K.~Tsumura,
  Phys.\ Rev.\  D {\bf 72}, 055012 (2005)
  [arXiv:hep-ph/0507264];
  {\bf 72}, 059904(E) (2005);
  S.~T.~Petcov, T.~Shindou and Y.~Takanishi,
  Nucl.\ Phys.\  B {\bf 738}, 219 (2006)
  [arXiv:hep-ph/0508243];
  S.~T.~Petcov and T.~Shindou,
  Phys.\ Rev.\  D {\bf 74}, 073006 (2006)
  [arXiv:hep-ph/0605151].

\bibitem{ref:directsearch}
  A.~Heister {\it et al.}  [ALEPH Collaboration],
  Phys.\ Lett.\  B {\bf 526}, 206 (2002)
  [arXiv:hep-ex/0112011];
  A.~Heister {\it et al.}  [ALEPH Collaboration],
  Phys.\ Lett.\  B {\bf 583}, 247 (2004);
  J.~Abdallah {\it et al.}  [DELPHI Collaboration],
  Eur.\ Phys.\ J.\  C {\bf 31}, 421 (2004)
  [arXiv:hep-ex/0311019];
  P.~Achard {\it et al.}  [L3 Collaboration],
  Phys.\ Lett.\  B {\bf 580}, 37 (2004)
  [arXiv:hep-ex/0310007];
  G.~Abbiendi {\it et al.}  [OPAL Collaboration],
  Eur.\ Phys.\ J.\  C {\bf 32}, 453 (2004)
  [arXiv:hep-ex/0309014];
  S.~Schael {\it et al.}  [The LEP Collaborations ALEPH, DELPHI, L3 and
  OPAL and The LEP Working Group for Higgs Boson searches],
  Eur.\ Phys.\ J.\  C {\bf 47}, 547 (2006)
  [arXiv:hep-ex/0602042];
  A.~A.~Affolder {\it et al.}  [CDF Collaboration],
  Phys.\ Rev.\ Lett.\  {\bf 88}, 041801 (2002)
  [arXiv:hep-ex/0106001];
  M.~Martinez-Perez  [CDF Collaboration],
  AIP Conf.\ Proc.\  {\bf 903}, 189 (2007);
  V.~M.~Abazov {\it et al.}  [D0 Collaboration],
  Phys.\ Lett.\  B {\bf 638}, 119 (2006)
  [arXiv:hep-ex/0604029].

\bibitem{ref:bsgamma}
  P.~Koppenburg {\it et al.}  [Belle Collaboration],
  Phys.\ Rev.\ Lett.\  {\bf 93}, 061803 (2004)
  [arXiv:hep-ex/0403004];
  B.~Aubert {\it et al.}  [BABAR Collaboration],
  Phys.\ Rev.\  D {\bf 72}, 052004 (2005)
  [arXiv:hep-ex/0508004];
  S.~Chen {\it et al.}  [CLEO Collaboration],
  Phys.\ Rev.\ Lett.\  {\bf 87}, 251807 (2001)
  [arXiv:hep-ex/0108032].

\bibitem{Ahmed:2001eh}
  M.~Ahmed {\it et al.}  [MEGA Collaboration],
  Phys.\ Rev.\  D {\bf 65}, 112002 (2002)
  [arXiv:hep-ex/0111030].

\bibitem{Aubert:2005ye}
  B.~Aubert {\it et al.}  [BABAR Collaboration],
  Phys.\ Rev.\ Lett.\  {\bf 95}, 041802 (2005)
  [arXiv:hep-ex/0502032].

\bibitem{Aubert:2005wa}
  B.~Aubert {\it et al.}  [BABAR Collaboration],
  Phys.\ Rev.\ Lett.\  {\bf 96}, 041801 (2006)
  [arXiv:hep-ex/0508012].

\bibitem{Romalis:2000mg}
  M.~V.~Romalis, W.~C.~Griffith and E.~N.~Fortson,
  Phys.\ Rev.\ Lett.\  {\bf 86}, 2505 (2001)
  [arXiv:hep-ex/0012001].

\bibitem{Baker:2006ts}
  C.~A.~Baker {\it et al.},
  Phys.\ Rev.\ Lett.\  {\bf 97}, 131801 (2006)
  [arXiv:hep-ex/0602020].

\bibitem{Regan:2002ta}
  B.~C.~Regan, E.~D.~Commins, C.~J.~Schmidt and D.~DeMille,
  Phys.\ Rev.\ Lett.\  {\bf 88}, 071805 (2002).

\bibitem{Barberio:2007cr}
  E.~Barberio {\it et al.}
  [Heavy Flavor Averaging Group (HFAG) Collaboration],
  arXiv:0704.3575 [hep-ex].

\bibitem{Abulencia:2006ze}
  A.~Abulencia {\it et al.}  [CDF Collaboration],
  Phys.\ Rev.\ Lett.\  {\bf 97}, 242003 (2006)
  [arXiv:hep-ex/0609040].

\bibitem{Feng:1999zg}
  J.~L.~Feng, K.~T.~Matchev and T.~Moroi,
  Phys.\ Rev.\  D {\bf 61}, 075005 (2000)
  [arXiv:hep-ph/9909334].

\bibitem{ref:DMfocuspoint}
  M.~Drees and M.~M.~Nojiri,
  Phys.\ Rev.\  D {\bf 47}, 376 (1993)
  [arXiv:hep-ph/9207234];
  S.~Mizuta and M.~Yamaguchi,
  Phys.\ Lett.\  B {\bf 298}, 120 (1993)
  [arXiv:hep-ph/9208251];
  J.~Edsj\"{o} and P.~Gondolo,
  Phys.\ Rev.\  D {\bf 56}, 1879 (1997)
  [arXiv:hep-ph/9704361].
  J.~L.~Feng, K.~T.~Matchev and F.~Wilczek,
  Phys.\ Lett.\  B {\bf 482}, 388 (2000)
  [arXiv:hep-ph/0004043];

\bibitem{Ellis:1998kh}
  J.~R.~Ellis, T.~Falk and K.~A.~Olive,
  Phys.\ Lett.\  B {\bf 444}, 367 (1998)
  [arXiv:hep-ph/9810360].

\bibitem{Browder:2007gg}
  T.~E.~Browder, M.~Ciuchini, T.~Gershon, M.~Hazumi, T.~Hurth, Y.~Okada
  and A.~Stocchi,
  arXiv:0710.3799 [hep-ph].

\bibitem{ref:SCPKstargamma}
  B.~Grinstein, Y.~Grossman, Z.~Ligeti and D.~Pirjol,
  Phys.\ Rev.\  D {\bf 71}, 011504 (2005)
  [arXiv:hep-ph/0412019].
  M.~Matsumori and A.~I.~Sanda,
  Phys.\ Rev.\  D {\bf 73}, 114022 (2006)
  [arXiv:hep-ph/0512175].
  P.~Ball and R.~Zwicky,
  Phys.\ Lett.\  B {\bf 642}, 478 (2006)
  [arXiv:hep-ph/0609037].

\bibitem{ref:SCPphiK}
  Y.~Grossman, G.~Isidori and M.~P.~Worah,
  Phys.\ Rev.\  D {\bf 58}, 057504 (1998)
  [arXiv:hep-ph/9708305].
  M.~Beneke,
  Phys.\ Lett.\  B {\bf 620}, 143 (2005)
  [arXiv:hep-ph/0505075].
  A.~R.~Williamson and J.~Zupan,
  Phys.\ Rev.\  D {\bf 74}, 014003 (2006)
  [Erratum-ibid.\  D {\bf 74}, 03901 (2006)]
  [arXiv:hep-ph/0601214].

\bibitem{ref:SUSYYukawacorr}
  L.~J.~Hall, R.~Rattazzi and U.~Sarid,
  Phys.\ Rev.\  D {\bf 50}, 7048 (1994)
  [arXiv:hep-ph/9306309].
  K.~S.~Babu and C.~F.~Kolda,
  Phys.\ Lett.\  B {\bf 451}, 77 (1999)
  [arXiv:hep-ph/9811308].
  M.~S.~Carena, D.~Garcia, U.~Nierste and C.~E.~M.~Wagner,
  Nucl.\ Phys.\  B {\bf 577}, 88 (2000)
  [arXiv:hep-ph/9912516].

\bibitem{ref:Higgs2B}
  C.~S.~Huang, W.~Liao and Q.~S.~Yan,
  Phys.\ Rev.\  D {\bf 59}, 011701 (1999)
  [arXiv:hep-ph/9803460];
  C.~Hamzaoui, M.~Pospelov and M.~Toharia,
  Phys.\ Rev.\  D {\bf 59}, 095005 (1999)
  [arXiv:hep-ph/9807350];
  S.~R.~Choudhury and N.~Gaur,
  Phys.\ Lett.\  B {\bf 451}, 86 (1999)
  [arXiv:hep-ph/9810307];
  K.~S.~Babu and C.~F.~Kolda,
  Phys.\ Rev.\ Lett.\  {\bf 84}, 228 (2000)
  [arXiv:hep-ph/9909476];

\end{thebibliography}
\end{document}